# Tertiary particle production and target optimization of the H2 beam line in the SPS North Area

F. Tellander[1, 2,*] and N. Charitonidis[2]

[1]Departement of Astronomy and Theoretical Physics, Lund University, SE-223 62 Lund, Sweden
[2] CERN, 1211 Geneva 23, Switzerland

*Corresponding author: felix@tellander.se

**Abstract**

In this note, the tertiary particle yield from secondary targets of different materials placed at the 'filter' position of the H2 beam line of SPS North Area are presented. The production is studied for secondary beams of different momenta in the range of 50-250 GeV/c. More specifically, we studied six different targets: two copper cylinders with a radius of 40 mm and lengths of 100 and 300 mm, one solid tungsten cylinder with a radius of 40 mm and a length of 150 mm and three polyethylene cylinders with radius of 40 mm and lengths of 550, 700 and 1000 mm. Eight different momenta of the secondary beam (50, 60, 70, 100, 120, 150, 200 and 250 GeV/c) as well as two different physics lists (QGSP_BIC and FTFP_BERT) have been extensively studied. The purpose of this study is (a) to optimize (using the appropriate filter target) the particle production from the secondary targets as demanded by the experiments (b) investigate the proton production (with respect to the pion production) in the produced tertiary beams, a fact interesting for cross-section measuring experiments (e.g. NA61), (c) provide an expected beam-composition database for each target and energy that will act as a reference for the test-beam users of the North Area and (d) demonstrate the differences between the different GEANT-4 physics lists. Moreover this work constitutes a starting point for a more detailed benchmark of the different available Monte-Carlo models and codes in this momentum range.

# 1. Introduction

## 1.1 The North Area H2 beam line

H2 beam line as well as the layout of the SPS North Area facility has been described in detail elsewhere (see e.g. [1]). H2 constitutes a multi-purpose beam line, able to transport hadrons in a very wide momentum range, from ~10 up to ~400 GeV/c, depending on the need of each installed experiment. This secondary beam is produced by the impingement of a primary, 400 GeV/c proton beam, slowly extracted from SPS on a thin beryllium plate.

The hadronic particle yield from the primary beryllium target has been studied in detail in [1] and [2]. As visualized in Figure 1, the secondary hadron beam produced by the primary beam is a mixture of protons, kaons and pions depending on the secondary momentum chosen.

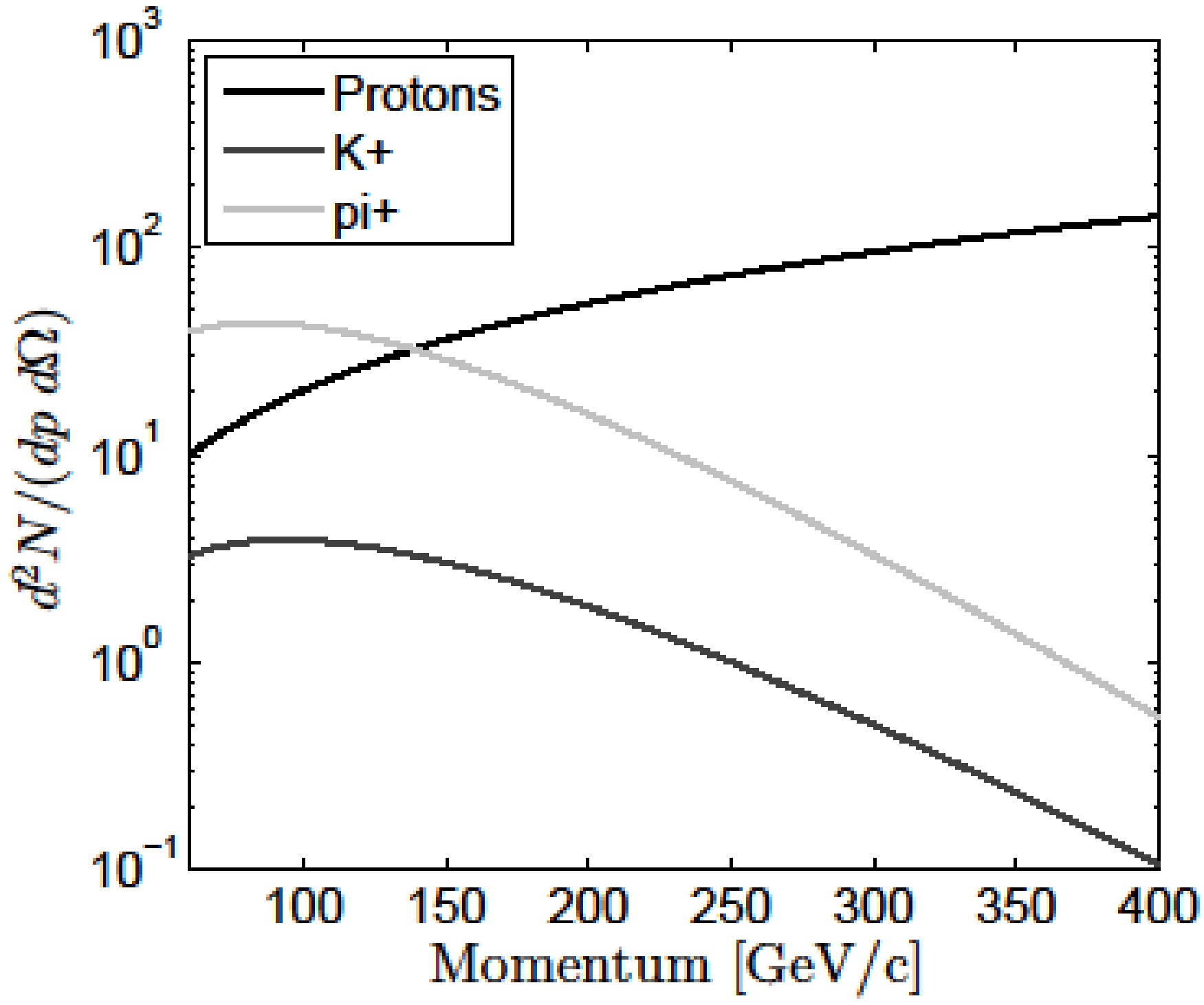

**Figure 1. Particle production from beryllium target. Production of protons, K+ and π+ from 400 GeV/c proton beam impinging on beryllium as a function of secondary momentum [1].**

However, depending on the experiment installed in the line, it is often favorable to suppress or enhance a specific particle type. This 'beam enrichment technique' by differential absorption has been applied in the past with success. The principle consists of inserting a secondary 'filter' target of length $L$ in the beam trajectory. This target is placed at a double focal point of the beam, specifically 131 meter downstream from the primary target. The fractional content of a given kind of particles at production before the secondary target, $a_i$, will become after passage through the filter

$$a_i' = \frac{a_i e^{-\frac{L}{\lambda_i}}}{\sum_\iota a_\iota e^{-\frac{L}{\lambda_\iota}}} \tag{3}$$

Where $\lambda_i$ is the nuclear interaction length for the filter material, and the denominator represents the factor by which the total beam flux is attenuated [3]. In this note, we summarize the results of a systematic Monte-Carlo study investigating the composition of tertiary beams produced in various targets and starting from various secondary beams. The electron production on the primary target is ignored in this study.

## 1.2 Electron beam content

The main contribution to the electron/positron yield from a target irradiated by a high energy proton beam is the production of $\pi^0$'s, leading to an electron-positron pair. Other electron sources as Dalitz electrons or the conversion of photons produced by the neutral decay of η particles, can contribute to the total yield of electrons. However, the first mechanism is the dominant. In the North Area secondary beam lines, the transport of middle-to-low energy electrons in the range of 10 – 60 GeV/c is based on the production of middle-to-high secondary electrons in the primary target, their subsequent transport on the secondary target where, through radiative losses, an amount of their energy is lost. These lower energy electrons are subsequently transported to the experiments. In the present work, the secondary beam impinging on the filter target is assumed to be electron-free. Even if the simulated beam composition is not realistic, this study is intended to give a first approximation of the contamination of a tertiary hadron beam with tertiary electrons, due to electron-production processes directly in the filter target.

## 1.3 Simulation model and parameters

The software used in the present study is the GEANT-4 based framework G4BeamLine [4]. The beam line model, as visualized in the program, is shown in Figure 2A. The beam line as a whole, starting from the primary Be target up to the experimental zones, located approx. 600 m downstream the target has been precisely modelled. In order to reduce the computational time, the interaction of the primary SPS beam with the Be-target is not modelled, and the formulas governing the particle production presented in ref [1] are used instead to generate the hadronic secondary beam composition from the target. Only the hadronic composition is modelled. The proton production is described by Equation (1) and the $K^{\pm}$ , $\pi^{\pm}$ , anti-proton production is given by Equation (2).

$$\frac{d^2N}{dp\, d\Omega} = A\left[\frac{(B+1)}{p_0}\left(\frac{p}{p_0}\right)^2\right]\left[\frac{2Cp^2}{2\pi}\,e^{-C(p\theta)^2}\right] \tag{1}$$

$$\frac{d^2N}{dp\, d\Omega} = A\left[\frac{B}{p}\,e^{-pB/p_0}\right]\left[\frac{2Cp^2}{2\pi}\,e^{-C(p\theta)^2}\right] \tag{2}$$

These formulas are visualized in Figure 1 as a function of the secondary momentum. The parameters $A, B,$ and $C$ depend on the particle type and are shown in Table 1. The angle θ is the particle production angle, selected for the simplicity to be 0.

**Table 1. Particle production parameters. The particle type depending parameters used for calculating the particle production from beryllium in Equation (1) and Equation (2).22211111**

| | $A$ | $B$ | $C$ |
|---|---|---|---|
| $\pi^+$ | 1.2 | 9.5 | 5.0 |
| $\pi^-$ | 0.8 | 11.5 | 5.0 |
| $K^+$ | 0.16 | 8.5 | 3.0 |
| $K^-$ | 0.10 | 13.0 | 3.5 |
| p | 0.8 | -0.6 | 3.5 |
| anti-p | 0.06 | 16.0 | 3.0 |

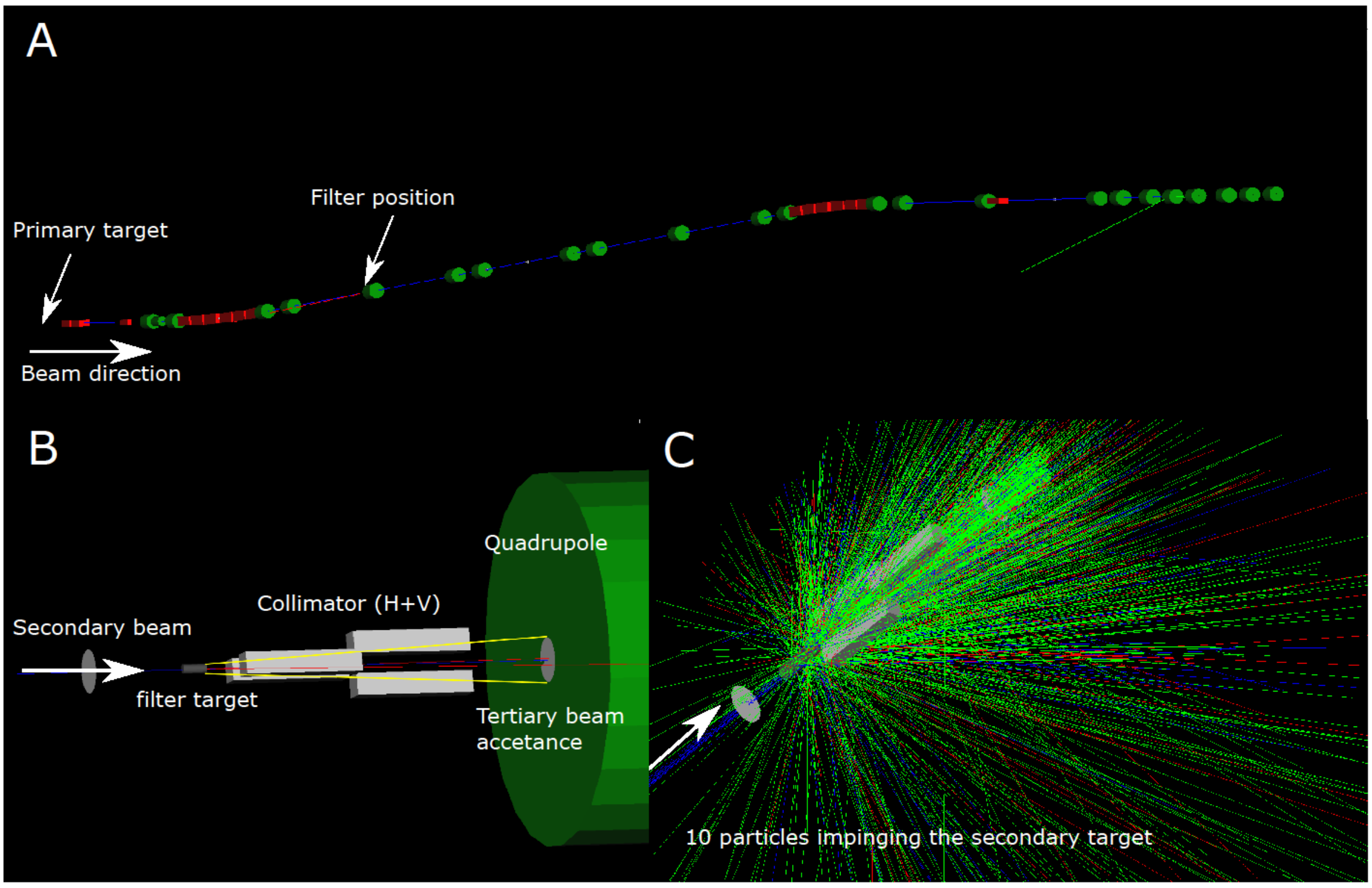

**Figure 2. The beam line as modeled in G4BeamLine. (A): Beam layout as modelled in G4beamline. One secondary particle is transported up to the filter target (red line) and the tertiary particle is then transported downstream to the experiment (green line). (B): A close view of the secondary target just before the momentum-selection collimator of the line. (C): Visualization of 10 particles impinging the secondary target.**

### 1.4 Secondary target and simulation setup

The secondary (filter) target has been modeled as a cylinder with a radius of 4 cm and with variable length. To sample the produced particles, a virtual scoring detector was placed at the entrance of the first subsequent quadrupole as shown in Figure 2B. This scoring detector represents the maximum geometrical acceptance of the beam downstream the target.

Massive Monte-Carlo simulations have been performed with the different targets. The choice and dimensions of materials was done based on the availability of those targets. The different materials and lengths studied are: pure Cu with length of 100 mm and 300 mm, polyethylene with length of 550, 700 and 1000 mm and pure W with a length of 150 mm, with their nominal densities.

In the results section, the total counts for protons, anti-protons and $\pi^{\pm}$ are shown for two different physics lists, FTFP_BERT and QGSP_BIC. These counts are normalized to $10^6$ incident particles while the momentum uncertainty is assumed to be equal to the line momentum byte, i.e. ±2 %. The purity of the beam in protons is described by the ratio p/π

## 2. Results

In the following plots, each plot title declares the secondary beam momentum and the target length, the $x$-axis indicates the momentum of the produced particles (‘tertiary’), while the $y$-axis shows the sampled number of particles normalized to $10^6$ incident particles. The vertical error bars are calculated by the Agresti-Coull method [5], using the binomial distribution and the horizontal bars indicate the ±2 % momentum sampling range.

### 2.1 50 – 70 GeV/c secondary beam regime

In the momentum range of 50 - 70 GeV/c, the proton production remains approximately the same in all targets, all momentum bins 20 – 45 GeV/c and all materials. The number of produced $\pi^+$ is always bigger than the number of produced protons whereas the number of $\pi^-$ is comparable with the number of protons. The optimal proton production in the region below the secondary momenta, 20 – 45 GeV/c, is given by the 700 mm polyethylene target with a secondary beam at 70 GeV/c. The peaks at the secondary momentum are due to non-interacting particles passing through the target. Finally, note that there are some notable differences between the physics lists.

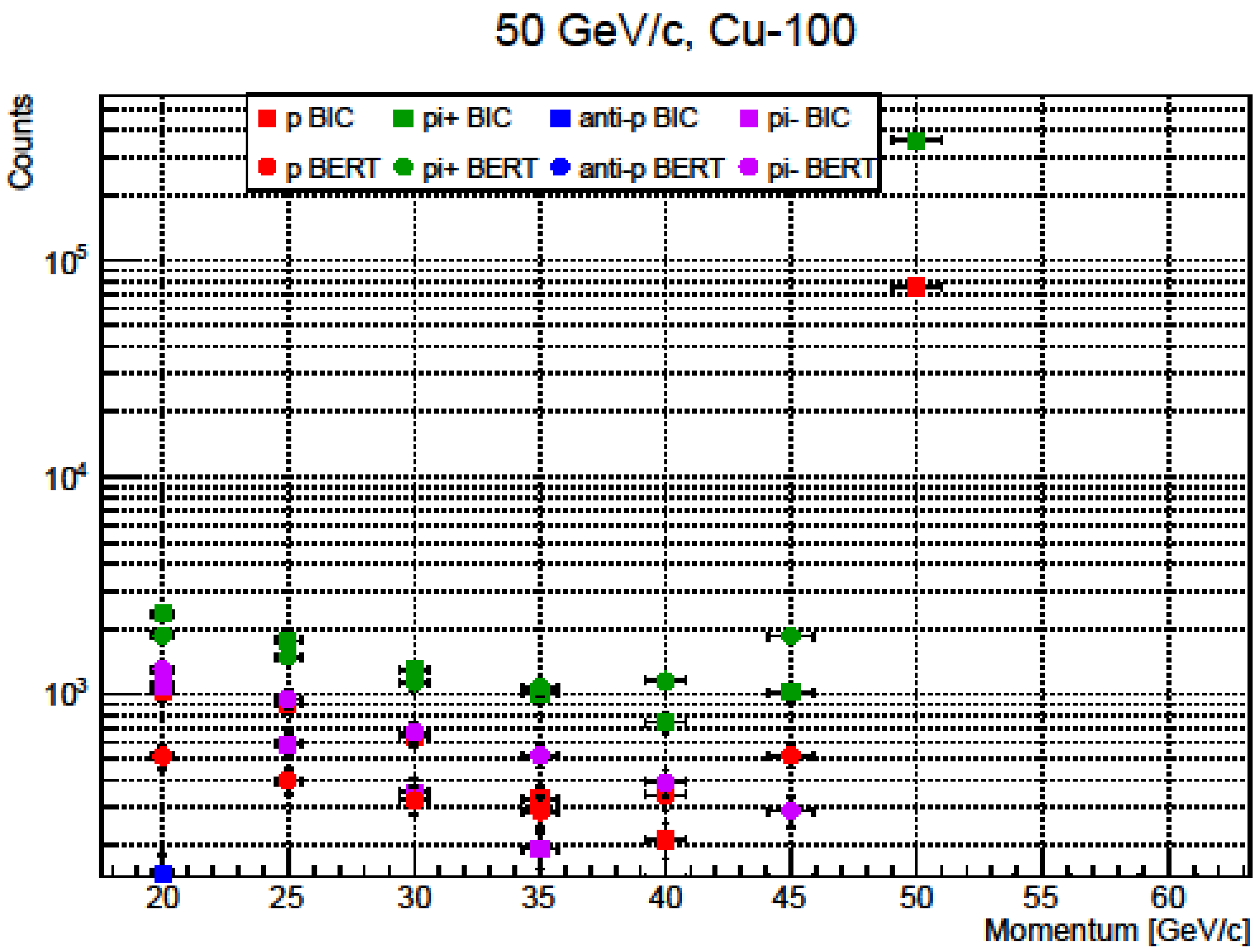

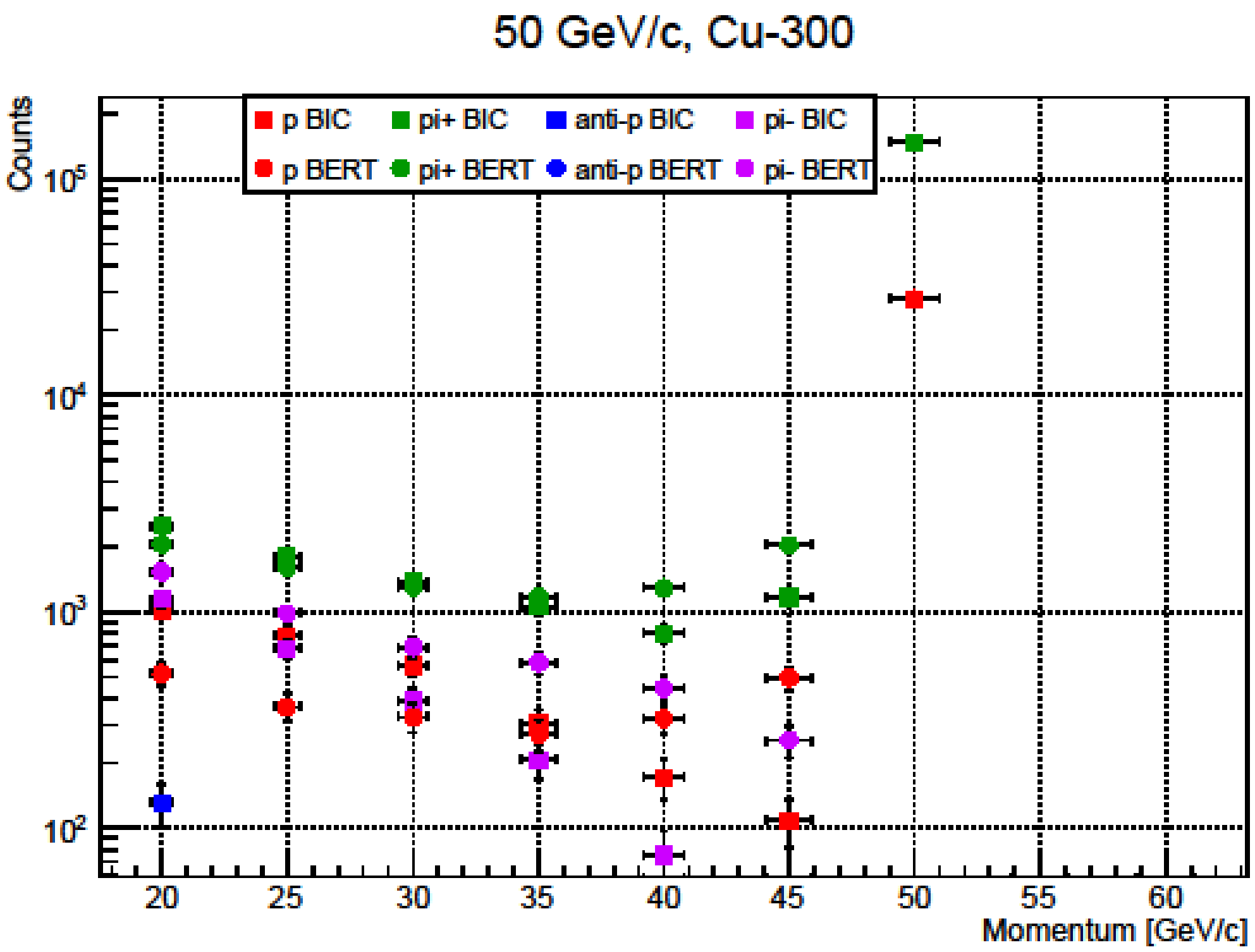
50 GeV/c, Cu-300
Counts
Momentum [GeV/c]
p BIC
pi+ BIC
anti-p BIC
pi- BIC
p BERT
pi+ BERT
anti-p BERT
pi- BERT
20
25
30
35
40
45
50
55
60

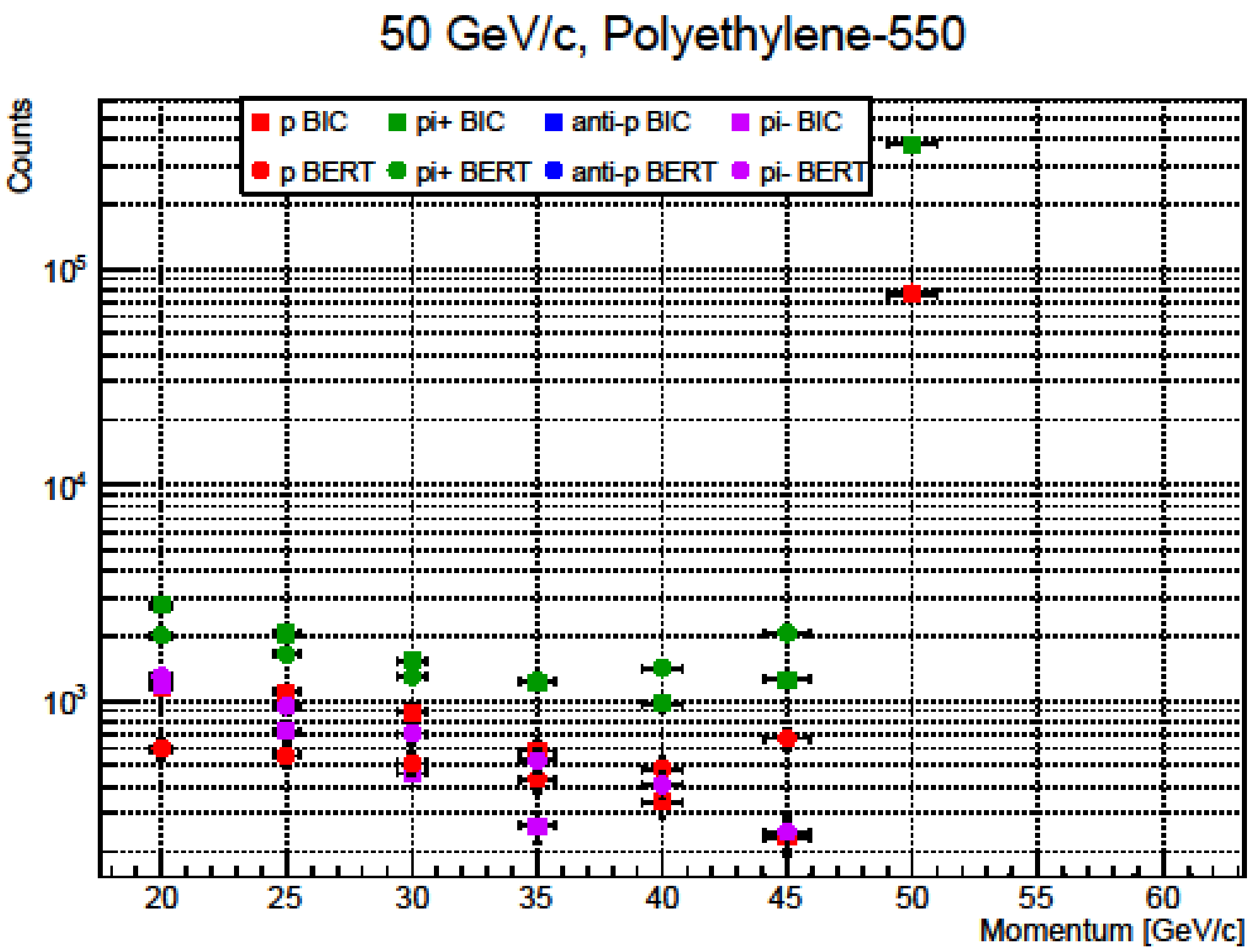
50 GeV/c, Polyethylene-550
Counts
Momentum [GeV/c]
p BIC
pi+ BIC
anti-p BIC
pi- BIC
p BERT
pi+ BERT
anti-p BERT
pi- BERT
20
25
30
35
40
45
50
55
60

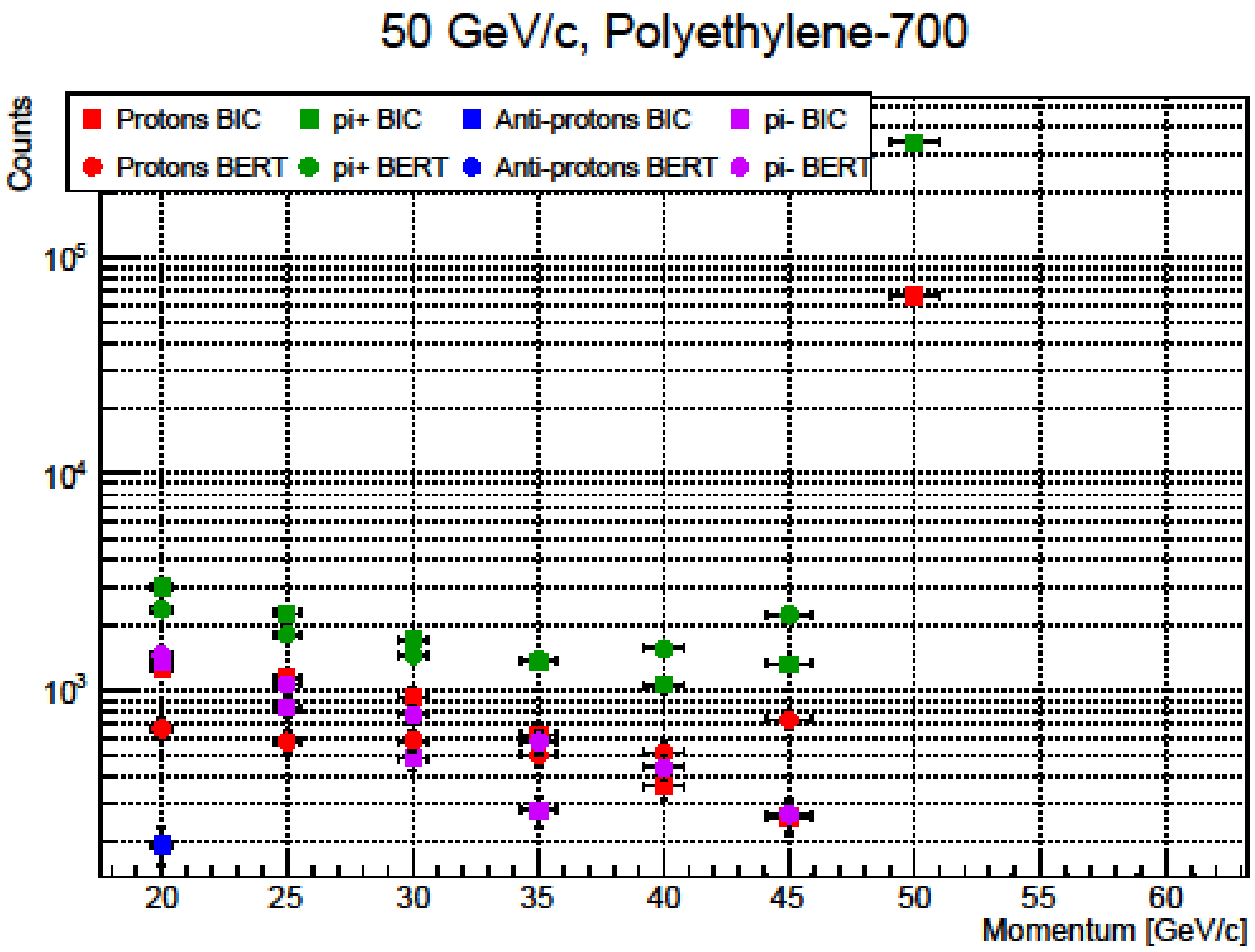

50 GeV/c, Polyethylene-700
Protons BIC
pi+ BIC
Anti-protons BIC
pi- BIC
Protons BERT
pi+ BERT
Anti-protons BERT
pi- BERT
Counts
Momentum [GeV/c]

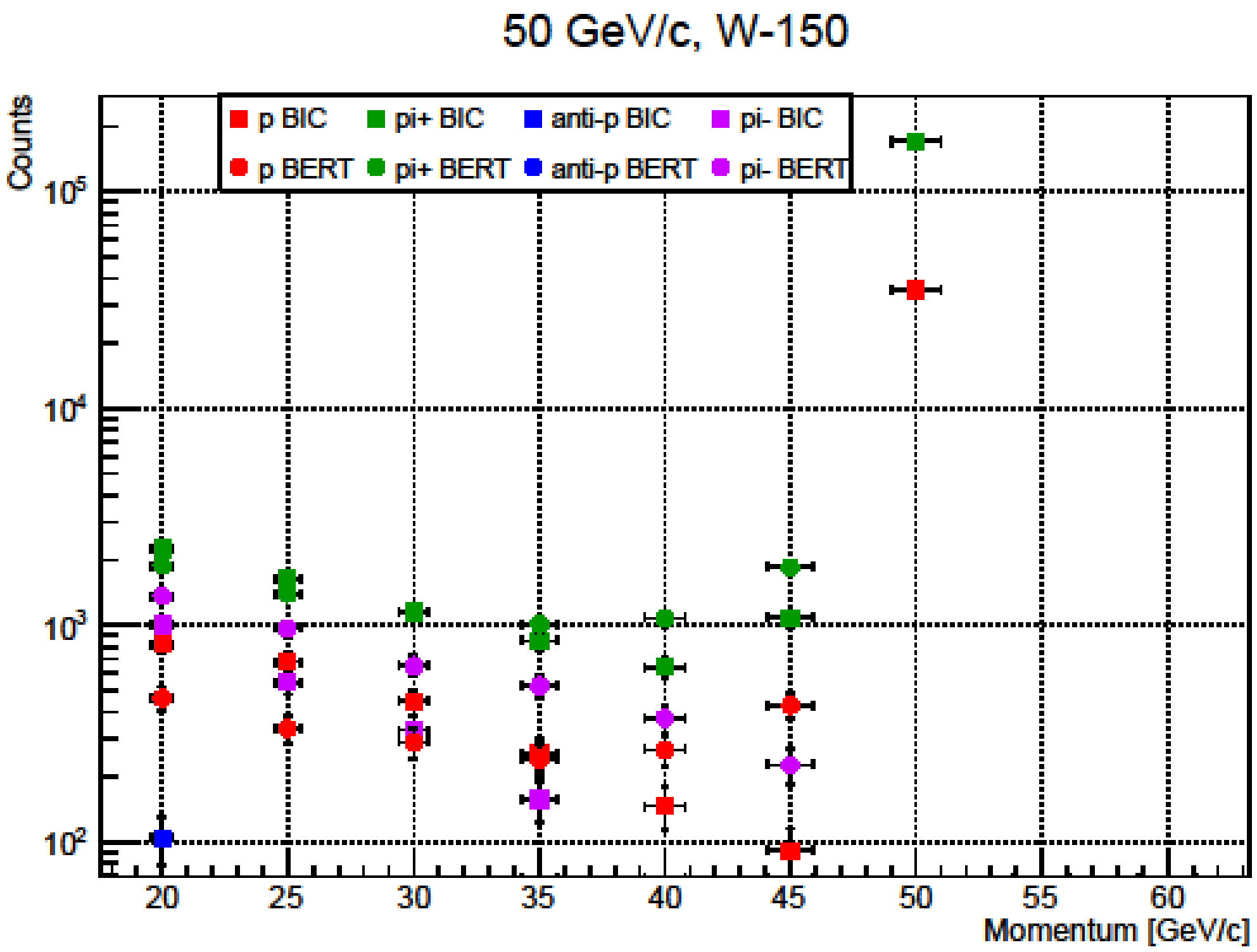

50 GeV/c, W-150
p BIC
pi+ BIC
anti-p BIC
pi- BIC
p BERT
pi+ BERT
anti-p BERT
pi- BERT
Counts
Momentum [GeV/c]

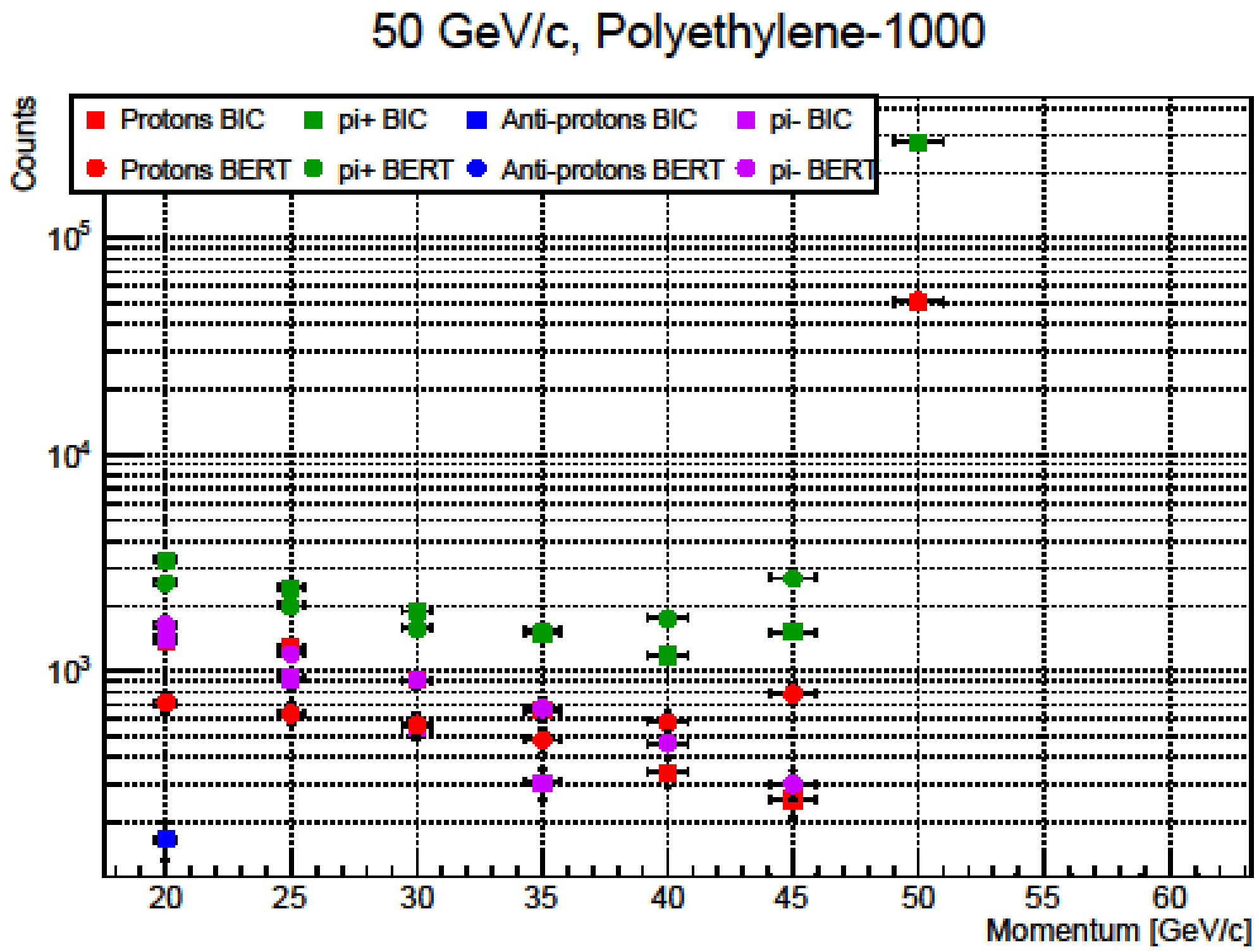
50 GeV/c, Polyethylene-1000
Counts
Protons BIC
pi+ BIC
Anti-protons BIC
pi- BIC
Protons BERT
pi+ BERT
Anti-protons BERT
pi- BERT
$10^5$
$10^4$
$10^3$
20
25
30
35
40
45
50
55
60
Momentum [GeV/c]

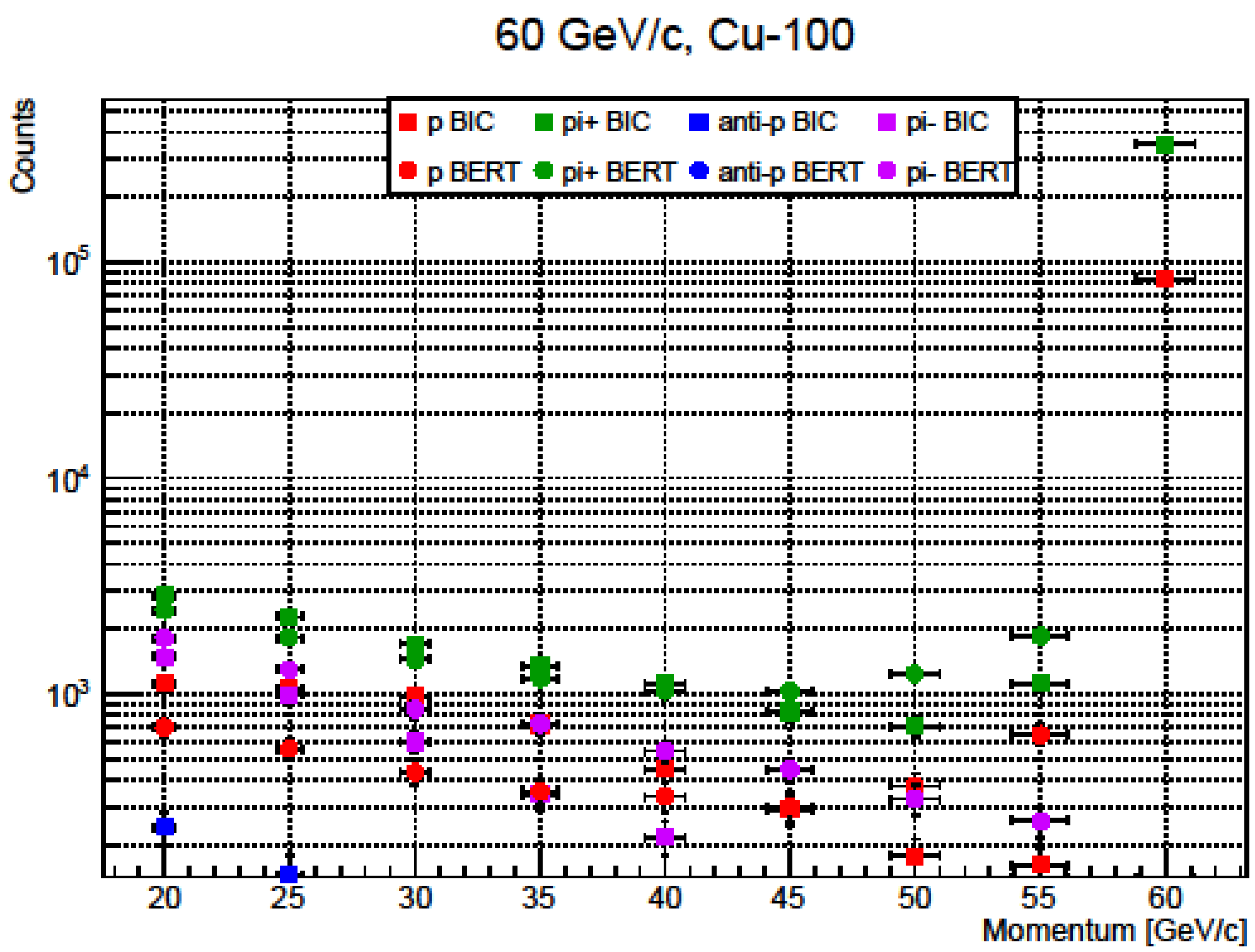
60 GeV/c, Cu-100
Counts
p BIC
pi+ BIC
anti-p BIC
pi- BIC
p BERT
pi+ BERT
anti-p BERT
pi- BERT
$10^5$
$10^4$
$10^3$
20
25
30
35
40
45
50
55
60
Momentum [GeV/c]

## 60 GeV/c, Cu-300

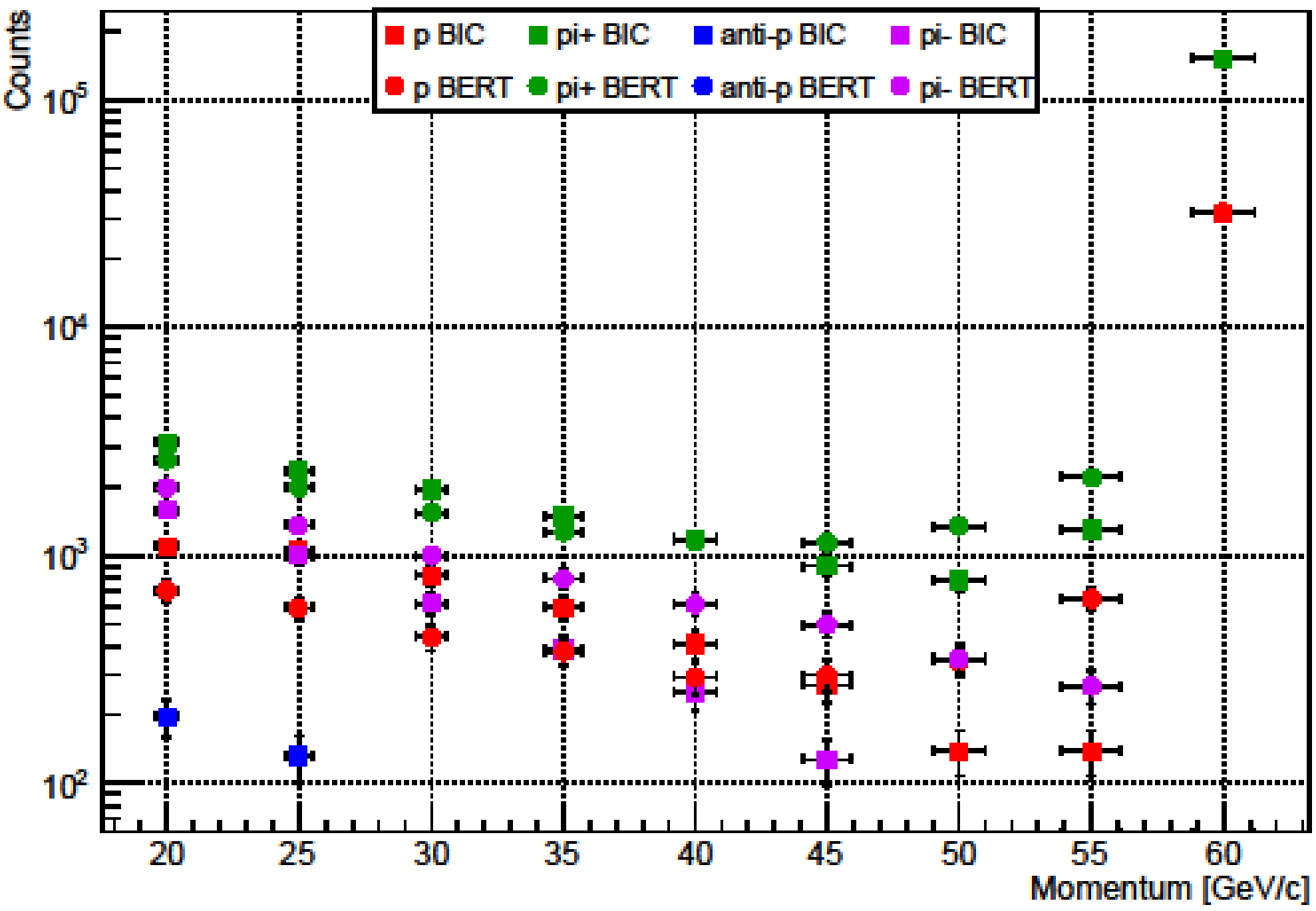

## 60 GeV/c, Polyethylene-550

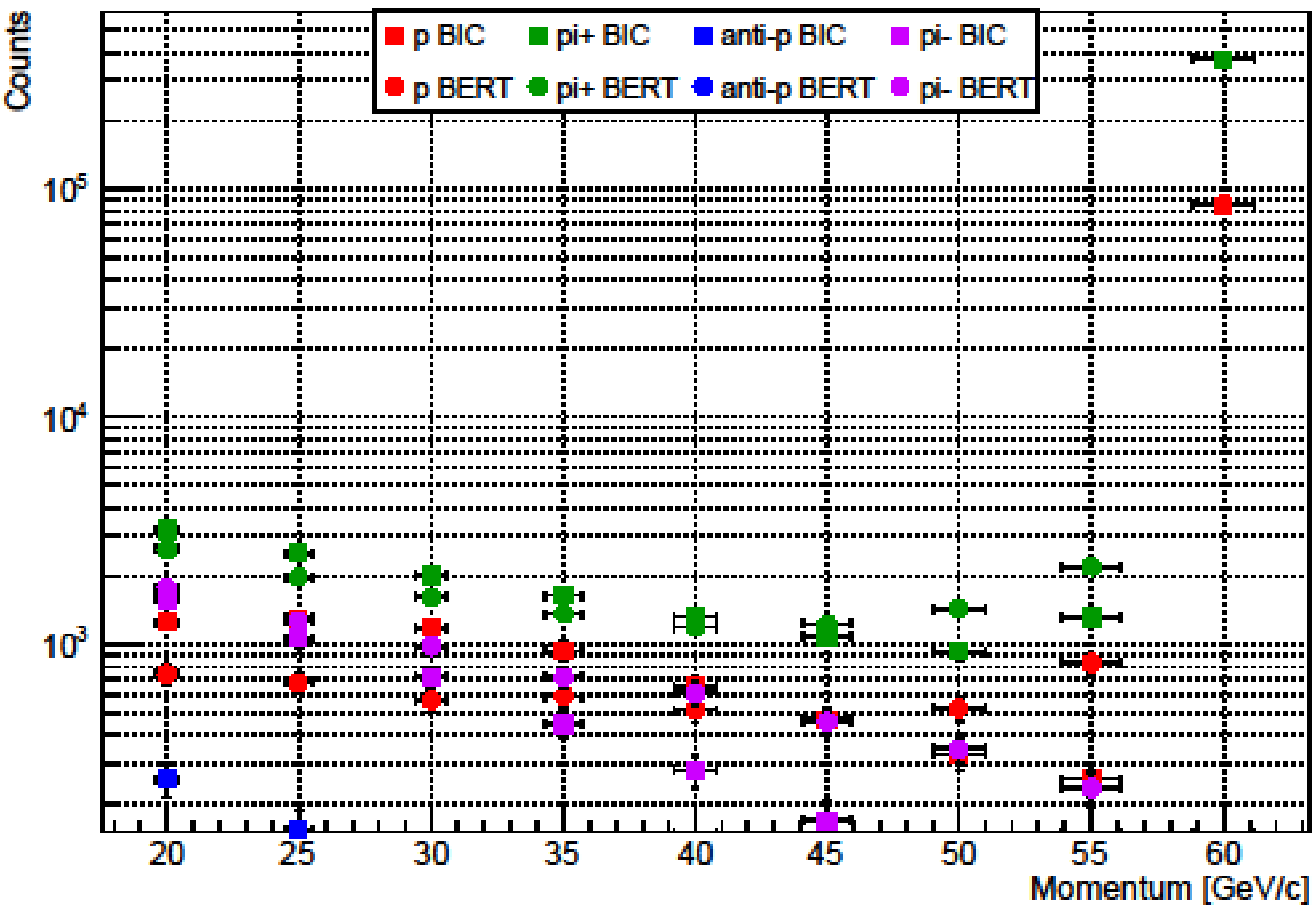

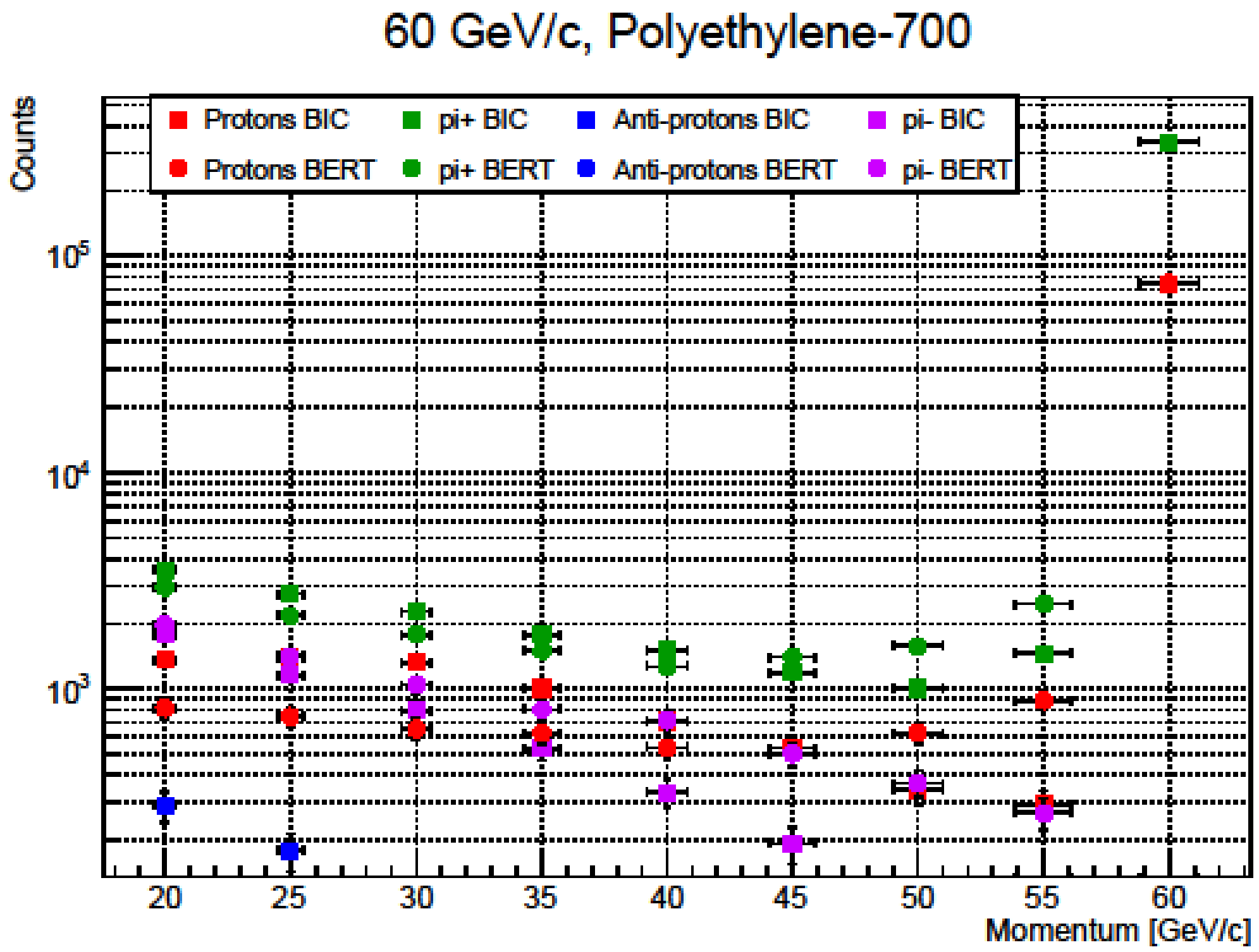
60 GeV/c, Polyethylene-700
Protons BIC
pi+ BIC
Anti-protons BIC
pi- BIC
Protons BERT
pi+ BERT
Anti-protons BERT
pi- BERT
Counts
$10^5$
$10^4$
$10^3$
20
25
30
35
40
45
50
55
60
Momentum [GeV/c]

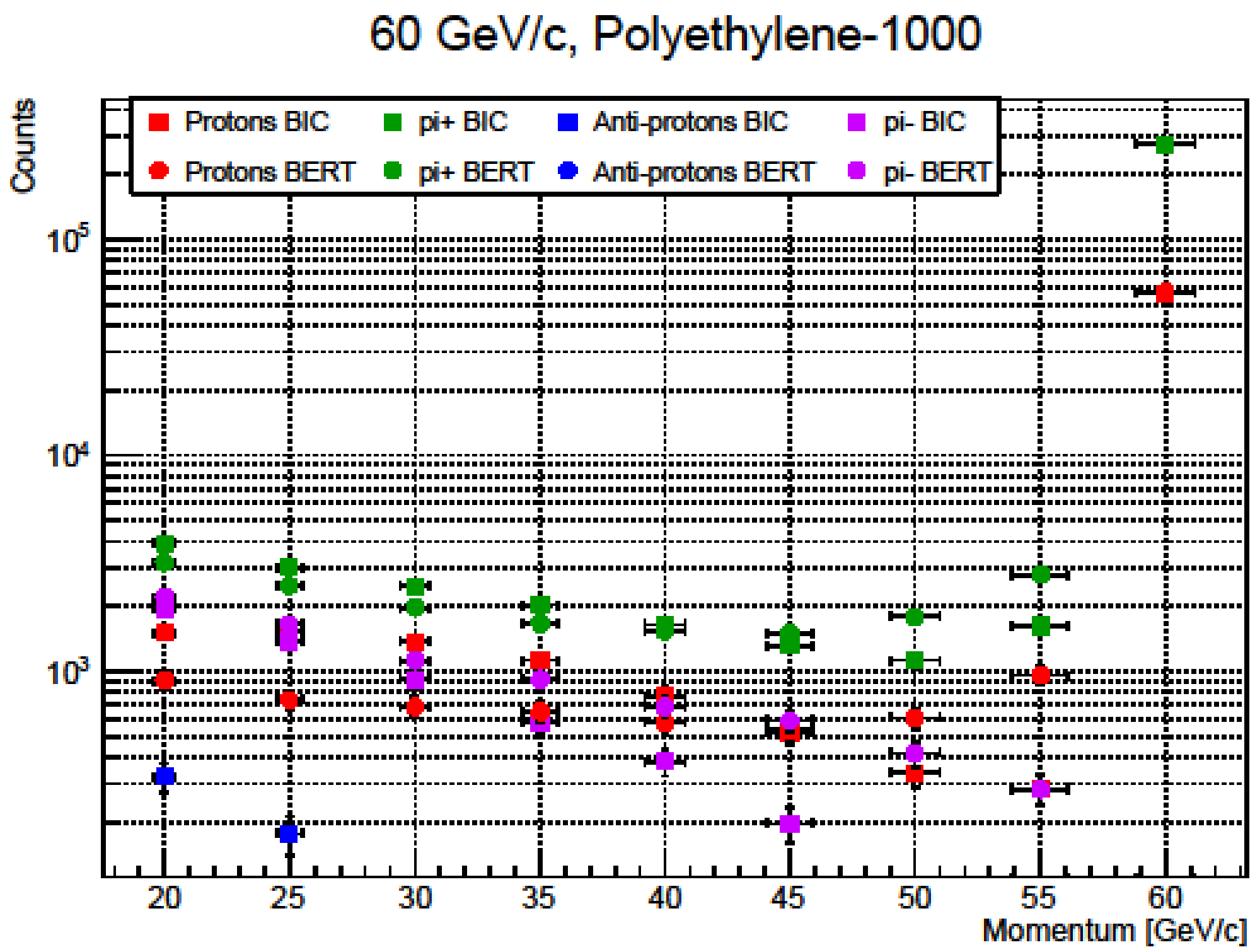
60 GeV/c, Polyethylene-1000
Protons BIC
pi+ BIC
Anti-protons BIC
pi- BIC
Protons BERT
pi+ BERT
Anti-protons BERT
pi- BERT
Counts
$10^5$
$10^4$
$10^3$
20
25
30
35
40
45
50
55
60
Momentum [GeV/c]

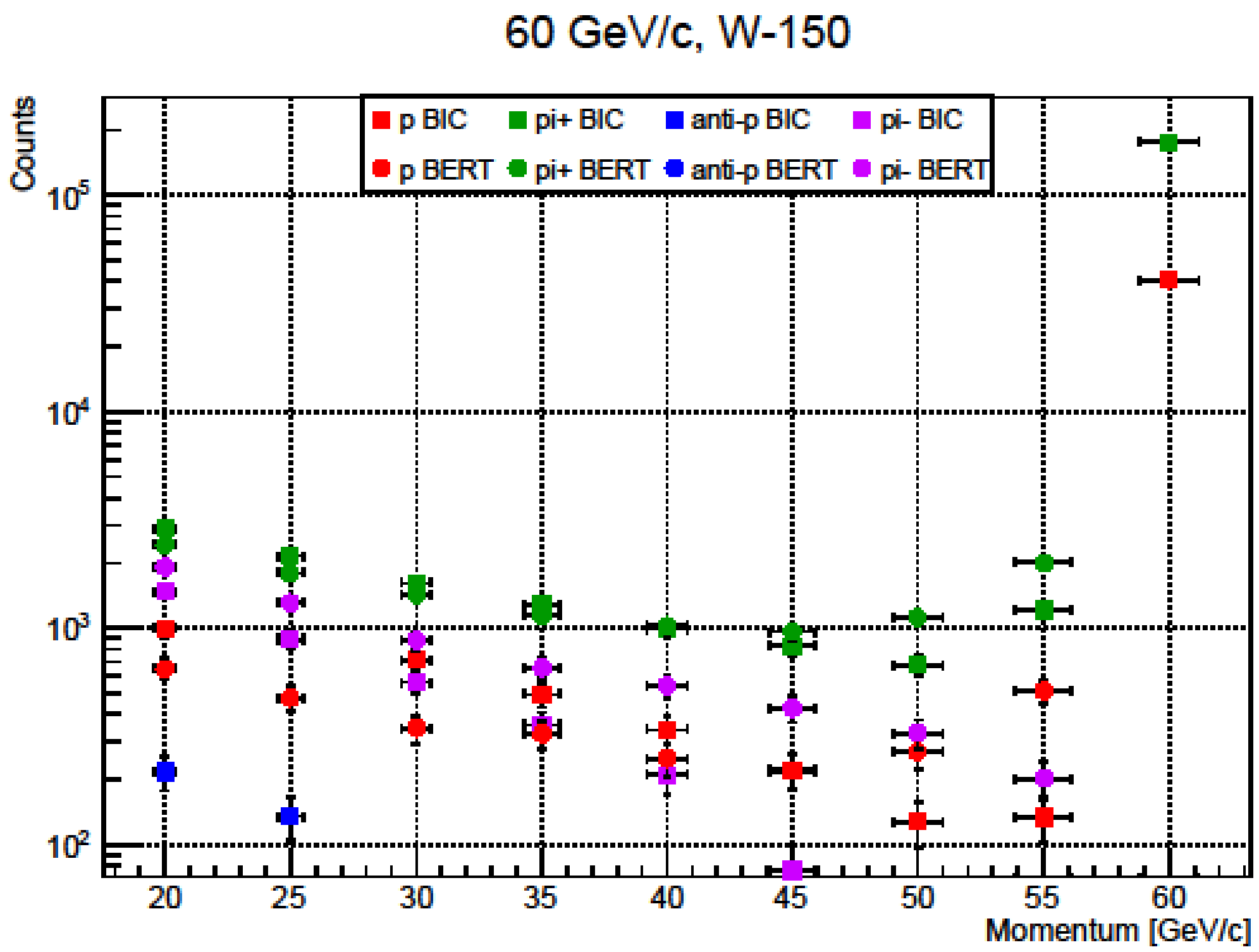
60 GeV/c, W-150
Counts
p BIC
pi+ BIC
anti-p BIC
pi- BIC
p BERT
pi+ BERT
anti-p BERT
pi- BERT
$10^5$
$10^4$
$10^3$
$10^2$
20
25
30
35
40
45
50
55
60
Momentum [GeV/c]

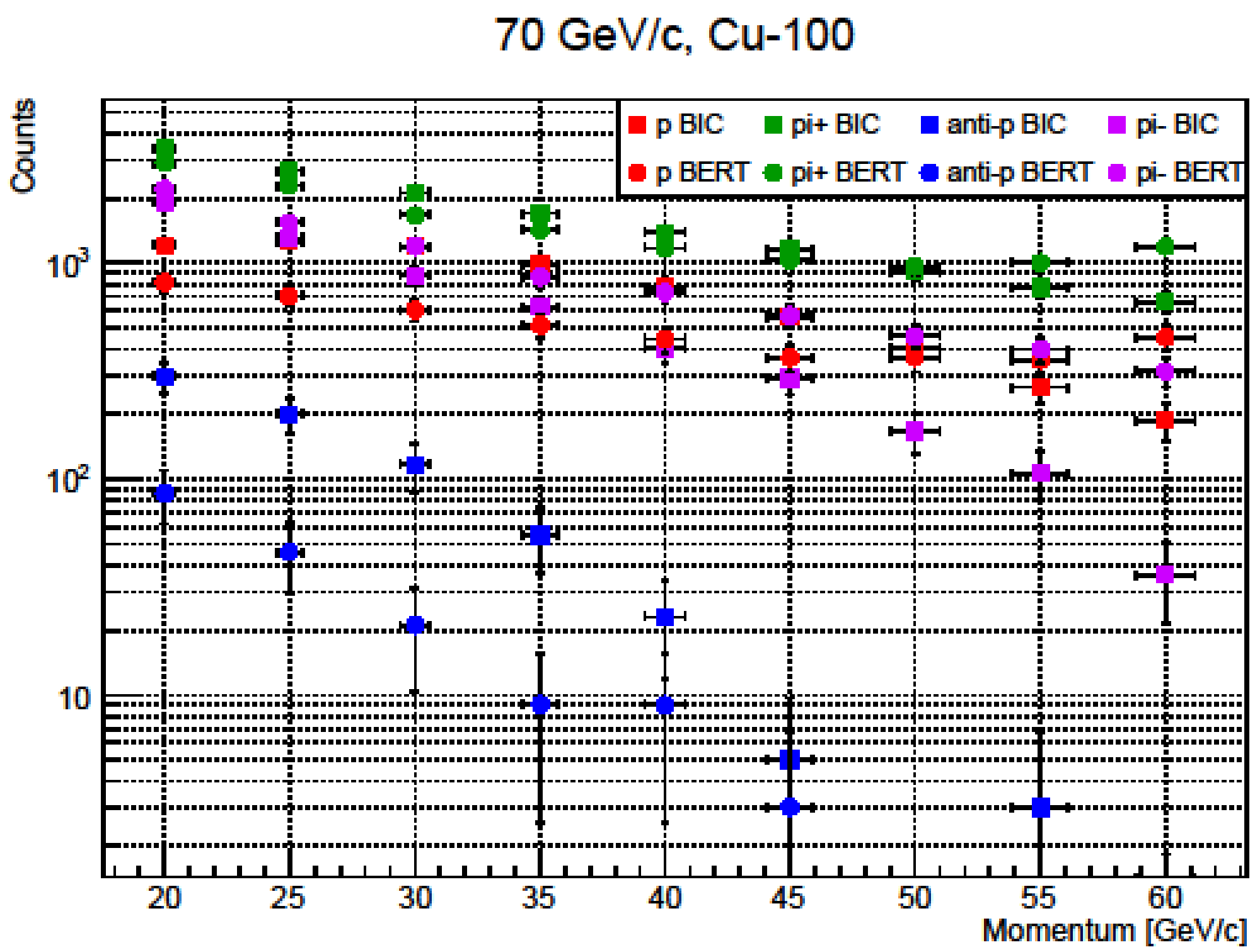
70 GeV/c, Cu-100
Counts
p BIC
pi+ BIC
anti-p BIC
pi- BIC
p BERT
pi+ BERT
anti-p BERT
pi- BERT
$10^3$
$10^2$
10
20
25
30
35
40
45
50
55
60
Momentum [GeV/c]

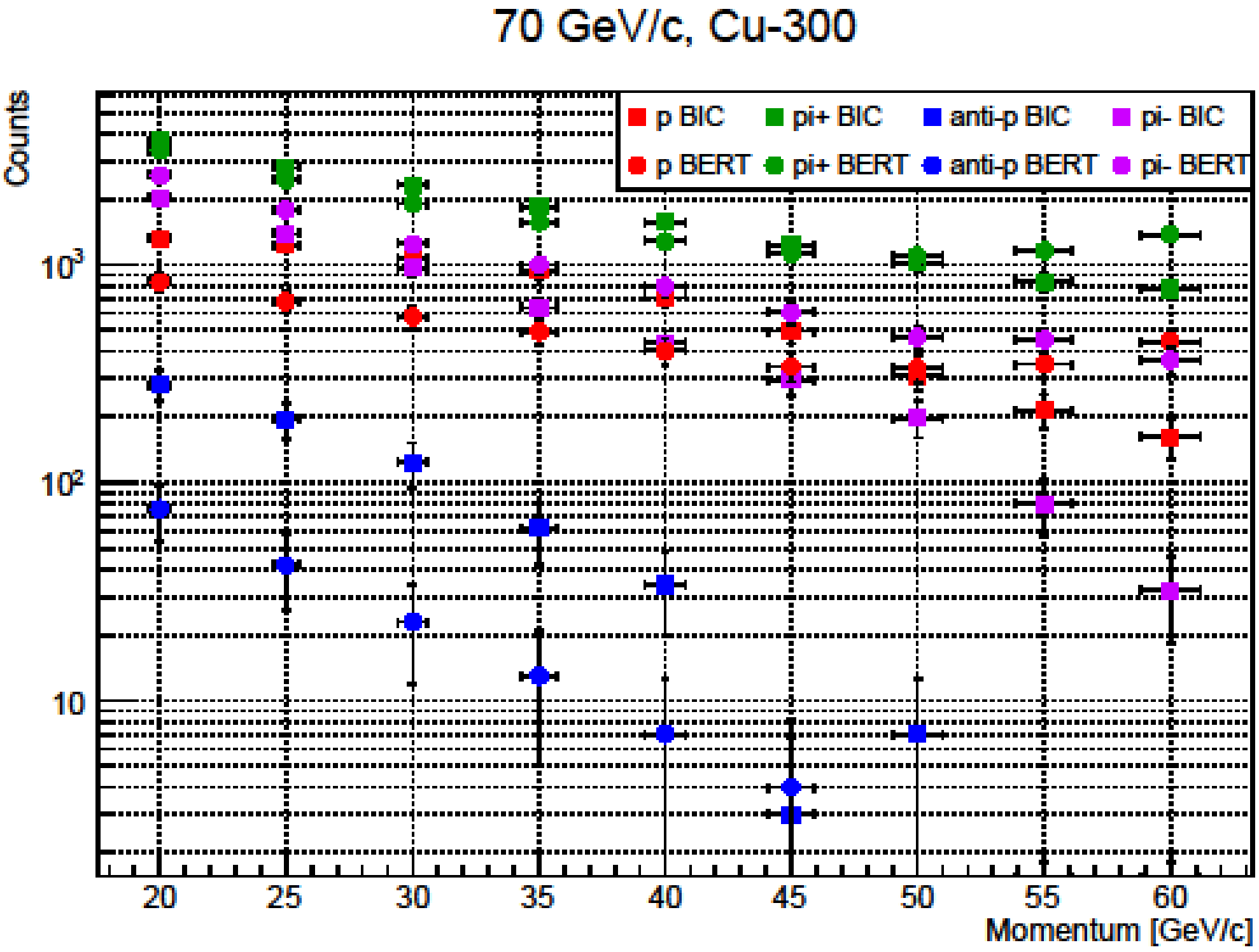

70 GeV/c, Cu-300
Counts
p BIC
pi+ BIC
anti-p BIC
pi- BIC
p BERT
pi+ BERT
anti-p BERT
pi- BERT
$10^3$
$10^2$
10
20
25
30
35
40
45
50
55
60
Momentum [GeV/c]

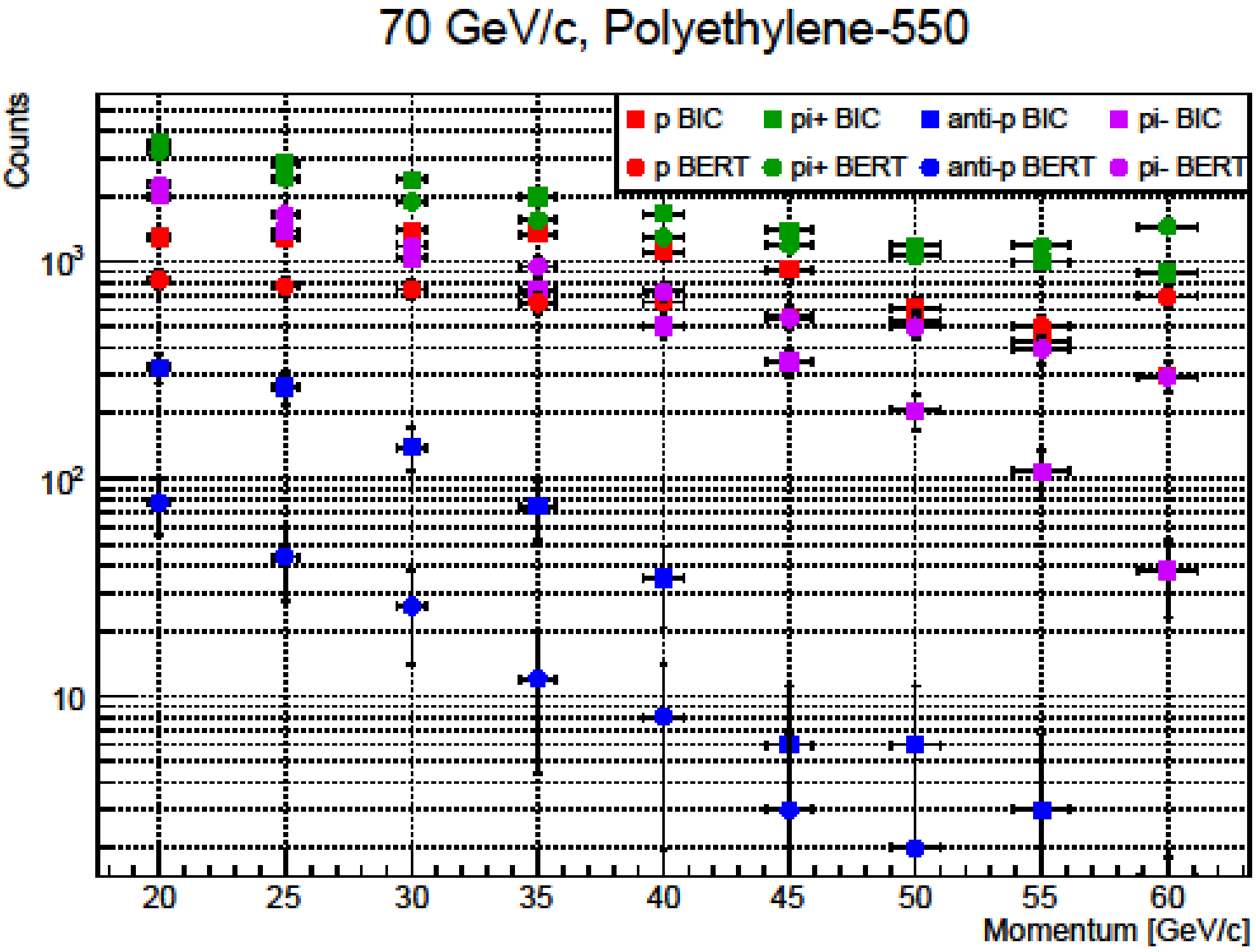

70 GeV/c, Polyethylene-550
Counts
p BIC
pi+ BIC
anti-p BIC
pi- BIC
p BERT
pi+ BERT
anti-p BERT
pi- BERT
$10^3$
$10^2$
10
20
25
30
35
40
45
50
55
60
Momentum [GeV/c]

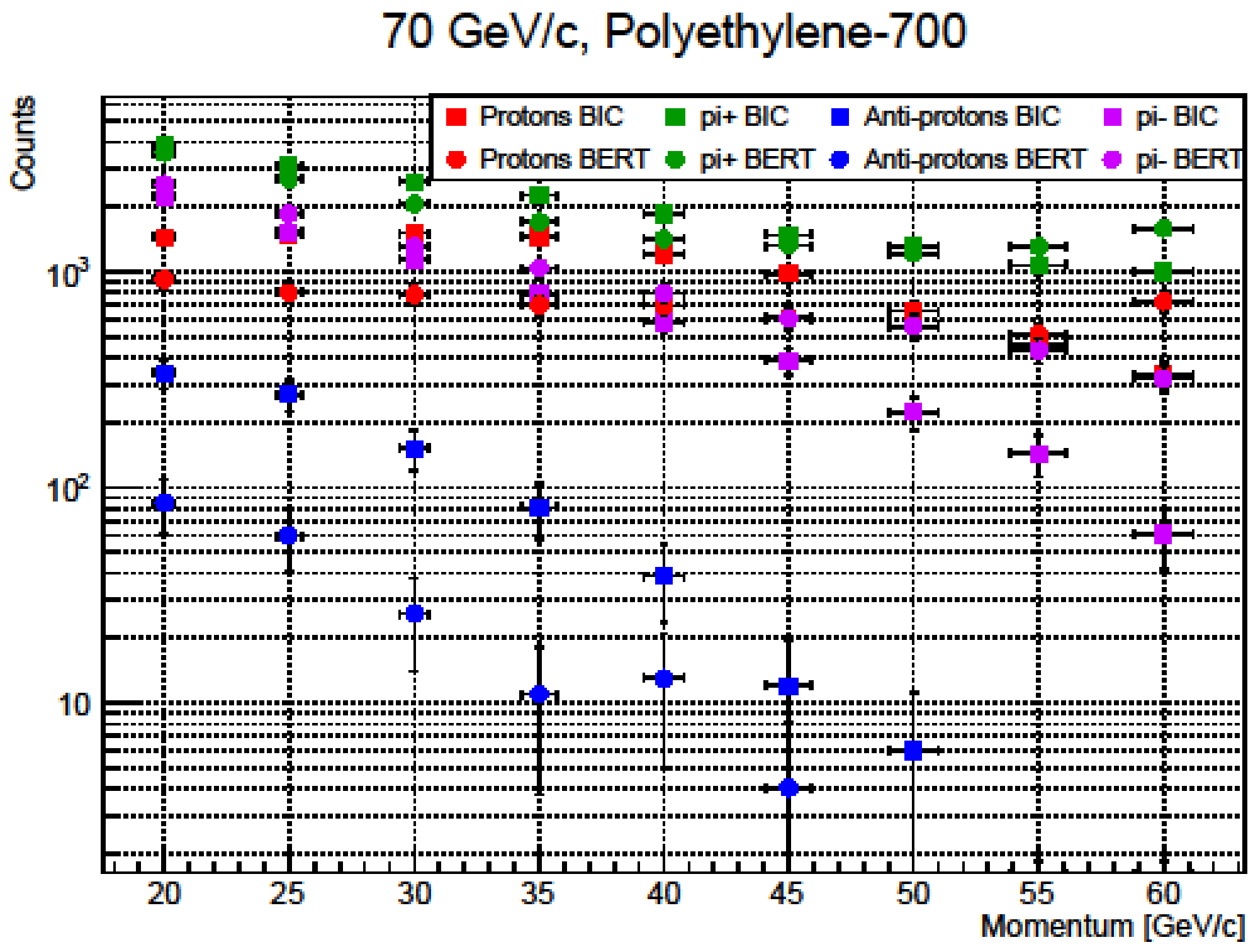
70 GeV/c, Polyethylene-700
Counts
Protons BIC
pi+ BIC
Anti-protons BIC
pi- BIC
Protons BERT
pi+ BERT
Anti-protons BERT
pi- BERT
$10^3$
$10^2$
10
20
25
30
35
40
45
50
55
60
Momentum [GeV/c]

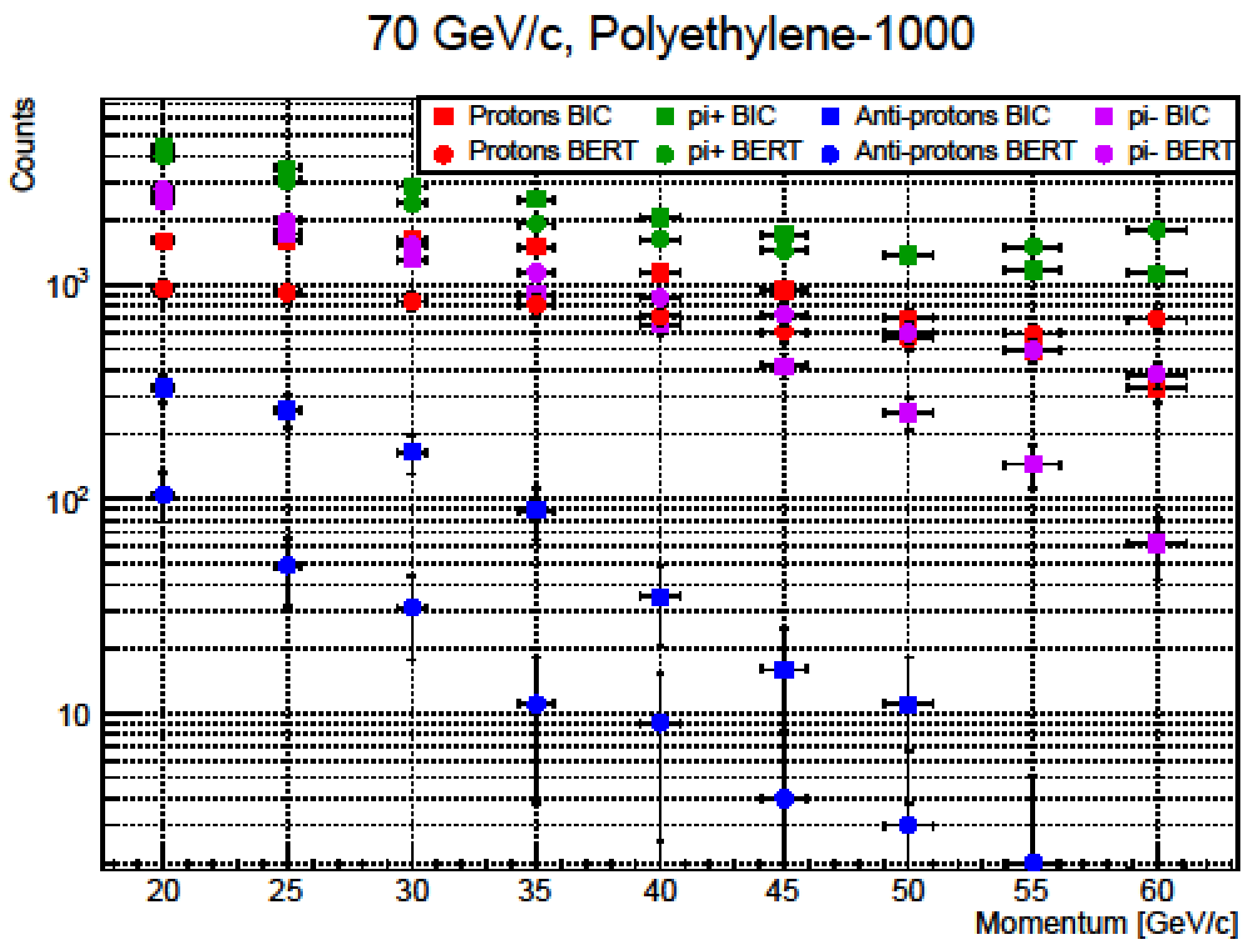
70 GeV/c, Polyethylene-1000
Counts
Protons BIC
pi+ BIC
Anti-protons BIC
pi- BIC
Protons BERT
pi+ BERT
Anti-protons BERT
pi- BERT
$10^3$
$10^2$
10
20
25
30
35
40
45
50
55
60
Momentum [GeV/c]

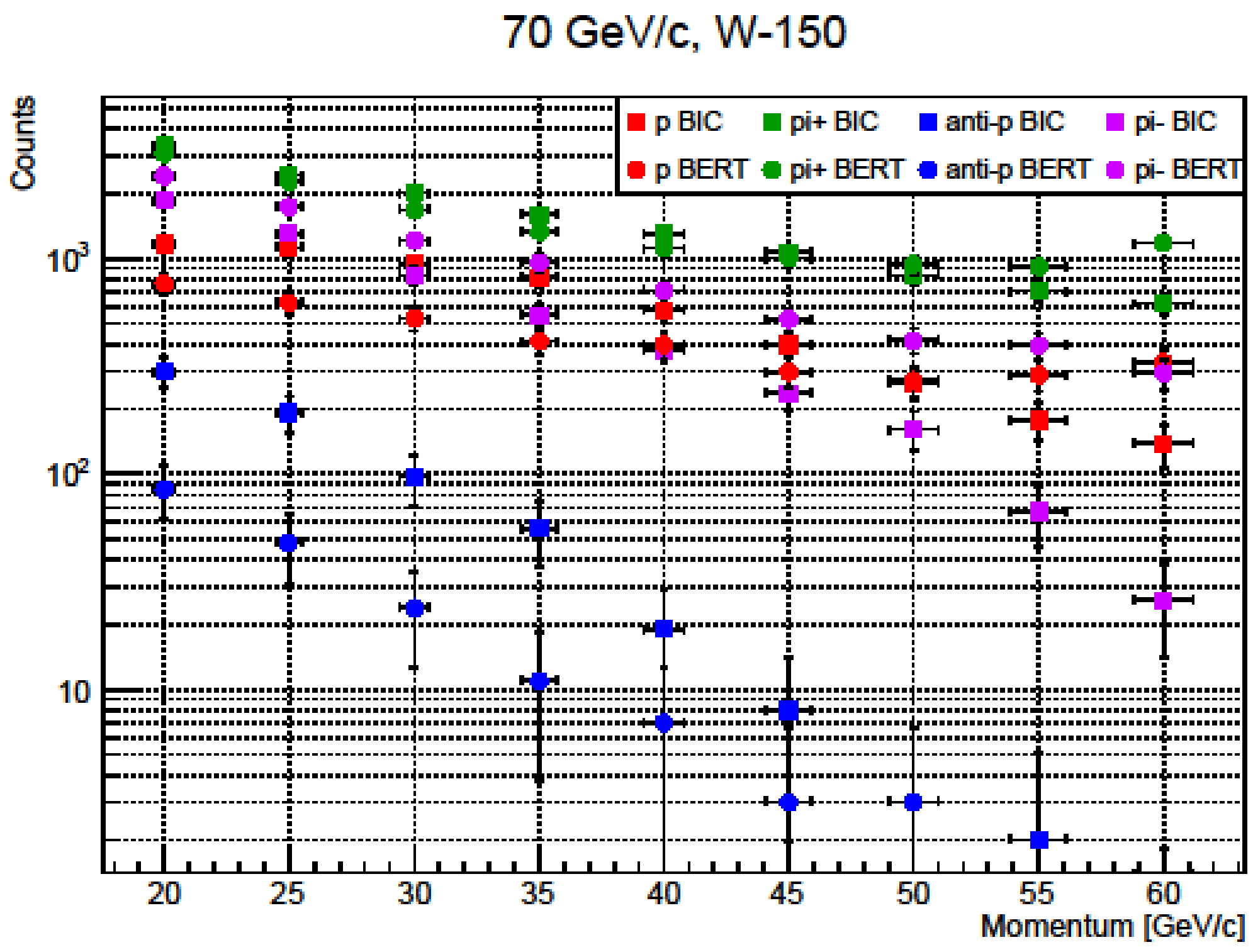

## 2.2 100 – 200 GeV/c secondary beam regime

In this higher-energy momentum regime, it is shown that for a certain secondary momentum and material the proton production remains relatively constant over the tertiary momenta. However, the pion production decreases as the tertiary momentum increases. It can be seen that in high enough tertiary momentum there will be more protons than pions in the tertiary beam, see e.g. 150 GeV/c Cu-100. Some differences in the proton production between the two physics lists used are more evident in this regime. Note especially in the aforementioned figure where only the QGSP_BIC protons are greater than the pions for the 50-60 GeV/c momentum bins, while this is not the case for FTFP_BERT for the same momenta. However, increasing the secondary momenta to 200 GeV/c both physics lists predict a higher protons than pions production at 60 GeV/c for 100 mm Cu. Differences between the list are evident also in the antiproton production, with the QGSP_BIC list to give always a higher yield.

100 GeV/c, Cu-100

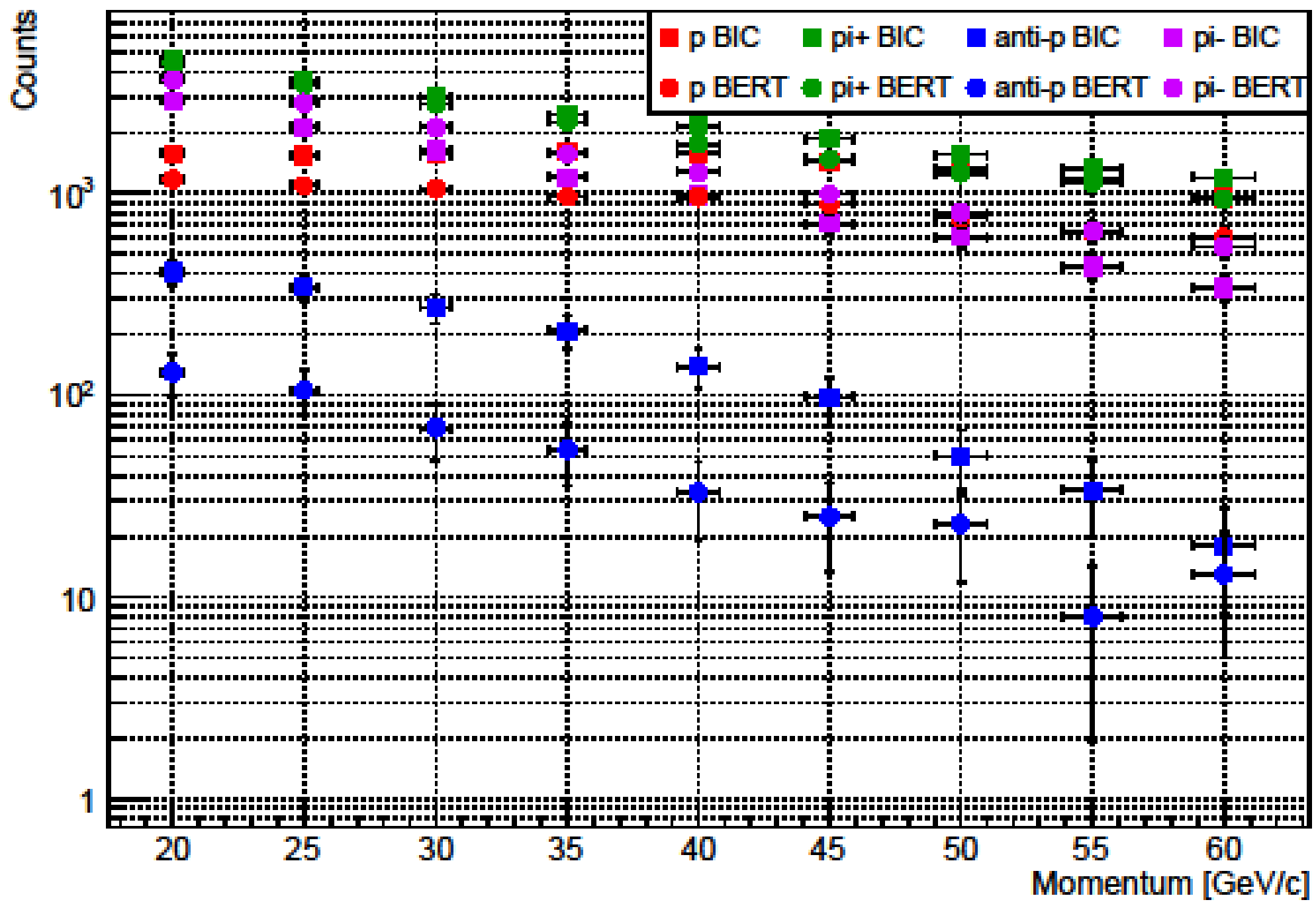

100 GeV/c, Cu-300

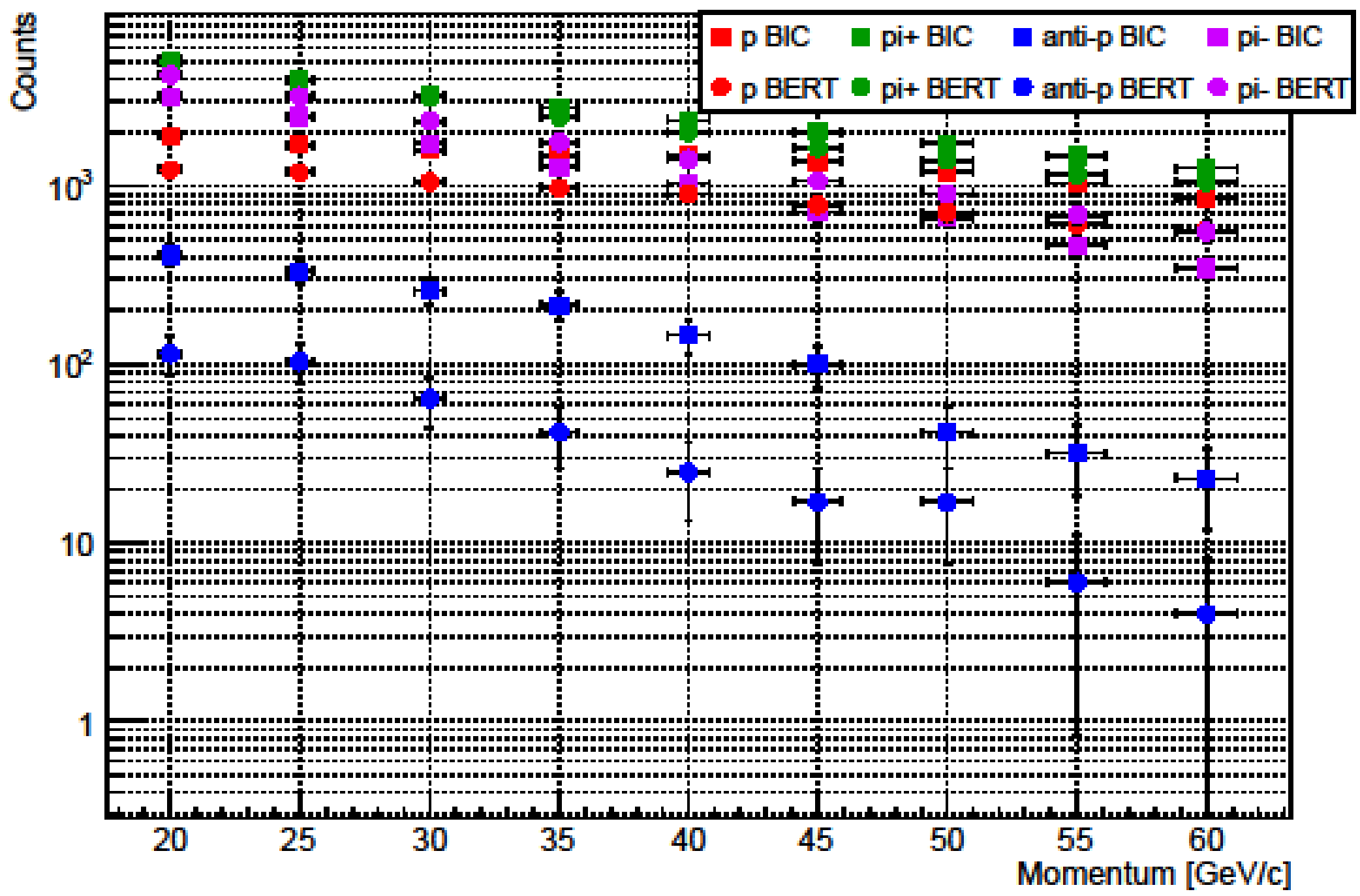

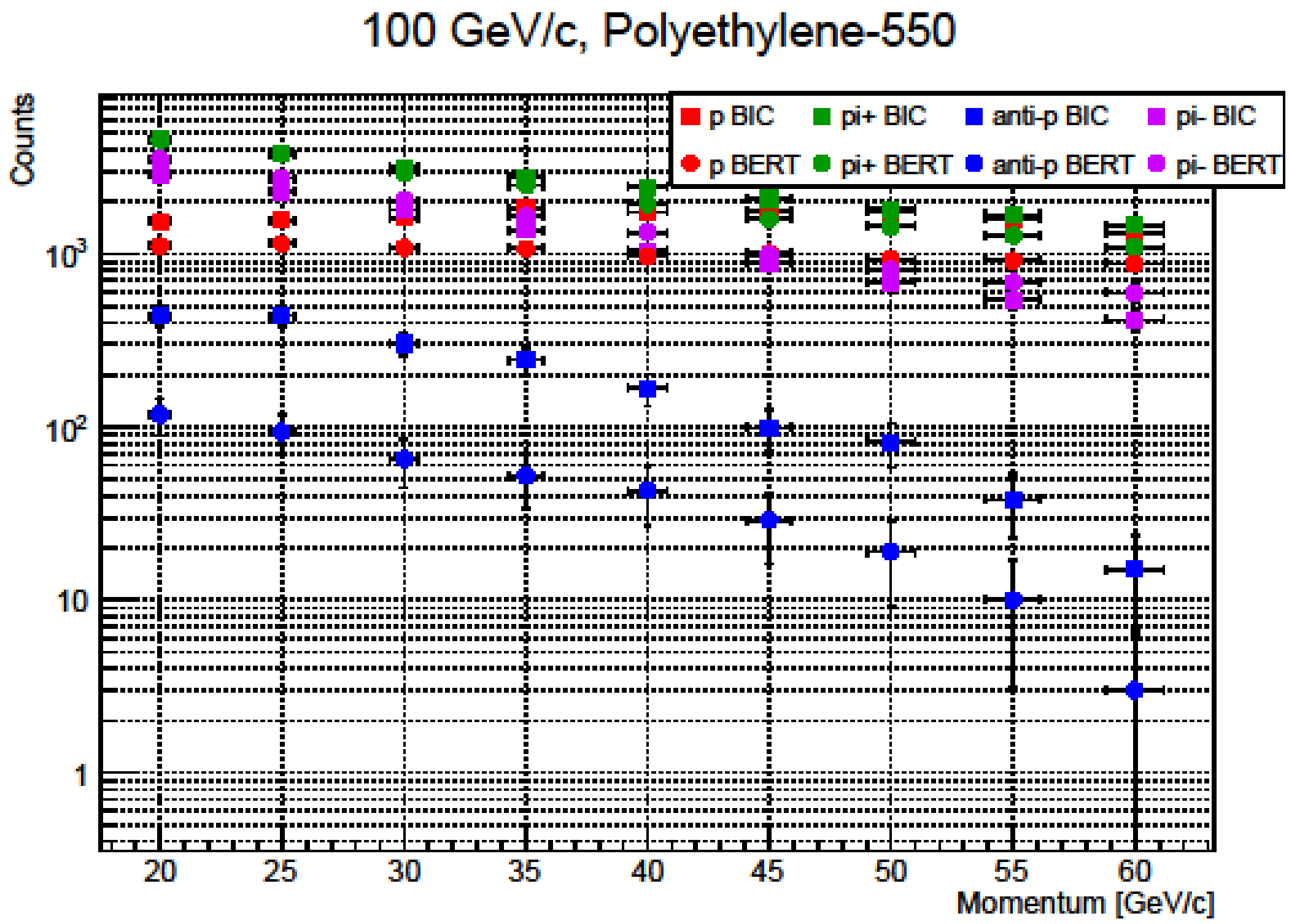

100 GeV/c, Polyethylene-550
Counts
Momentum [GeV/c]
p BIC
pi+ BIC
anti-p BIC
pi- BIC
p BERT
pi+ BERT
anti-p BERT
pi- BERT
1
10
$10^2$
$10^3$
20
25
30
35
40
45
50
55
60

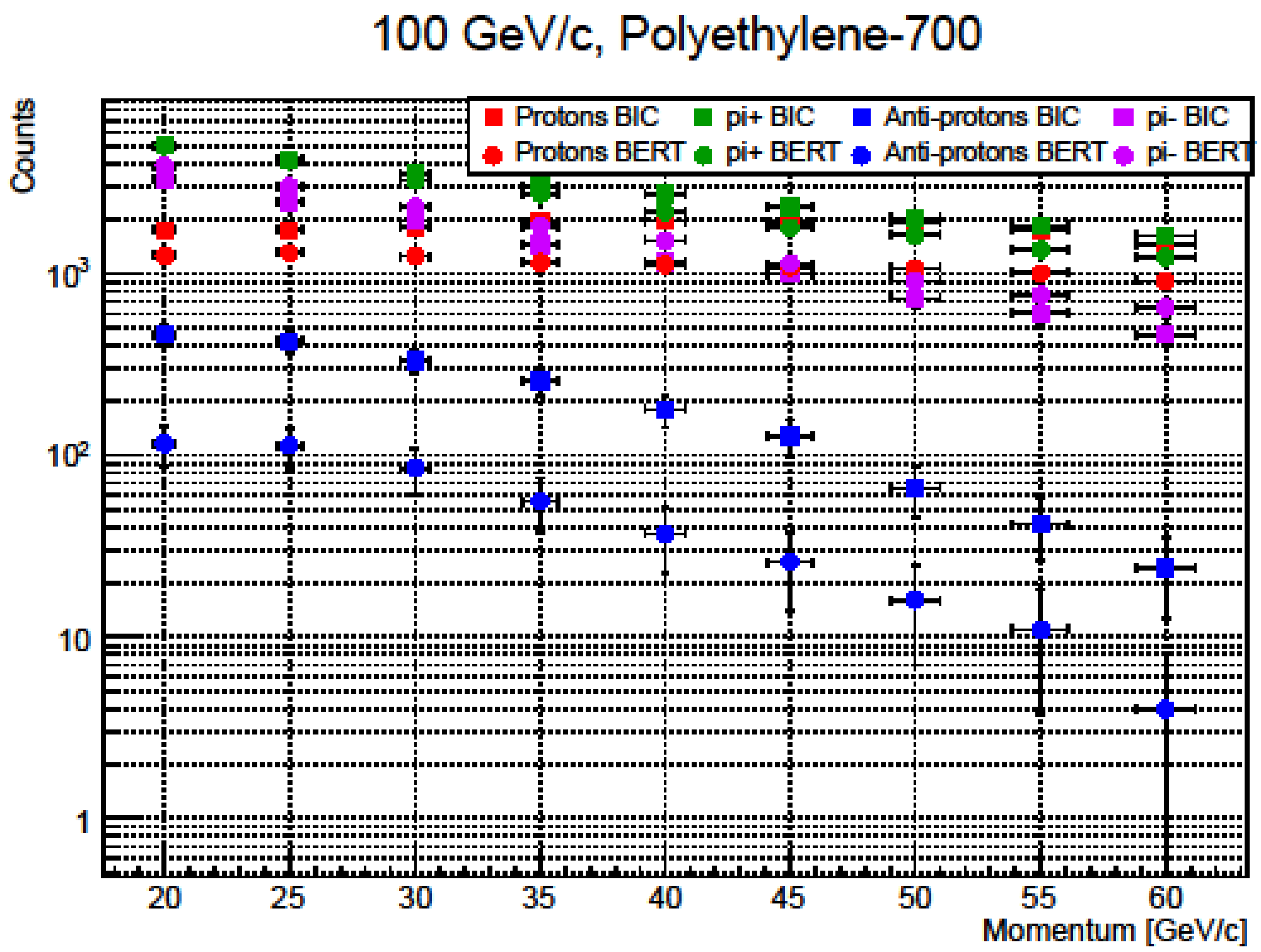

100 GeV/c, Polyethylene-700
Counts
Momentum [GeV/c]
Protons BIC
pi+ BIC
Anti-protons BIC
pi- BIC
Protons BERT
pi+ BERT
Anti-protons BERT
pi- BERT
1
10
$10^2$
$10^3$
20
25
30
35
40
45
50
55
60

## 100 GeV/c, Polyethylene-1000

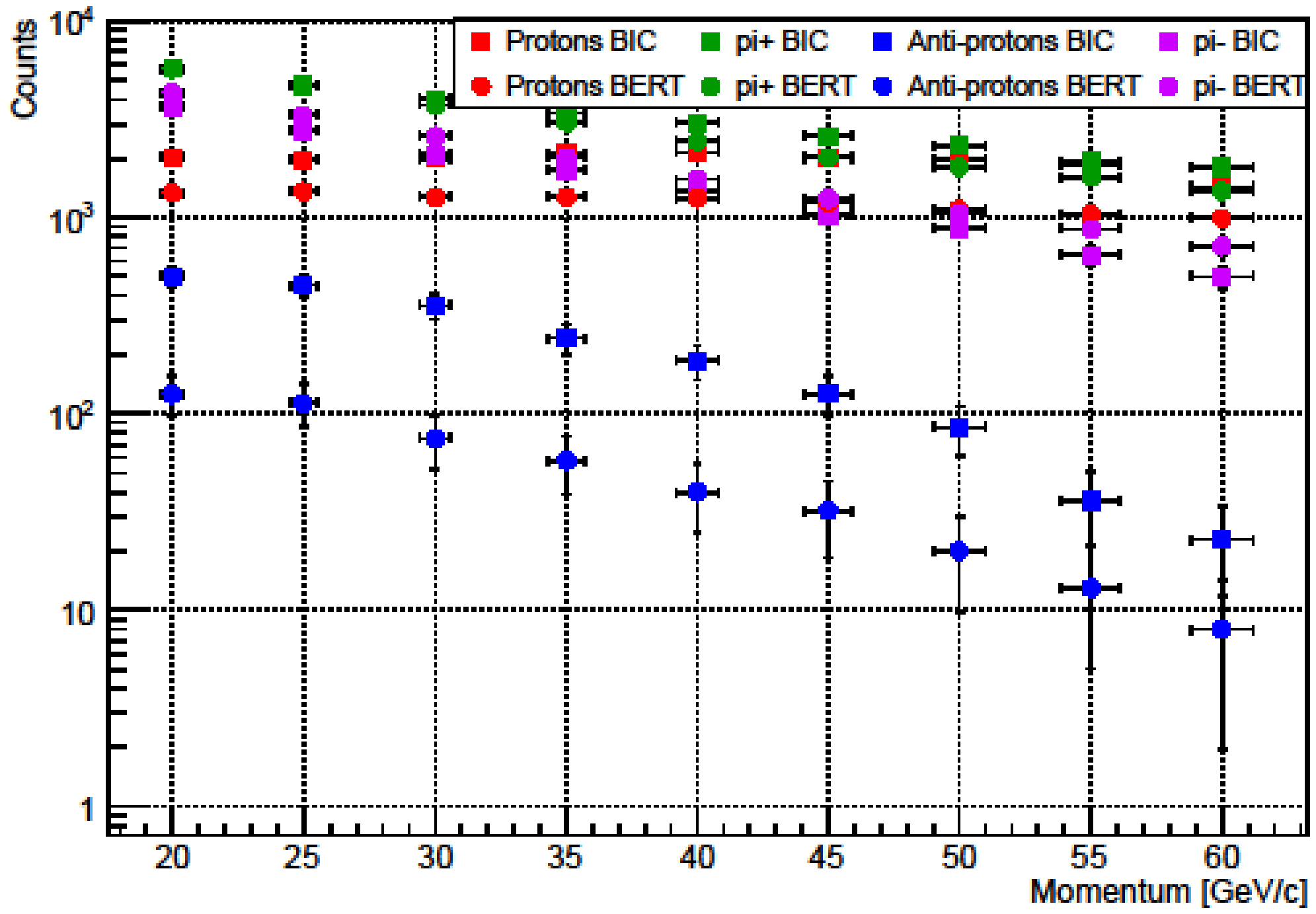

## 100 GeV/c, W-150

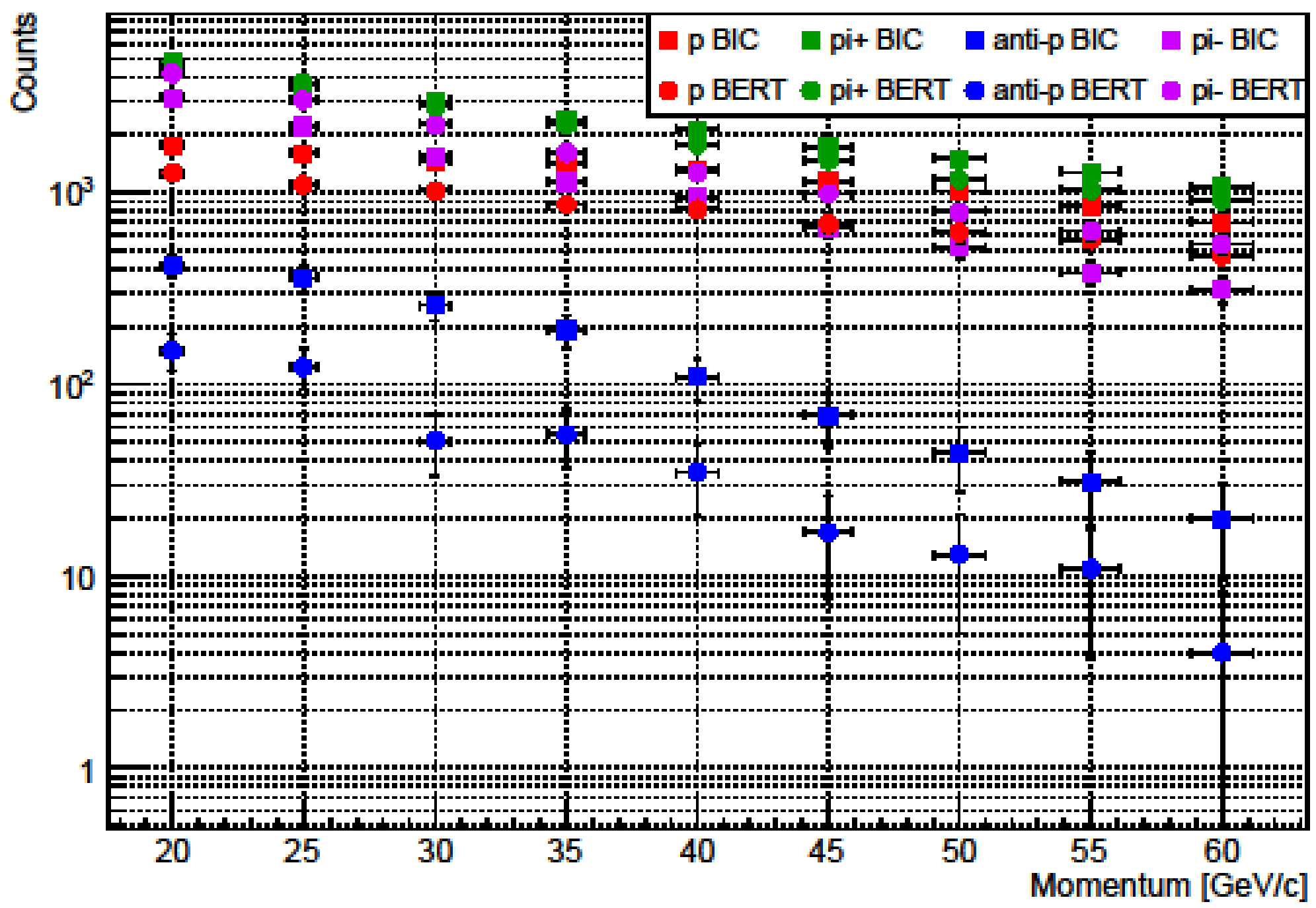

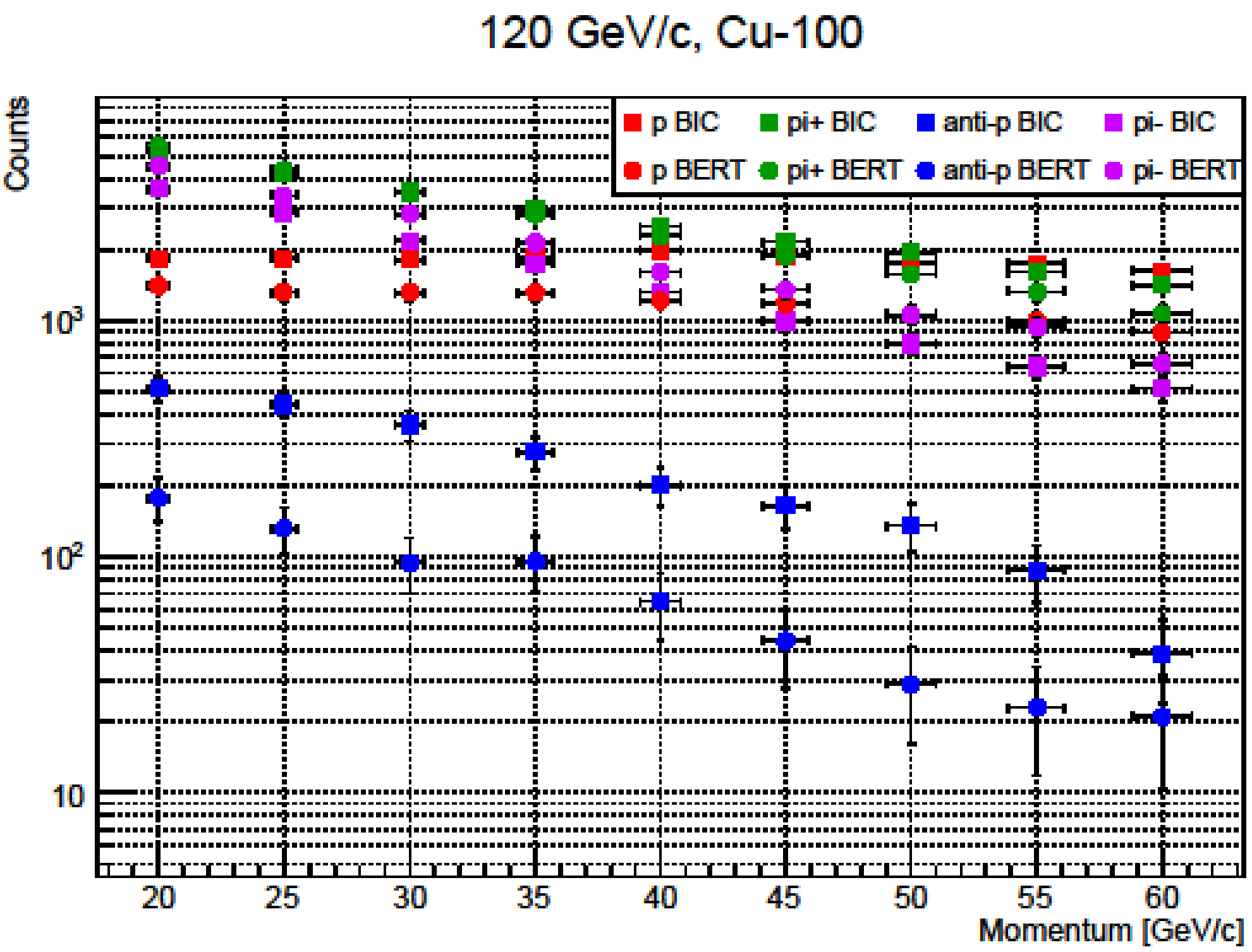
120 GeV/c, Cu-100
Counts
p BIC
pi+ BIC
anti-p BIC
pi- BIC
p BERT
pi+ BERT
anti-p BERT
pi- BERT
$10^3$
$10^2$
10
20
25
30
35
40
45
50
55
60
Momentum [GeV/c]

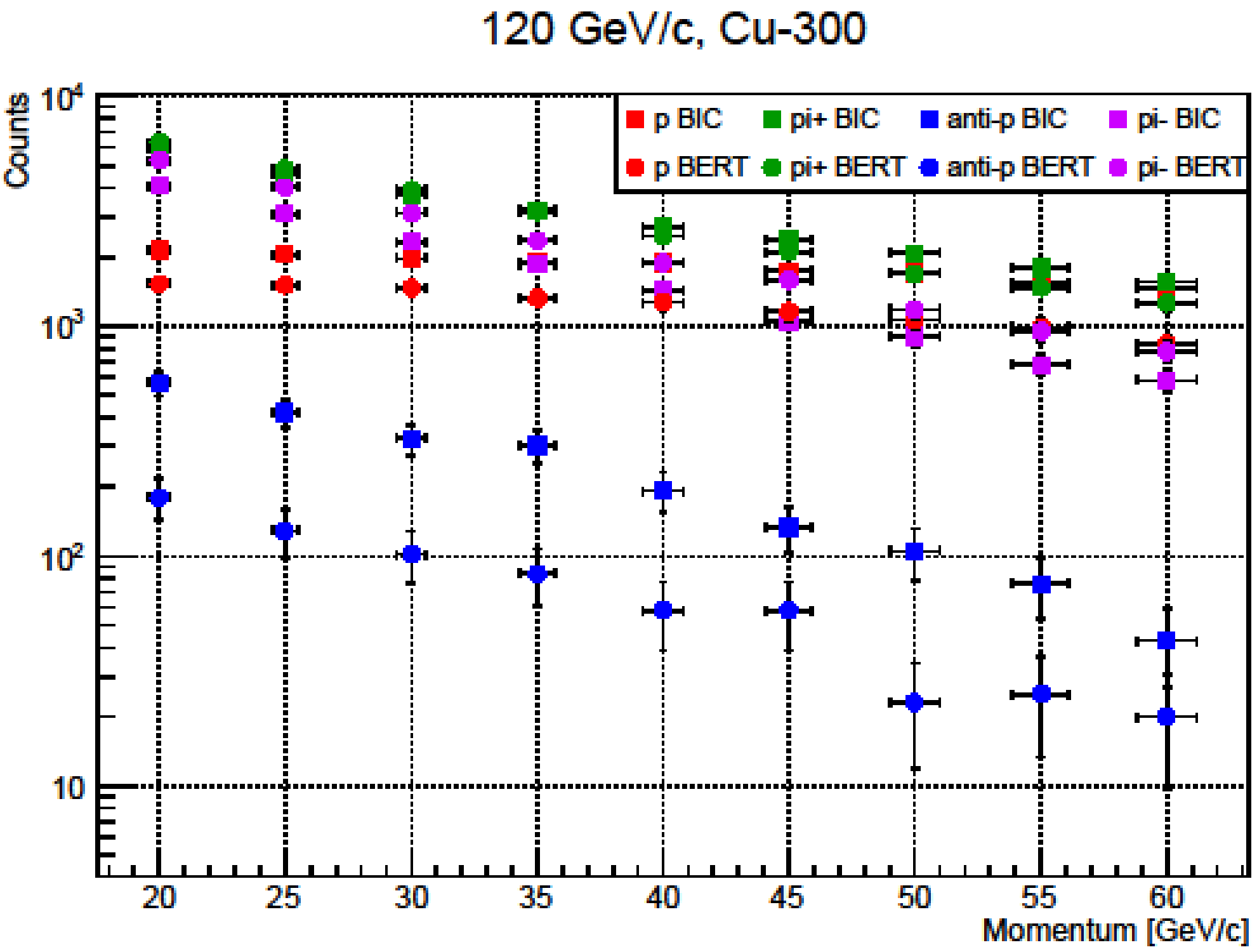
120 GeV/c, Cu-300
Counts
$10^4$
p BIC
pi+ BIC
anti-p BIC
pi- BIC
p BERT
pi+ BERT
anti-p BERT
pi- BERT
$10^3$
$10^2$
10
20
25
30
35
40
45
50
55
60
Momentum [GeV/c]

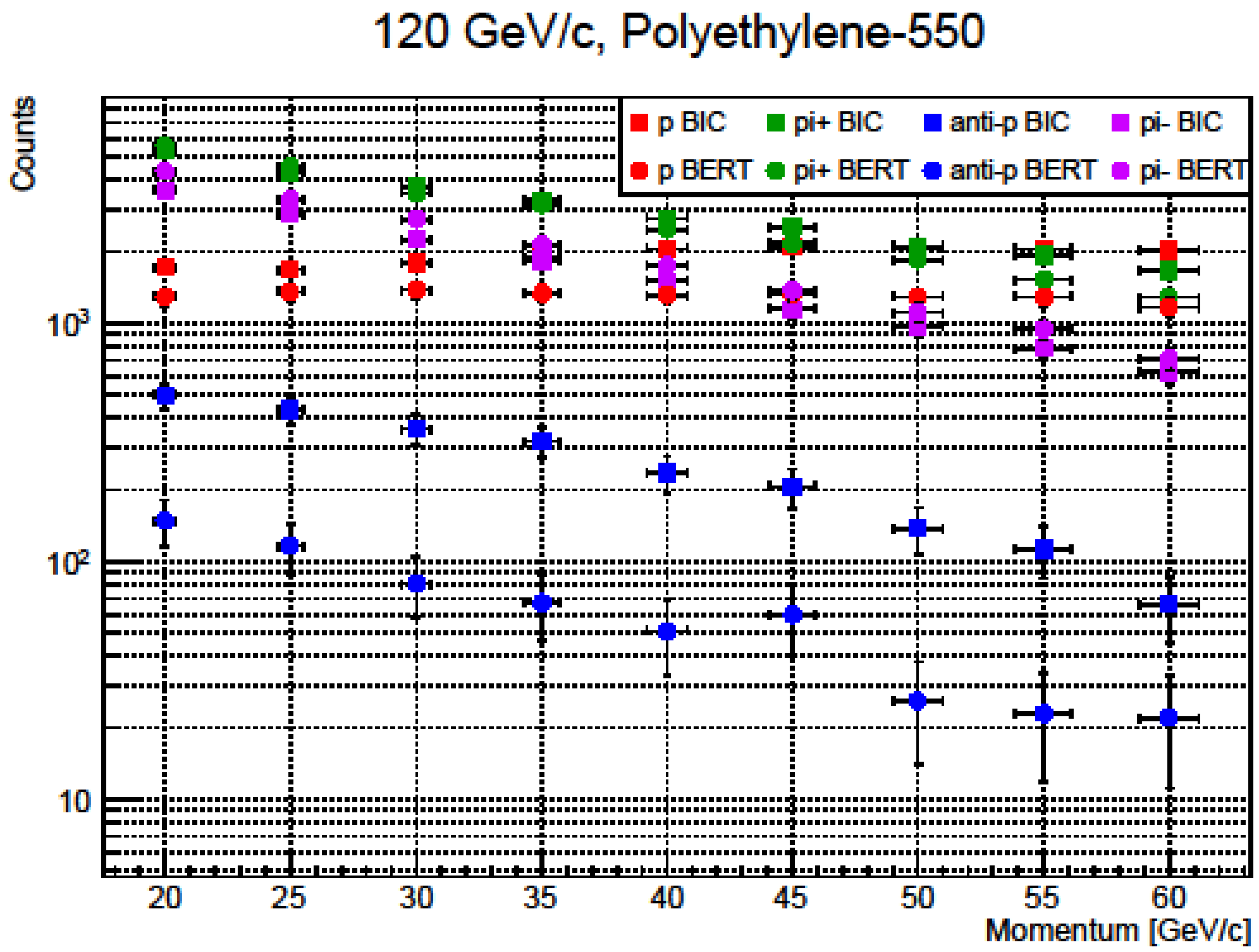
120 GeV/c, Polyethylene-550
Counts
p BIC
pi+ BIC
anti-p BIC
pi- BIC
p BERT
pi+ BERT
anti-p BERT
pi- BERT
$10^3$
$10^2$
10
20
25
30
35
40
45
50
55
60
Momentum [GeV/c]

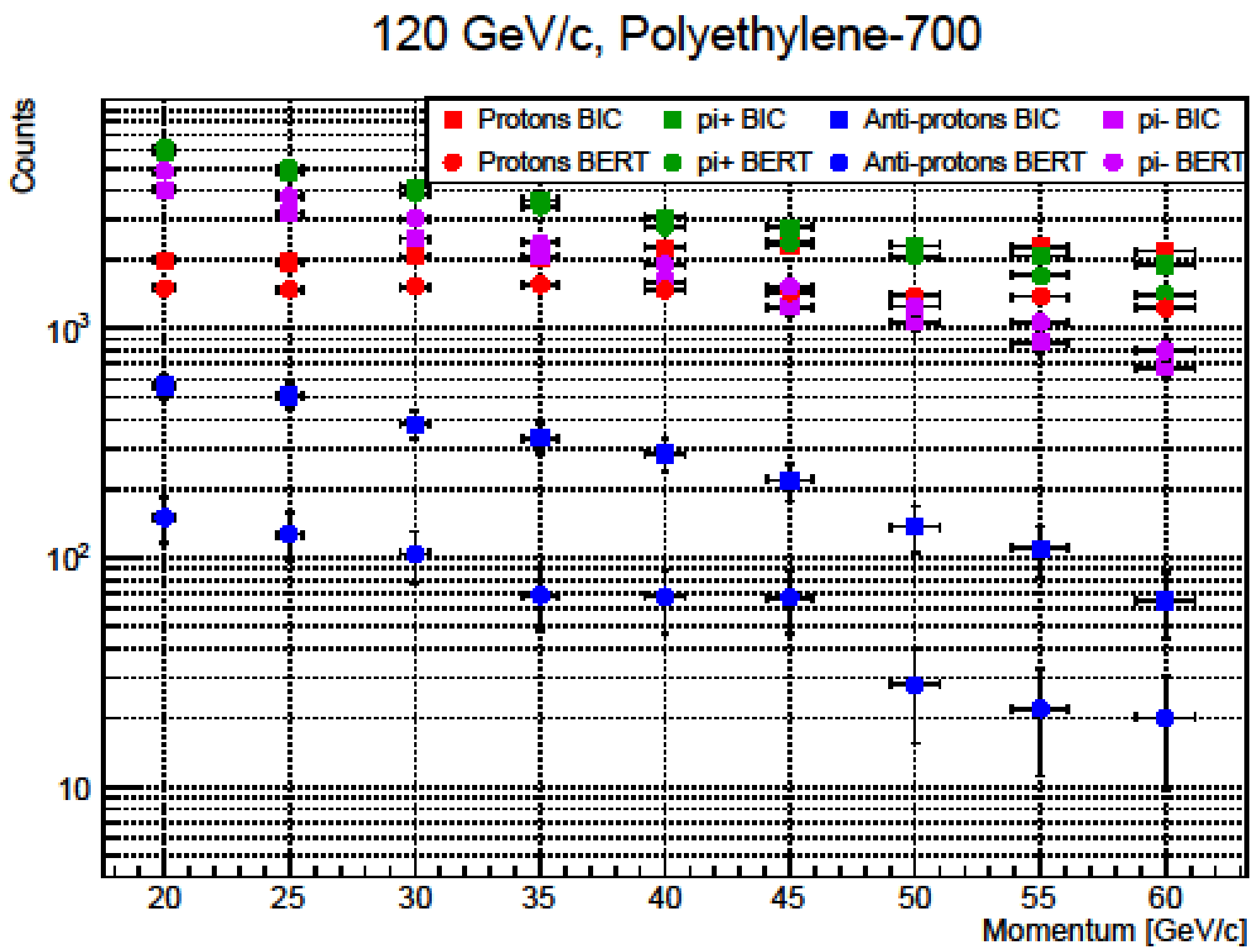
120 GeV/c, Polyethylene-700
Counts
Protons BIC
pi+ BIC
Anti-protons BIC
pi- BIC
Protons BERT
pi+ BERT
Anti-protons BERT
pi- BERT
$10^3$
$10^2$
10
20
25
30
35
40
45
50
55
60
Momentum [GeV/c]

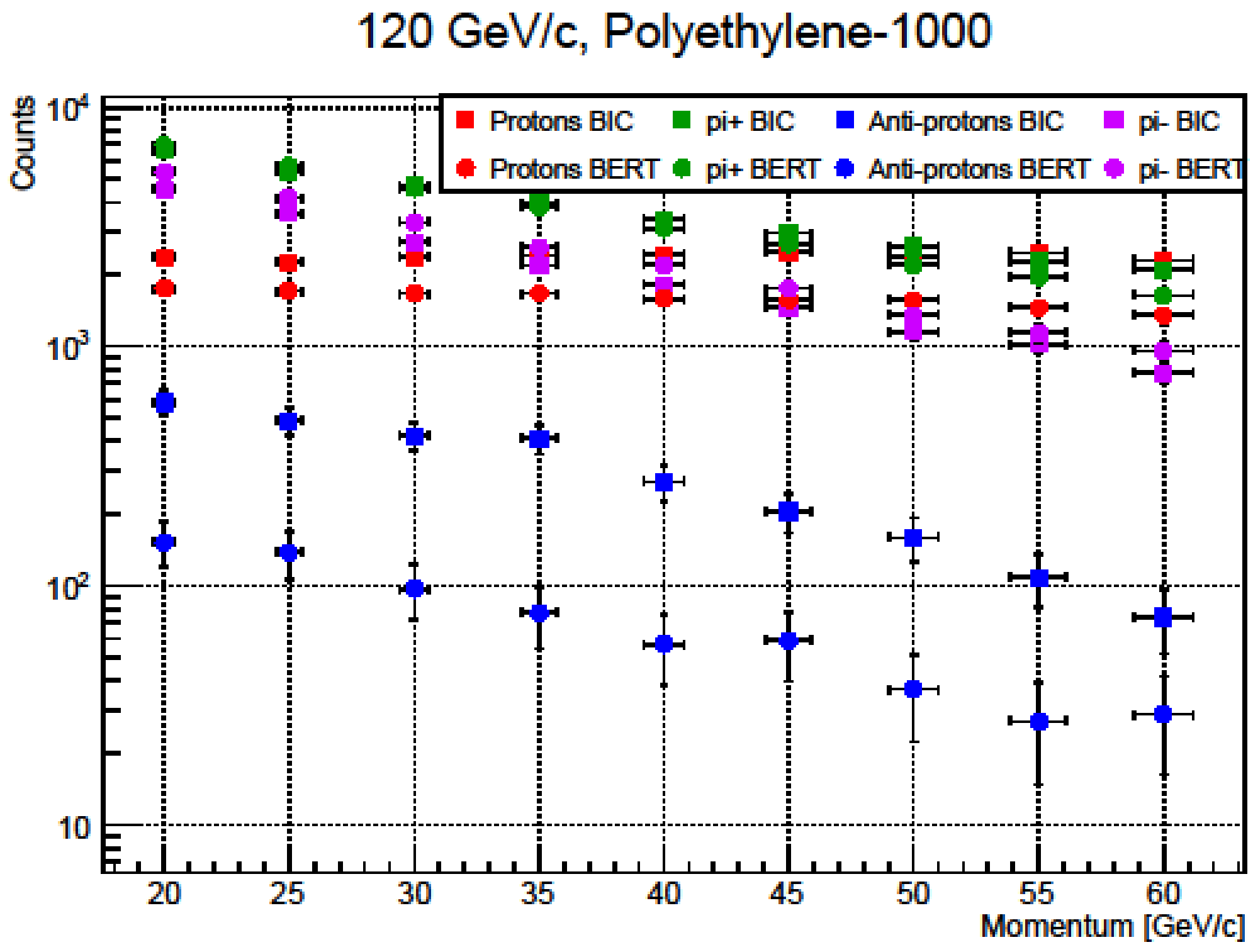
120 GeV/c, Polyethylene-1000
Counts
Protons BIC
pi+ BIC
Anti-protons BIC
pi- BIC
Protons BERT
pi+ BERT
Anti-protons BERT
pi- BERT
Momentum [GeV/c]

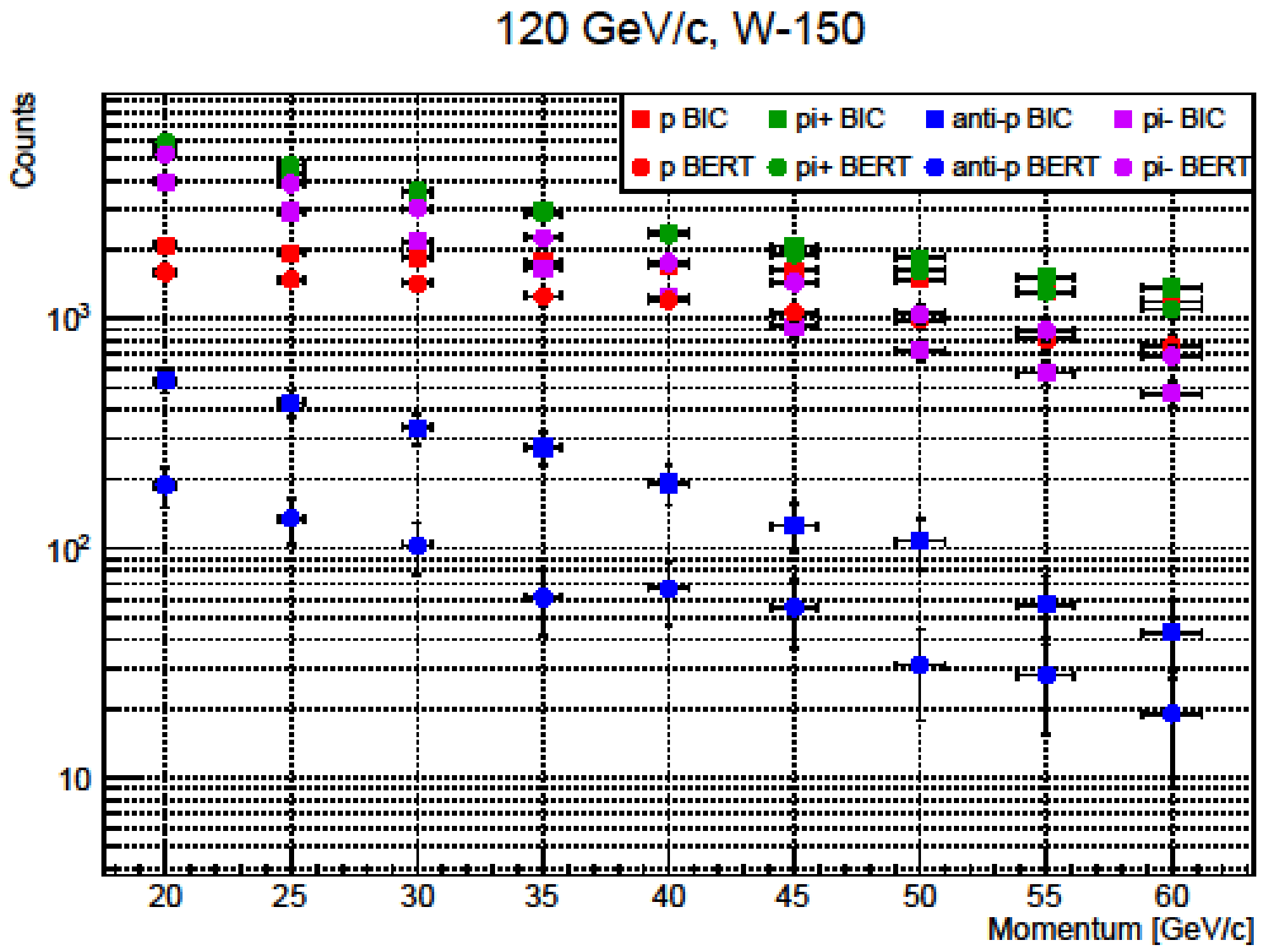
120 GeV/c, W-150
Counts
p BIC
pi+ BIC
anti-p BIC
pi- BIC
p BERT
pi+ BERT
anti-p BERT
pi- BERT
Momentum [GeV/c]

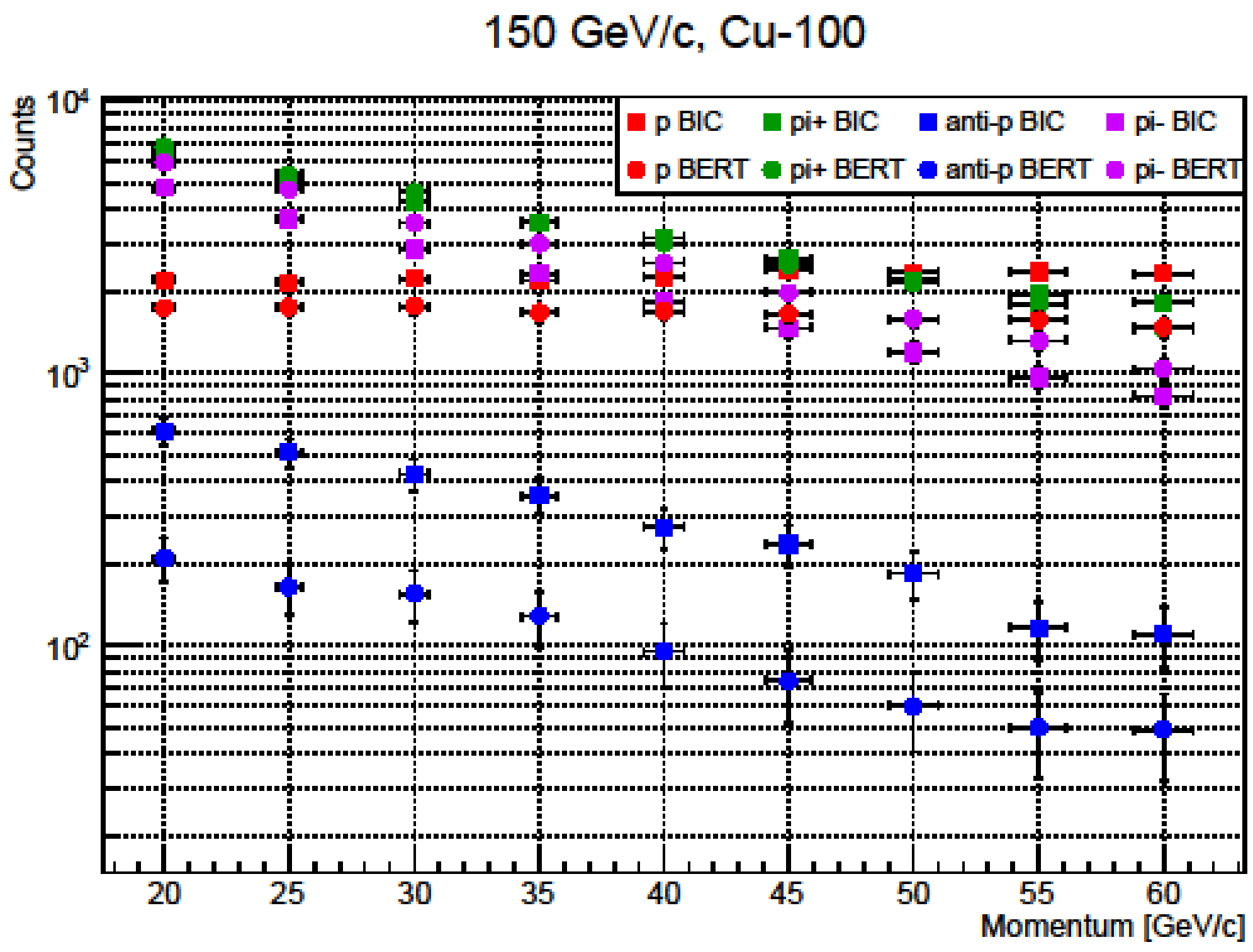
150 GeV/c, Cu-100
Counts
p BIC
pi+ BIC
anti-p BIC
pi- BIC
p BERT
pi+ BERT
anti-p BERT
pi- BERT
Momentum [GeV/c]

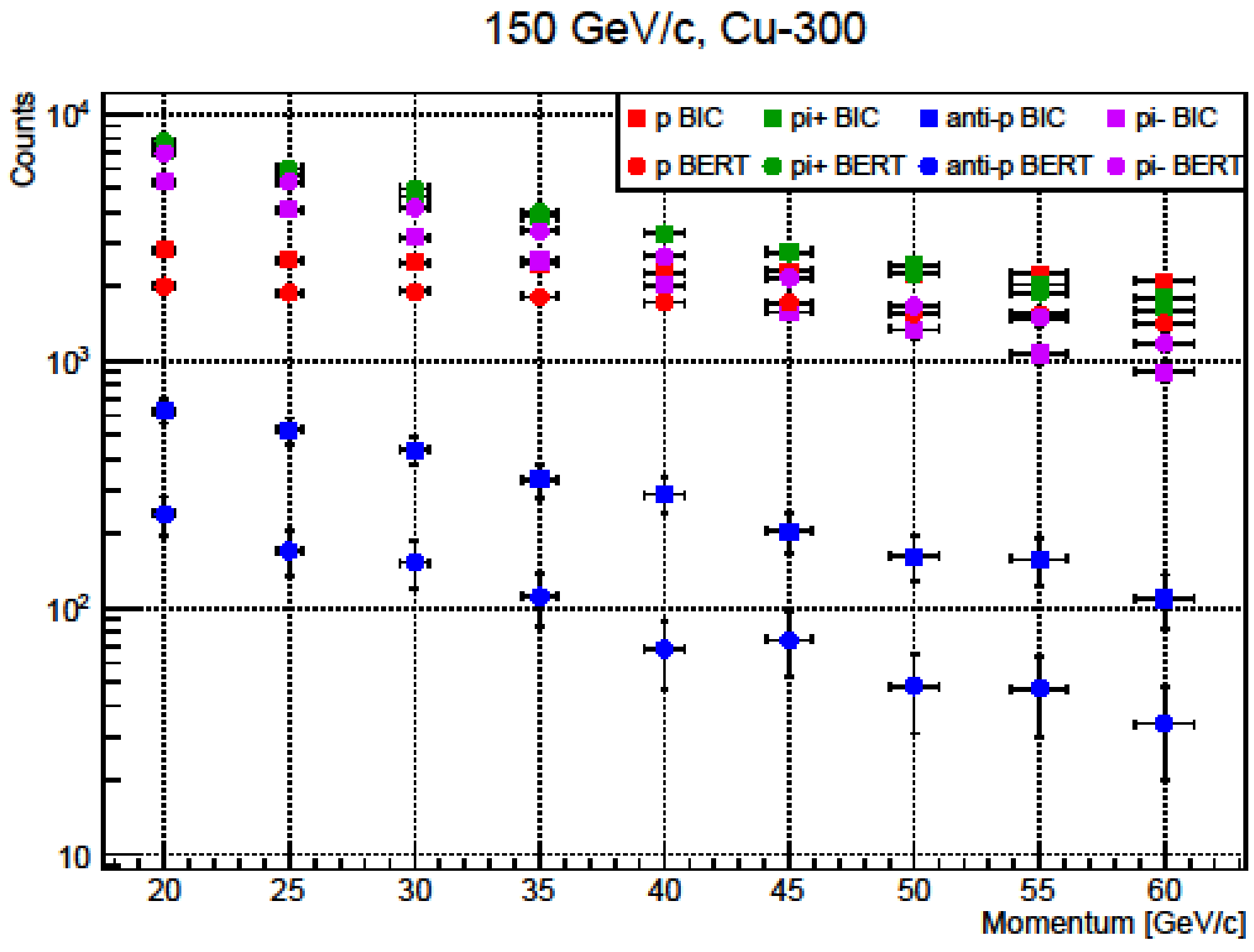
150 GeV/c, Cu-300
Counts
p BIC
pi+ BIC
anti-p BIC
pi- BIC
p BERT
pi+ BERT
anti-p BERT
pi- BERT
Momentum [GeV/c]

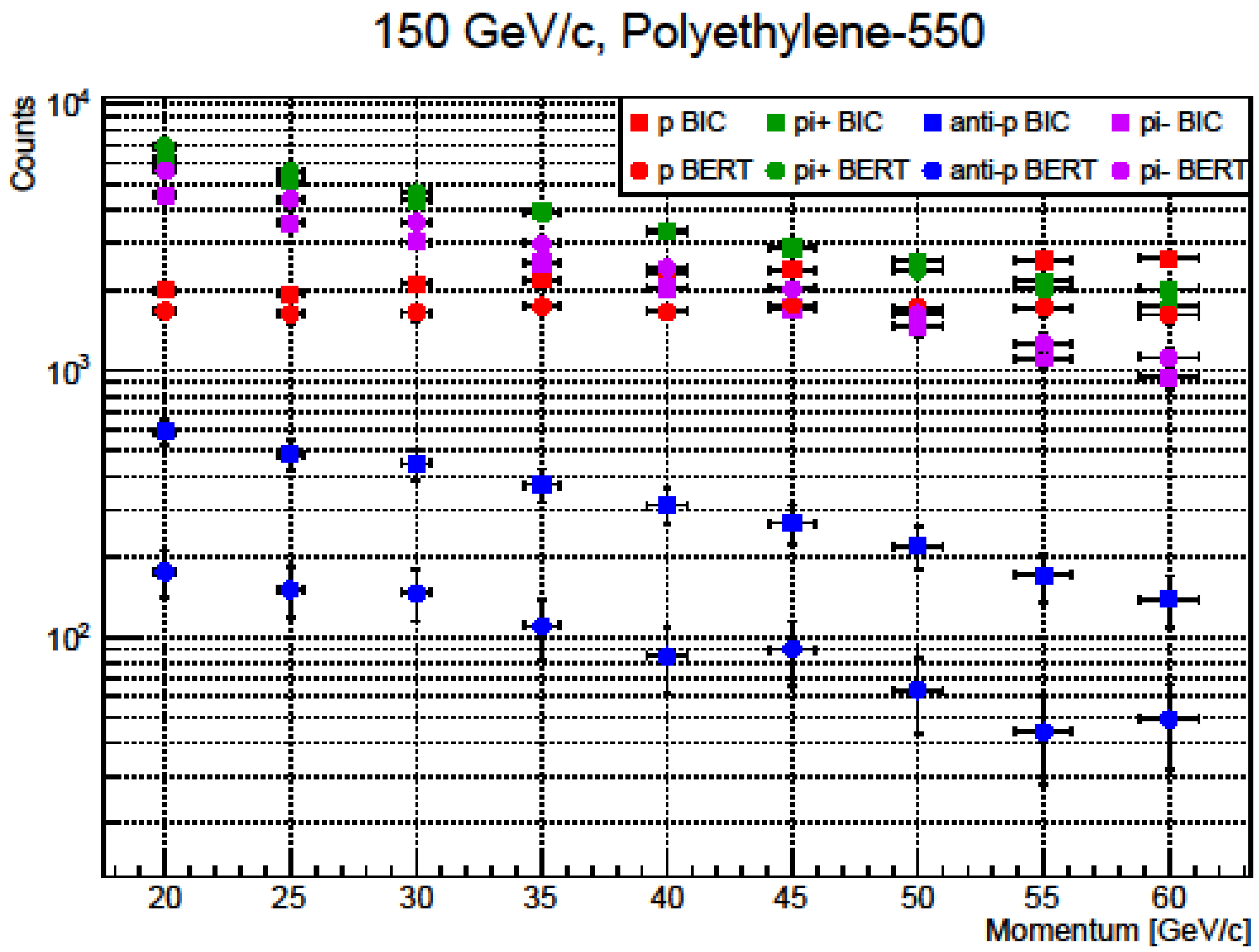
150 GeV/c, Polyethylene-550
Counts
$10^4$
$10^3$
$10^2$
p BIC
pi+ BIC
anti-p BIC
pi- BIC
p BERT
pi+ BERT
anti-p BERT
pi- BERT
20
25
30
35
40
45
50
55
60
Momentum [GeV/c]

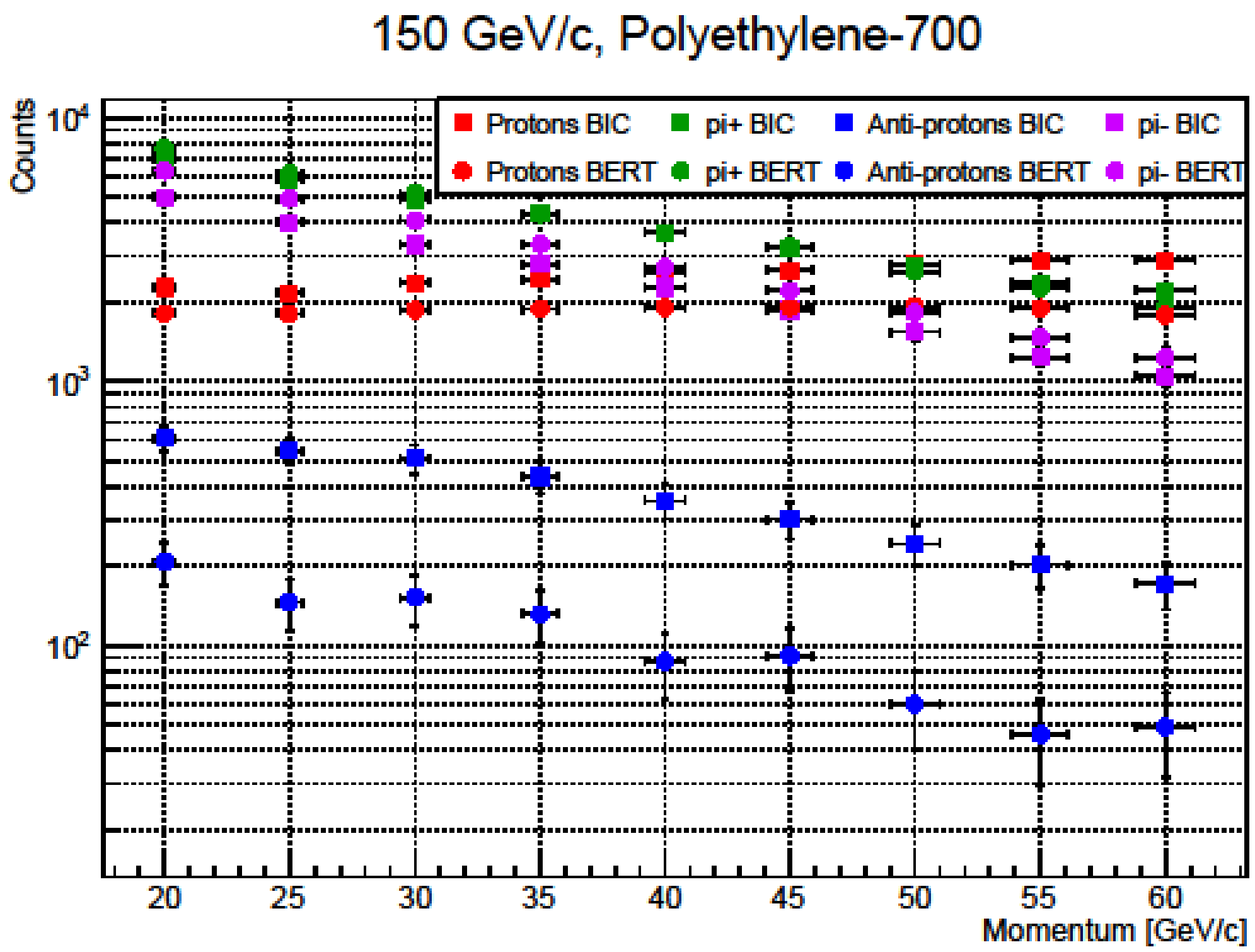
150 GeV/c, Polyethylene-700
Counts
$10^4$
$10^3$
$10^2$
Protons BIC
pi+ BIC
Anti-protons BIC
pi- BIC
Protons BERT
pi+ BERT
Anti-protons BERT
pi- BERT
20
25
30
35
40
45
50
55
60
Momentum [GeV/c]

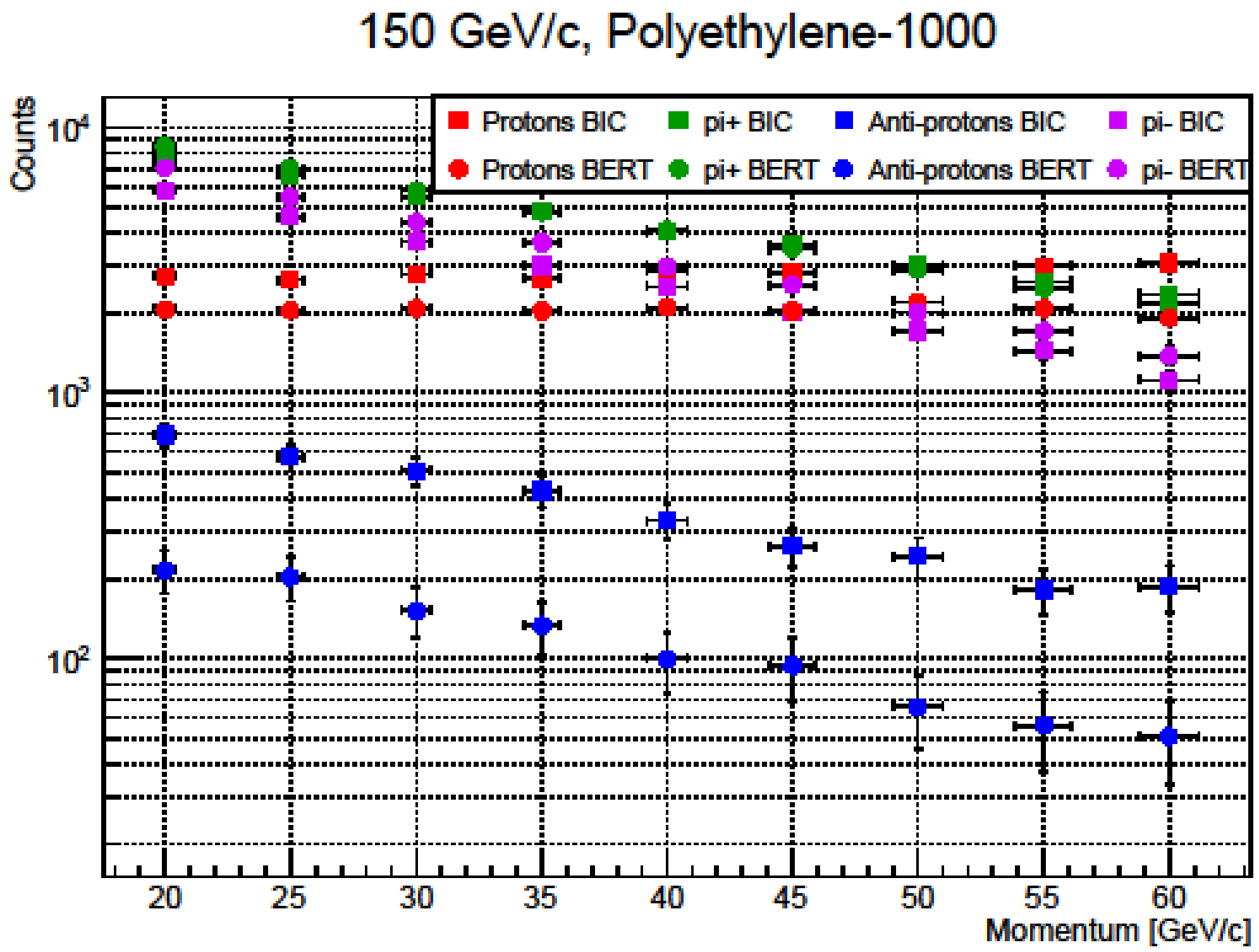
150 GeV/c, Polyethylene-1000
Counts
Protons BIC
pi+ BIC
Anti-protons BIC
pi- BIC
Protons BERT
pi+ BERT
Anti-protons BERT
pi- BERT
Momentum [GeV/c]

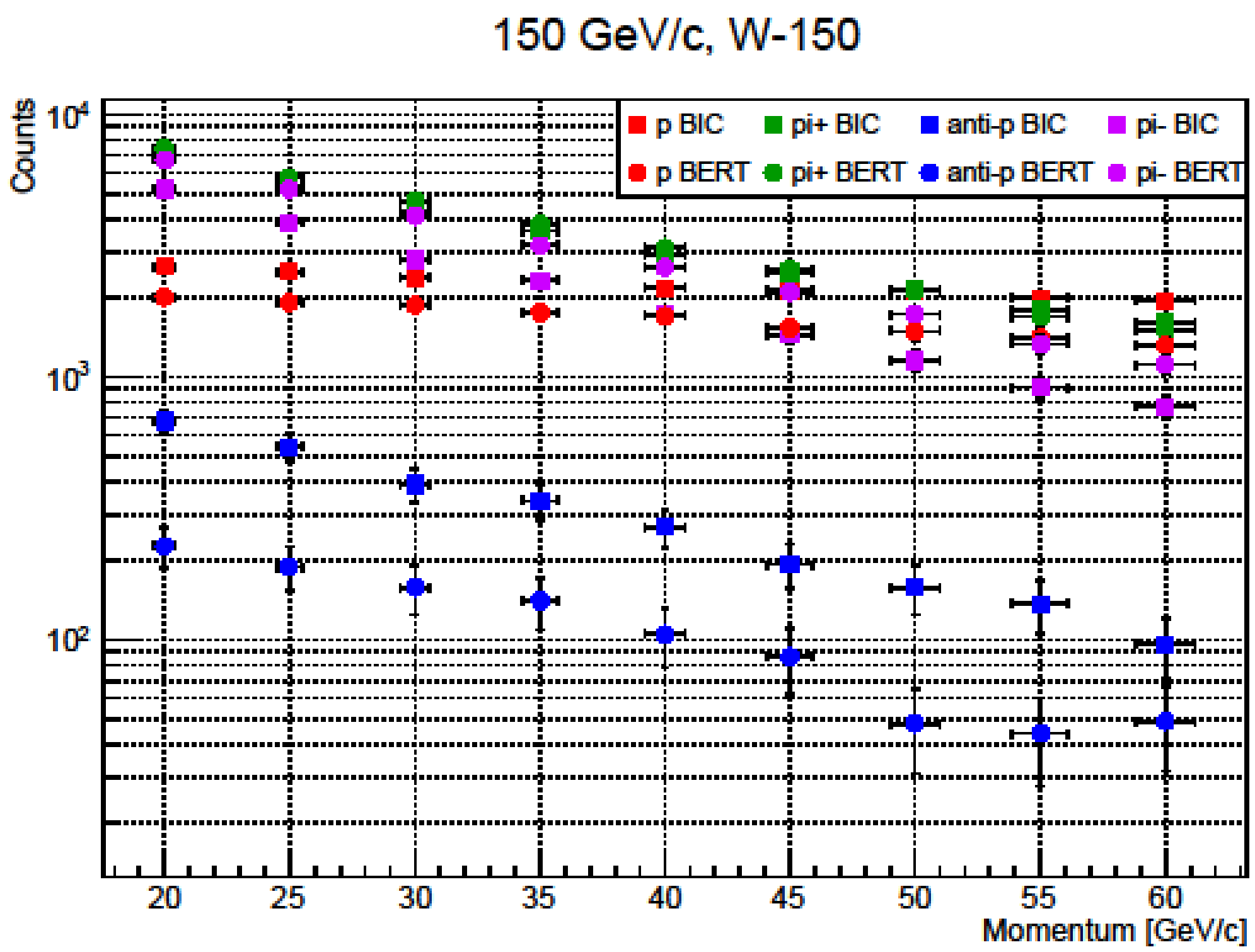
150 GeV/c, W-150
Counts
p BIC
pi+ BIC
anti-p BIC
pi- BIC
p BERT
pi+ BERT
anti-p BERT
pi- BERT
Momentum [GeV/c]

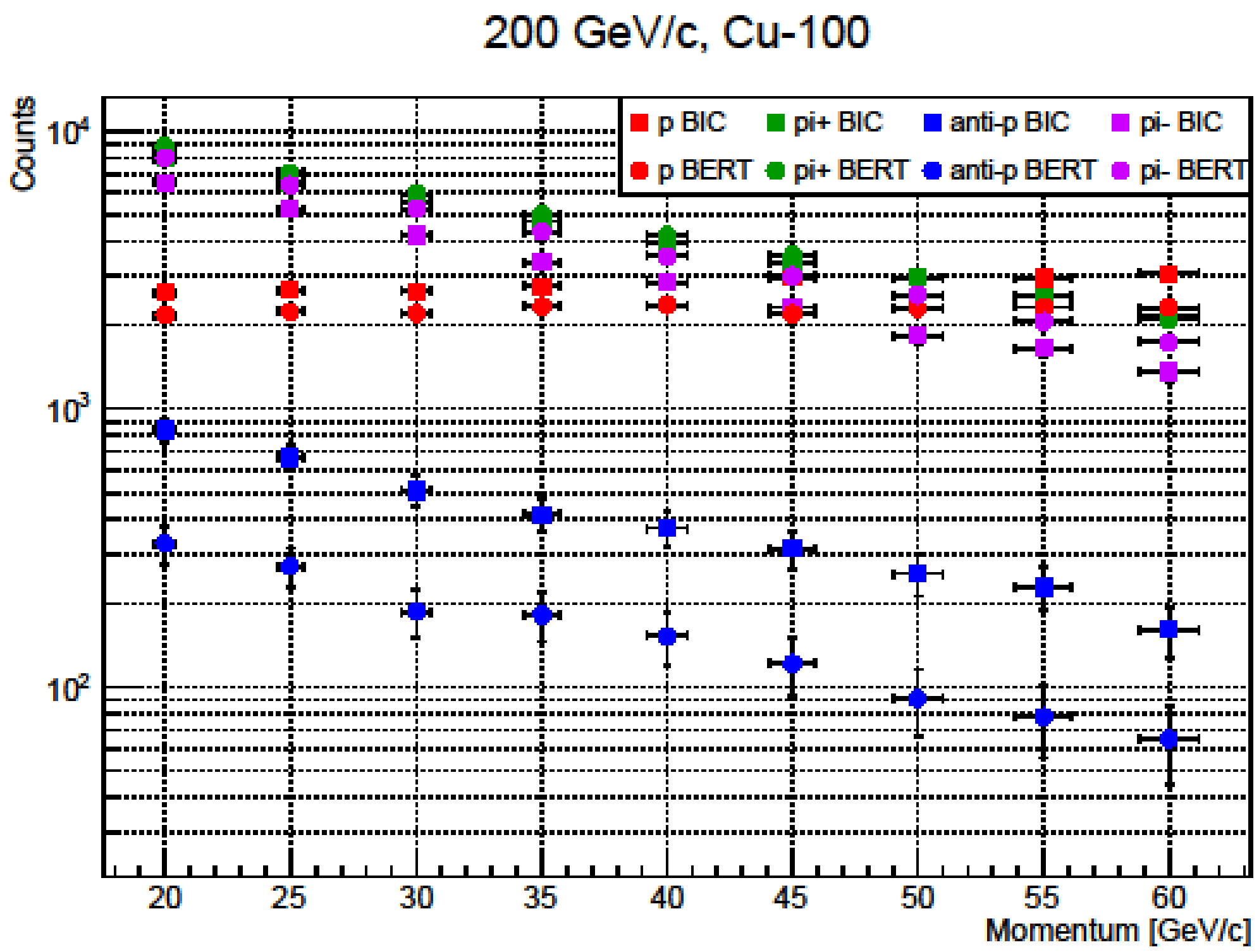
200 GeV/c, Cu-100
Counts
p BIC
pi+ BIC
anti-p BIC
pi- BIC
p BERT
pi+ BERT
anti-p BERT
pi- BERT
Momentum [GeV/c]

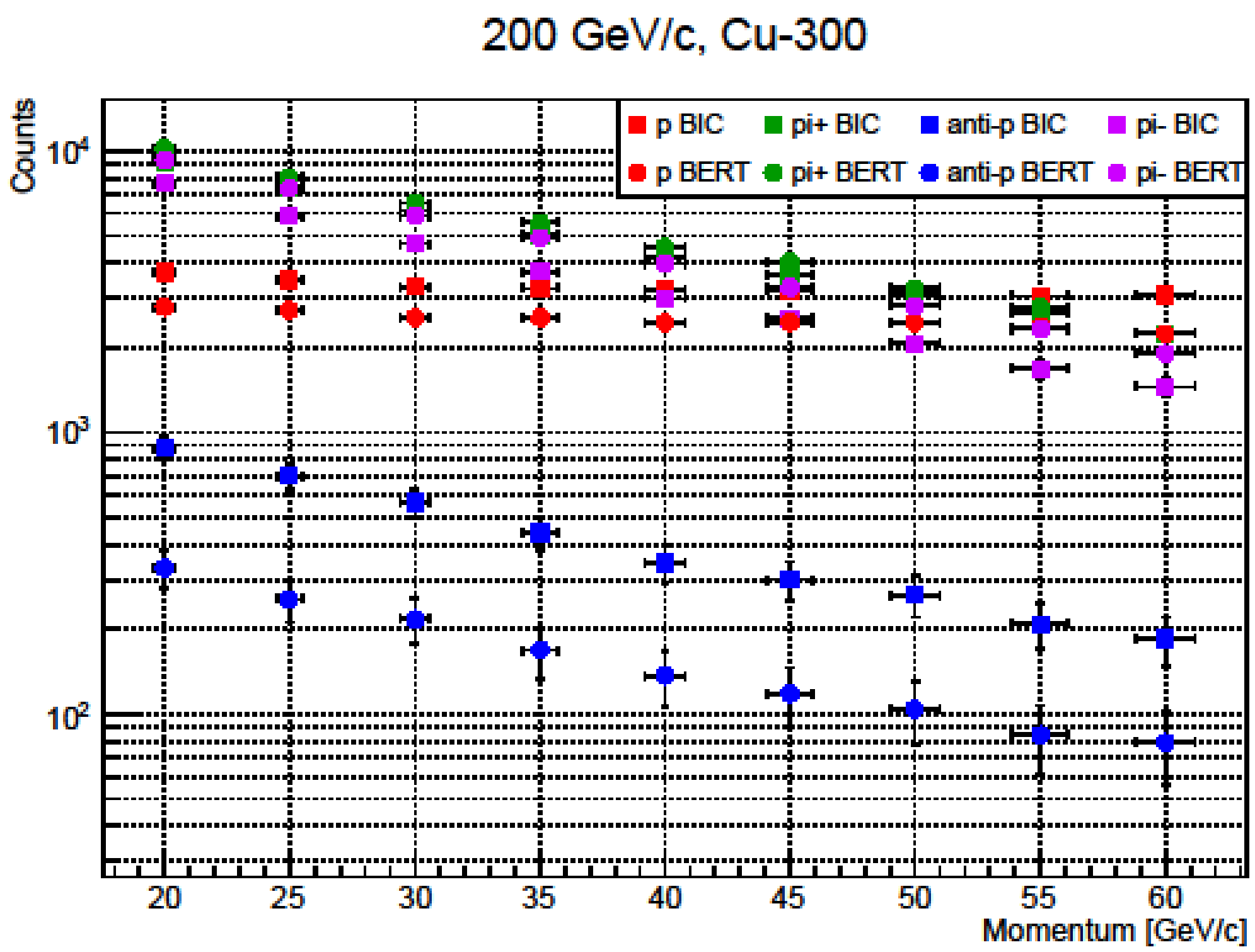
200 GeV/c, Cu-300
Counts
p BIC
pi+ BIC
anti-p BIC
pi- BIC
p BERT
pi+ BERT
anti-p BERT
pi- BERT
Momentum [GeV/c]

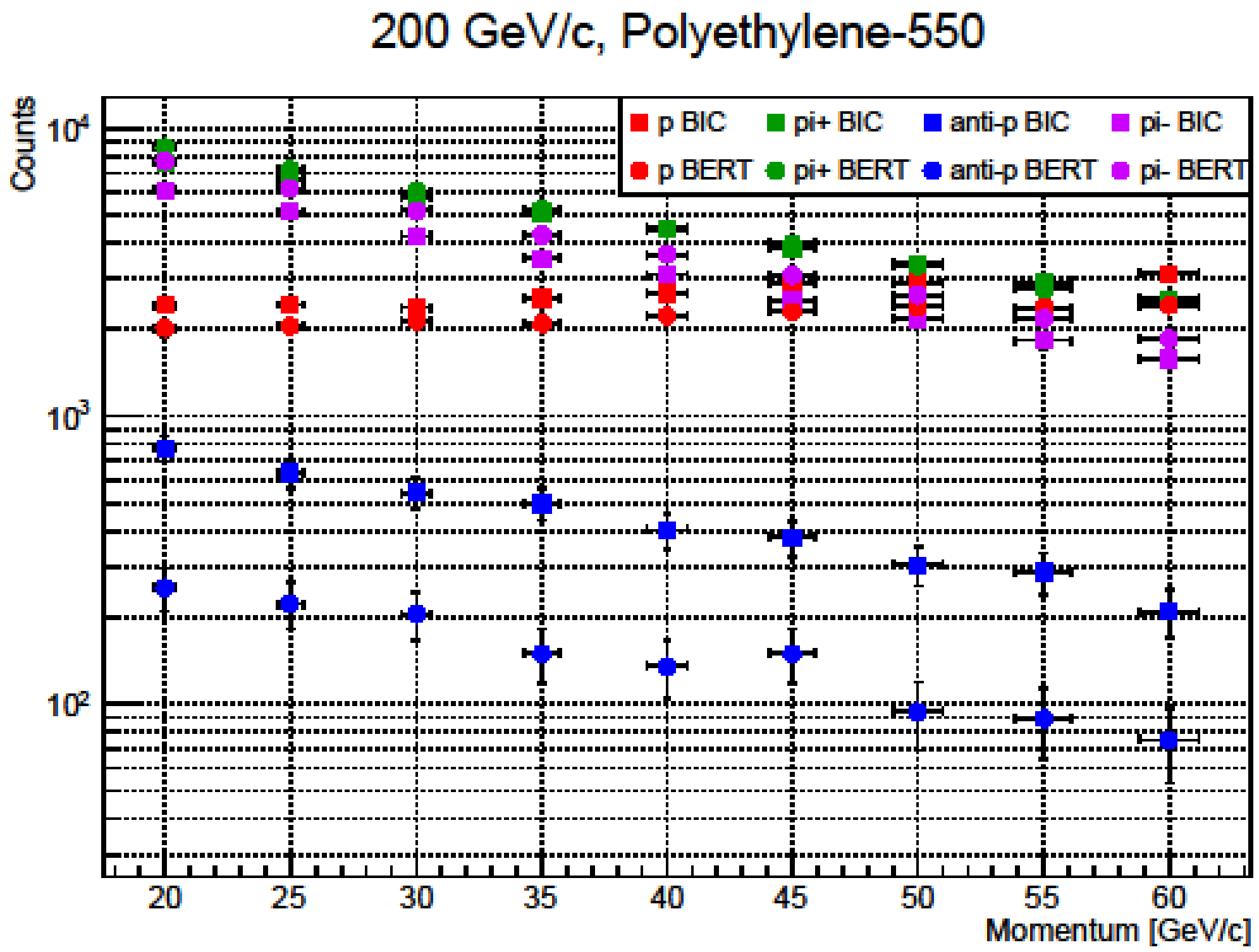
200 GeV/c, Polyethylene-550
Counts
p BIC
pi+ BIC
anti-p BIC
pi- BIC
p BERT
pi+ BERT
anti-p BERT
pi- BERT
$10^4$
$10^3$
$10^2$
20
25
30
35
40
45
50
55
60
Momentum [GeV/c]

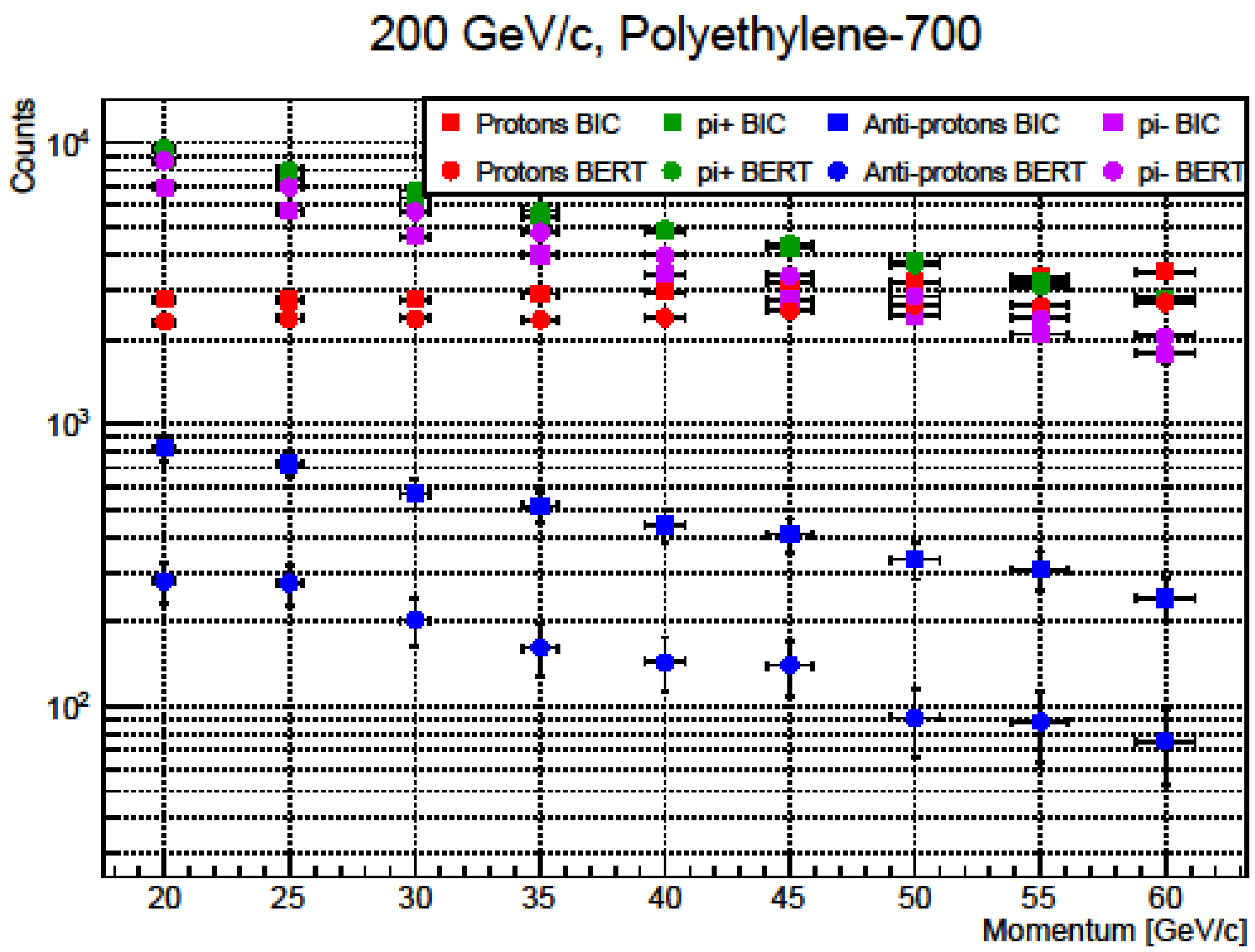
200 GeV/c, Polyethylene-700
Counts
Protons BIC
pi+ BIC
Anti-protons BIC
pi- BIC
Protons BERT
pi+ BERT
Anti-protons BERT
pi- BERT
$10^4$
$10^3$
$10^2$
20
25
30
35
40
45
50
55
60
Momentum [GeV/c]

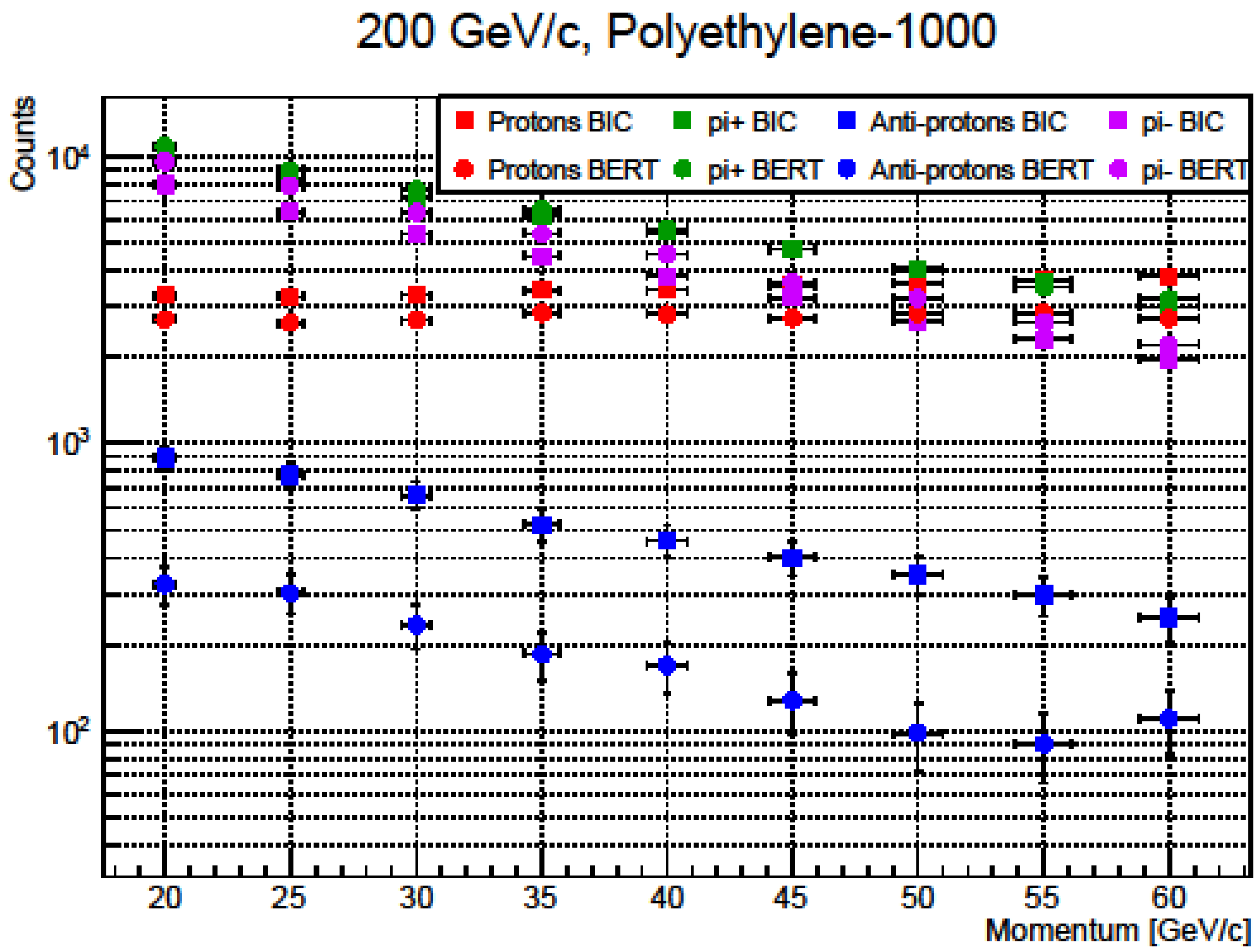
200 GeV/c, Polyethylene-1000
Counts
Protons BIC
pi+ BIC
Anti-protons BIC
pi- BIC
Protons BERT
pi+ BERT
Anti-protons BERT
pi- BERT
$10^4$
$10^3$
$10^2$
20
25
30
35
40
45
50
55
60
Momentum [GeV/c]

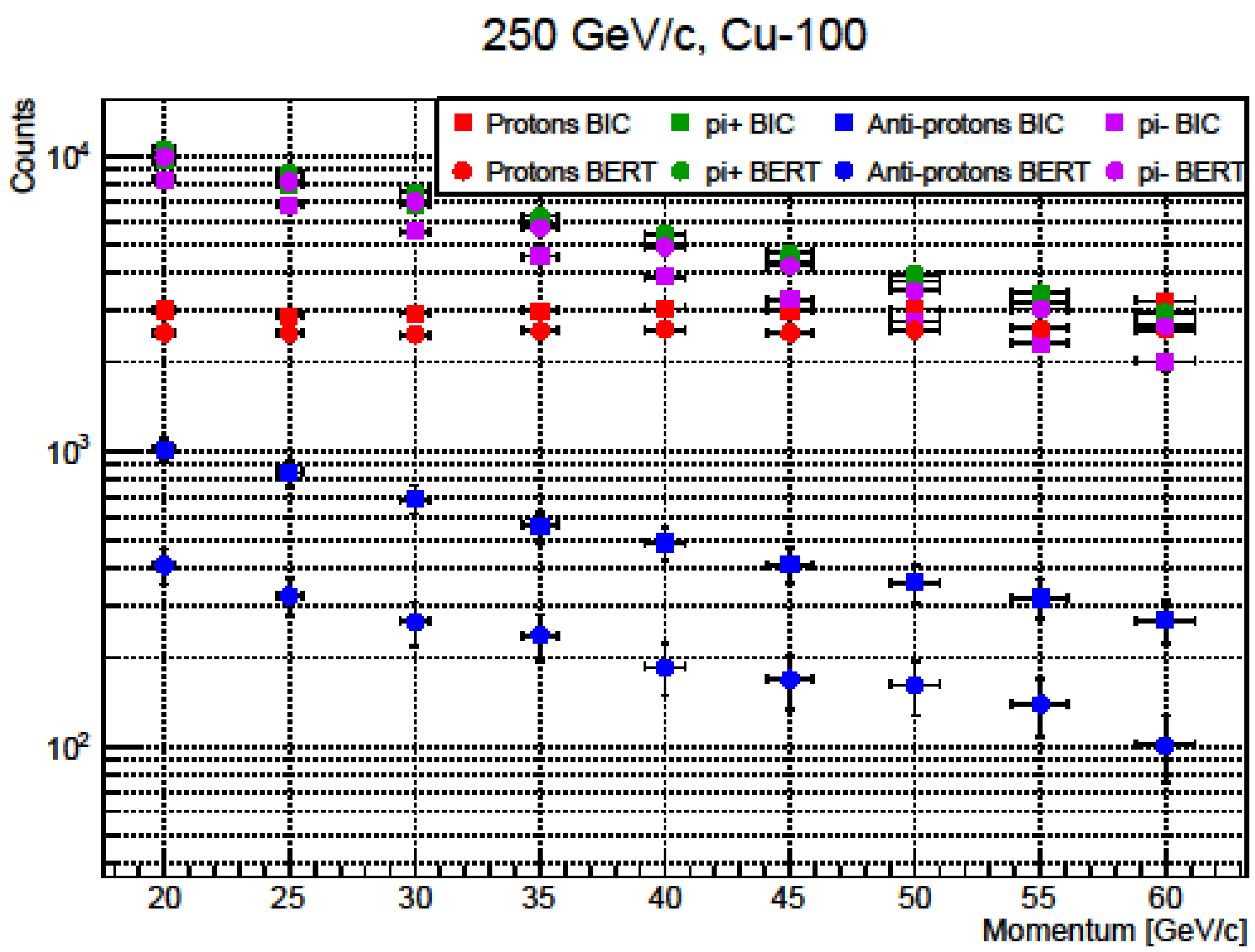
250 GeV/c, Cu-100
Counts
Protons BIC
pi+ BIC
Anti-protons BIC
pi- BIC
Protons BERT
pi+ BERT
Anti-protons BERT
pi- BERT
$10^4$
$10^3$
$10^2$
20
25
30
35
40
45
50
55
60
Momentum [GeV/c]

## 250 GeV/c, Cu-100

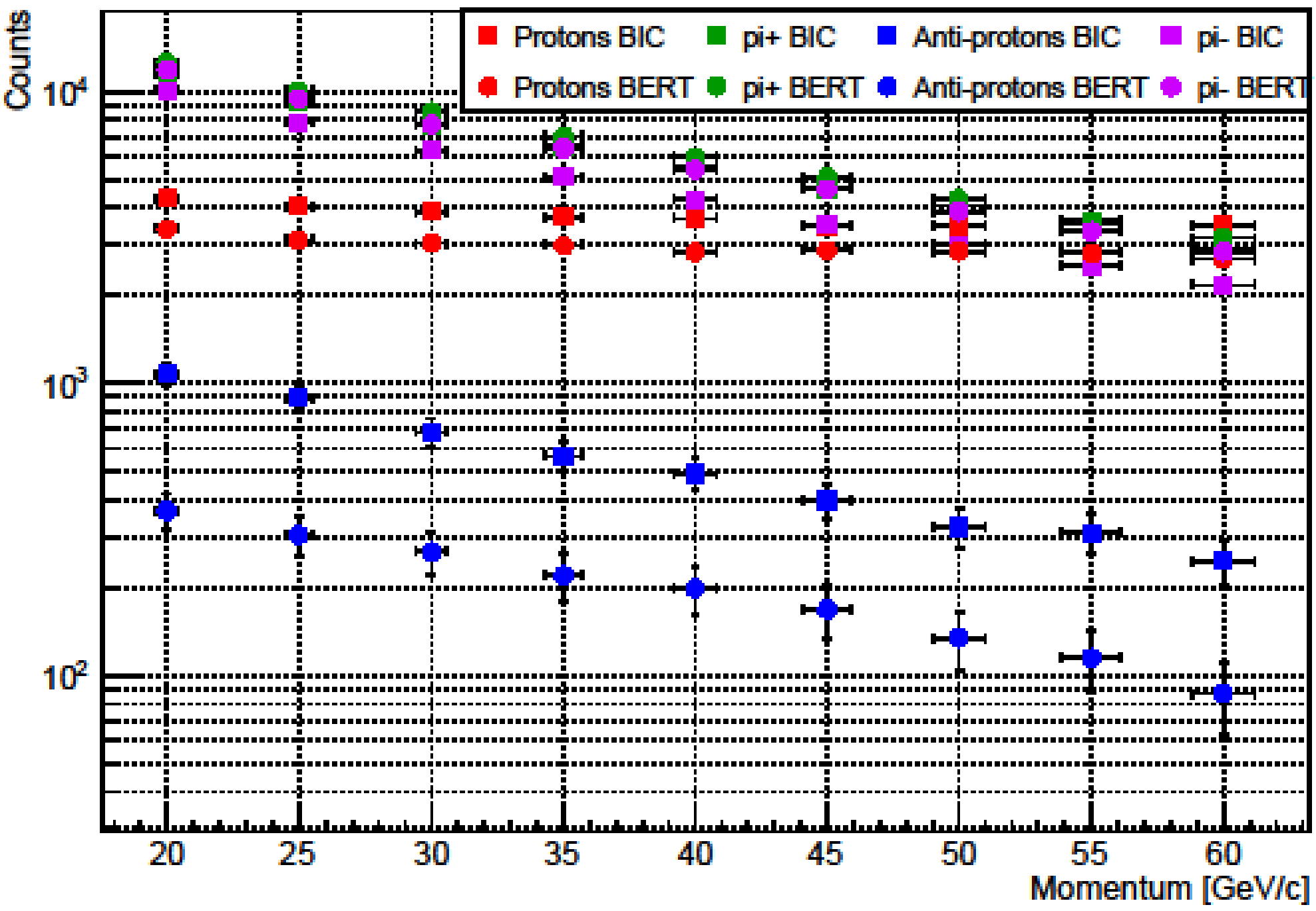

## 250 GeV/c, Polyethylene-550

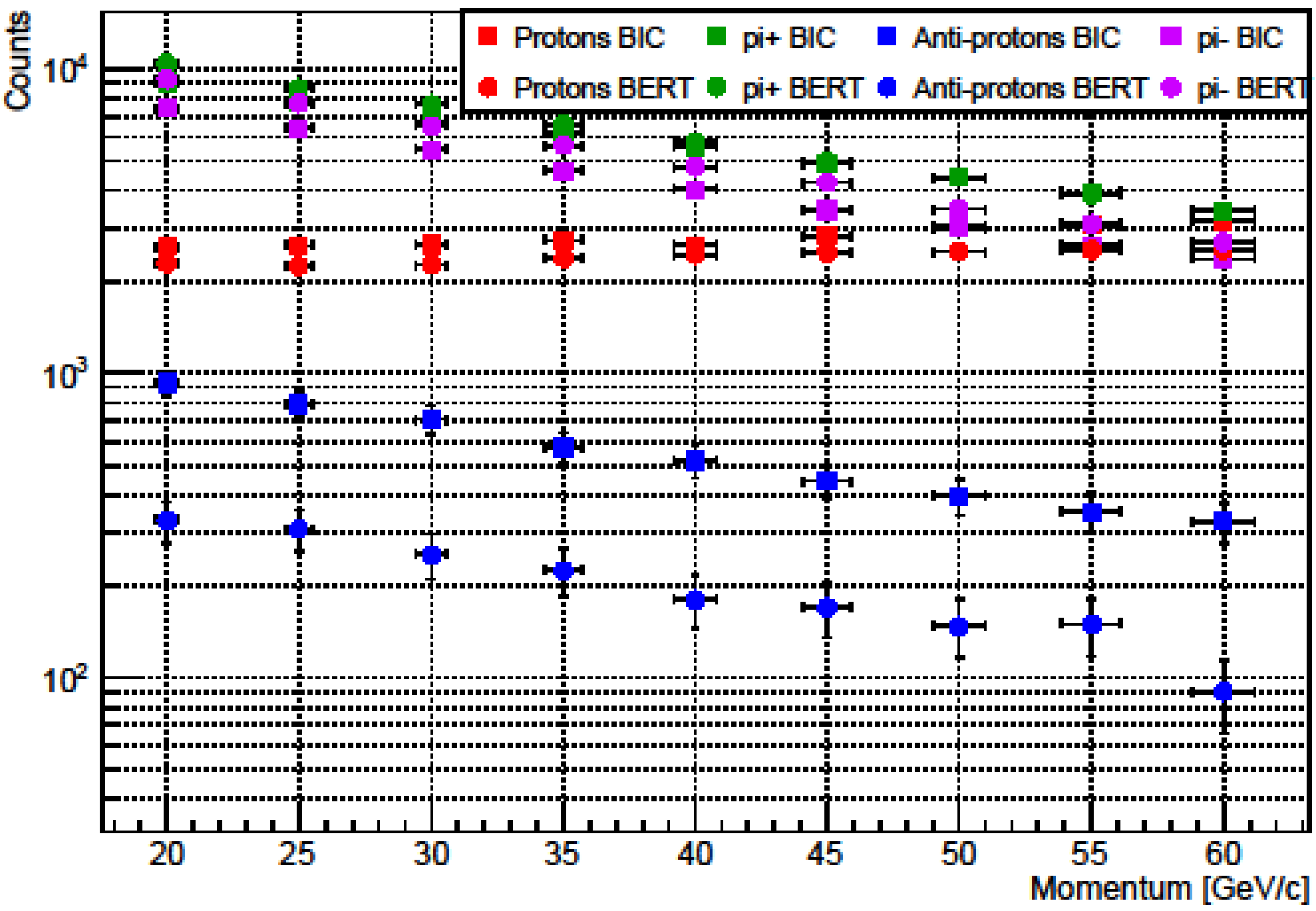

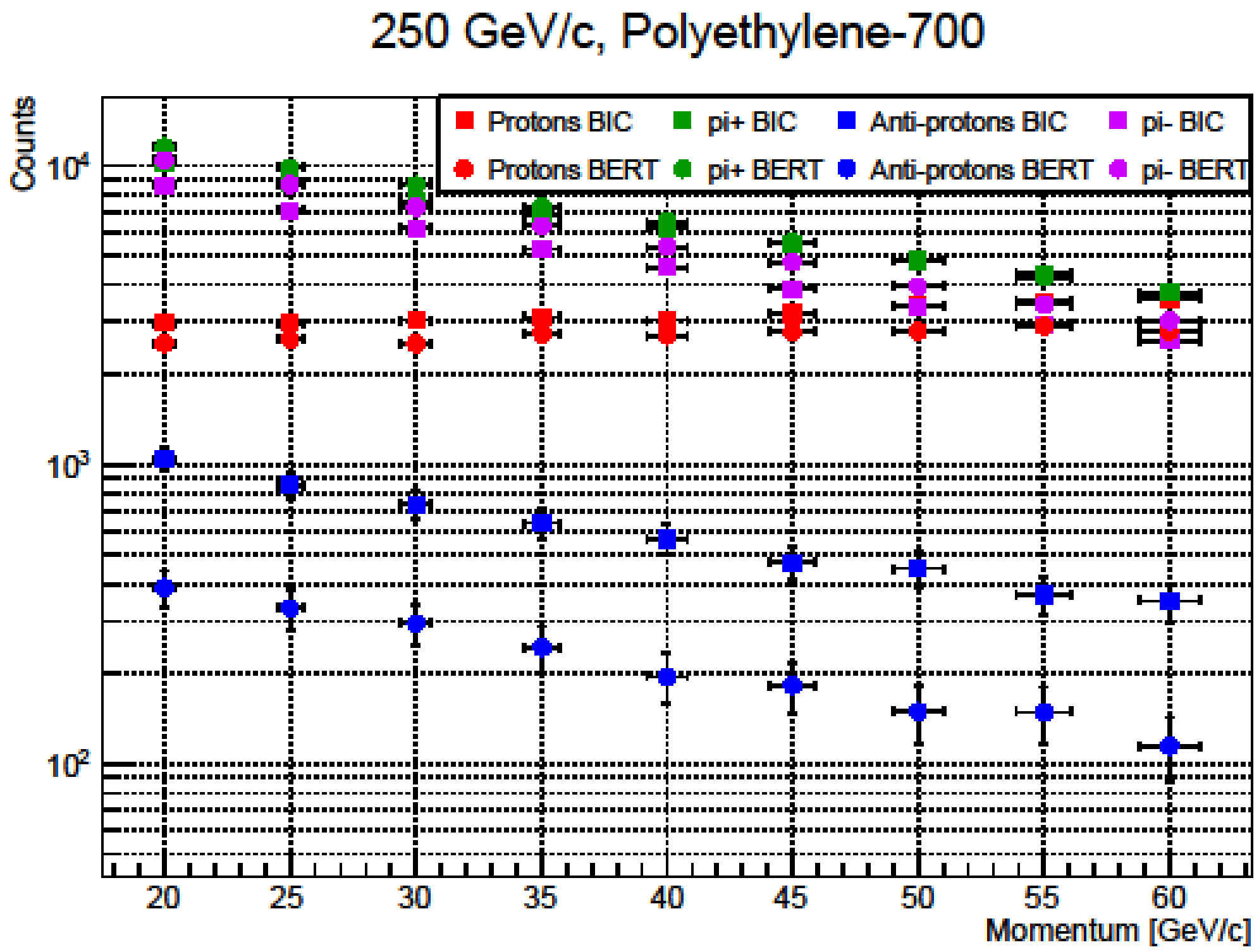
250 GeV/c, Polyethylene-700
Protons BIC
pi+ BIC
Anti-protons BIC
pi- BIC
Protons BERT
pi+ BERT
Anti-protons BERT
pi- BERT
Counts
Momentum [GeV/c]

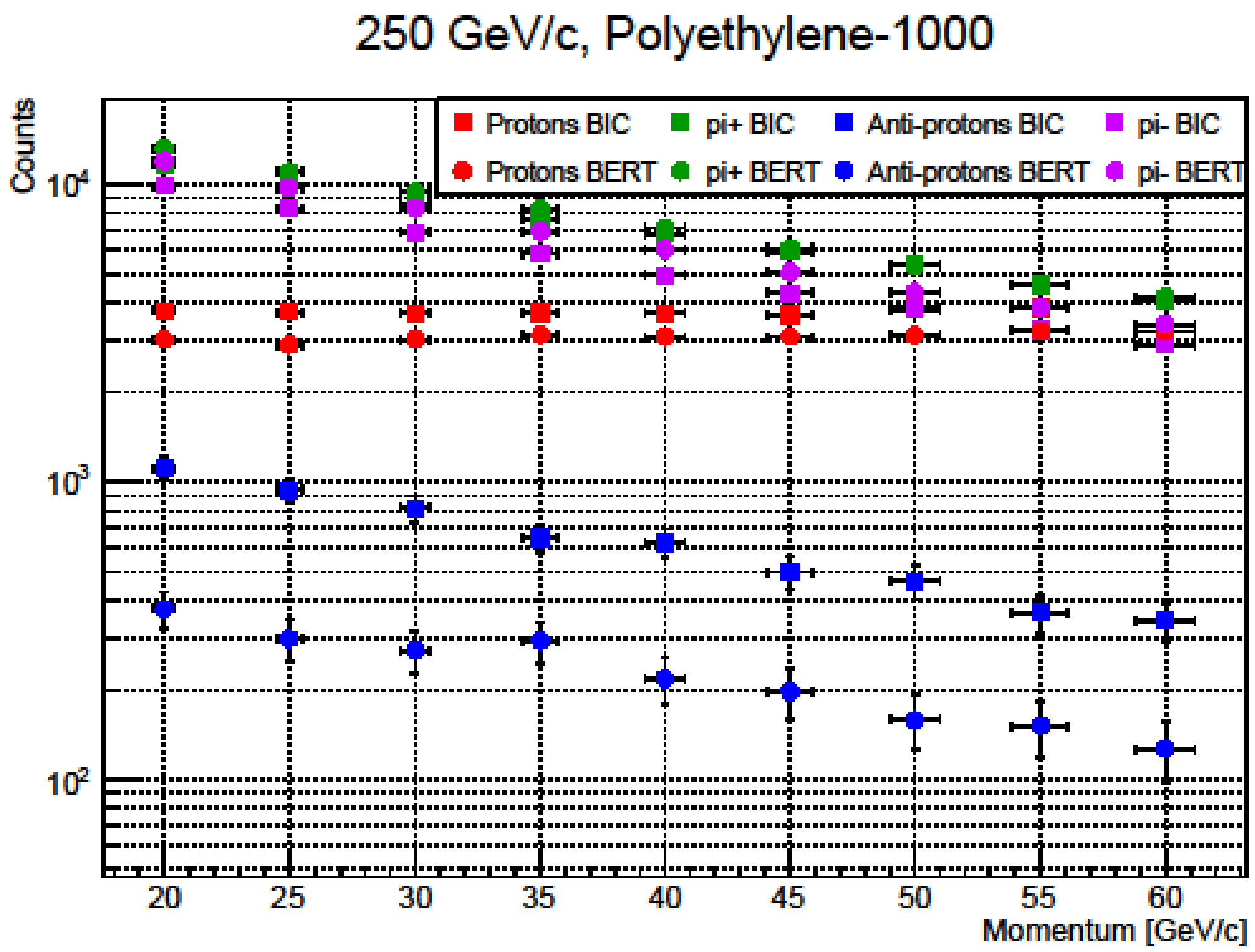
250 GeV/c, Polyethylene-1000
Protons BIC
pi+ BIC
Anti-protons BIC
pi- BIC
Protons BERT
pi+ BERT
Anti-protons BERT
pi- BERT
Counts
Momentum [GeV/c]

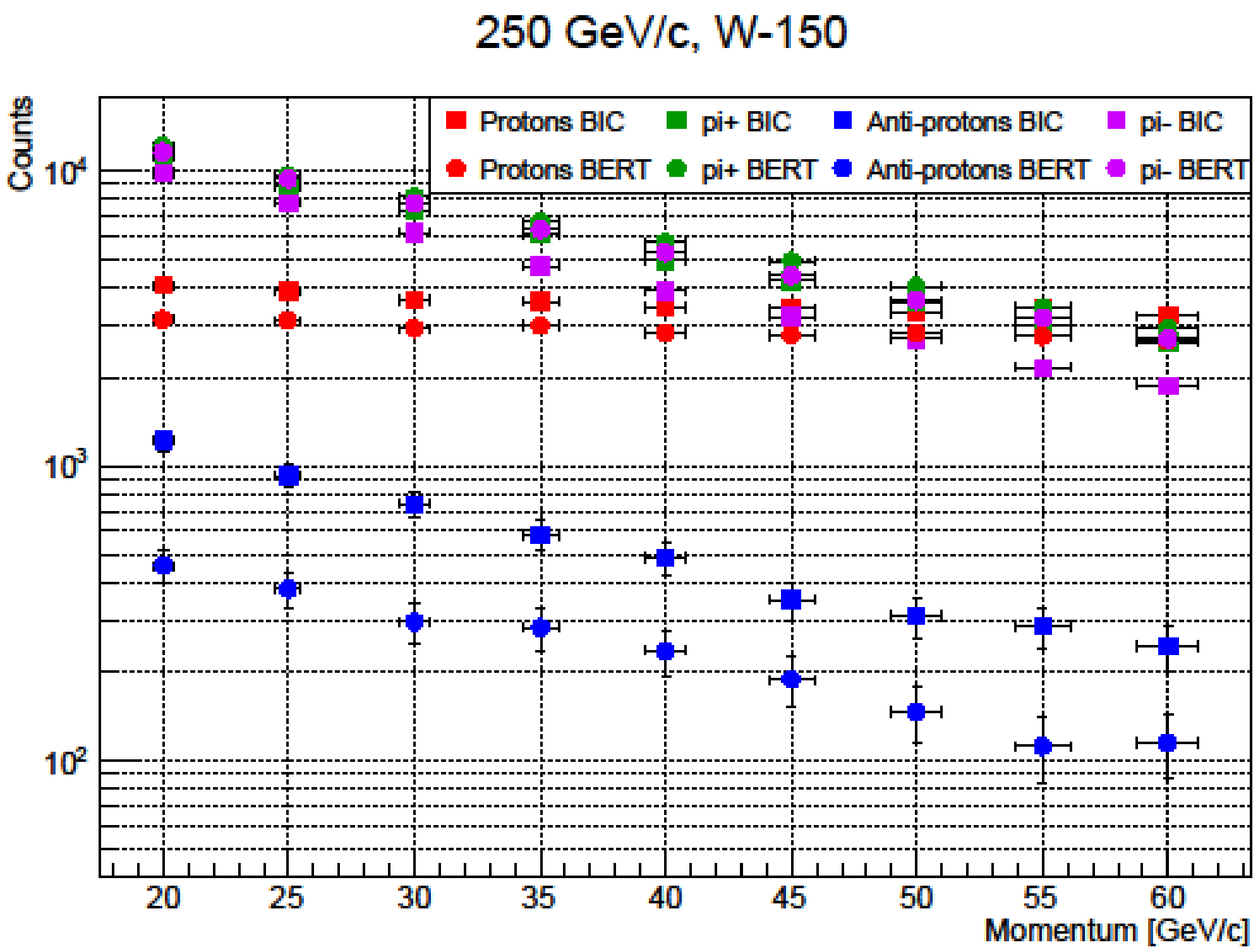

## 2.3 Electron contamination

As discussed in the introduction, the secondary beam is not electron free. In the framework of this study, the electron content has been ignored, with the purpose of investigating the direct electron production on the targets. From the figures below, it can be seen that for lower secondary momenta the electron production lies between two and three orders of magnitudes less than the proton production. As the secondary momentum increases, the difference between proton and electron production decreases. Due to this fact, there are cases where the electron contamination becomes significant for the lower momentum bins, e.g. 200 GeV/c, Polyethylene-1000. In this case the number of protons and electrons are of the same order of magnitude for a tertiary momentum of 20 GeV/c. However, at 60 GeV/c the number of protons are more than one order of magnitude greater.

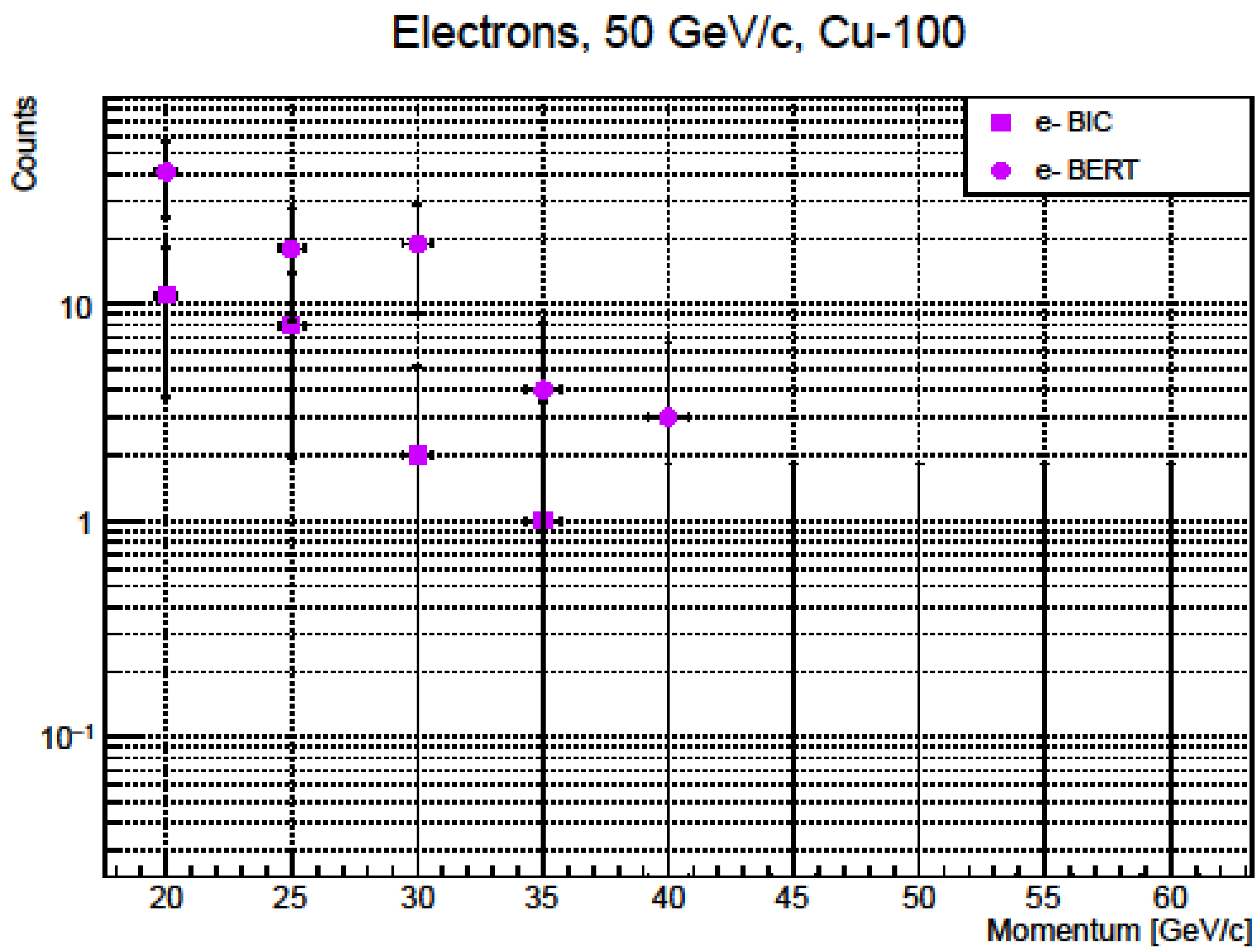
Electrons, 50 GeV/c, Cu-100
Counts
10
1
10⁻¹
20
25
30
35
40
45
50
55
60
Momentum [GeV/c]
e- BIC
e- BERT

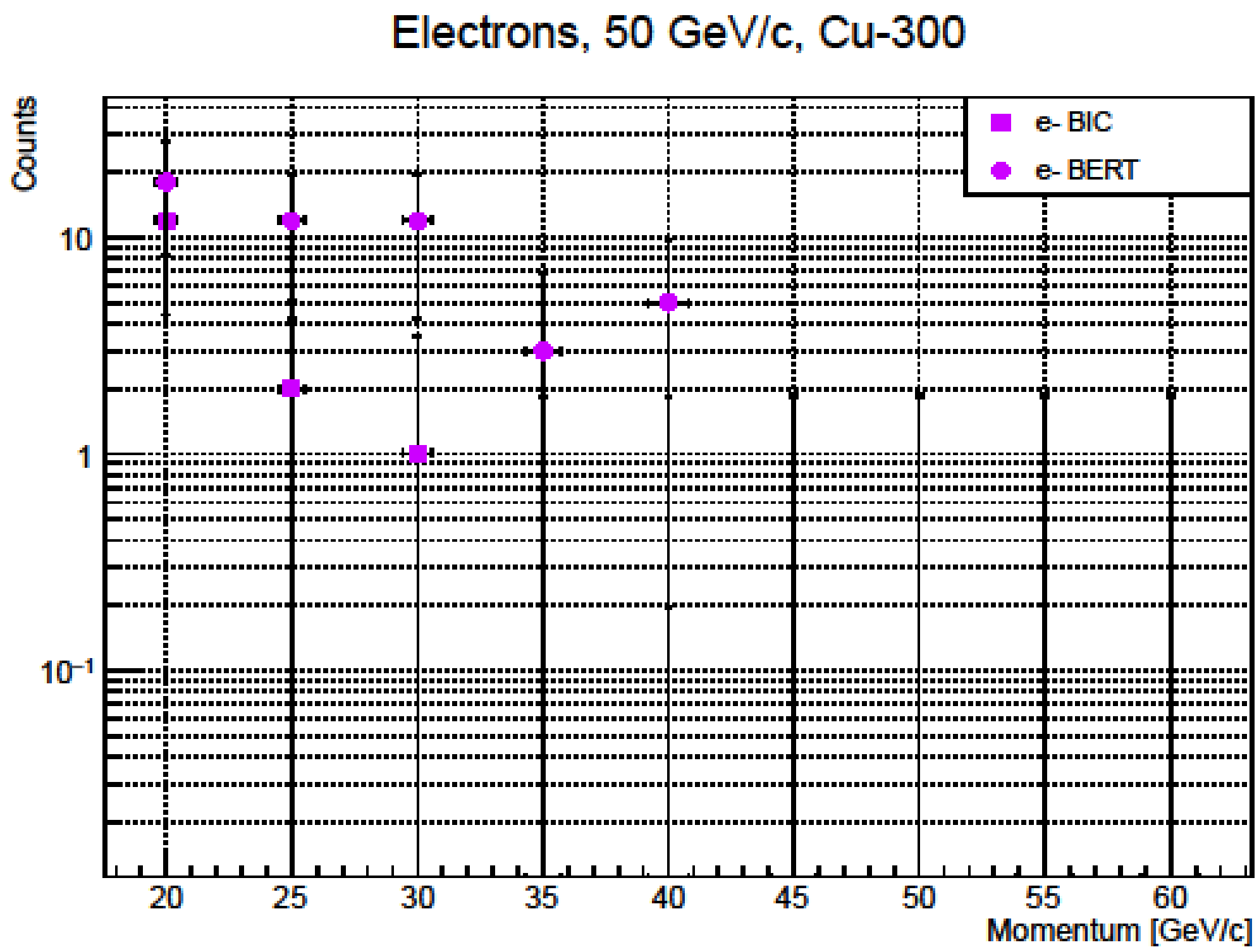
Electrons, 50 GeV/c, Cu-300
Counts
10
1
10⁻¹
20
25
30
35
40
45
50
55
60
Momentum [GeV/c]
e- BIC
e- BERT

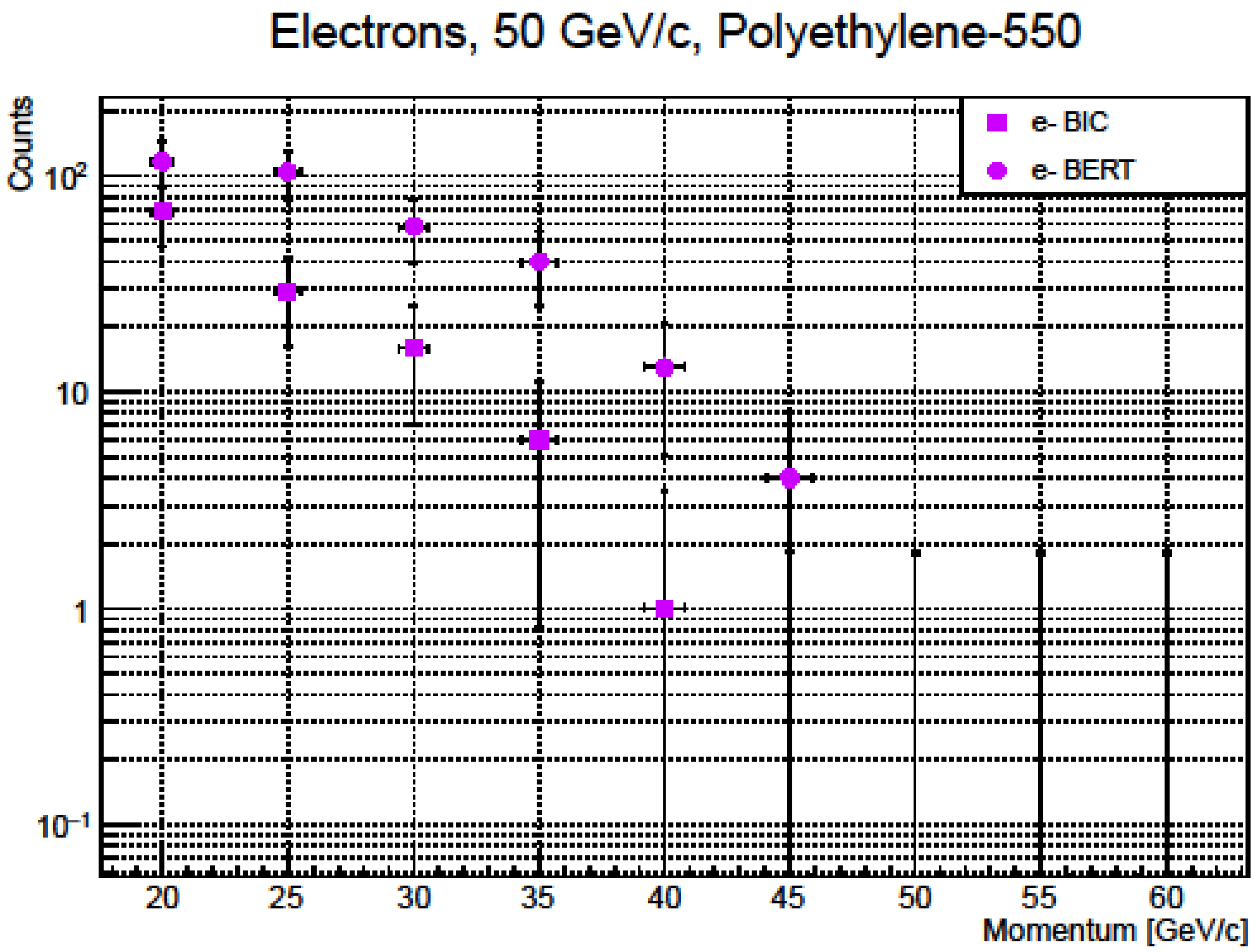
Electrons, 50 GeV/c, Polyethylene-550
Counts
$10^2$
10
1
$10^{-1}$
20
25
30
35
40
45
50
55
60
Momentum [GeV/c]
e- BIC
e- BERT

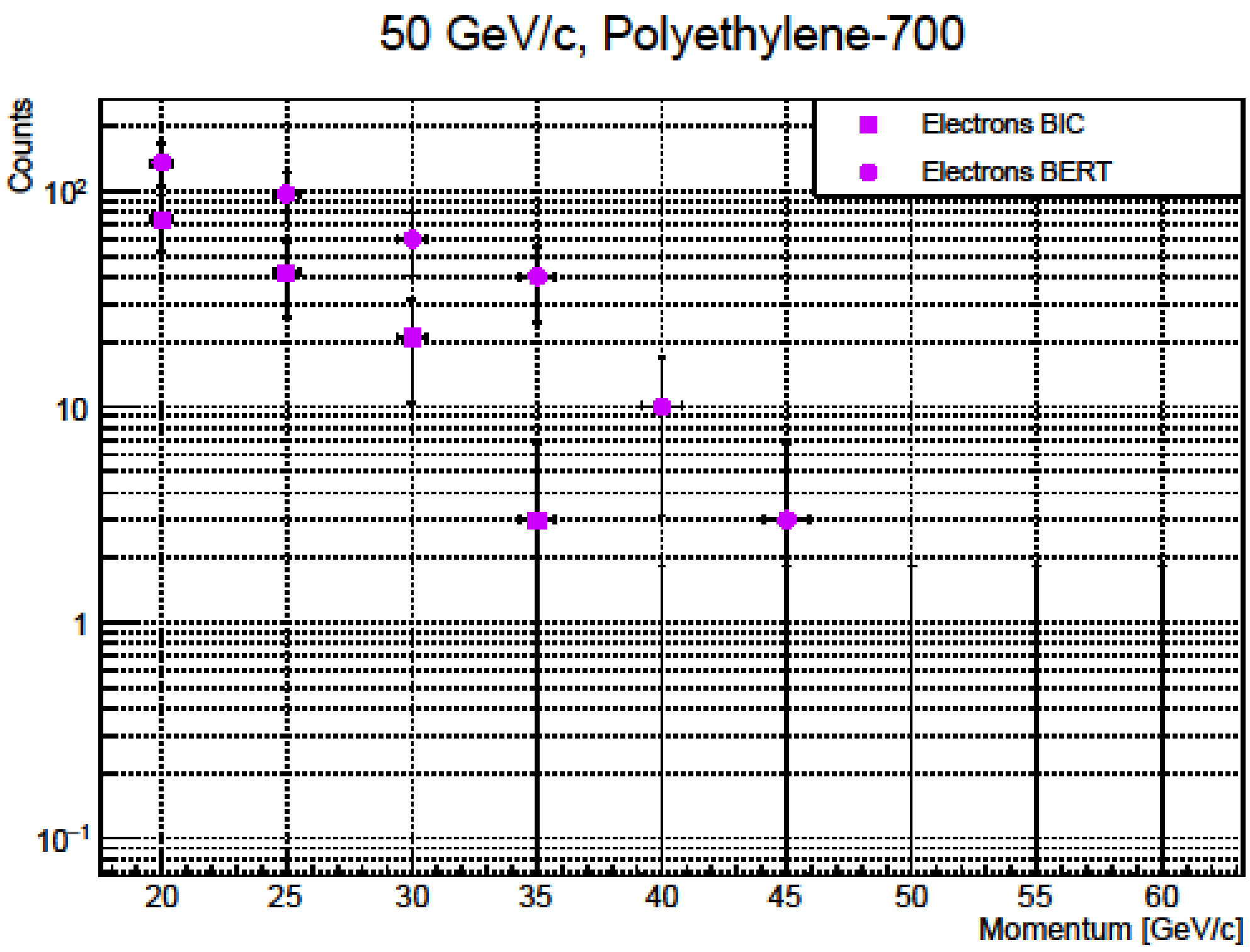
50 GeV/c, Polyethylene-700
Counts
$10^2$
10
1
$10^{-1}$
20
25
30
35
40
45
50
55
60
Momentum [GeV/c]
Electrons BIC
Electrons BERT

## 50 GeV/c, Polyethylene-1000

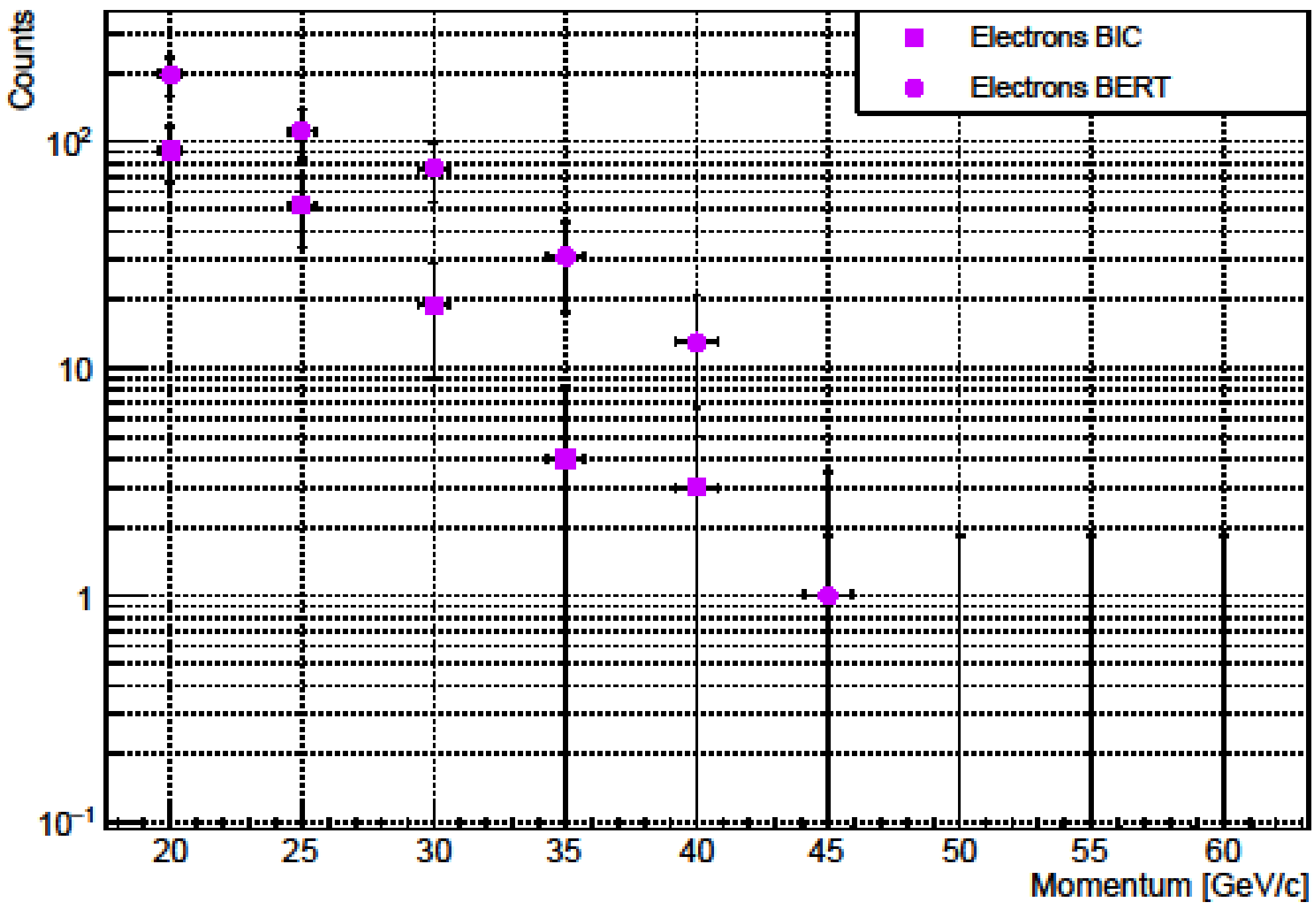

## Electrons, 50 GeV/c, W-150

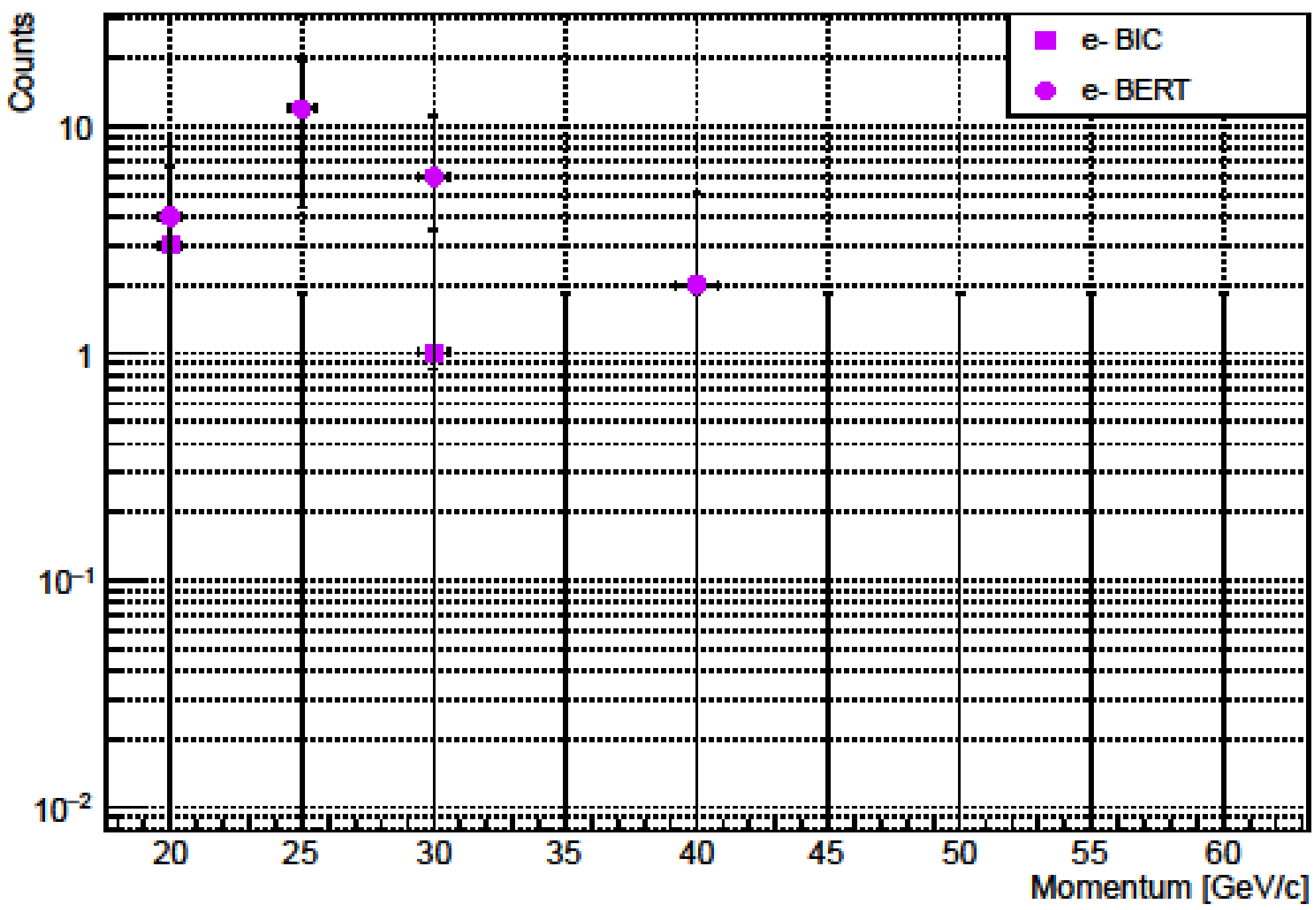

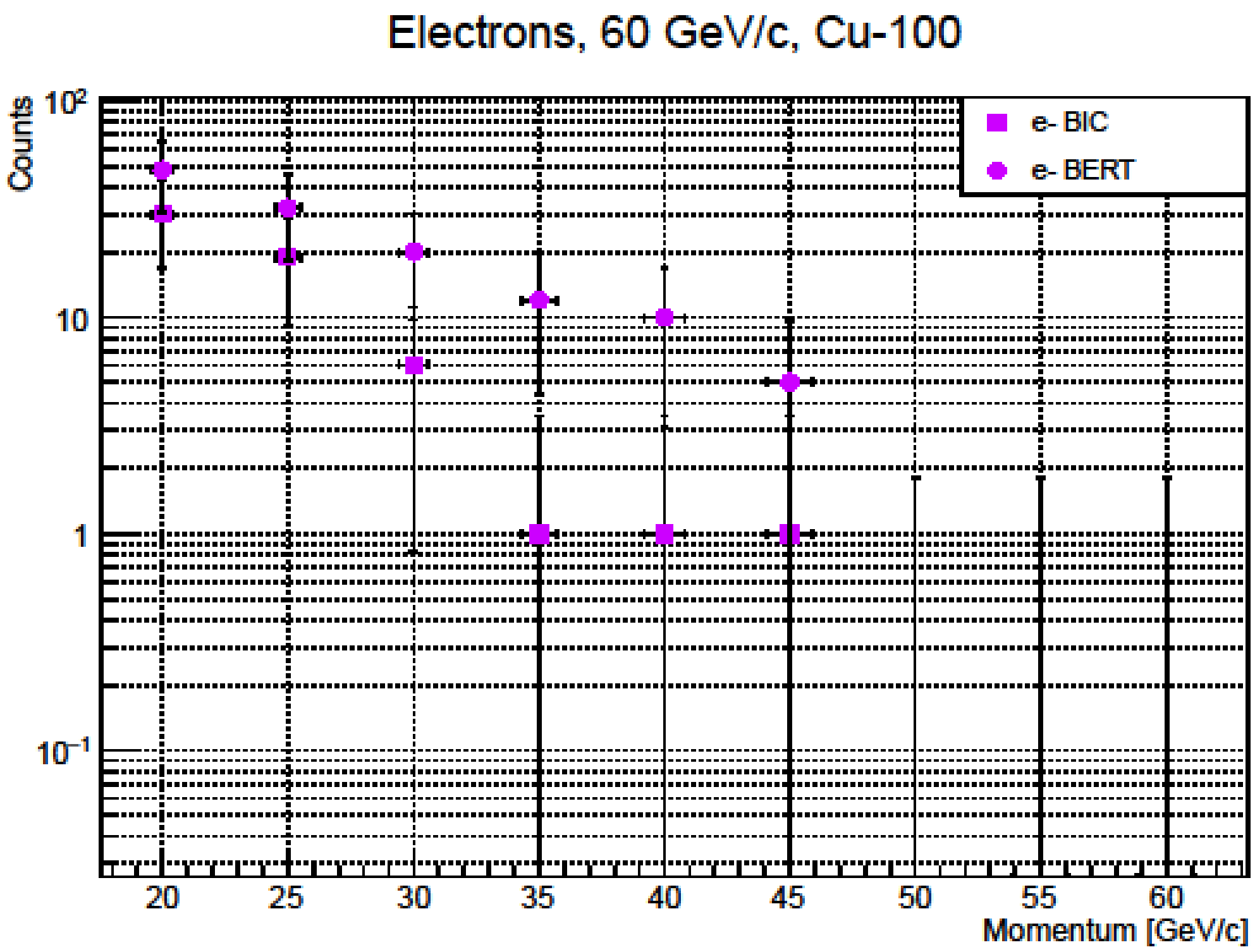
Electrons, 60 GeV/c, Cu-100
Counts
$10^2$
10
1
$10^{-1}$
20
25
30
35
40
45
50
55
60
Momentum [GeV/c]
e- BIC
e- BERT

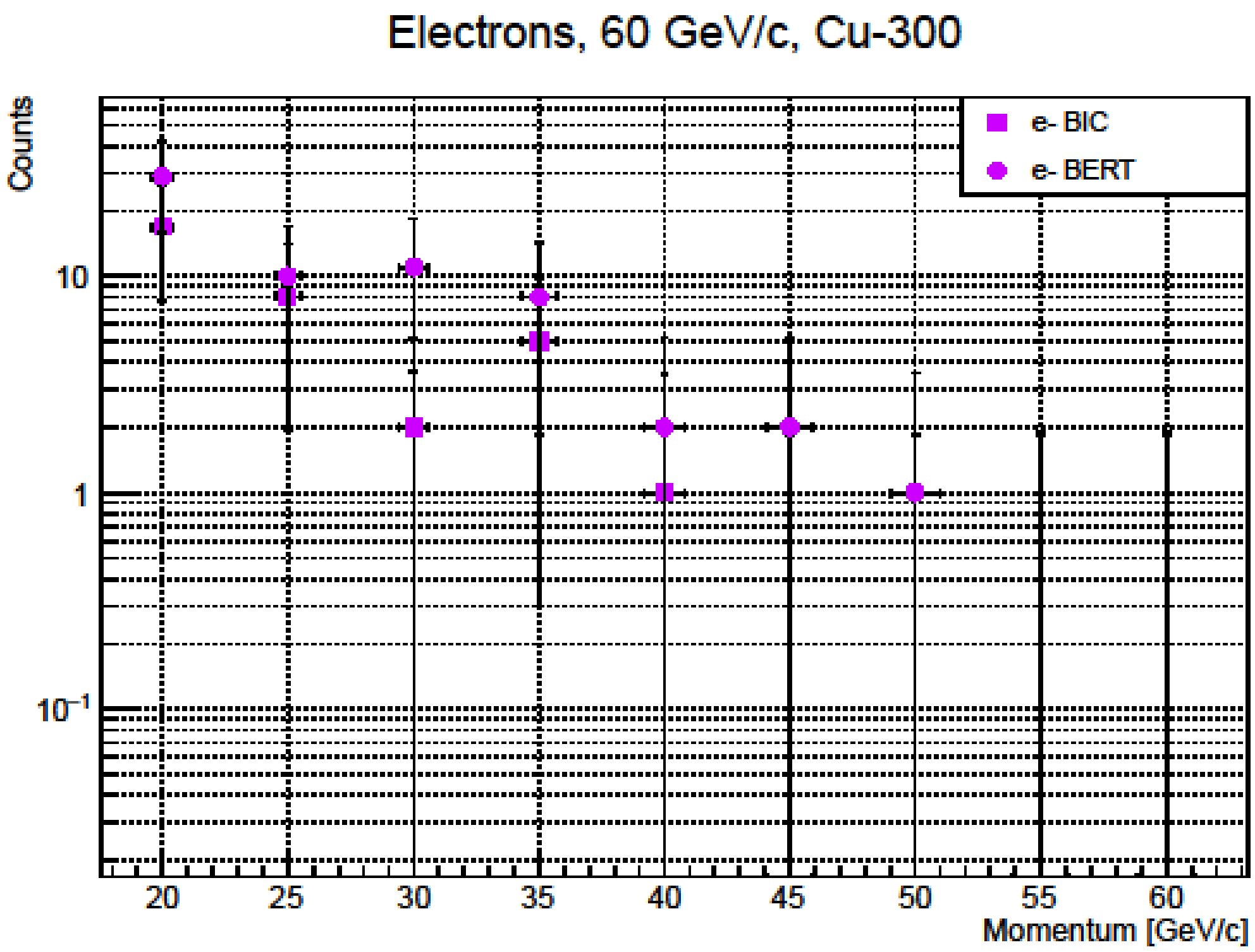
Electrons, 60 GeV/c, Cu-300
Counts
10
1
$10^{-1}$
20
25
30
35
40
45
50
55
60
Momentum [GeV/c]
e- BIC
e- BERT

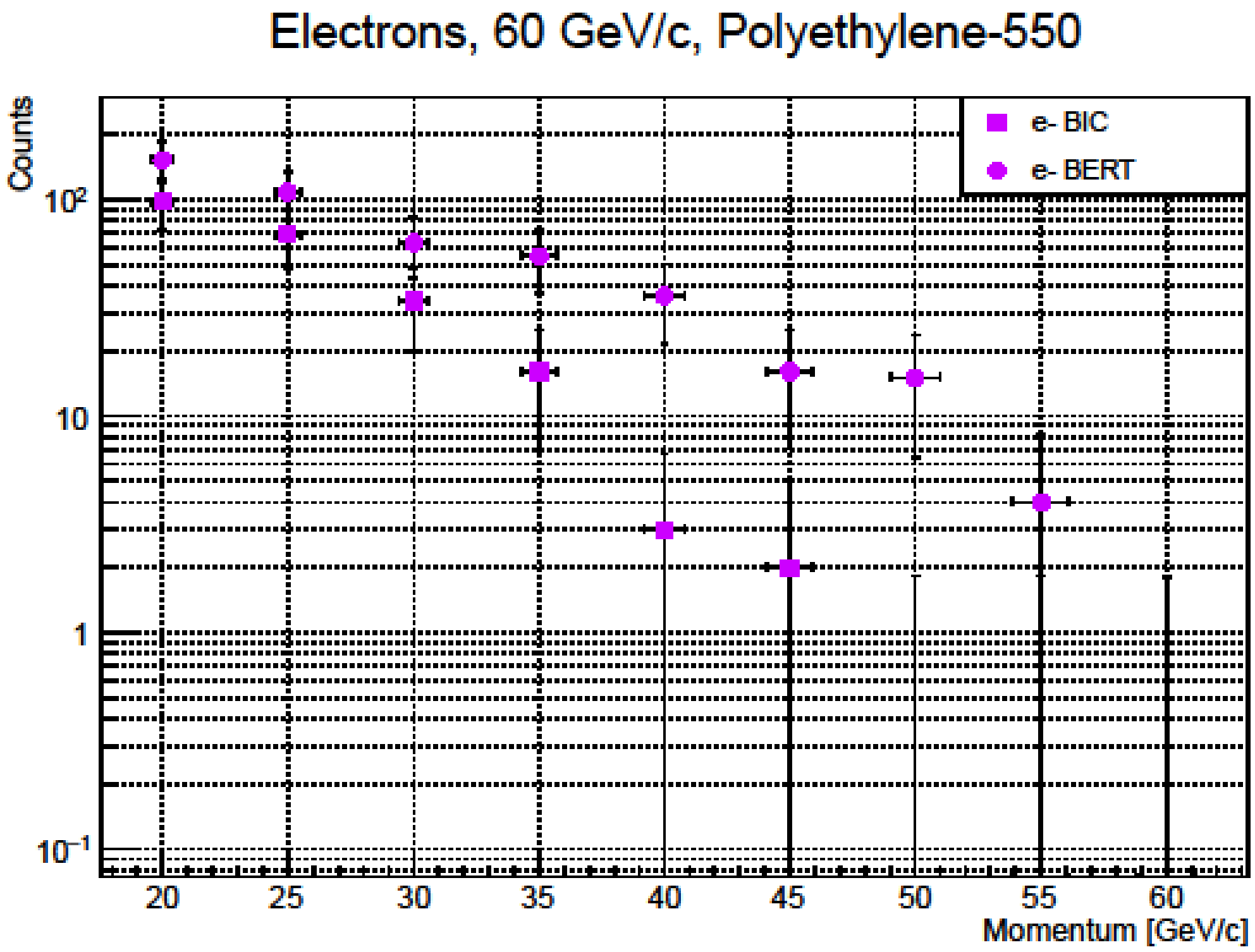
Electrons, 60 GeV/c, Polyethylene-550
Counts
$10^2$
10
1
$10^{-1}$
20
25
30
35
40
45
50
55
60
Momentum [GeV/c]
e- BIC
e- BERT

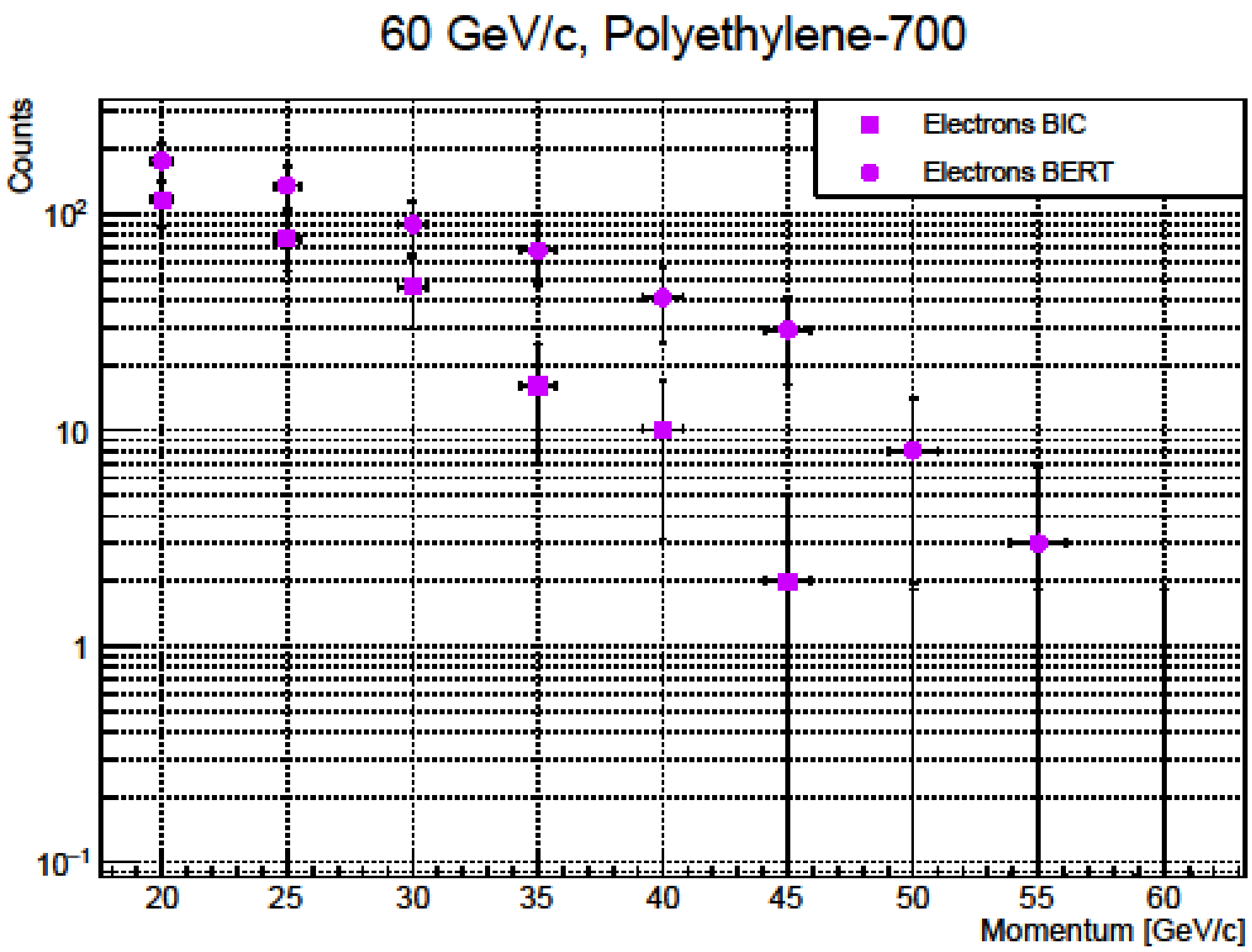
60 GeV/c, Polyethylene-700
Counts
$10^2$
10
1
$10^{-1}$
20
25
30
35
40
45
50
55
60
Momentum [GeV/c]
Electrons BIC
Electrons BERT

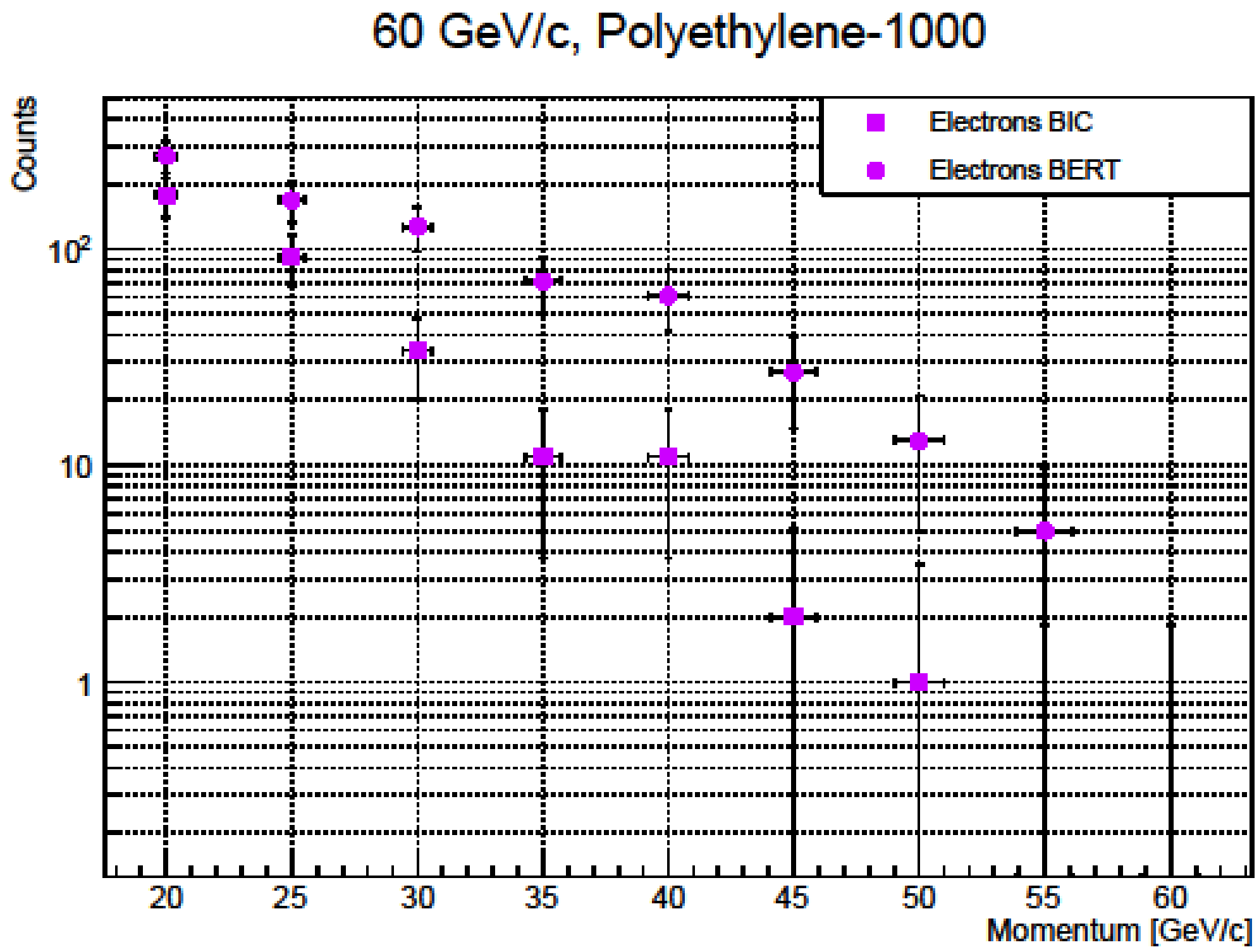

60 GeV/c, Polyethylene-1000
Electrons BIC
Electrons BERT
Counts
10²
10
1
20
25
30
35
40
45
50
55
60
Momentum [GeV/c]

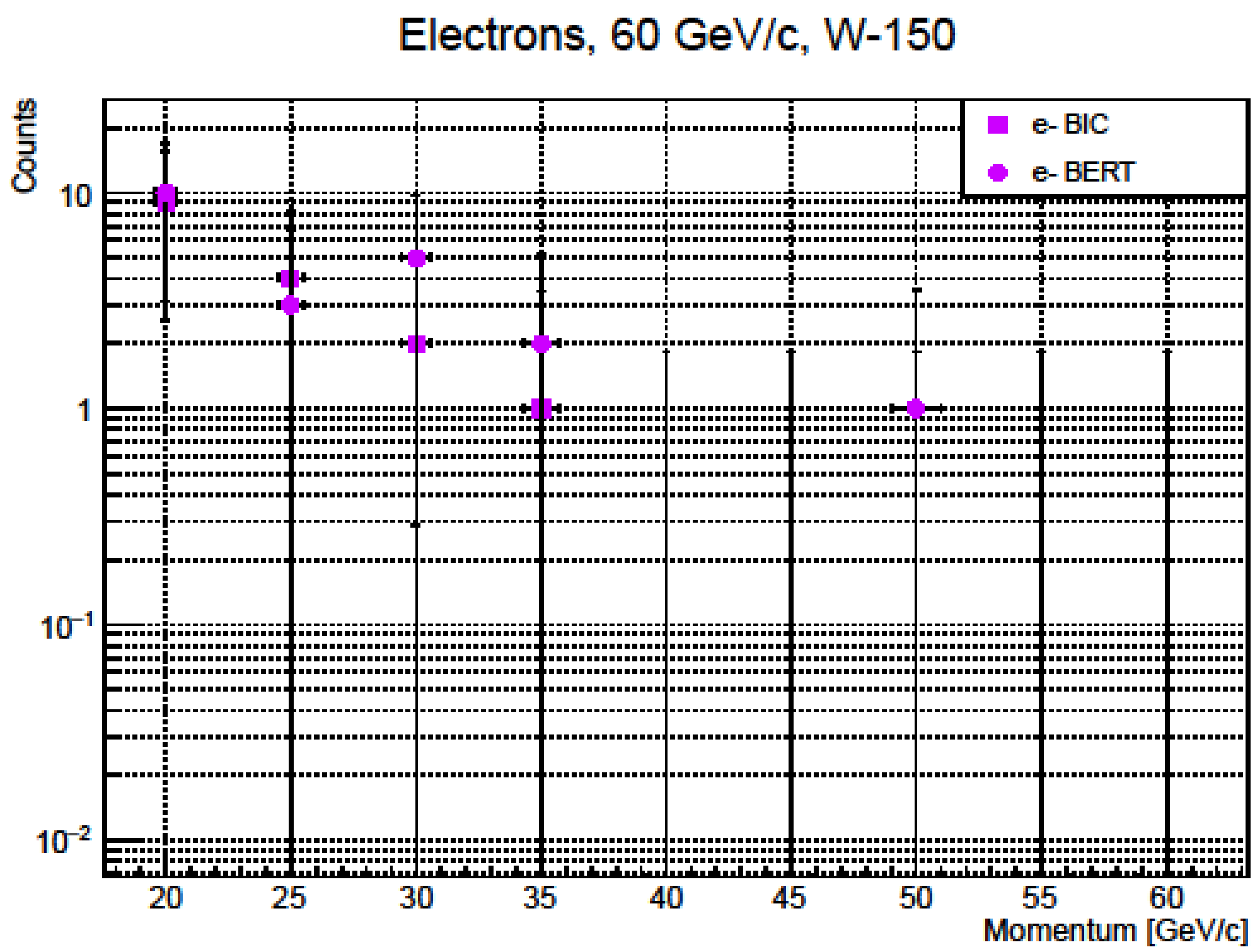

Electrons, 60 GeV/c, W-150
e- BIC
e- BERT
Counts
10
1
10⁻¹
10⁻²
20
25
30
35
40
45
50
55
60
Momentum [GeV/c]

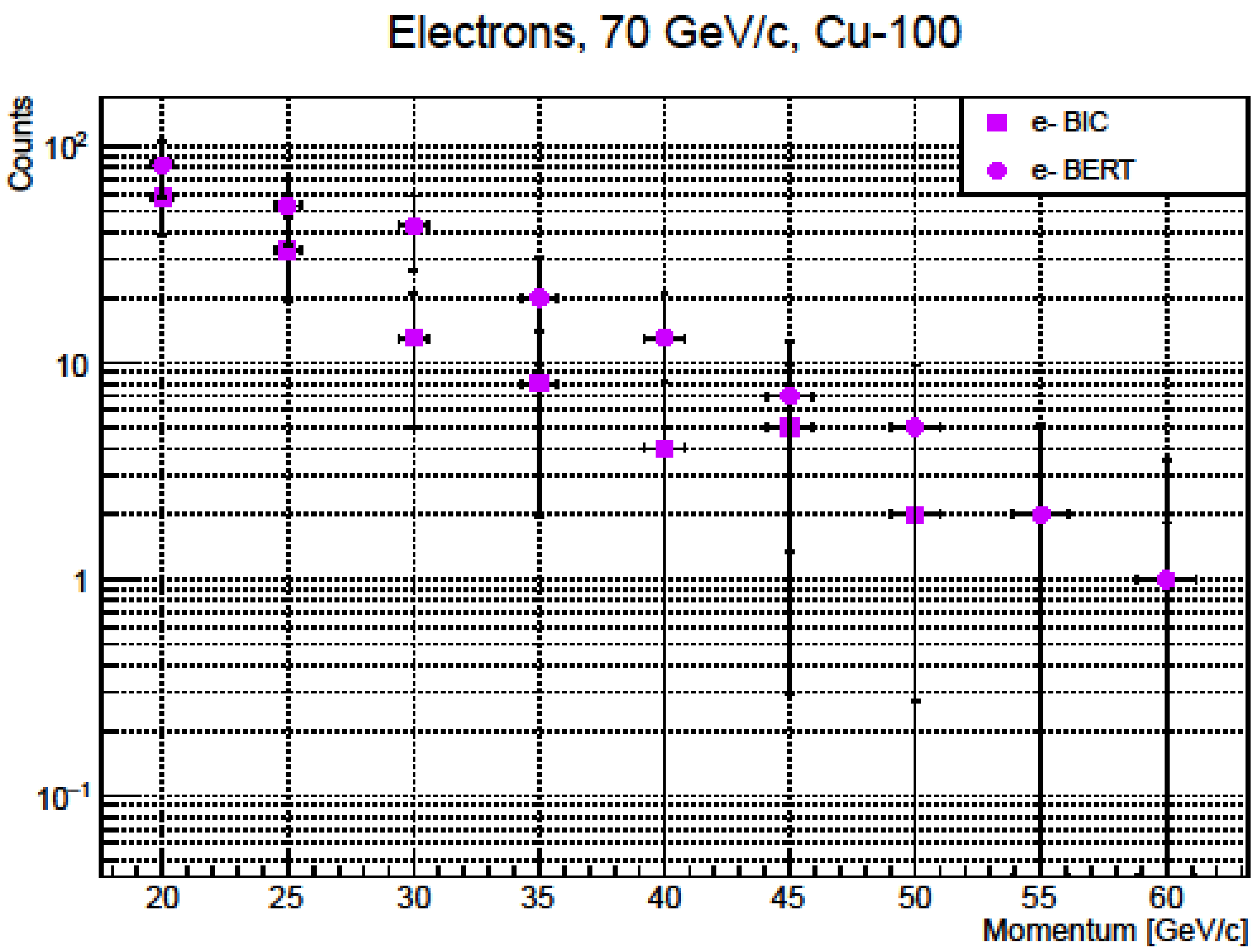
Electrons, 70 GeV/c, Cu-100
Counts
Momentum [GeV/c]
e- BIC
e- BERT

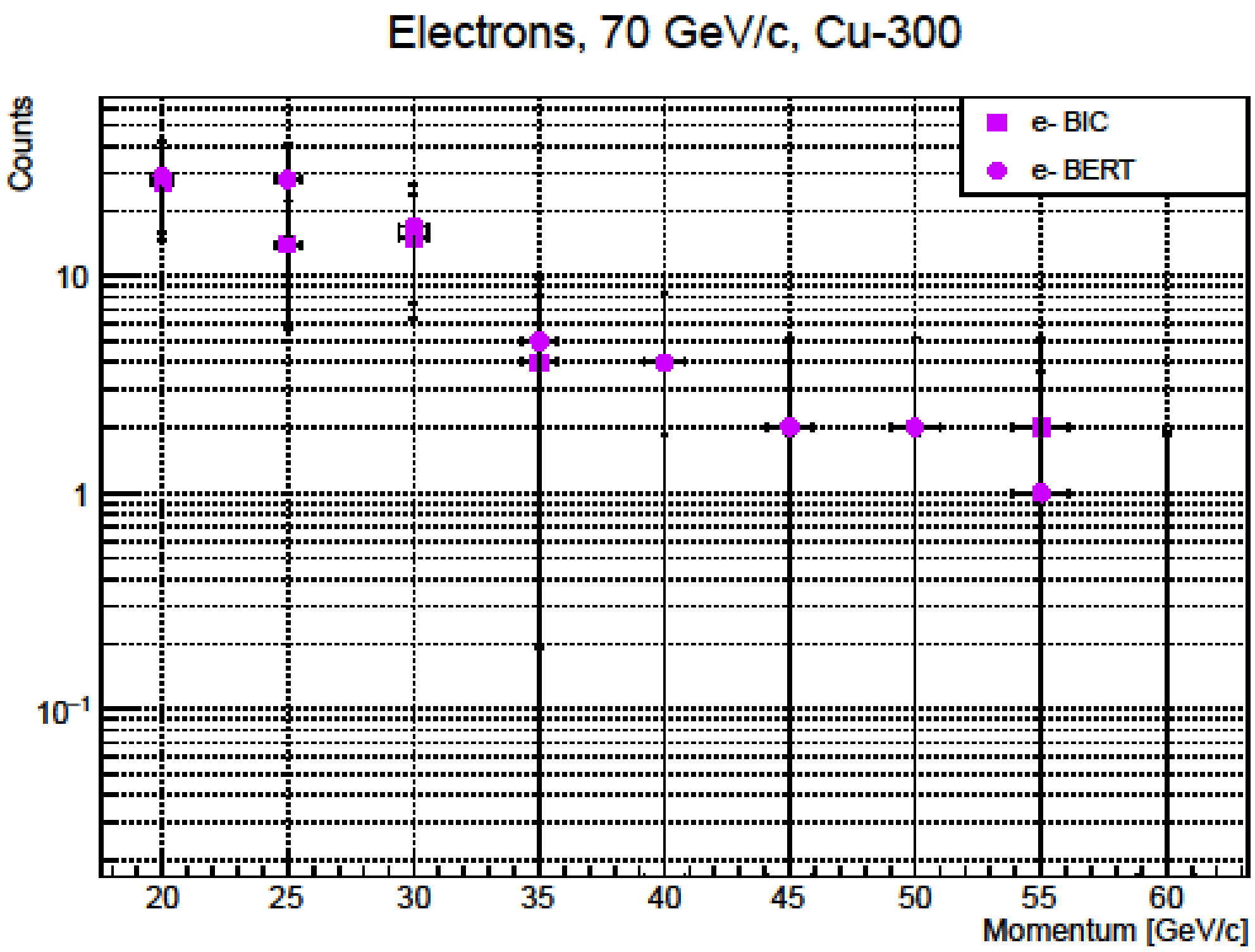
Electrons, 70 GeV/c, Cu-300
Counts
Momentum [GeV/c]
e- BIC
e- BERT

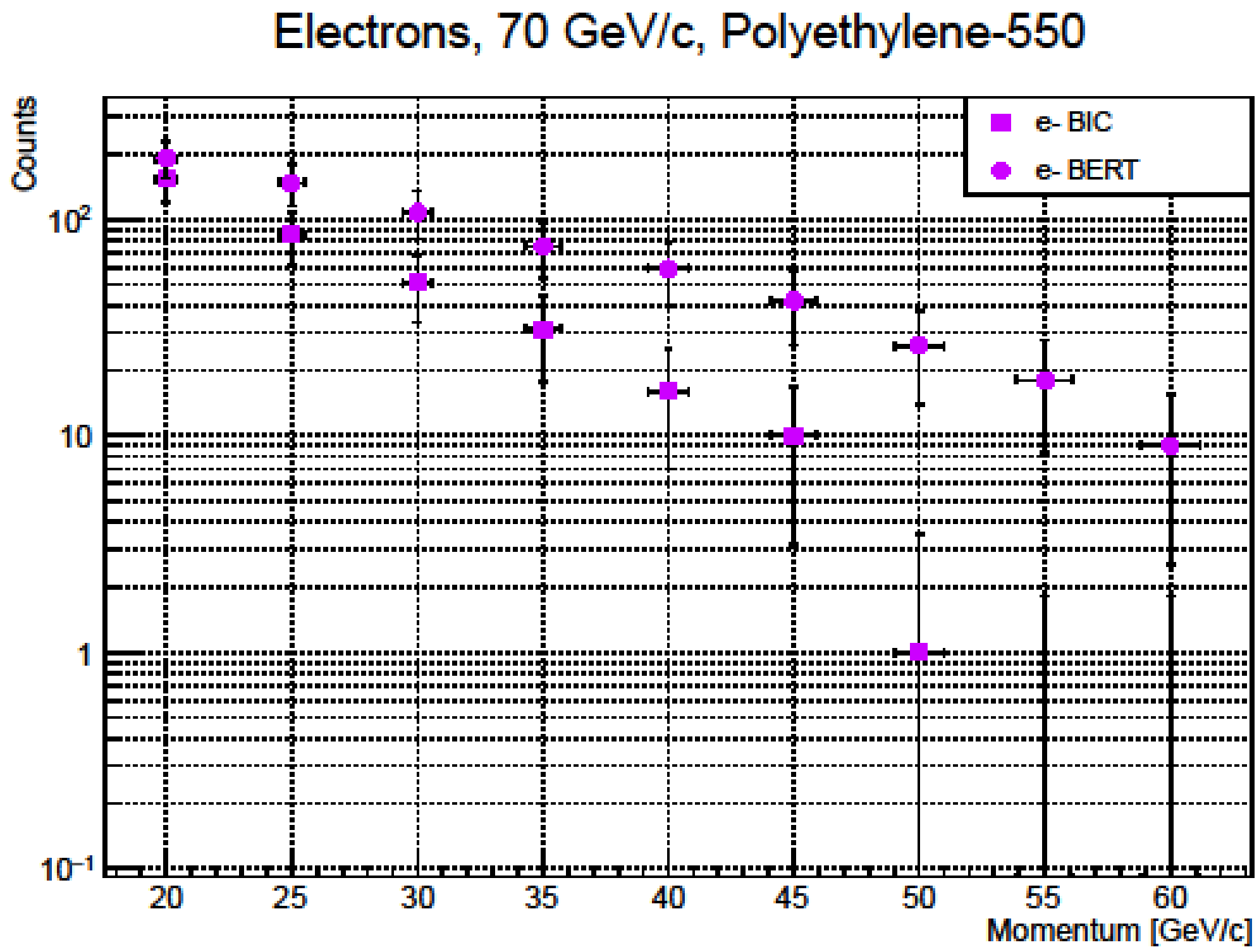
Electrons, 70 GeV/c, Polyethylene-550
Counts
Momentum [GeV/c]
e- BIC
e- BERT
$10^{-1}$
1
10
$10^2$
20
25
30
35
40
45
50
55
60

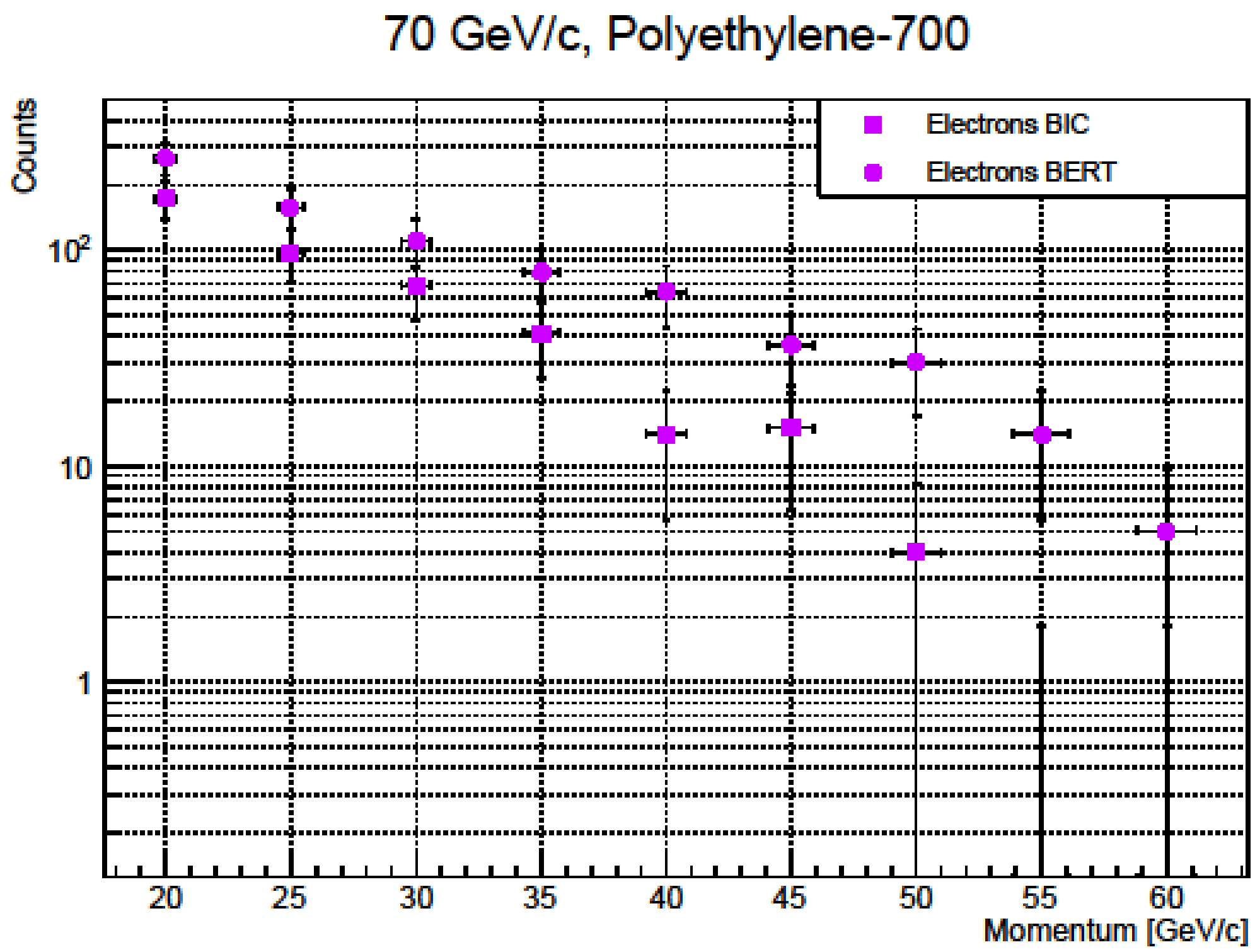
70 GeV/c, Polyethylene-700
Counts
Momentum [GeV/c]
Electrons BIC
Electrons BERT
1
10
$10^2$
20
25
30
35
40
45
50
55
60

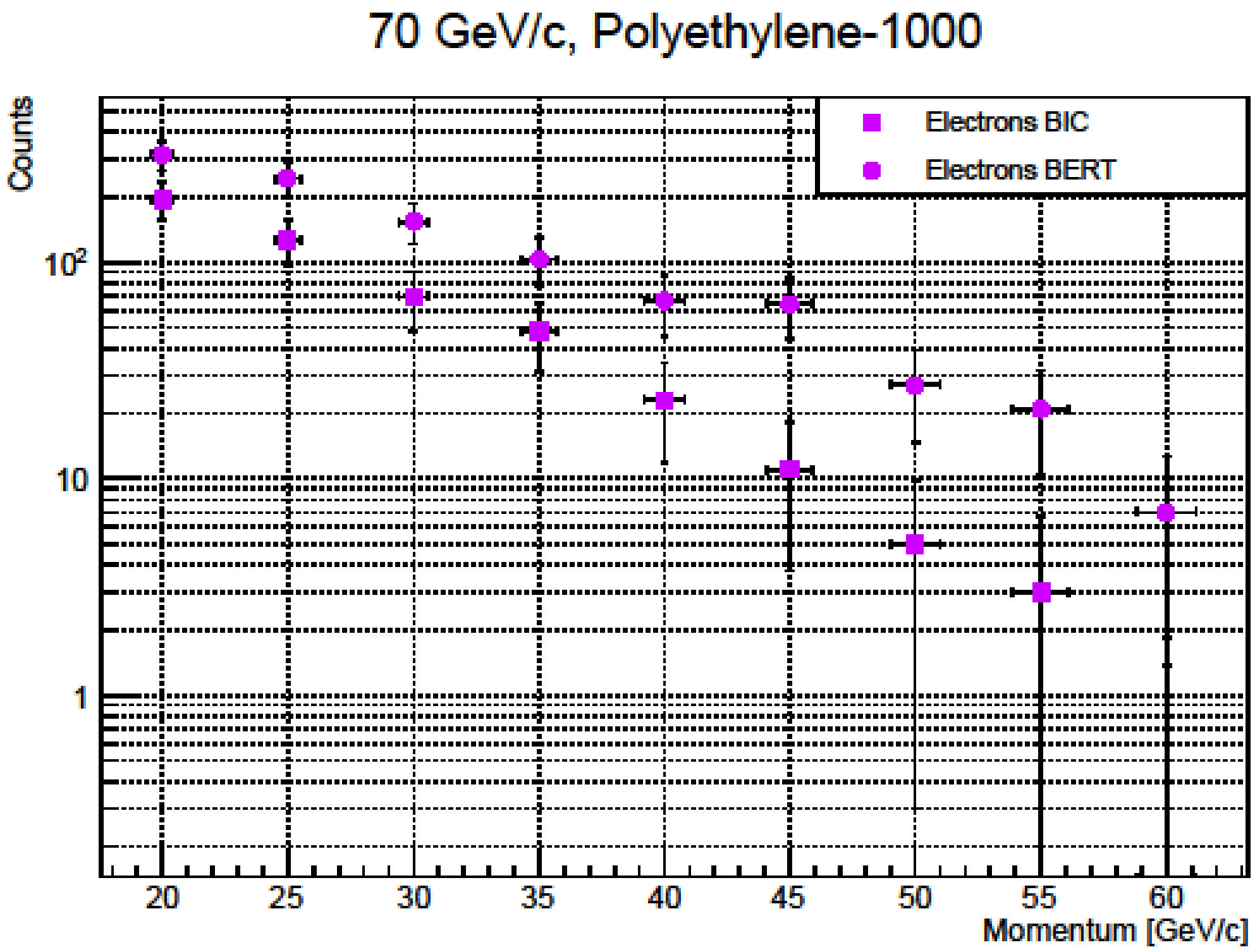

70 GeV/c, Polyethylene-1000
Counts
Electrons BIC
Electrons BERT
$10^2$
10
1
20
25
30
35
40
45
50
55
60
Momentum [GeV/c]

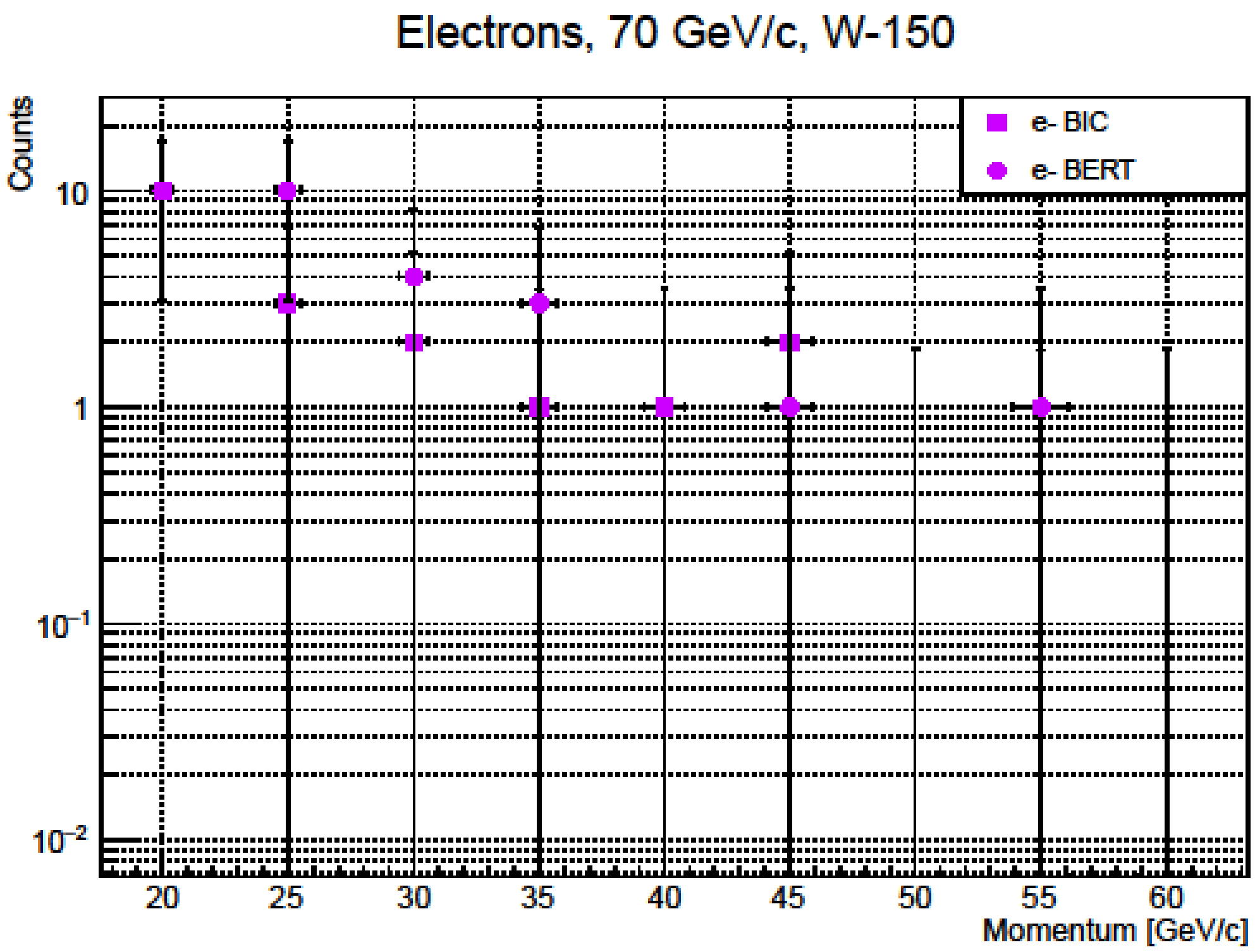

Electrons, 70 GeV/c, W-150
Counts
e- BIC
e- BERT
10
1
$10^{-1}$
$10^{-2}$
20
25
30
35
40
45
50
55
60
Momentum [GeV/c]

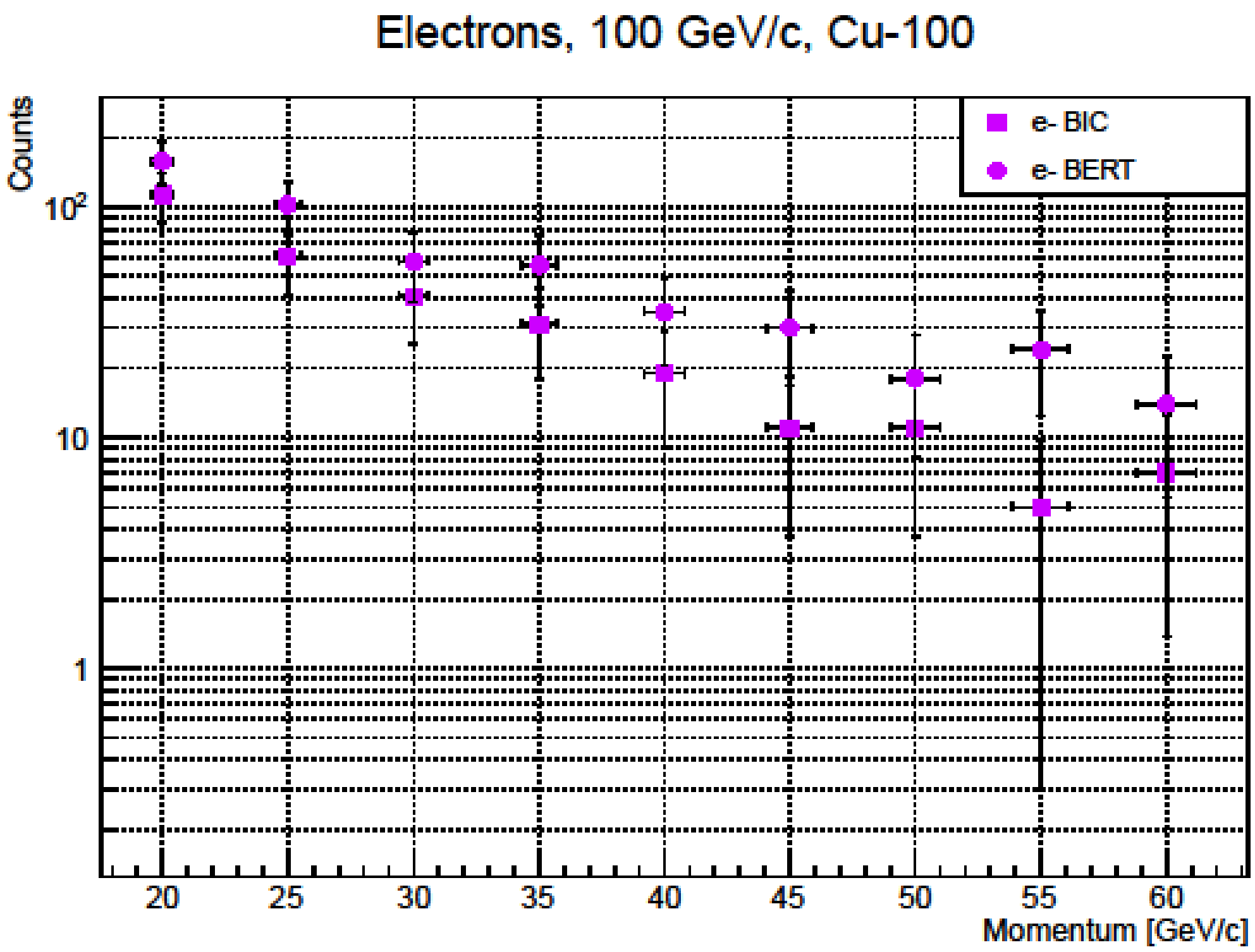

Electrons, 100 GeV/c, Cu-100
Counts
Momentum [GeV/c]
e- BIC
e- BERT

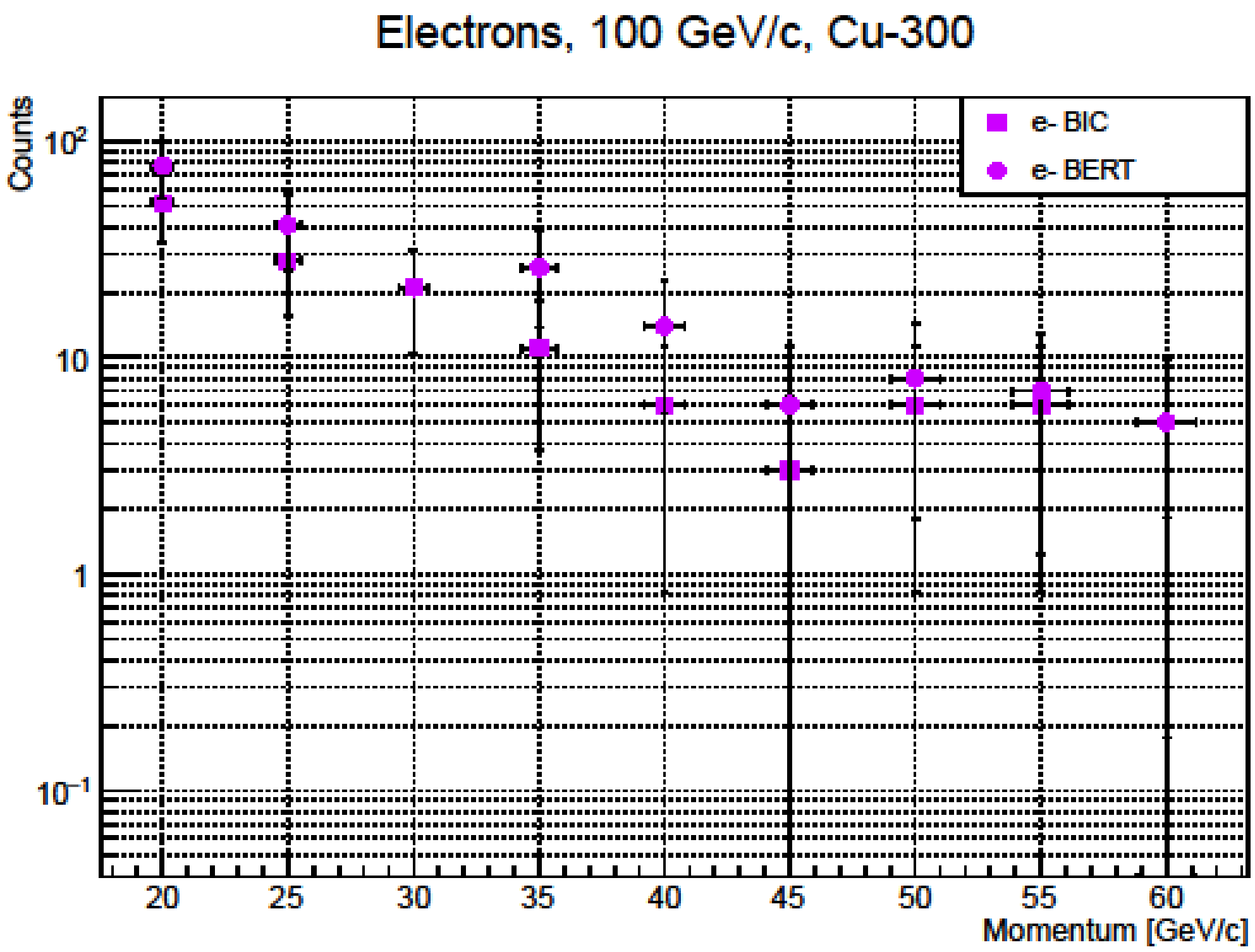

Electrons, 100 GeV/c, Cu-300
Counts
Momentum [GeV/c]
e- BIC
e- BERT

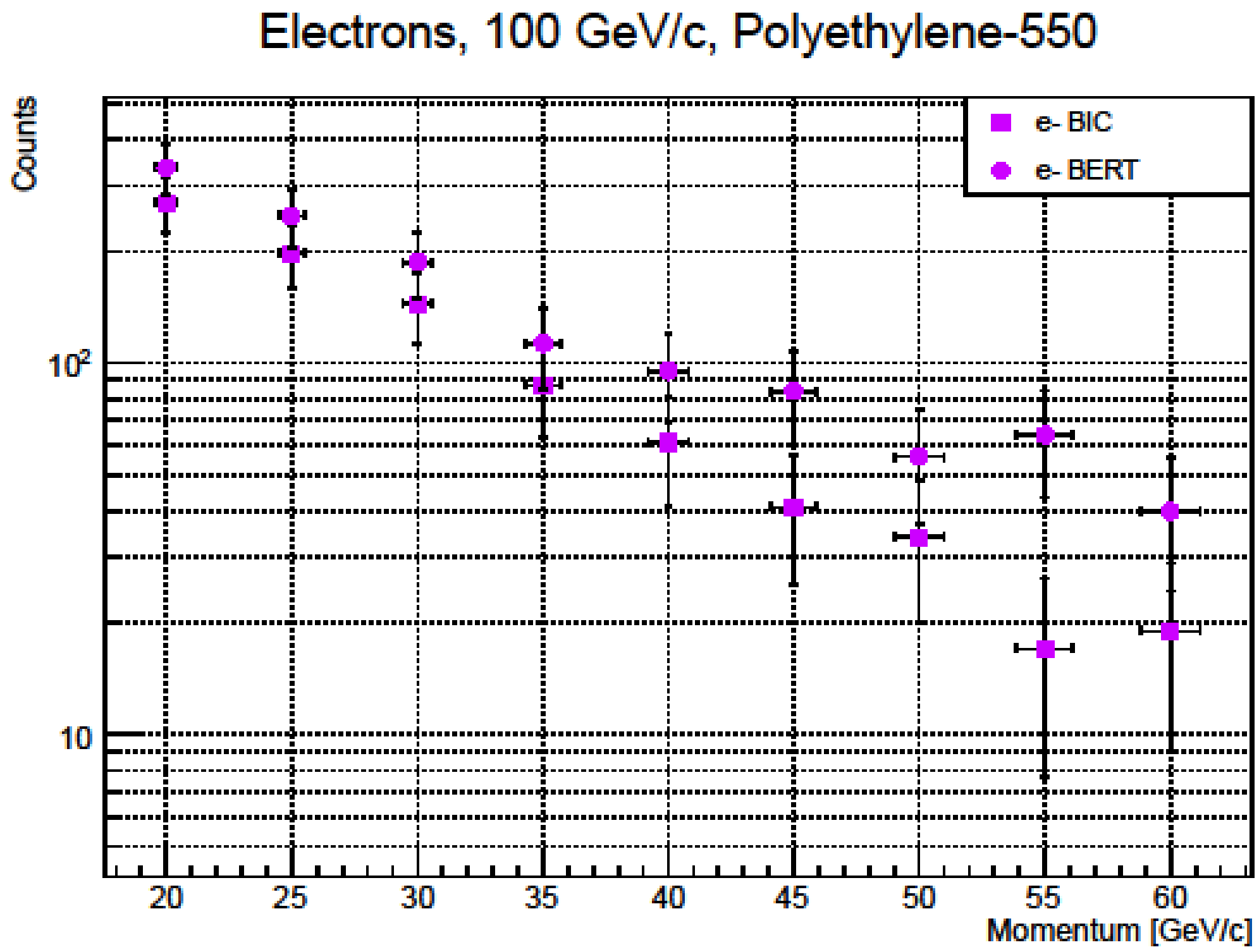
Electrons, 100 GeV/c, Polyethylene-550
Counts
e- BIC
e- BERT
10
$10^2$
20
25
30
35
40
45
50
55
60
Momentum [GeV/c]

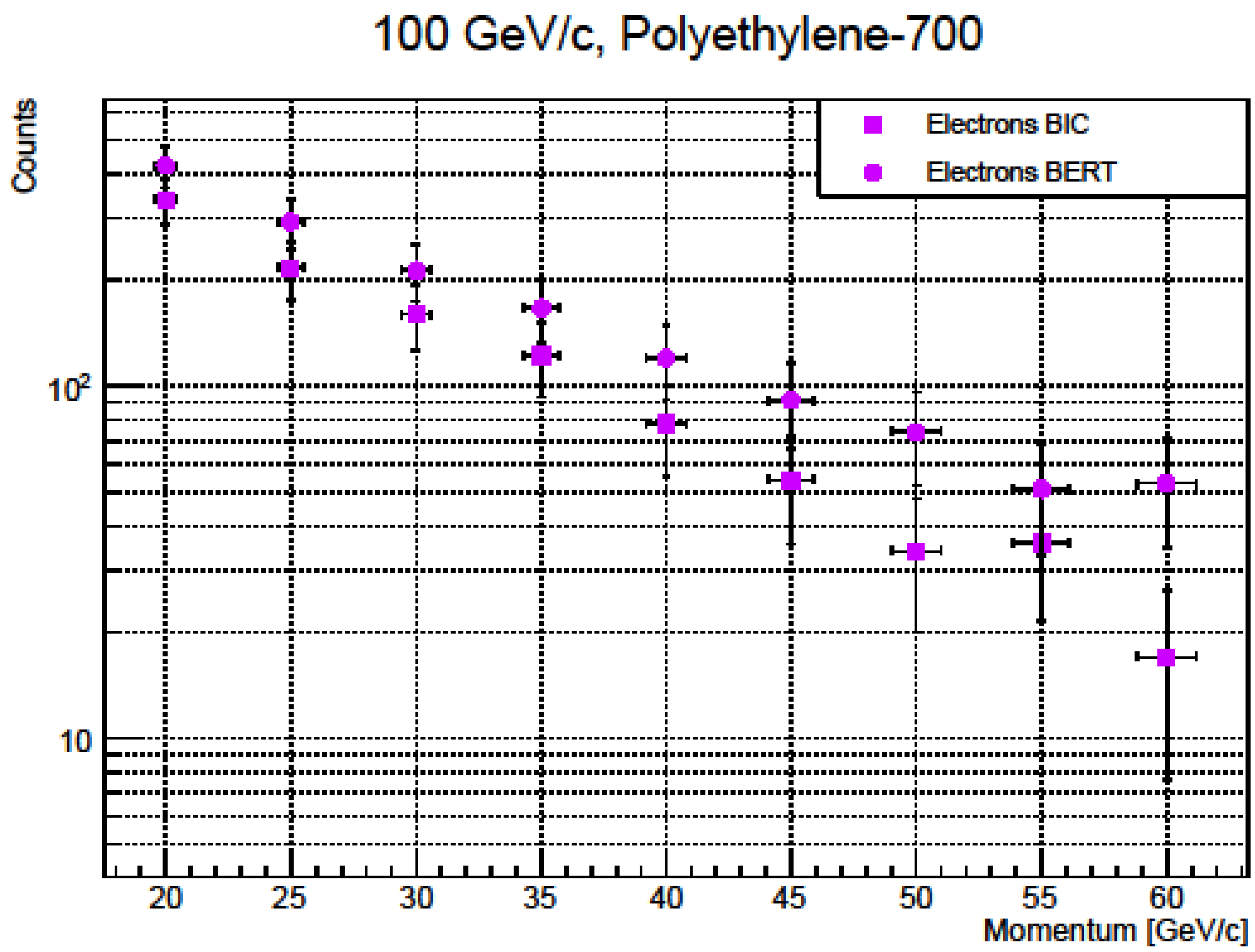
100 GeV/c, Polyethylene-700
Counts
Electrons BIC
Electrons BERT
10
$10^2$
20
25
30
35
40
45
50
55
60
Momentum [GeV/c]

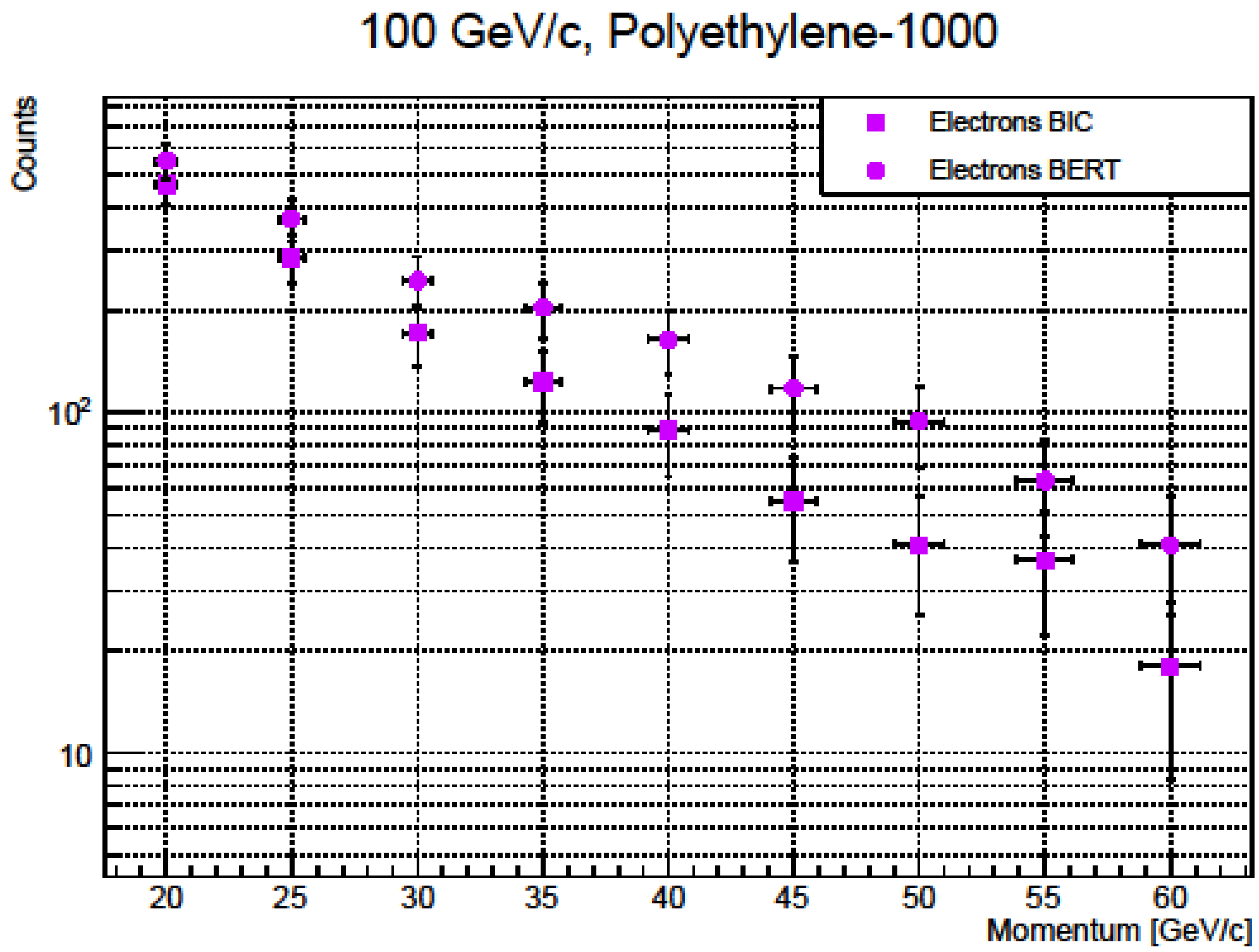

100 GeV/c, Polyethylene-1000
Counts
Electrons BIC
Electrons BERT
$10^2$
10
20
25
30
35
40
45
50
55
60
Momentum [GeV/c]

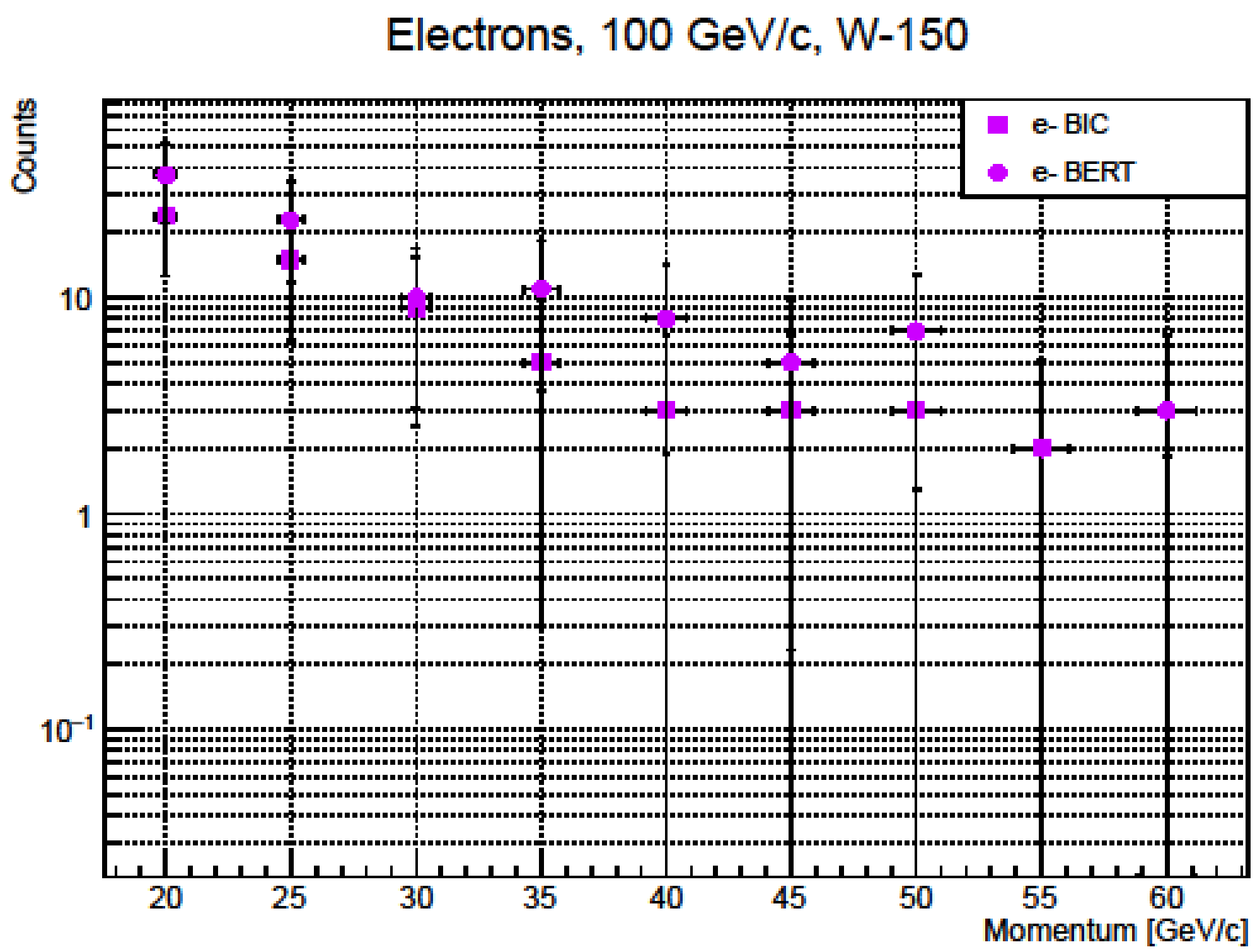

Electrons, 100 GeV/c, W-150
Counts
e- BIC
e- BERT
10
1
$10^{-1}$
20
25
30
35
40
45
50
55
60
Momentum [GeV/c]

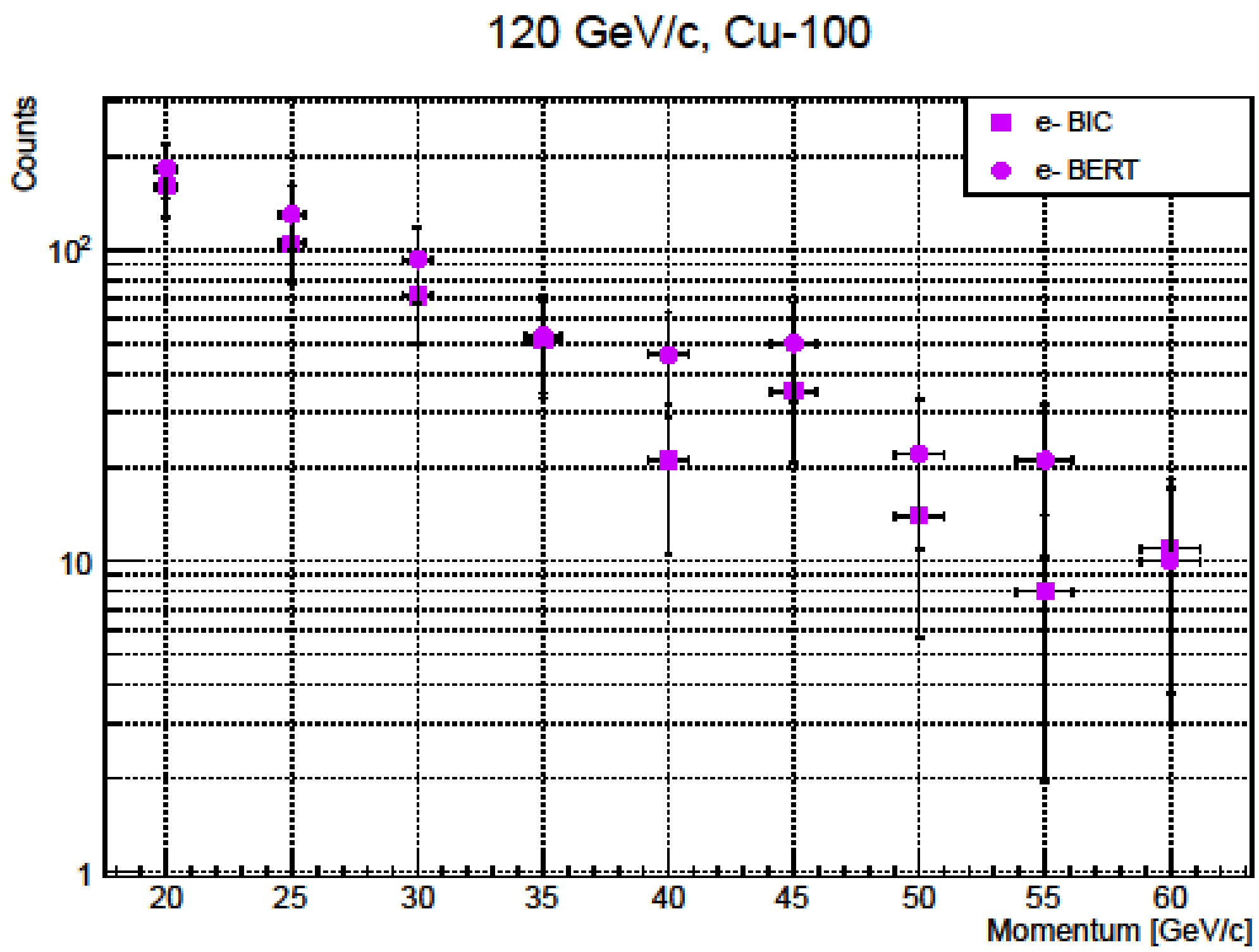
120 GeV/c, Cu-100
Counts
e- BIC
e- BERT
$10^2$
10
1
20
25
30
35
40
45
50
55
60
Momentum [GeV/c]

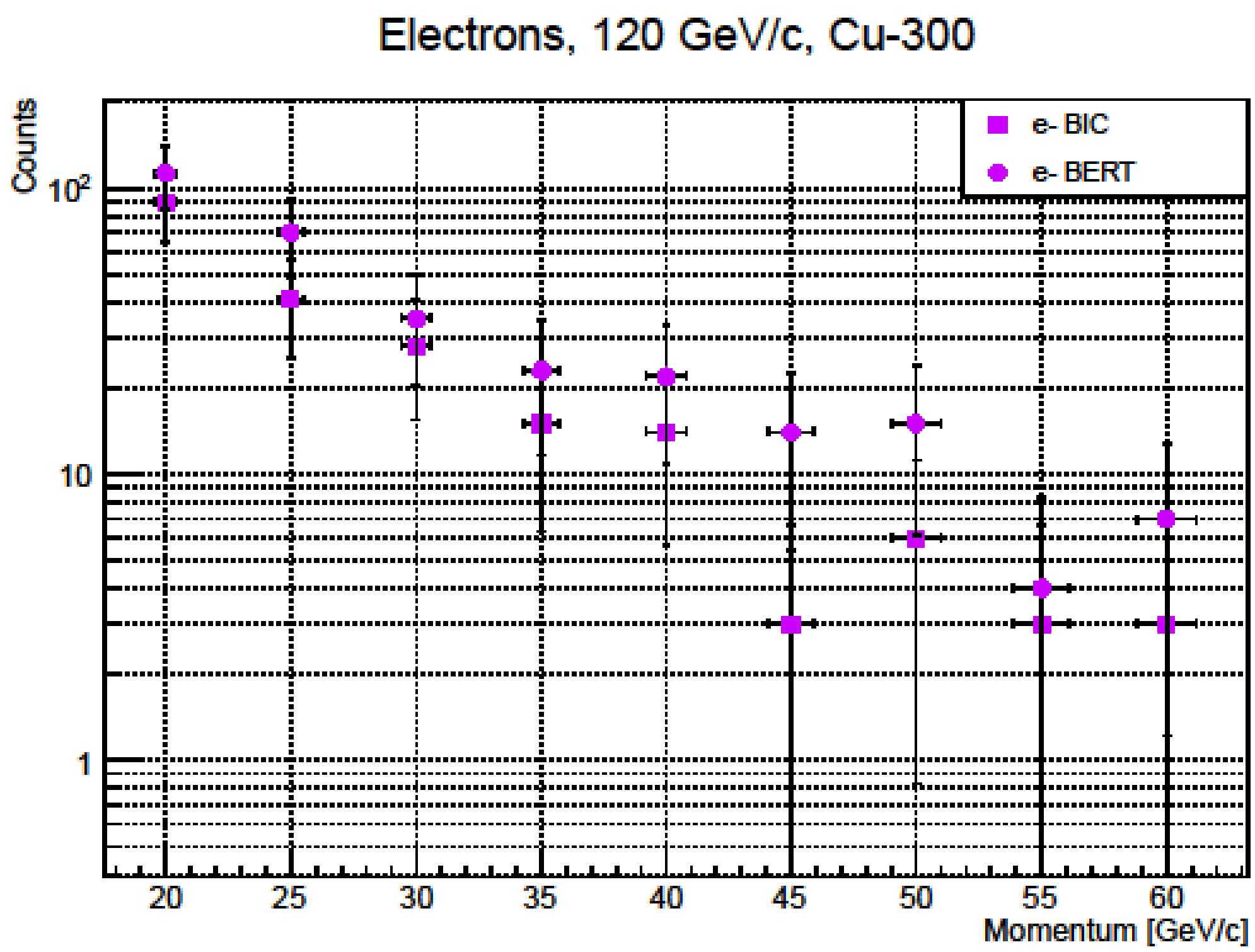
Electrons, 120 GeV/c, Cu-300
Counts
e- BIC
e- BERT
$10^2$
10
1
20
25
30
35
40
45
50
55
60
Momentum [GeV/c]

Electrons, 120 GeV/c, Polyethylene-550

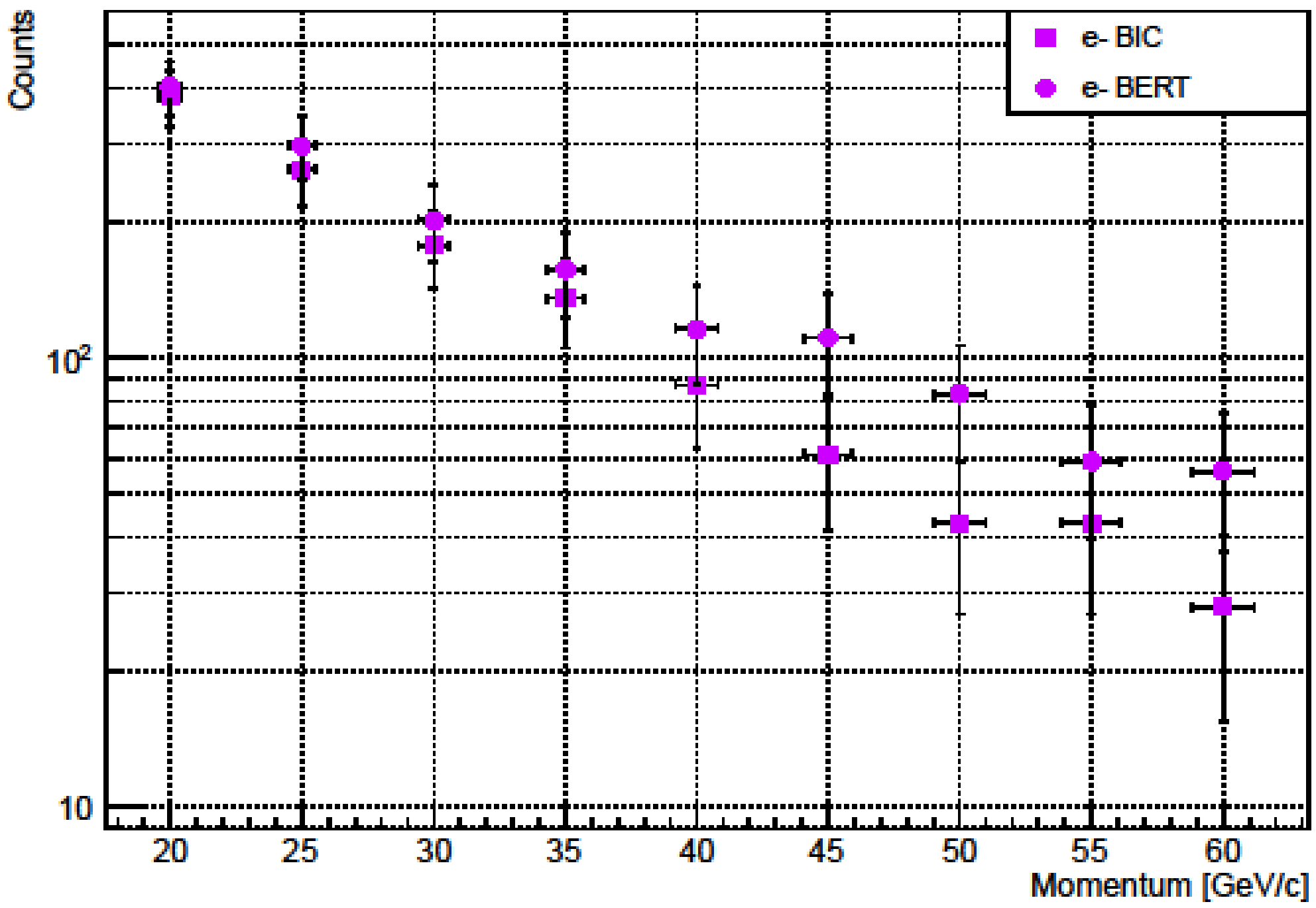

120 GeV/c, Polyethylene-700

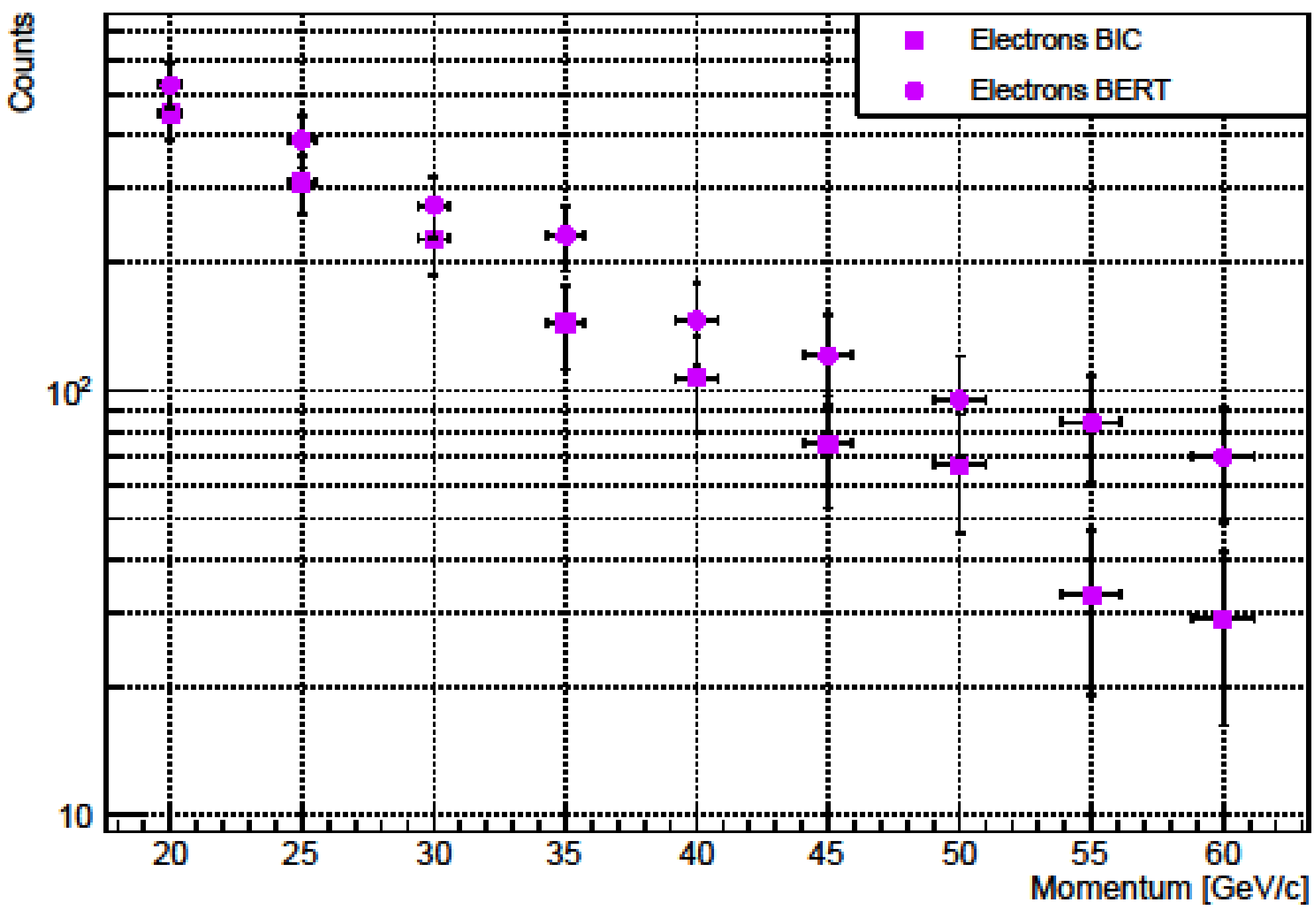

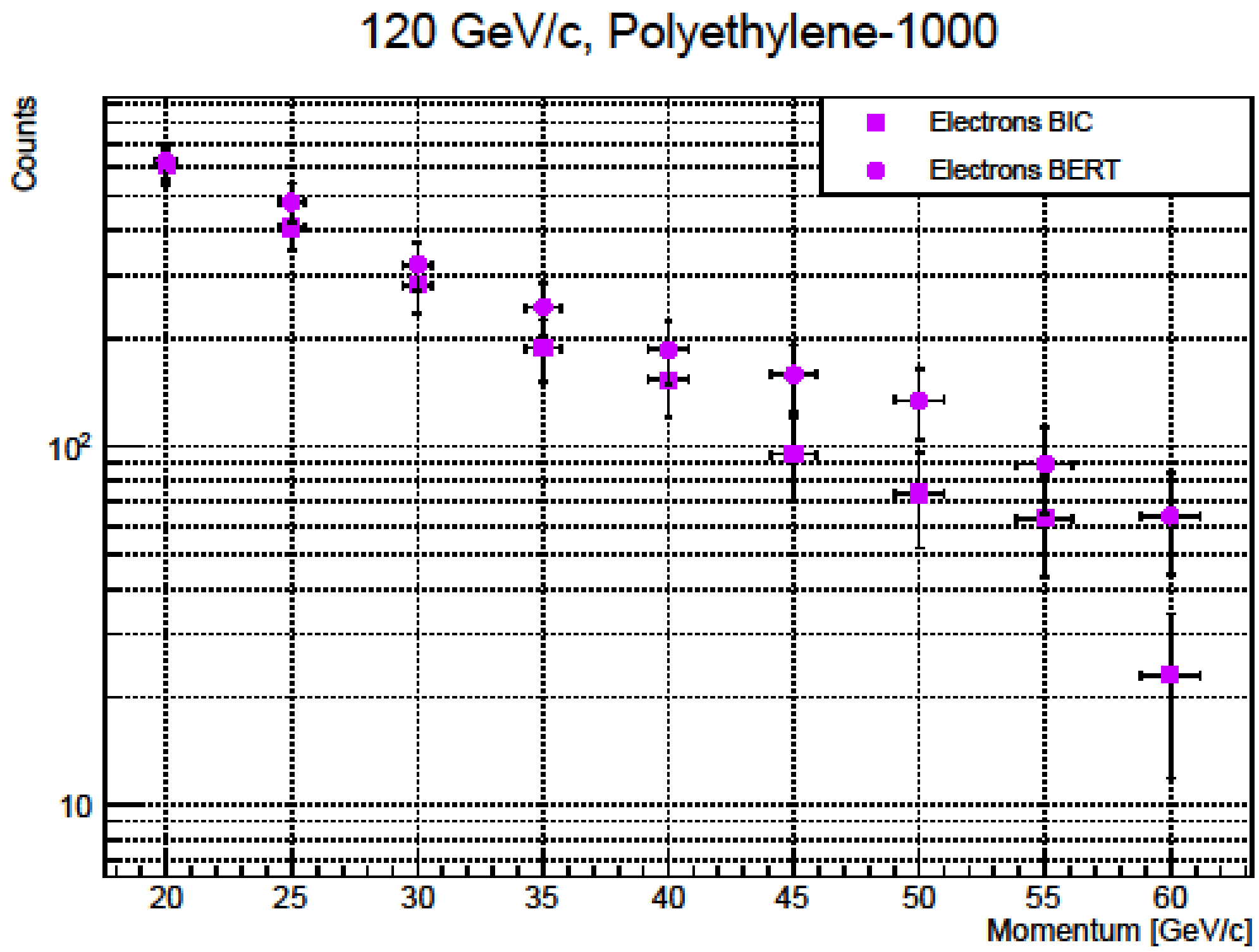
120 GeV/c, Polyethylene-1000
Electrons BIC
Electrons BERT
Counts
$10^2$
10
20
25
30
35
40
45
50
55
60
Momentum [GeV/c]

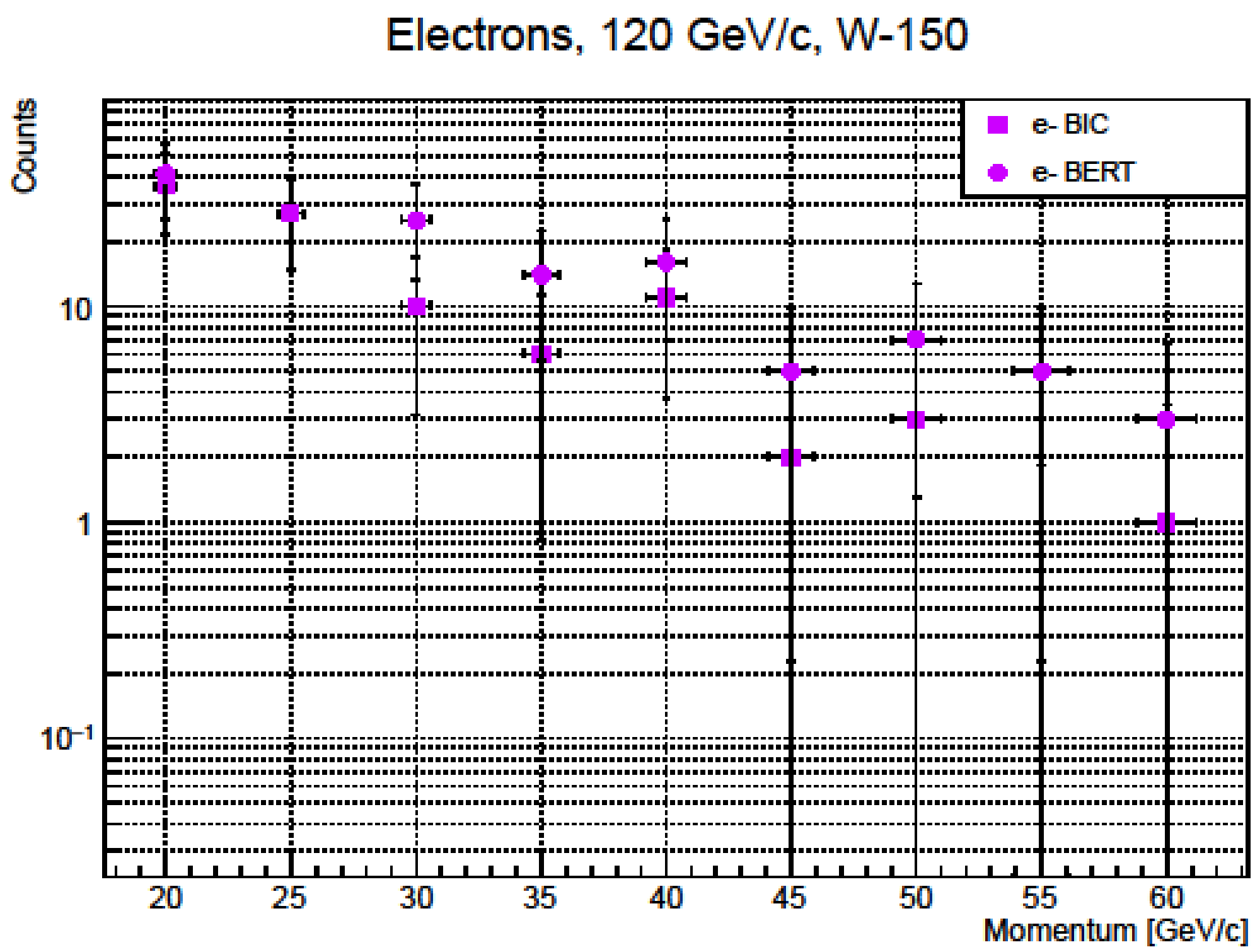
Electrons, 120 GeV/c, W-150
e- BIC
e- BERT
Counts
10
1
$10^{-1}$
20
25
30
35
40
45
50
55
60
Momentum [GeV/c]

Electrons, 150 GeV/c, Cu-100

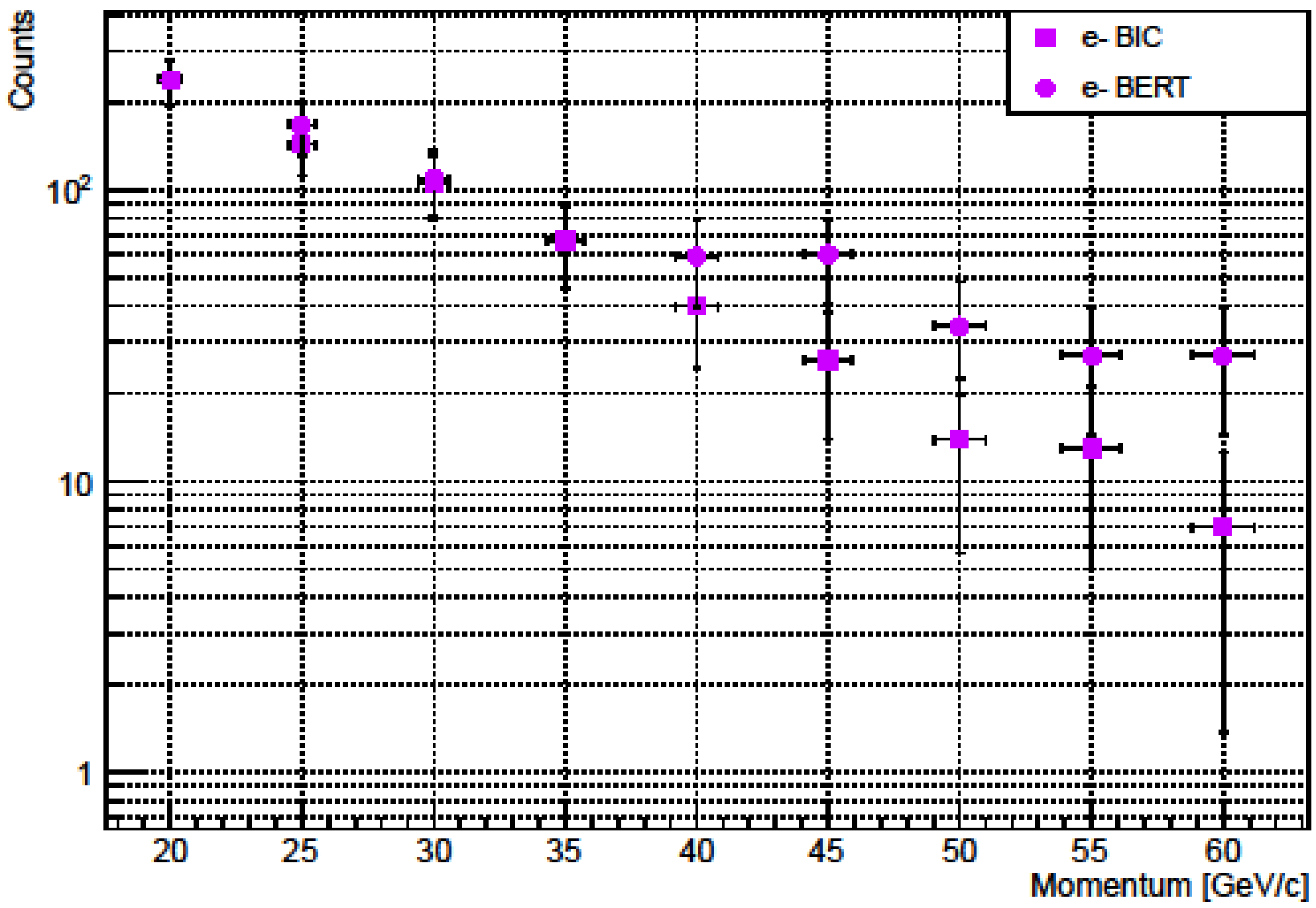

Electrons, 150 GeV/c, Cu-300

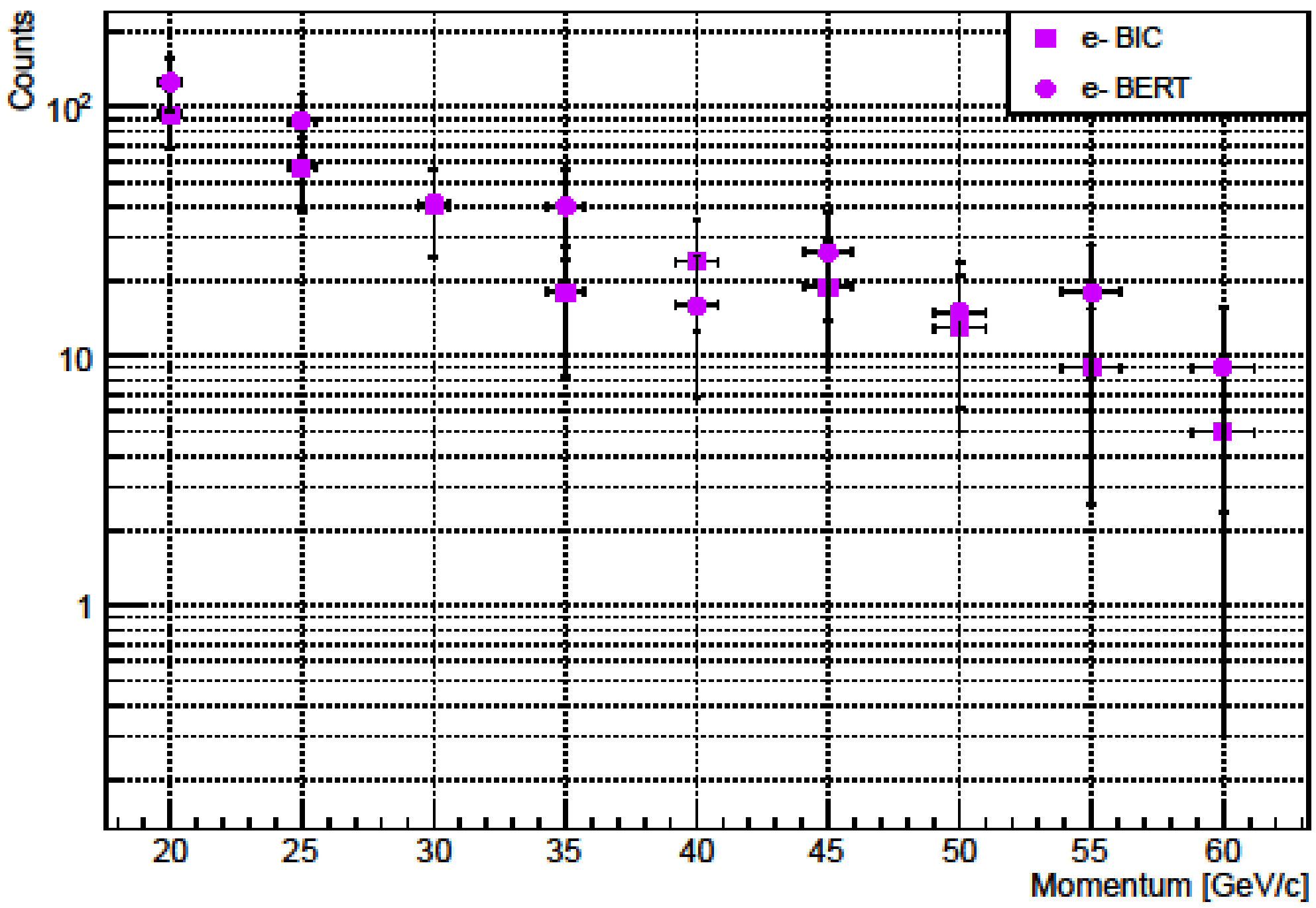

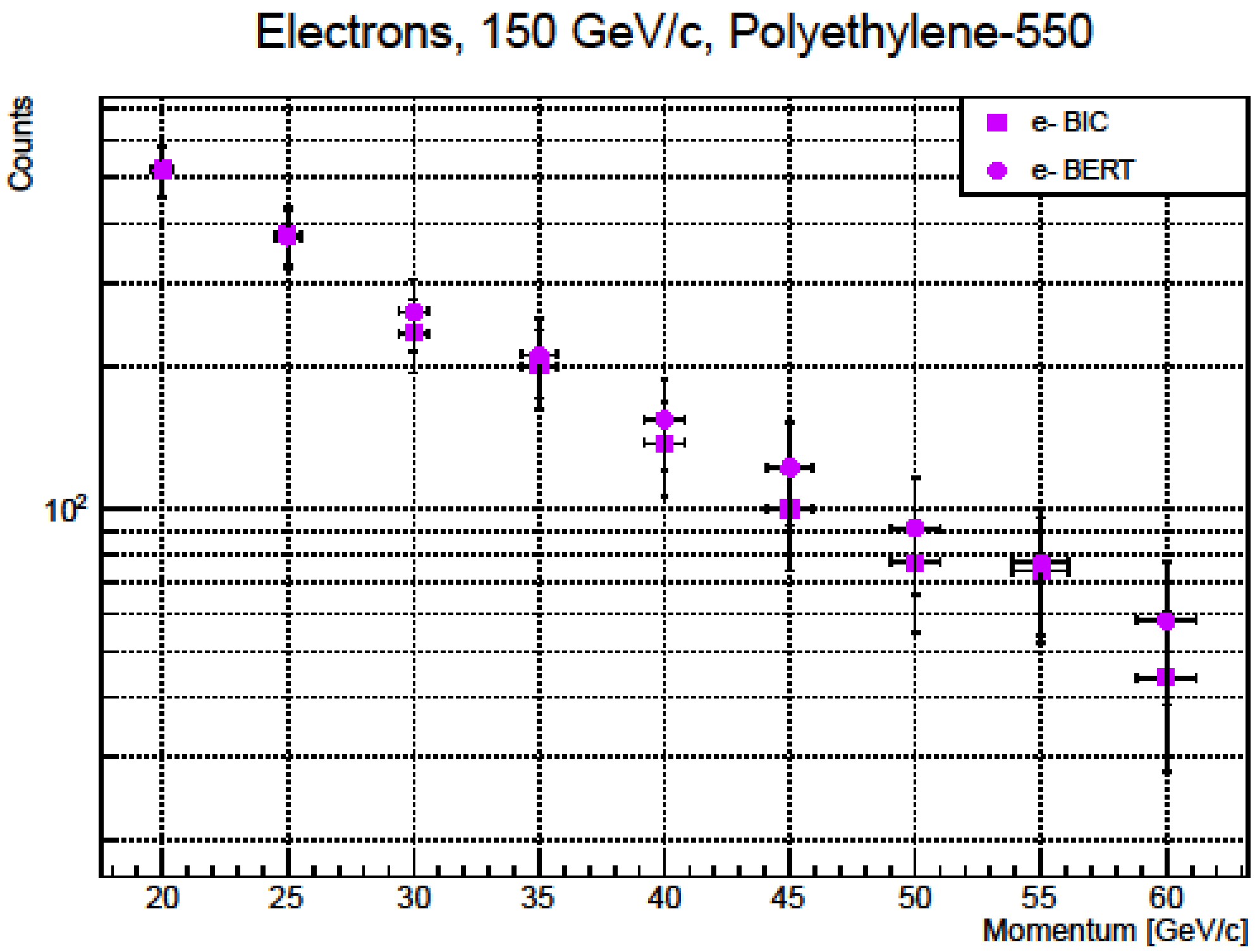
Electrons, 150 GeV/c, Polyethylene-550
Counts
e- BIC
e- BERT
$10^2$
20
25
30
35
40
45
50
55
60
Momentum [GeV/c]

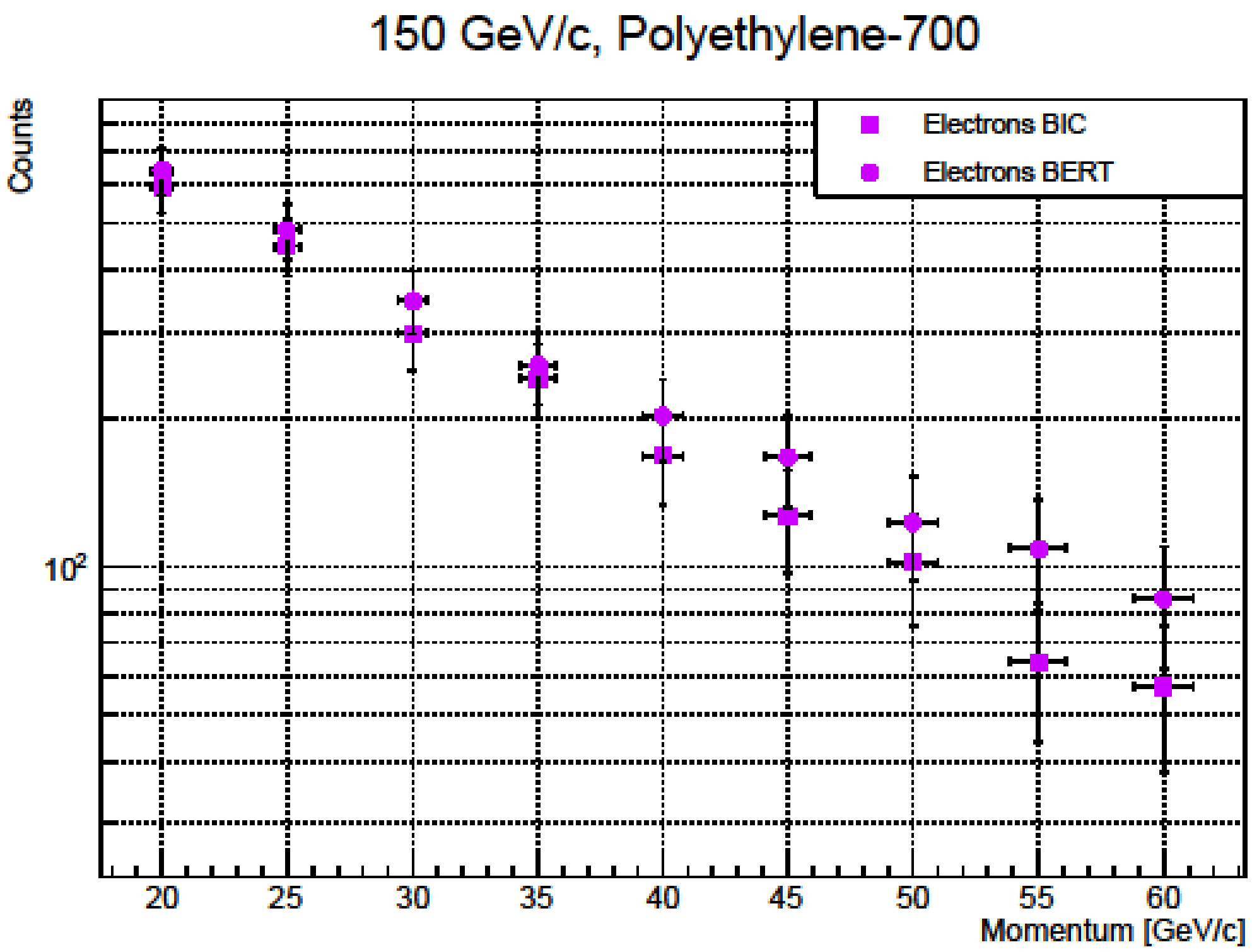
150 GeV/c, Polyethylene-700
Counts
Electrons BIC
Electrons BERT
$10^2$
20
25
30
35
40
45
50
55
60
Momentum [GeV/c]

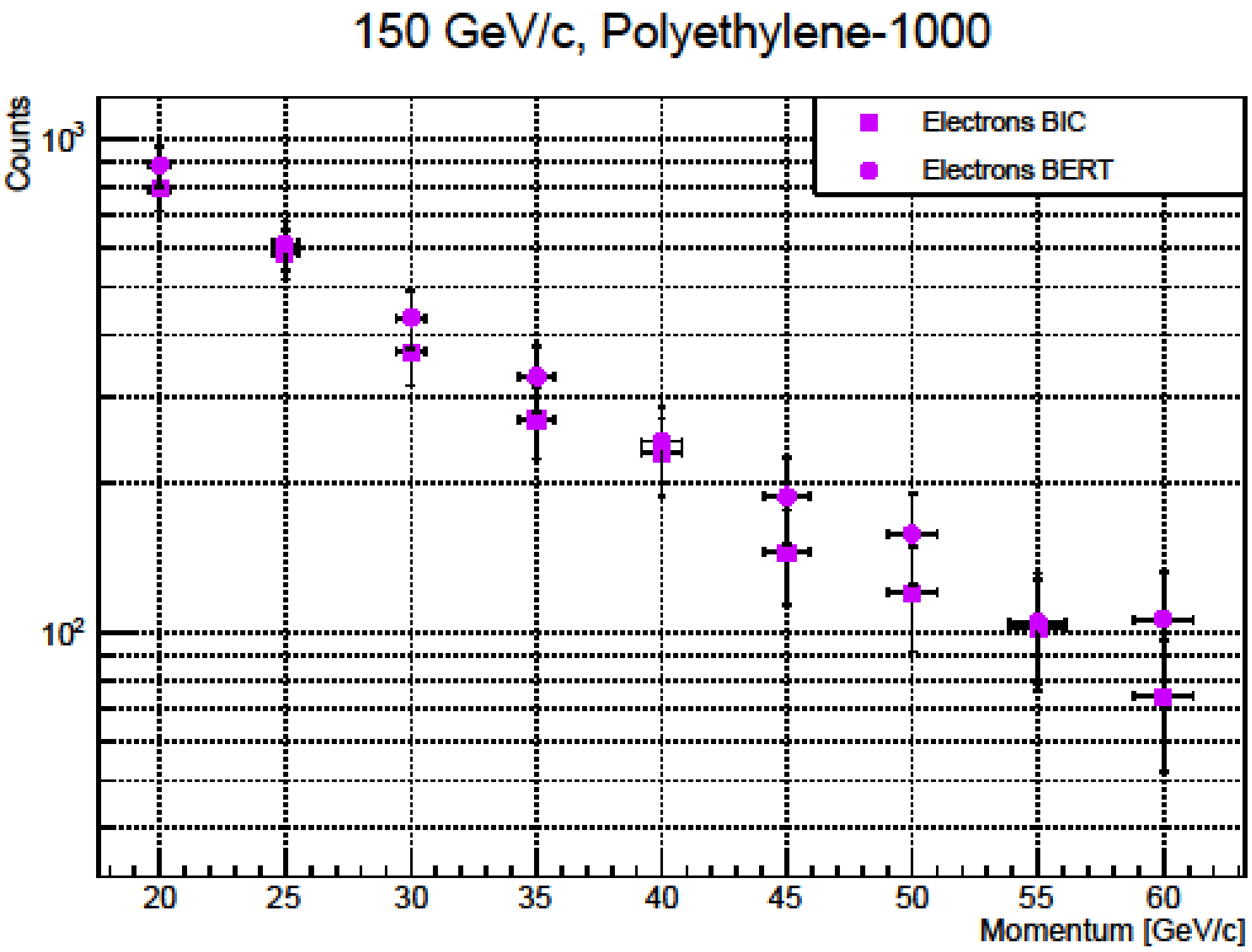
150 GeV/c, Polyethylene-1000
Counts
$10^3$
$10^2$
Electrons BIC
Electrons BERT
20
25
30
35
40
45
50
55
60
Momentum [GeV/c]

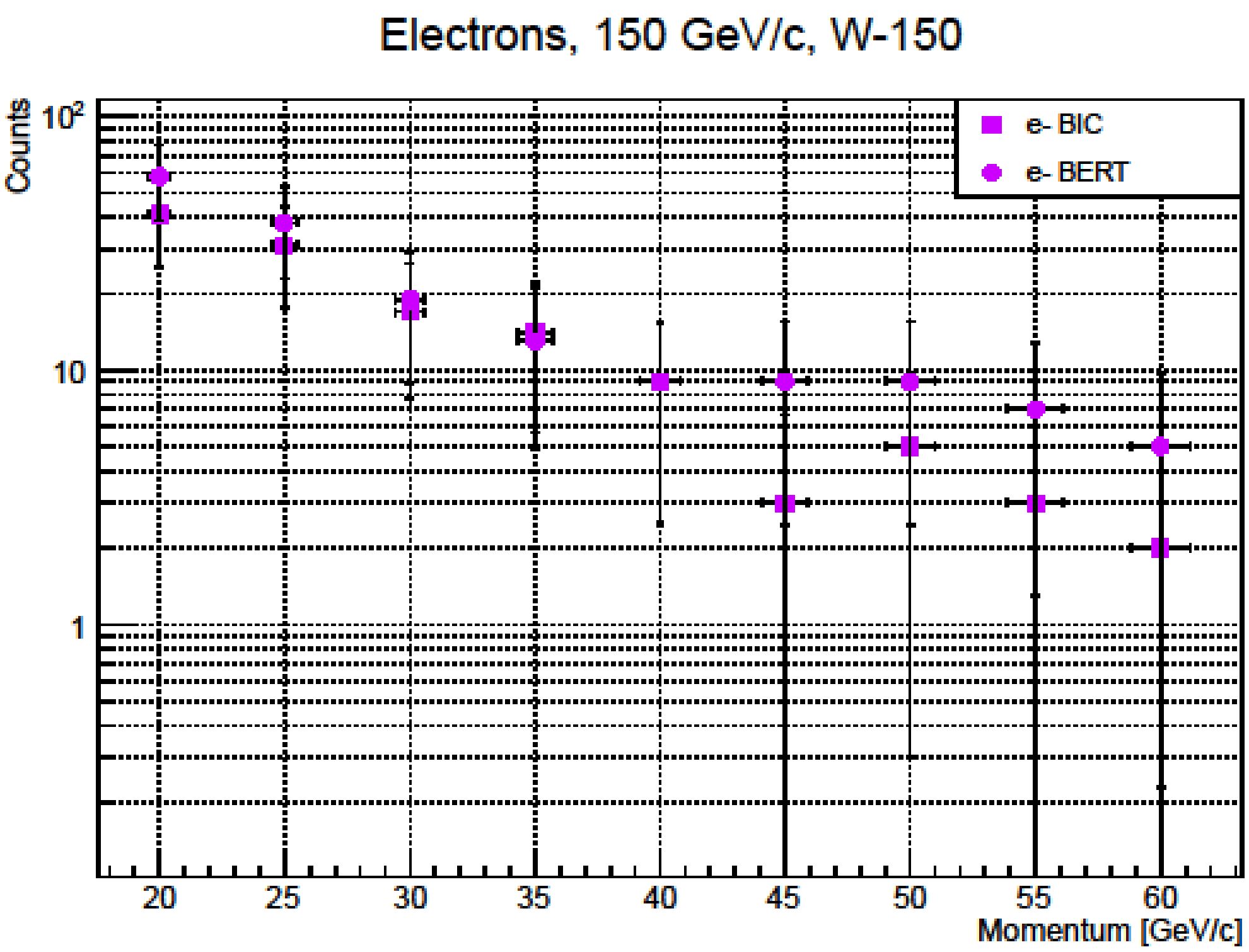
Electrons, 150 GeV/c, W-150
Counts
$10^2$
10
1
e- BIC
e- BERT
20
25
30
35
40
45
50
55
60
Momentum [GeV/c]

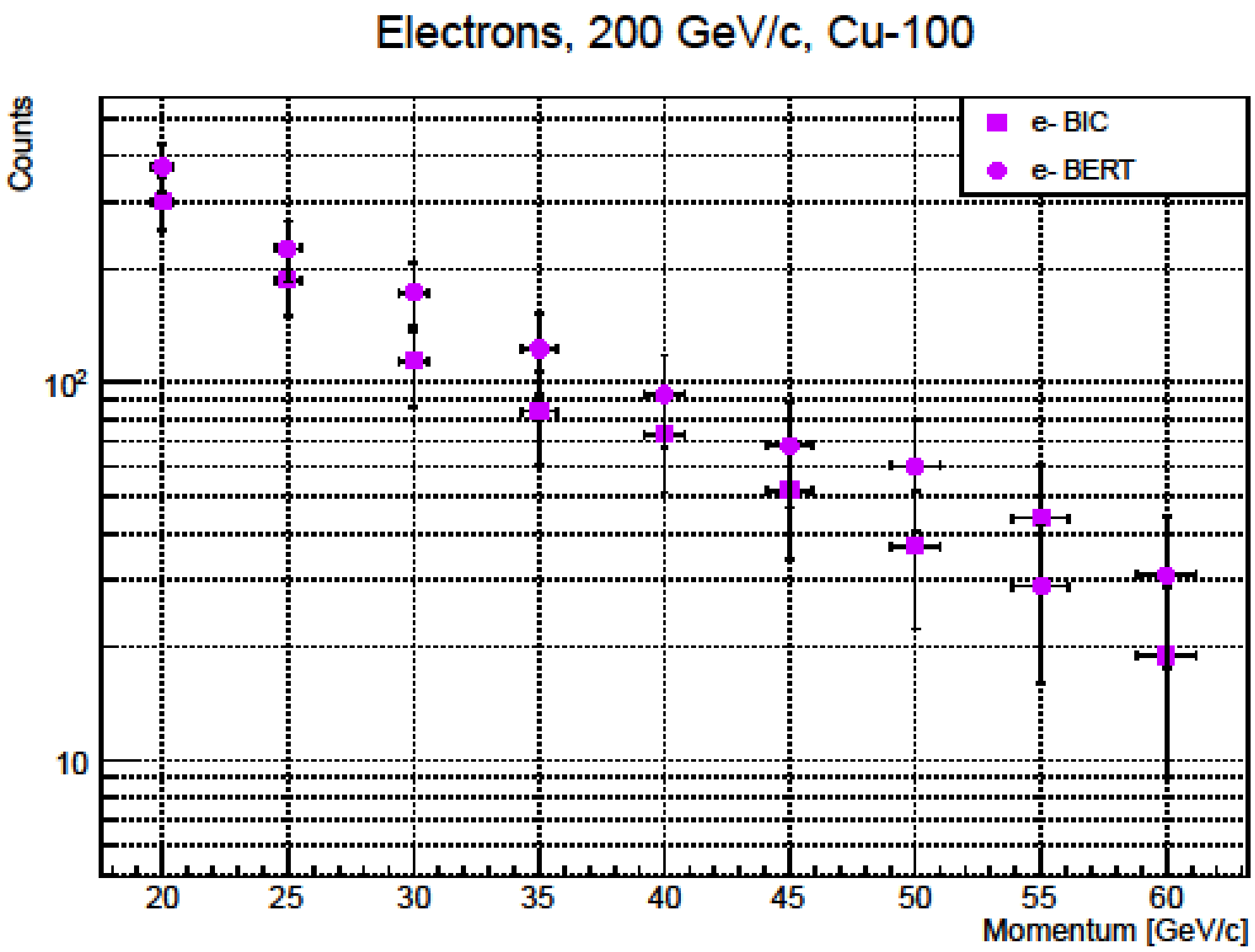

Electrons, 200 GeV/c, Cu-100
Counts
$10^2$
10
20
25
30
35
40
45
50
55
60
Momentum [GeV/c]
e- BIC
e- BERT

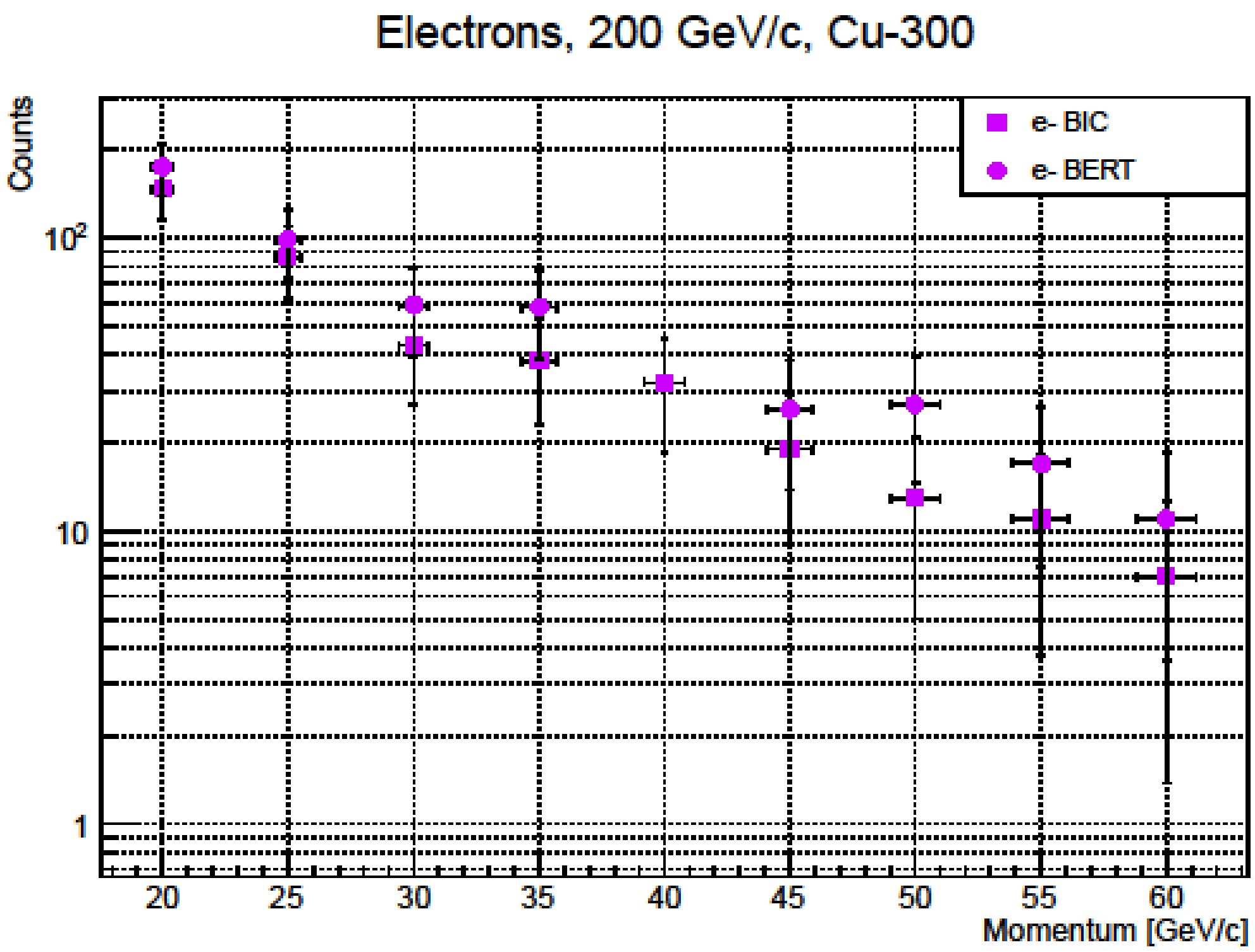

Electrons, 200 GeV/c, Cu-300
Counts
$10^2$
10
1
20
25
30
35
40
45
50
55
60
Momentum [GeV/c]
e- BIC
e- BERT

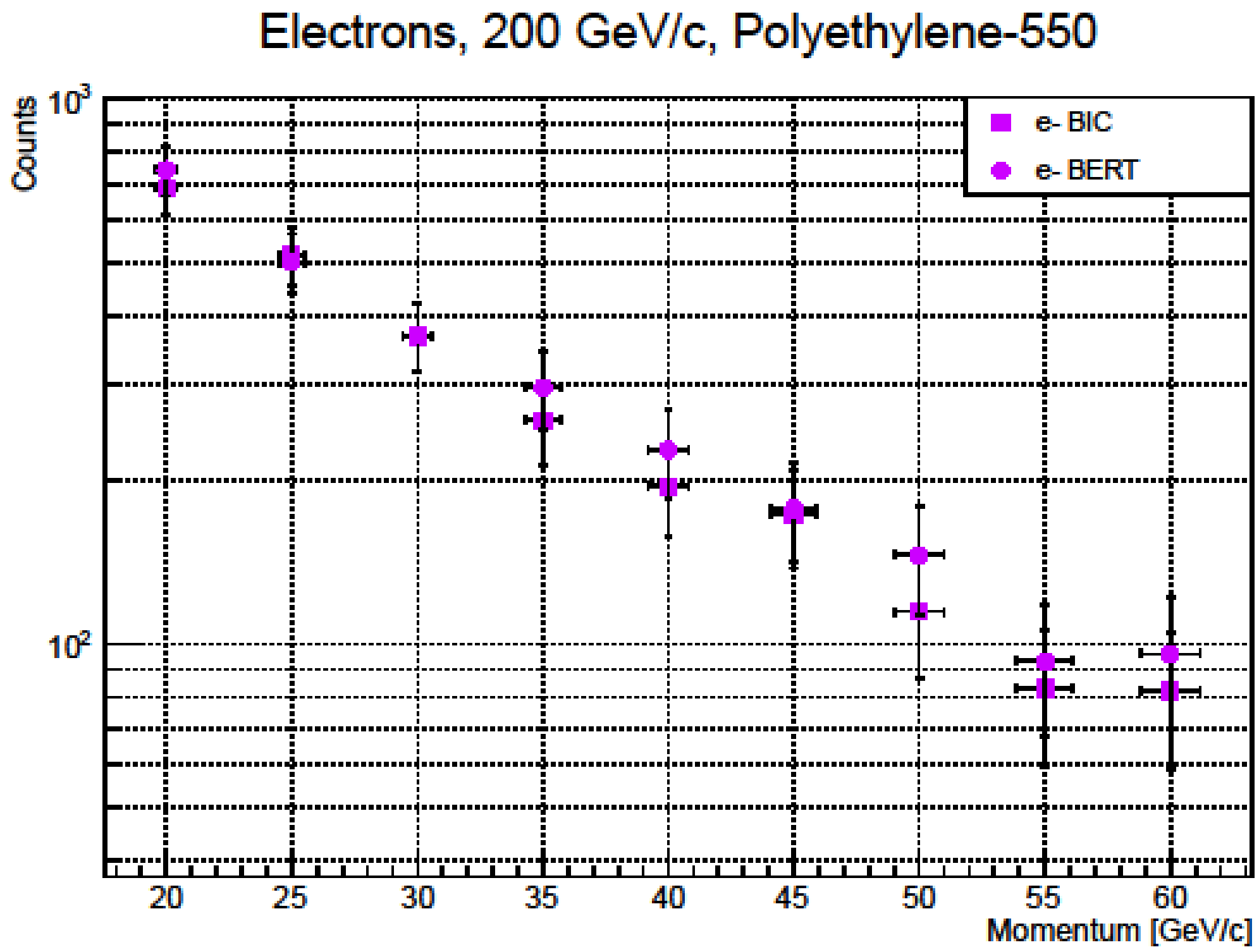
Electrons, 200 GeV/c, Polyethylene-550
Counts
$10^3$
$10^2$
e- BIC
e- BERT
20
25
30
35
40
45
50
55
60
Momentum [GeV/c]

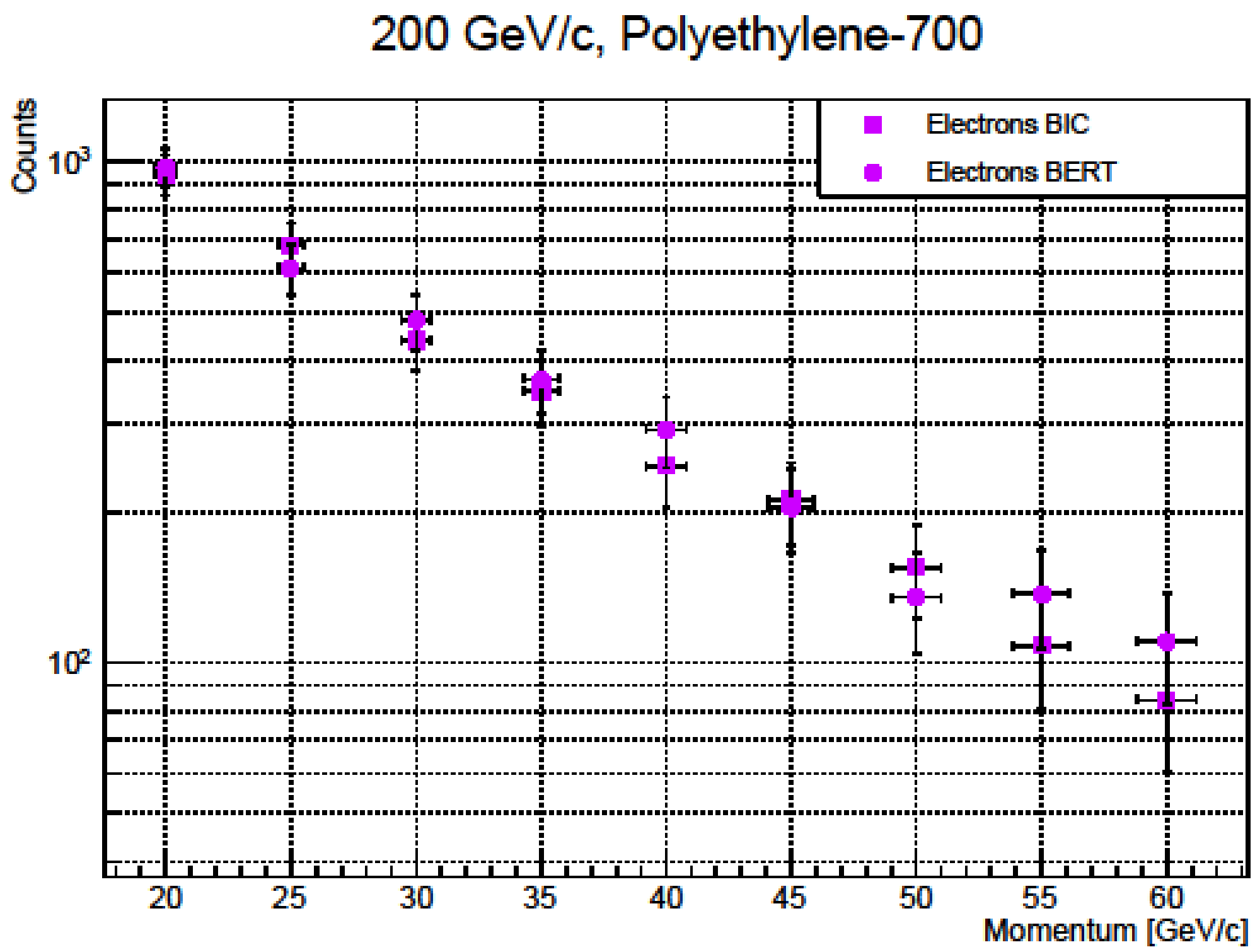
200 GeV/c, Polyethylene-700
Counts
$10^3$
$10^2$
Electrons BIC
Electrons BERT
20
25
30
35
40
45
50
55
60
Momentum [GeV/c]

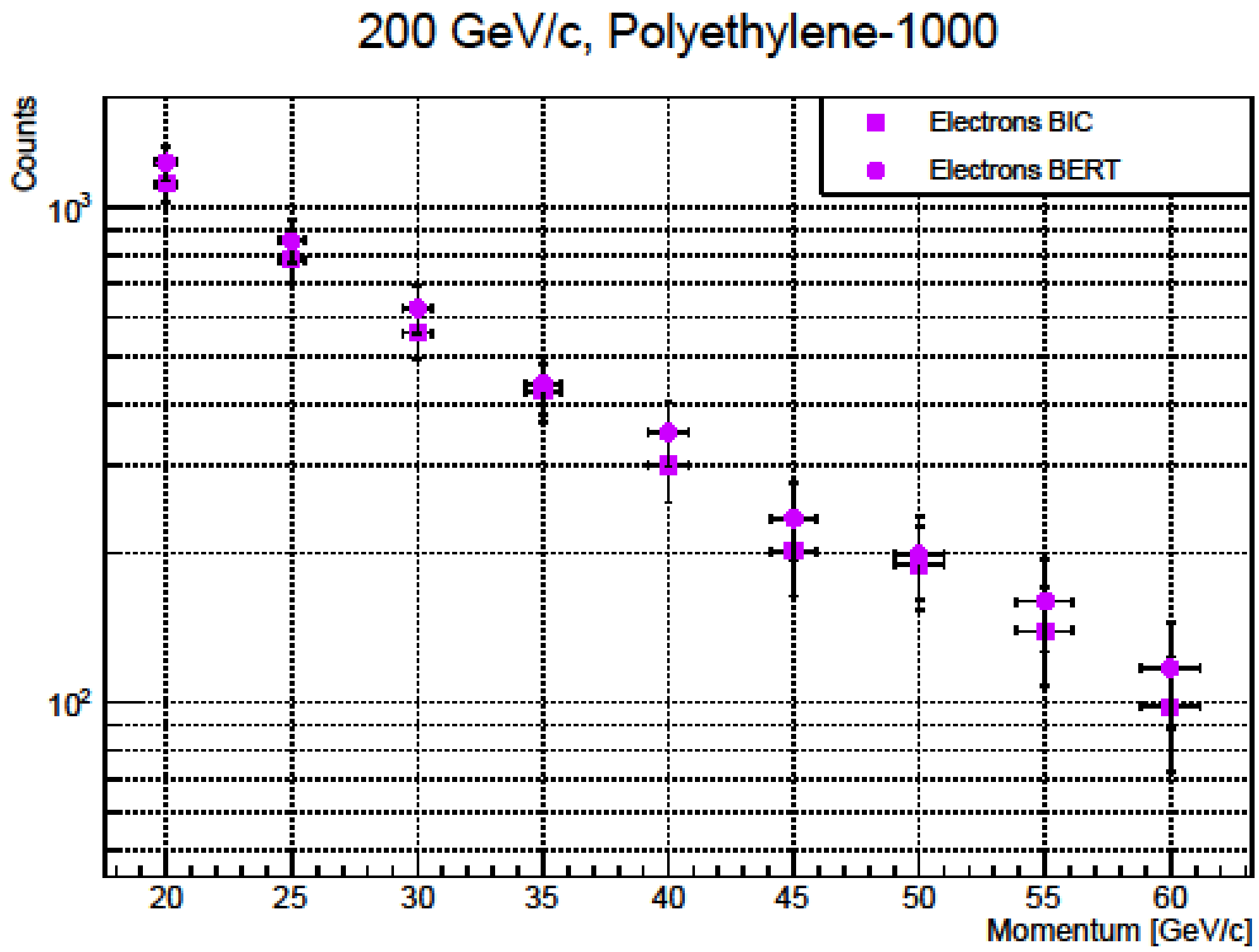

200 GeV/c, Polyethylene-1000
Electrons BIC
Electrons BERT
Counts
$10^3$
$10^2$
20
25
30
35
40
45
50
55
60
Momentum [GeV/c]

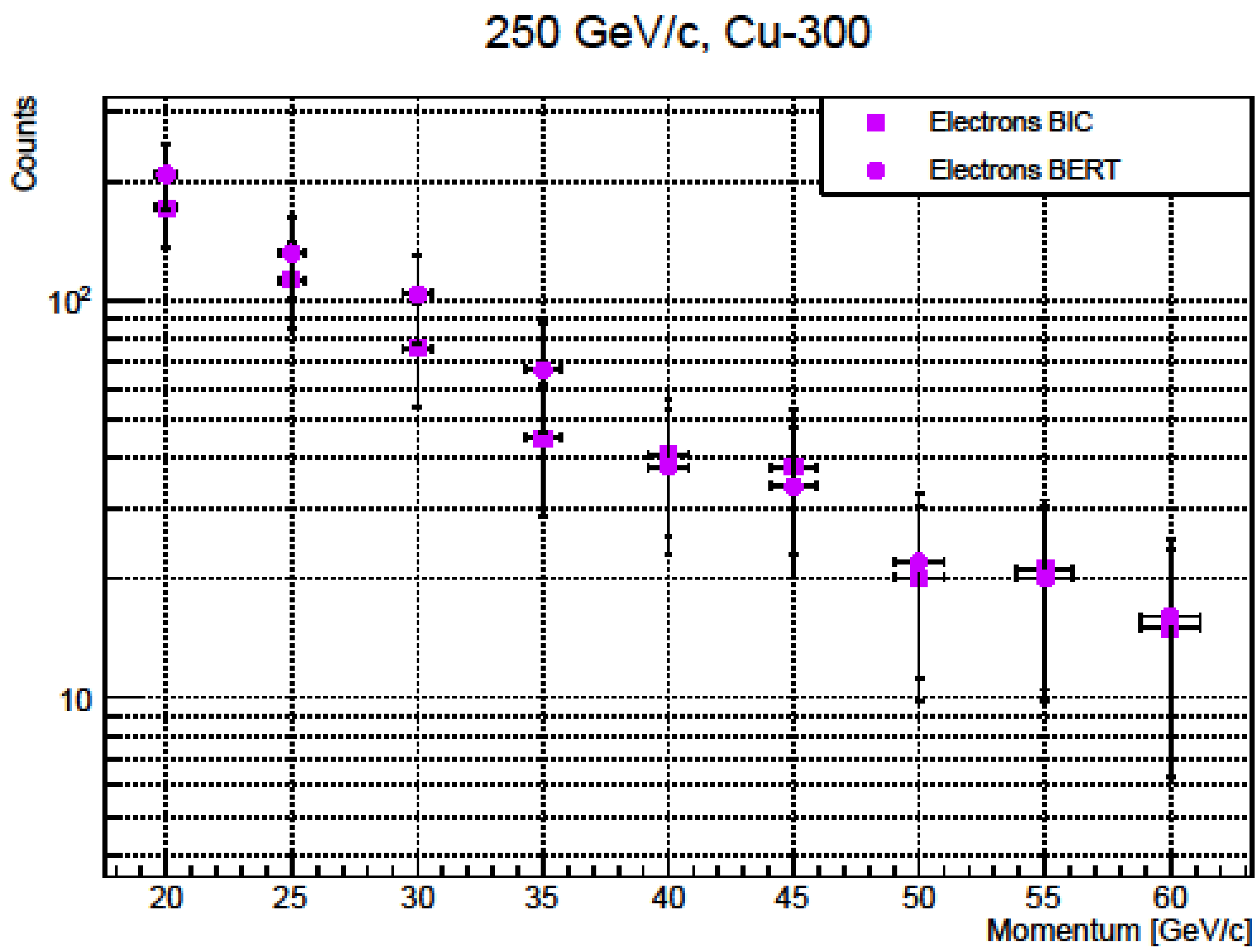

250 GeV/c, Cu-300
Electrons BIC
Electrons BERT
Counts
$10^2$
10
20
25
30
35
40
45
50
55
60
Momentum [GeV/c]

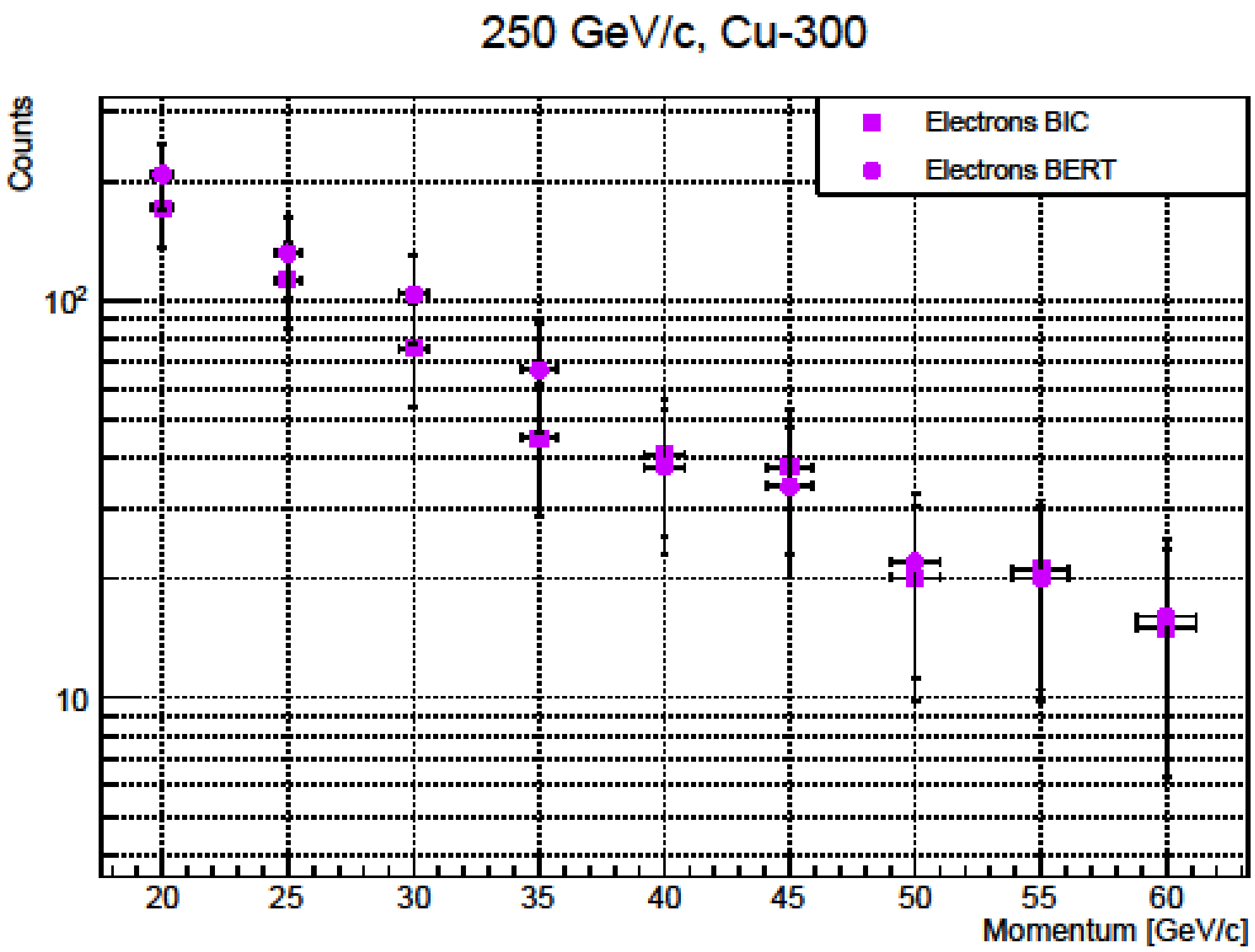
250 GeV/c, Cu-300
Counts
Electrons BIC
Electrons BERT
10
$10^2$
20
25
30
35
40
45
50
55
60
Momentum [GeV/c]

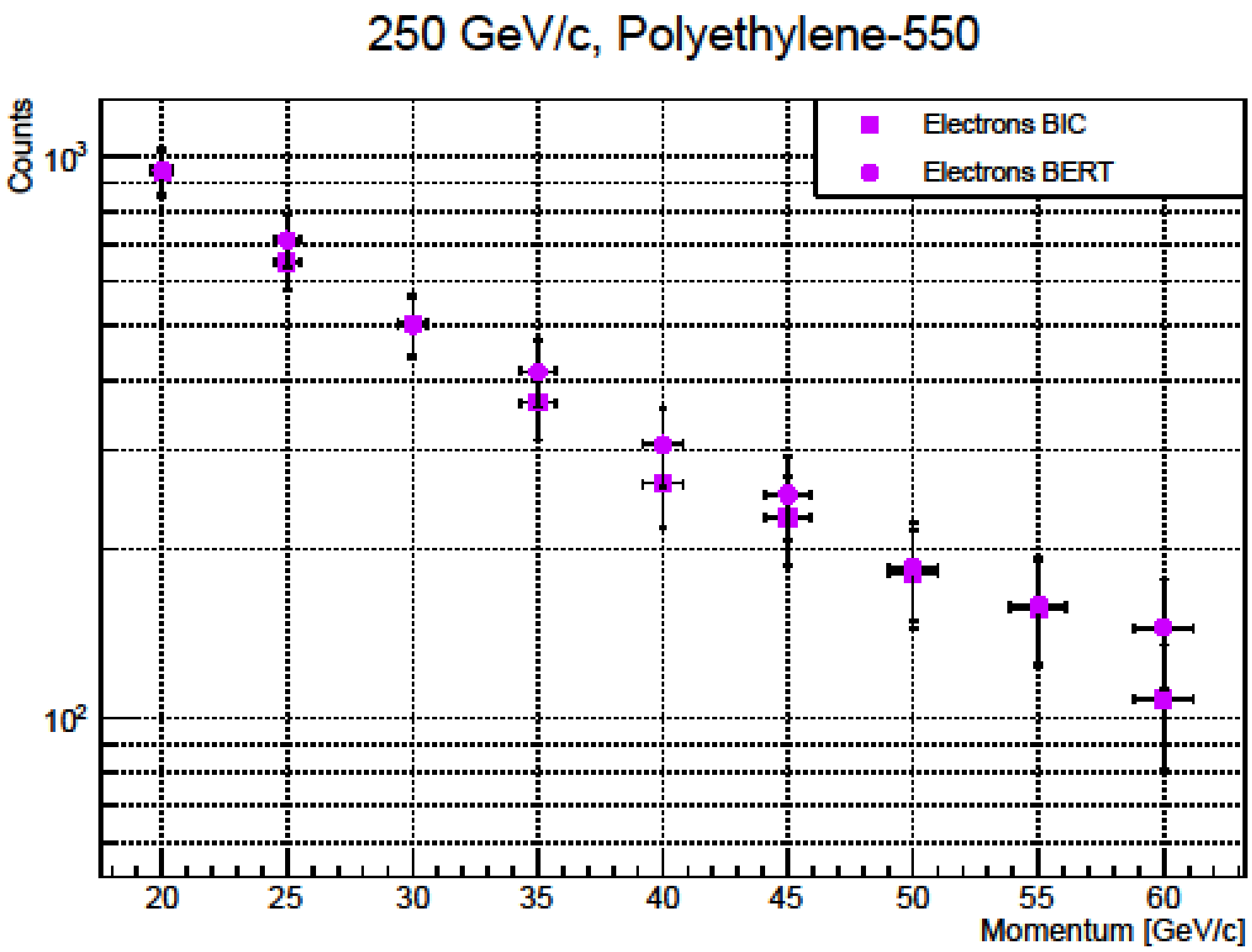
250 GeV/c, Polyethylene-550
Counts
Electrons BIC
Electrons BERT
$10^2$
$10^3$
20
25
30
35
40
45
50
55
60
Momentum [GeV/c]

250 GeV/c, Polyethylene-700

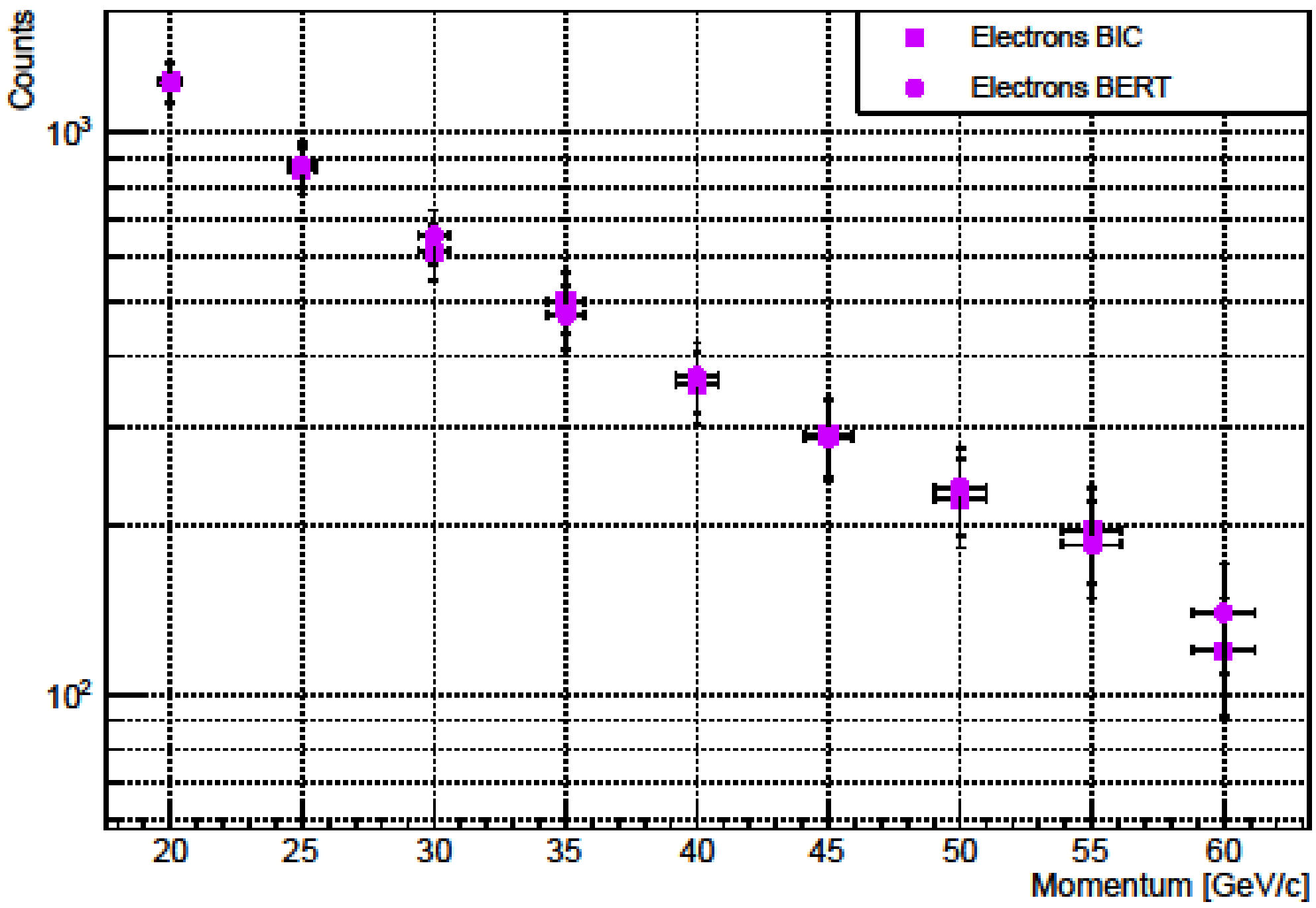

250 GeV/c, Polyethylene-1000

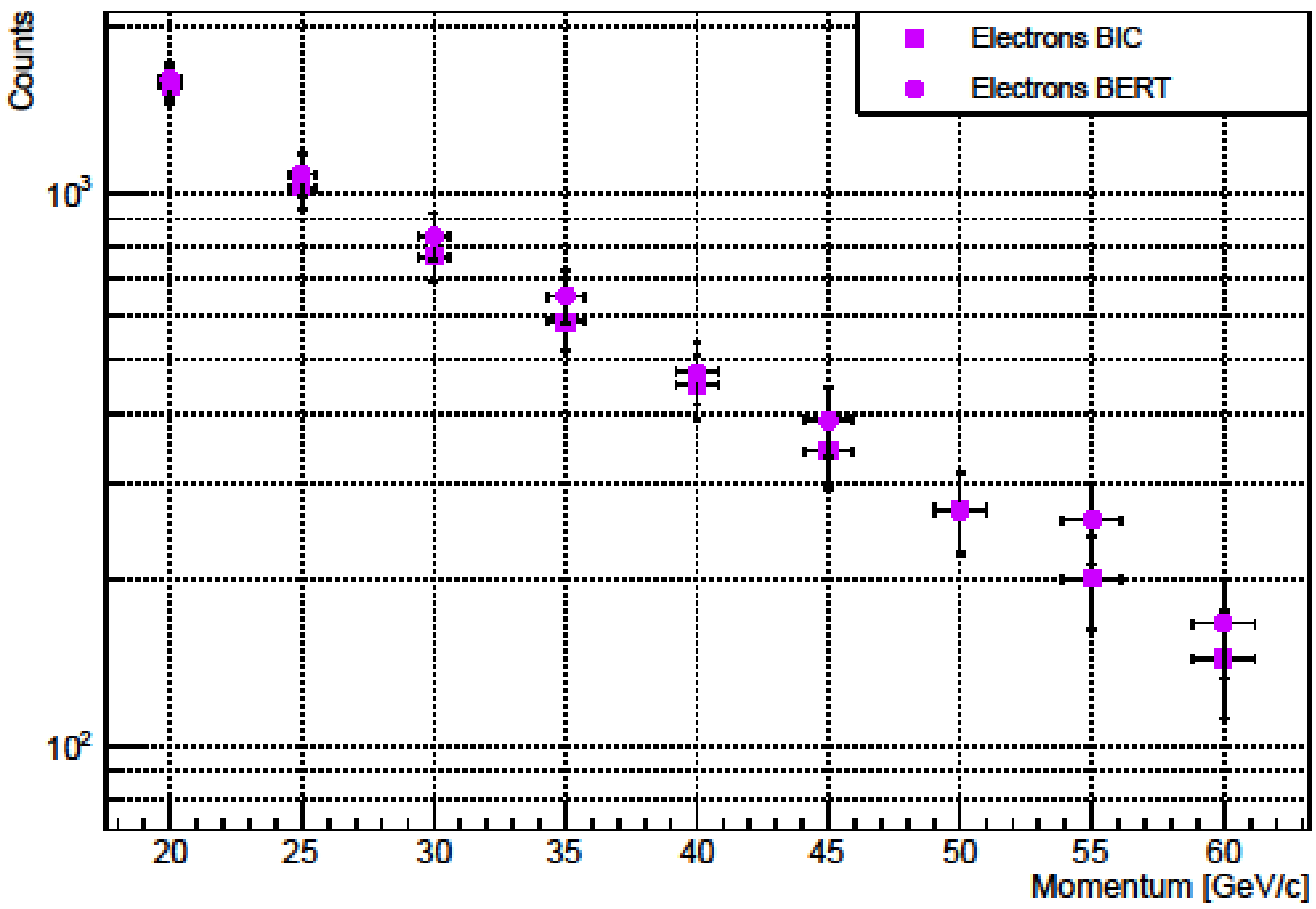

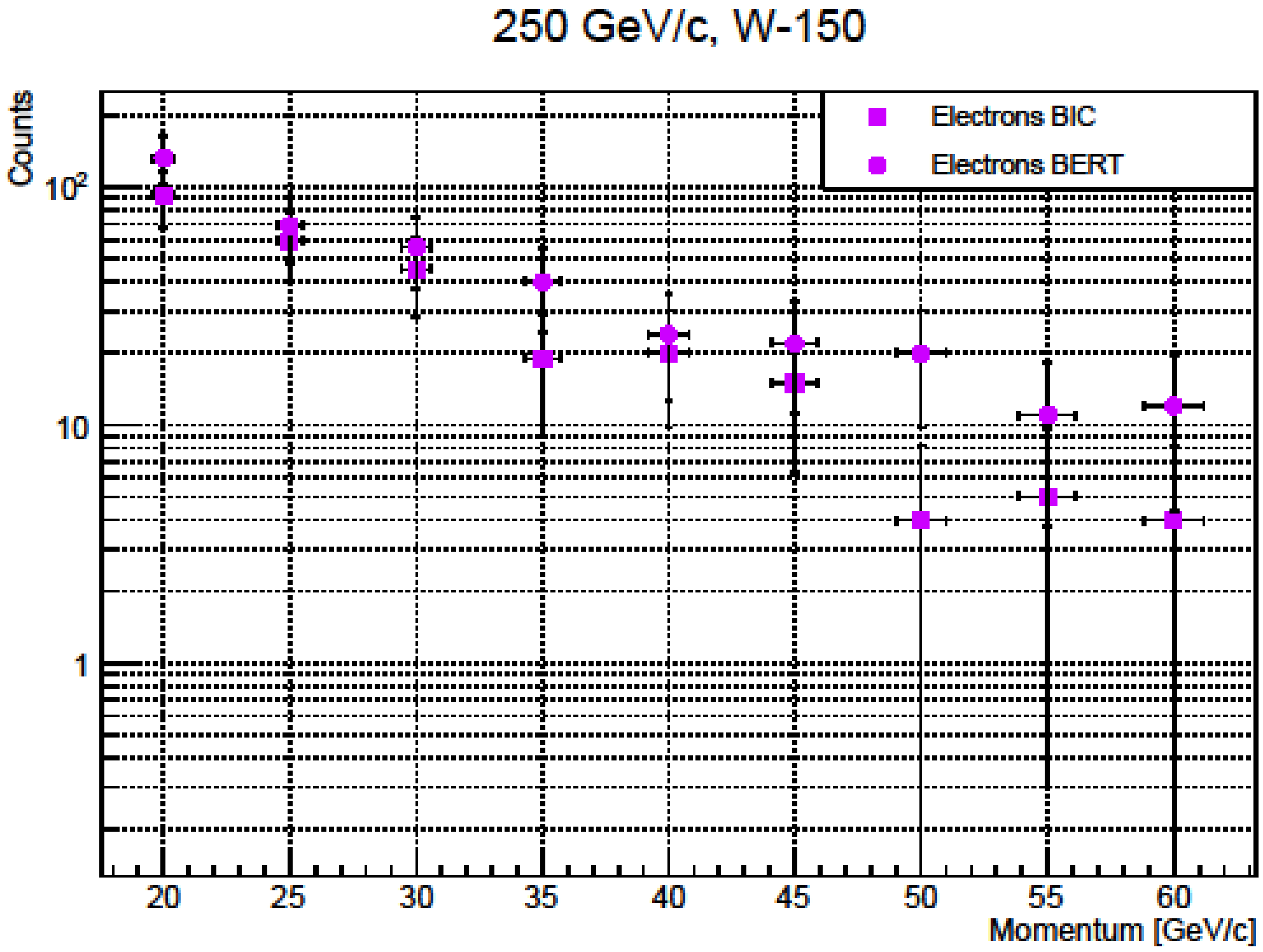

## 2.4 Ratio p/pi

In this section the ratio of protons over pions has been calculated. As already seen from section 2.1 and 2.2, the ratio is small for low secondary momenta. But from secondary momenta higher than 100 GeV/c the ratio starts to increase linearly as a function of the tertiary momentum. For high enough secondary momenta (≥120 GeV/c) and tertiary momenta (≥50 GeV/c) the beam in general contains more protons than pions, at least with QGSP_BIC. Significant differences are observed between the two physics lists, especially for momenta < 30 GeV/c.

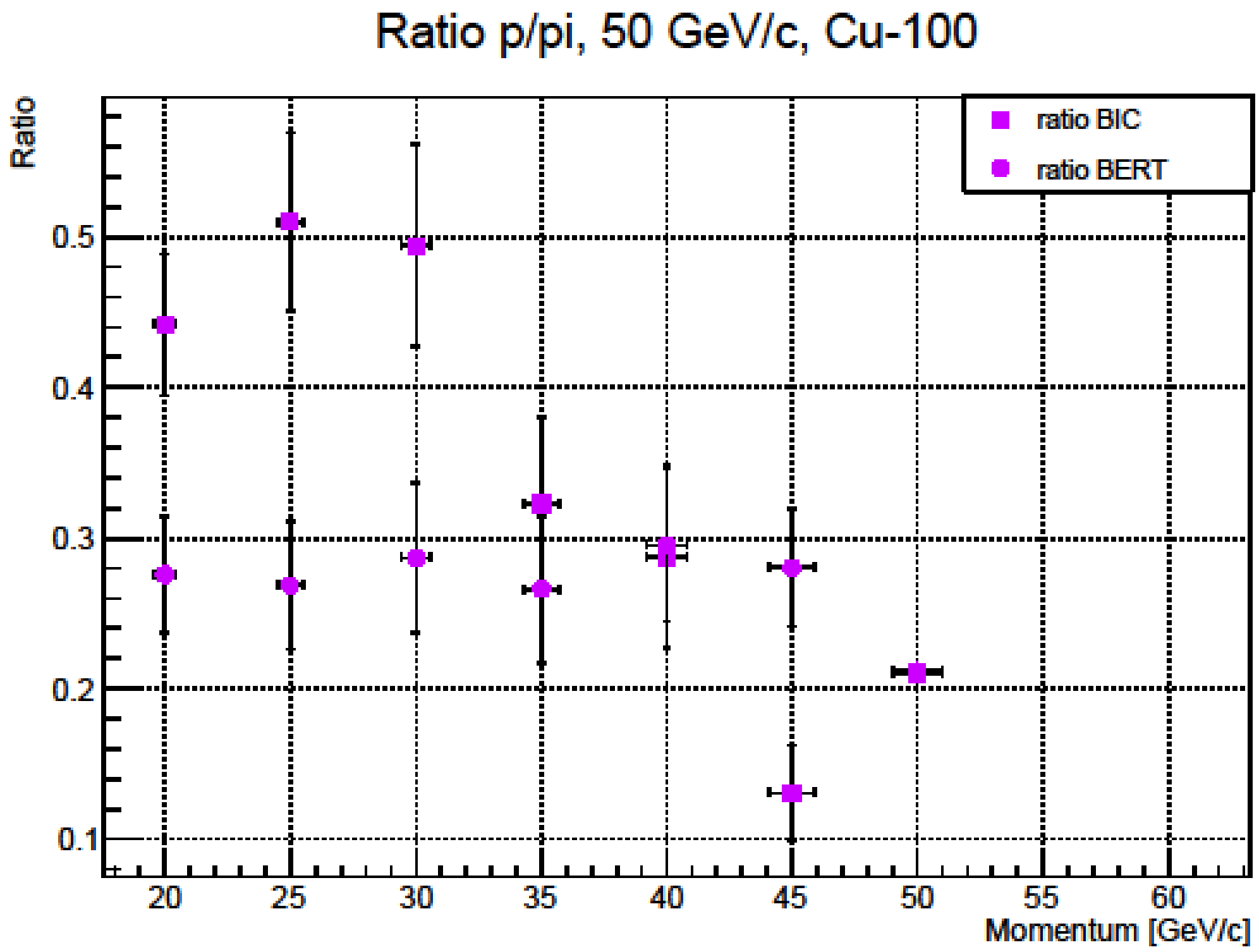
Ratio p/pi, 50 GeV/c, Cu-100
Ratio
0.5
0.4
0.3
0.2
0.1
20
25
30
35
40
45
50
55
60
Momentum [GeV/c]
ratio BIC
ratio BERT

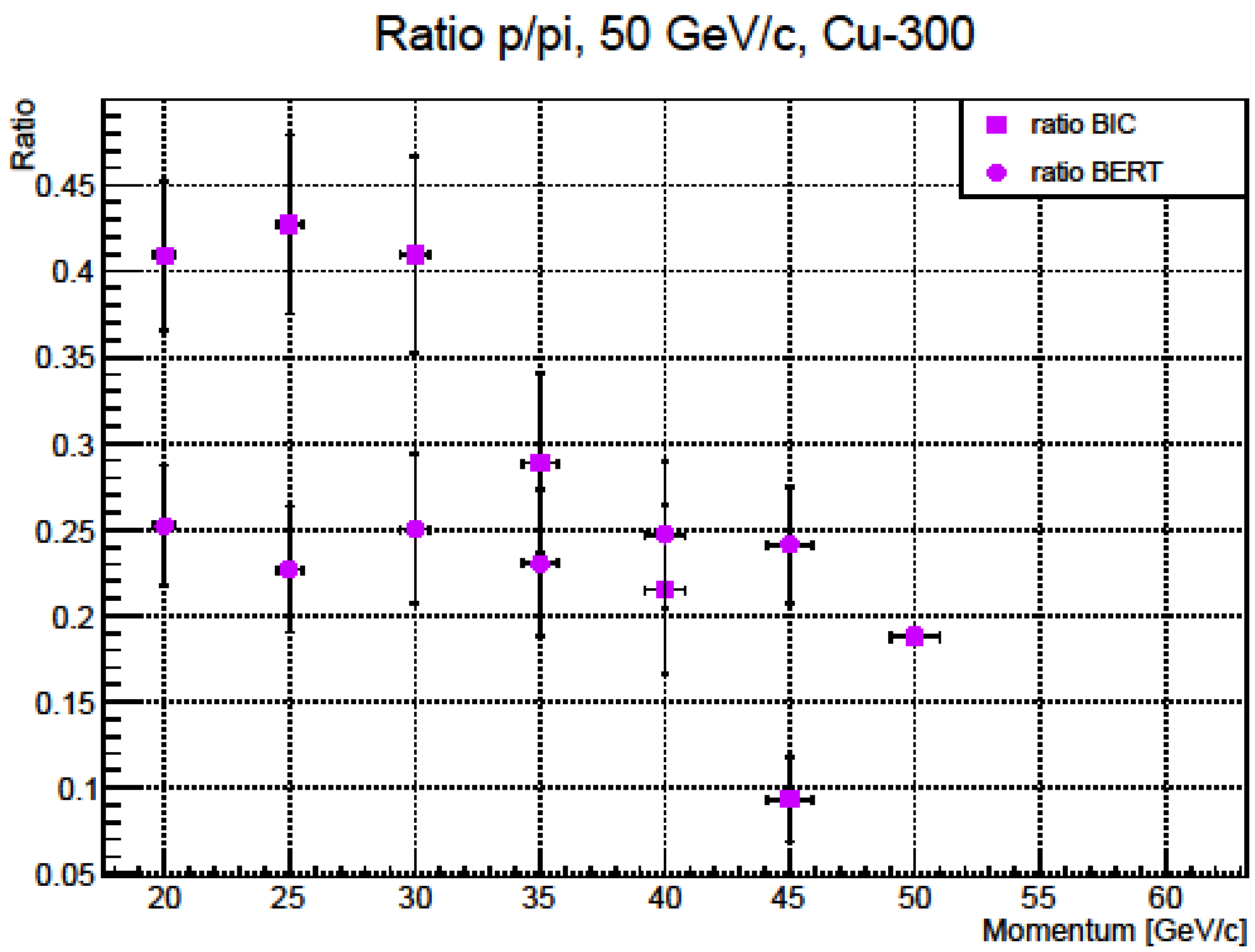
Ratio p/pi, 50 GeV/c, Cu-300
Ratio
0.45
0.4
0.35
0.3
0.25
0.2
0.15
0.1
0.05
20
25
30
35
40
45
50
55
60
Momentum [GeV/c]
ratio BIC
ratio BERT

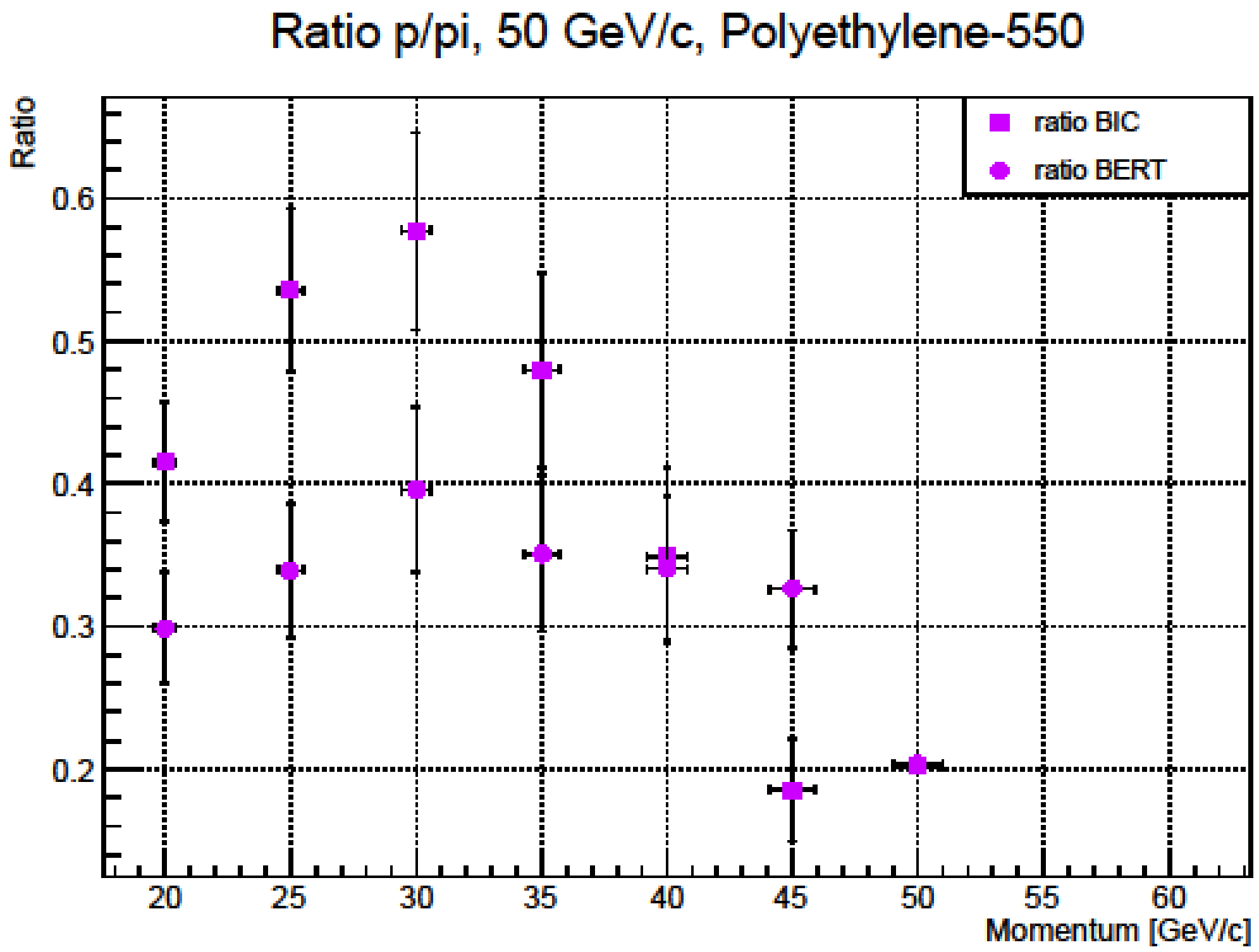
Ratio p/pi, 50 GeV/c, Polyethylene-550
Ratio
Momentum [GeV/c]
ratio BIC
ratio BERT

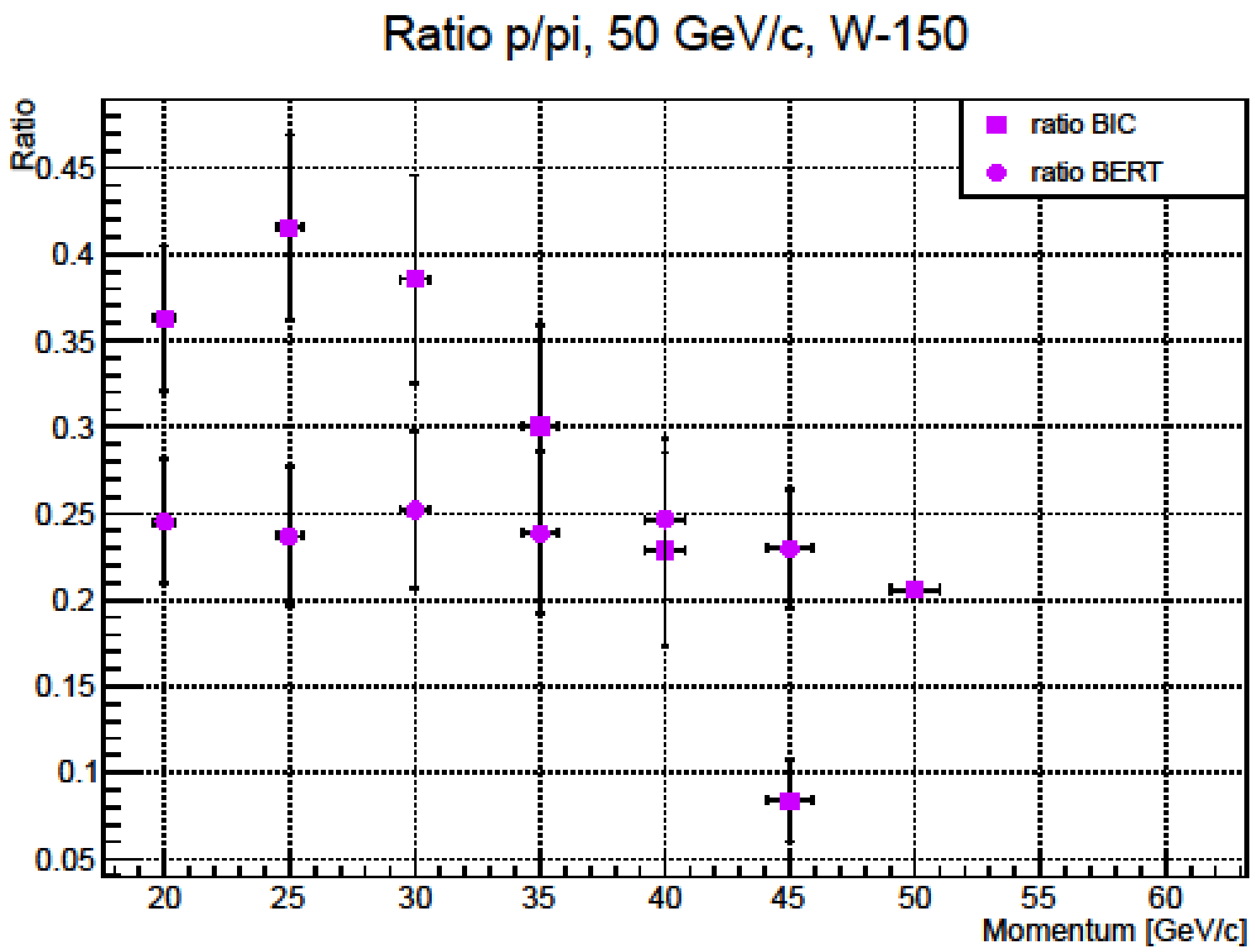
Ratio p/pi, 50 GeV/c, W-150
Ratio
Momentum [GeV/c]
ratio BIC
ratio BERT

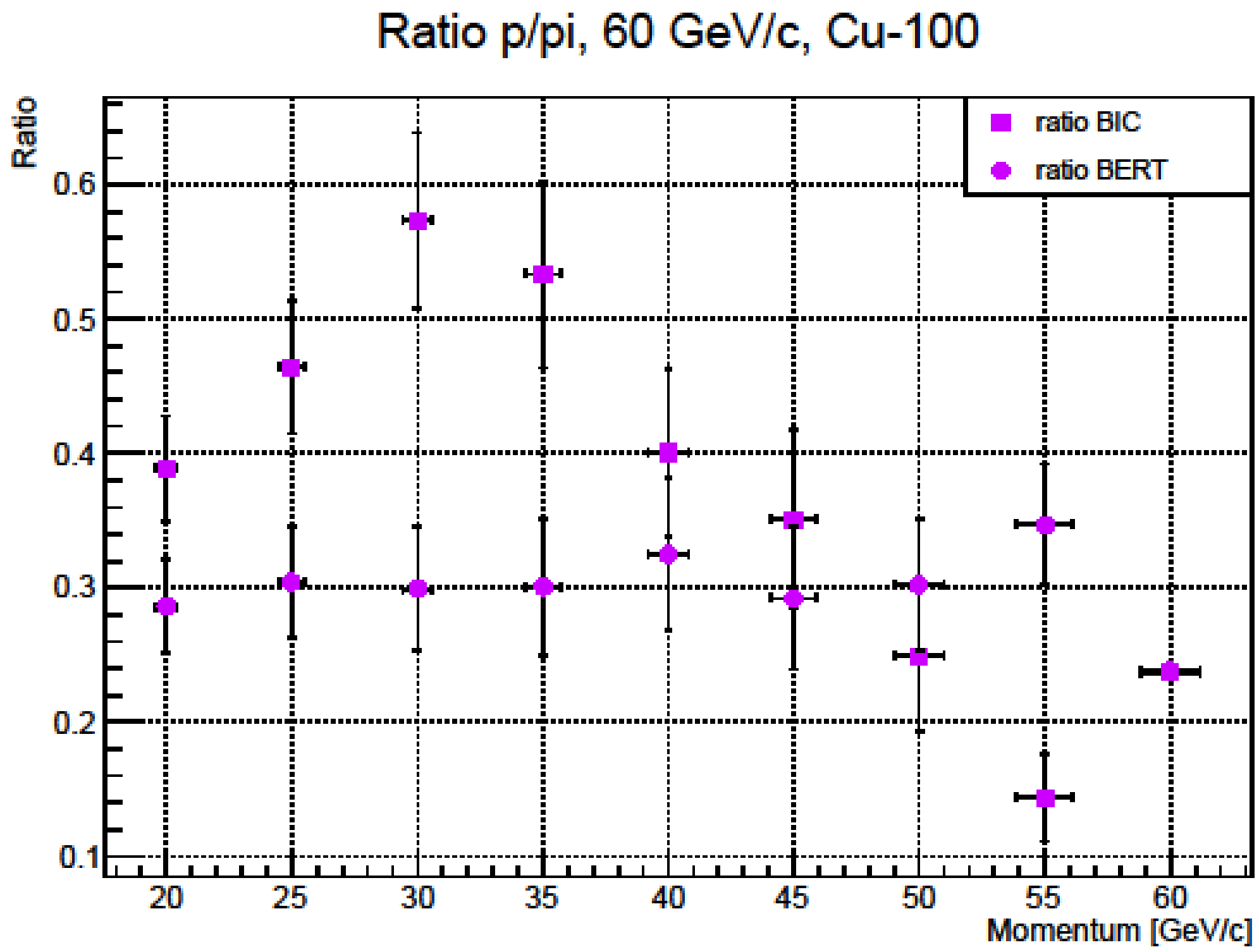
Ratio p/pi, 60 GeV/c, Cu-100
Ratio
Momentum [GeV/c]
ratio BIC
ratio BERT
0.1
0.2
0.3
0.4
0.5
0.6
20
25
30
35
40
45
50
55
60

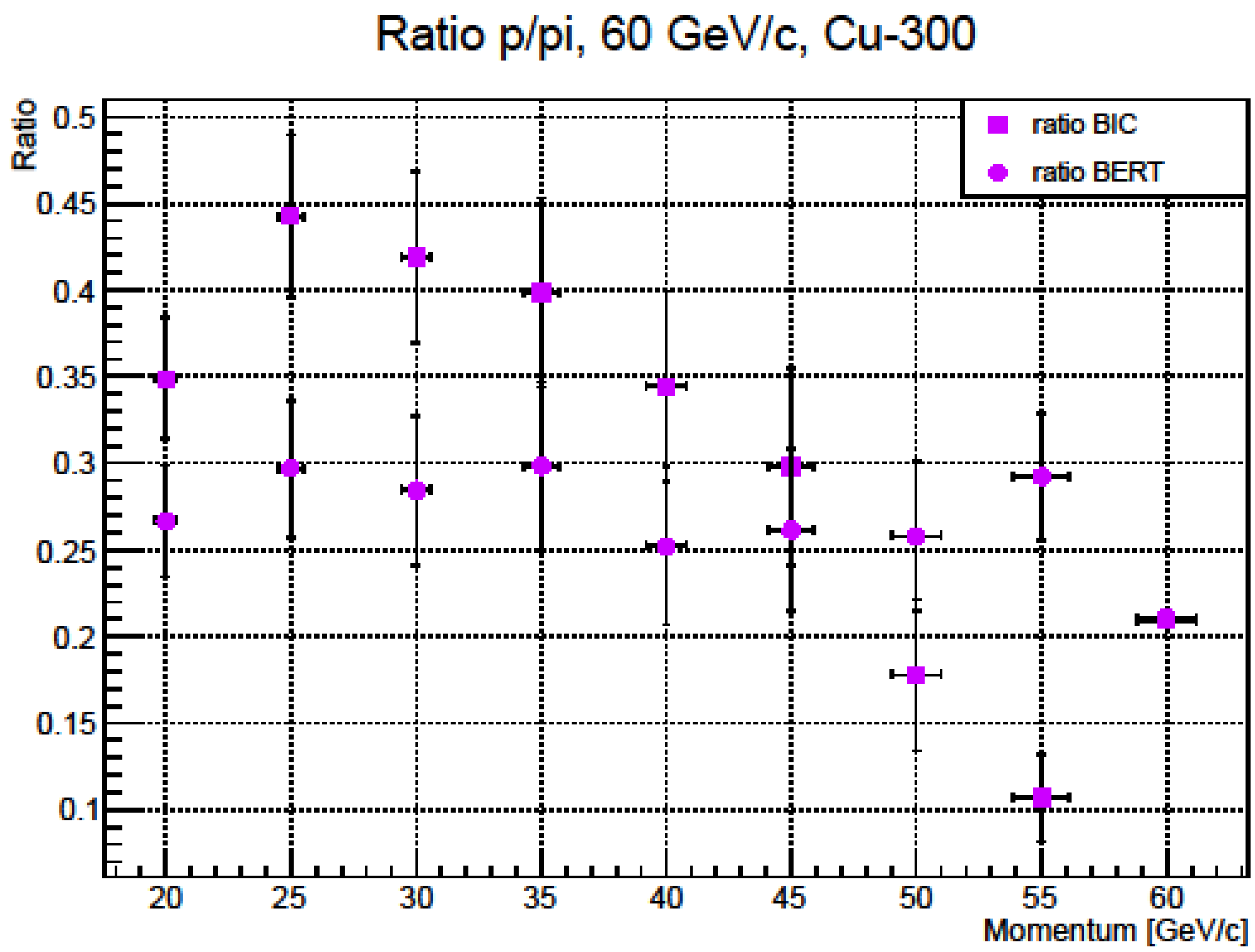
Ratio p/pi, 60 GeV/c, Cu-300
Ratio
Momentum [GeV/c]
ratio BIC
ratio BERT
0.1
0.15
0.2
0.25
0.3
0.35
0.4
0.45
0.5
20
25
30
35
40
45
50
55
60

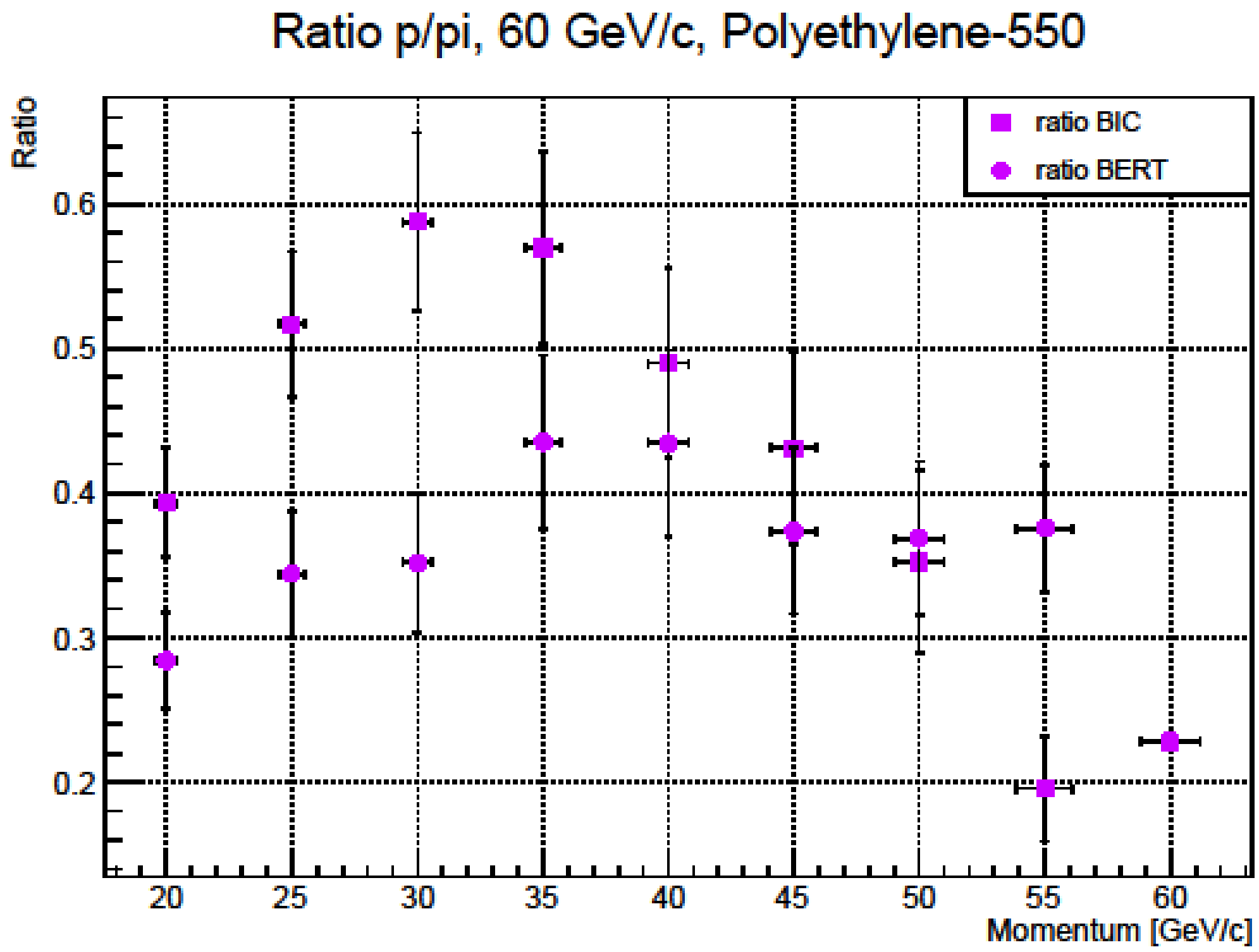
Ratio p/pi, 60 GeV/c, Polyethylene-550
Ratio
0.6
0.5
0.4
0.3
0.2
20
25
30
35
40
45
50
55
60
Momentum [GeV/c]
ratio BIC
ratio BERT

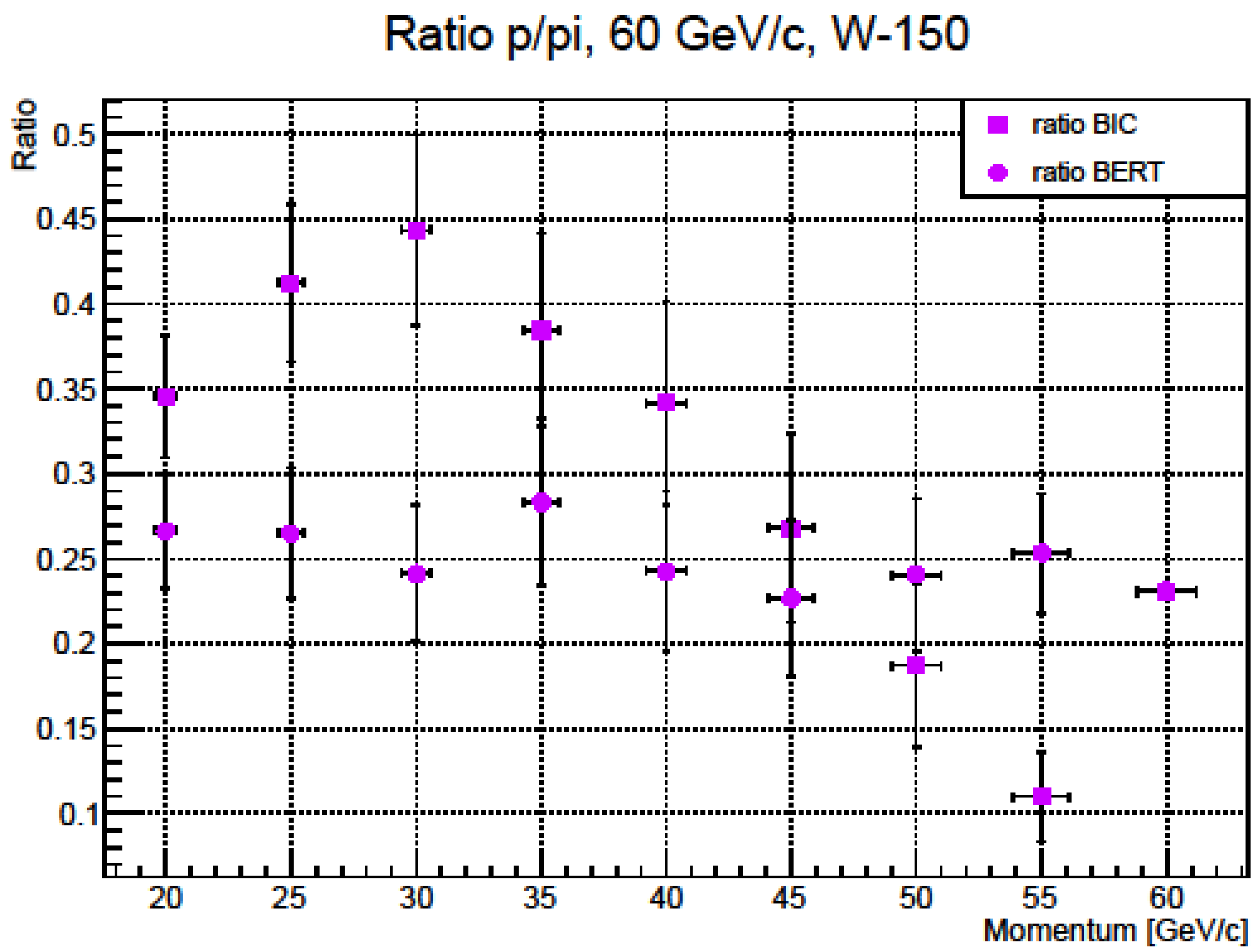
Ratio p/pi, 60 GeV/c, W-150
Ratio
0.5
0.45
0.4
0.35
0.3
0.25
0.2
0.15
0.1
20
25
30
35
40
45
50
55
60
Momentum [GeV/c]
ratio BIC
ratio BERT

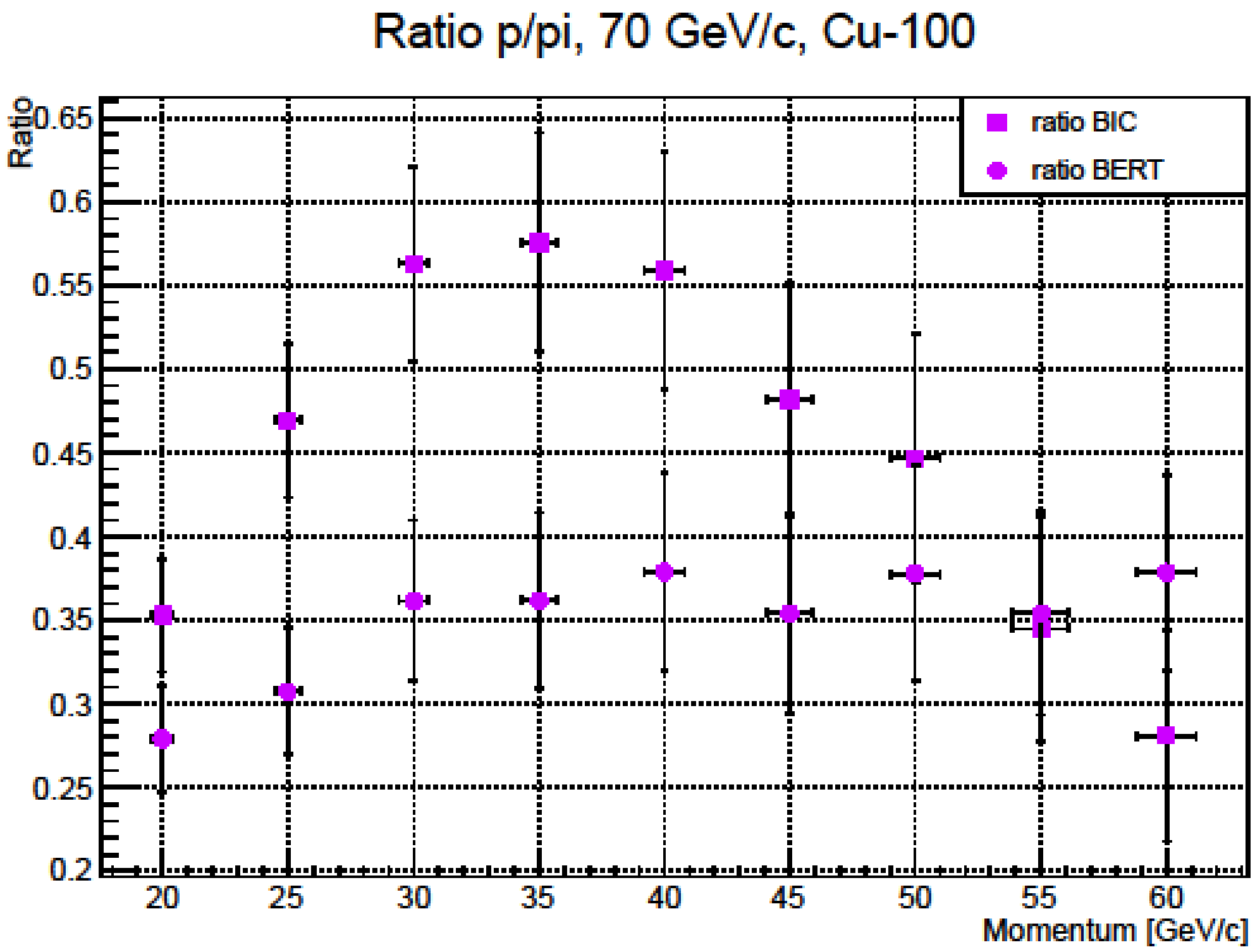
Ratio p/pi, 70 GeV/c, Cu-100
Ratio
0.65
0.6
0.55
0.5
0.45
0.4
0.35
0.3
0.25
0.2
20
25
30
35
40
45
50
55
60
Momentum [GeV/c]
ratio BIC
ratio BERT

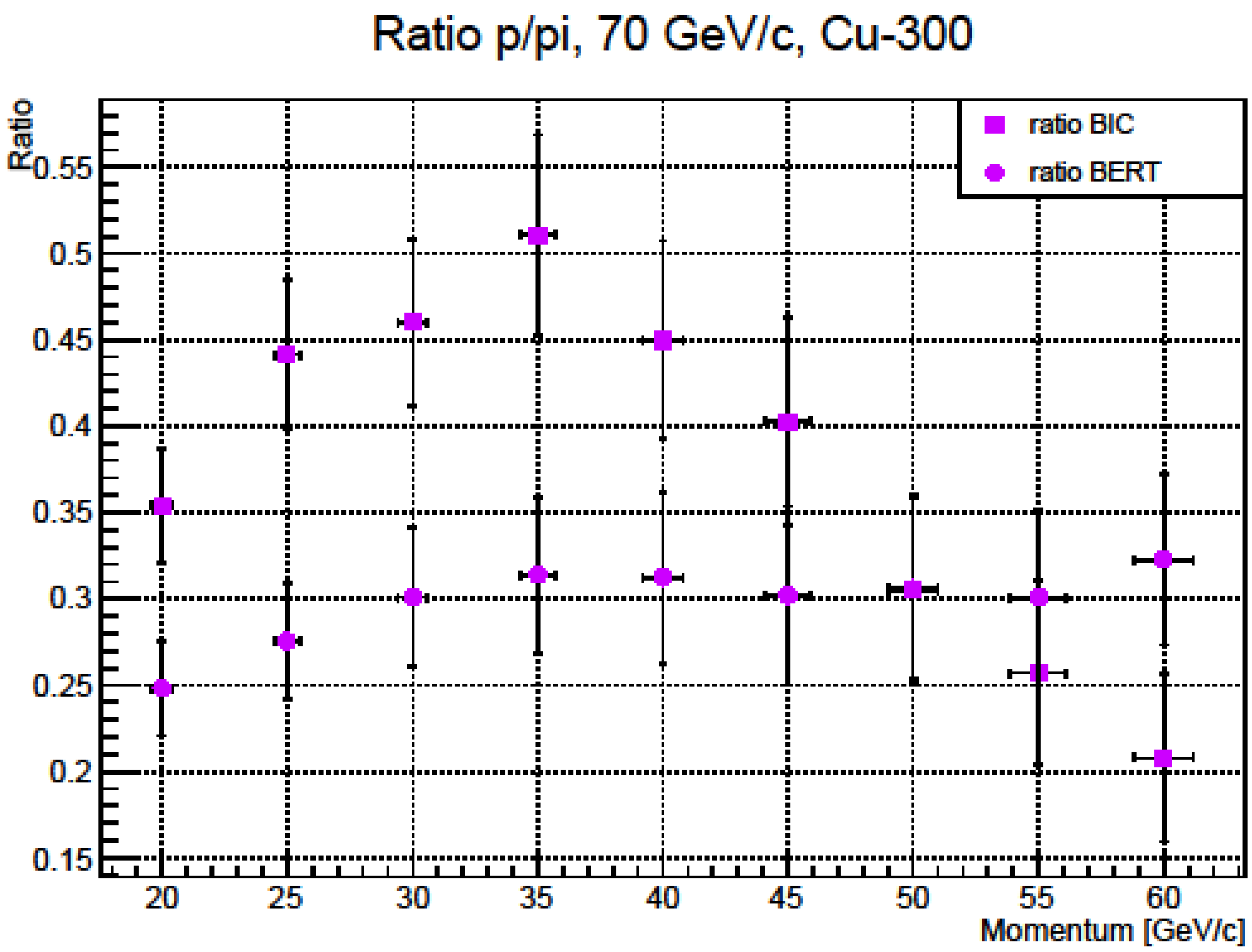
Ratio p/pi, 70 GeV/c, Cu-300
Ratio
0.55
0.5
0.45
0.4
0.35
0.3
0.25
0.2
0.15
20
25
30
35
40
45
50
55
60
Momentum [GeV/c]
ratio BIC
ratio BERT

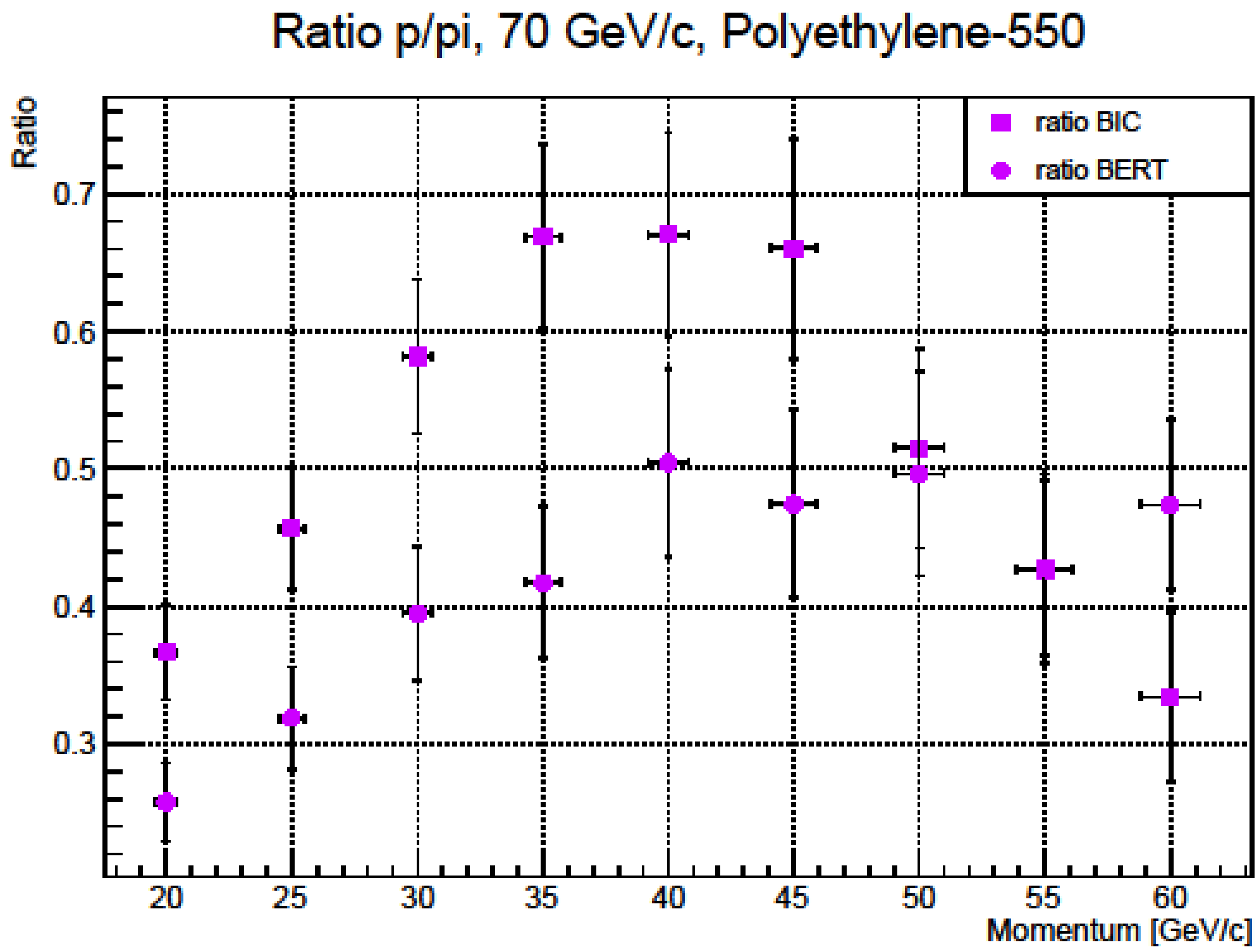

Ratio p/pi, 70 GeV/c, Polyethylene-550
Ratio
0.7
0.6
0.5
0.4
0.3
ratio BIC
ratio BERT
20
25
30
35
40
45
50
55
60
Momentum [GeV/c]

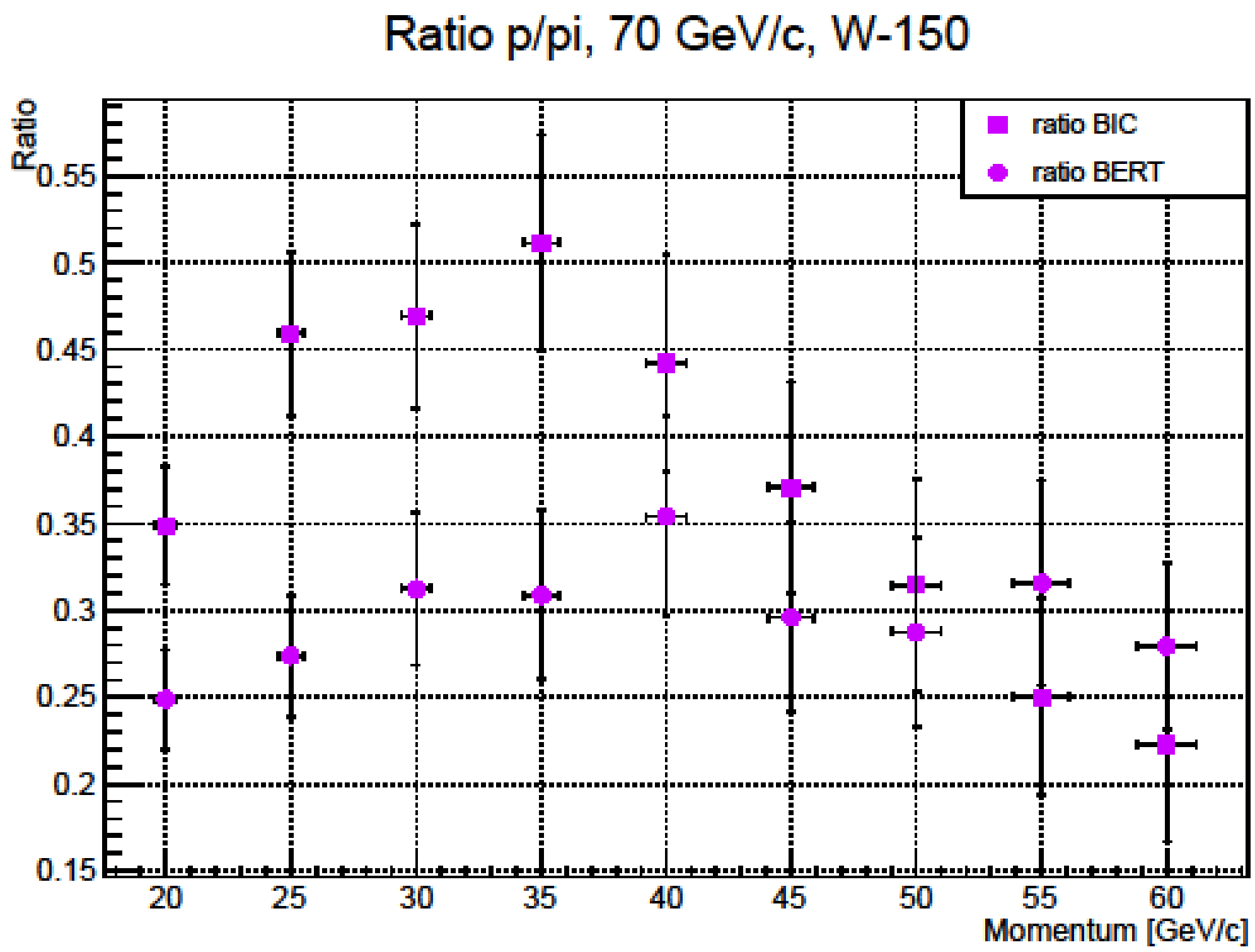

Ratio p/pi, 70 GeV/c, W-150
Ratio
0.55
0.5
0.45
0.4
0.35
0.3
0.25
0.2
0.15
ratio BIC
ratio BERT
20
25
30
35
40
45
50
55
60
Momentum [GeV/c]

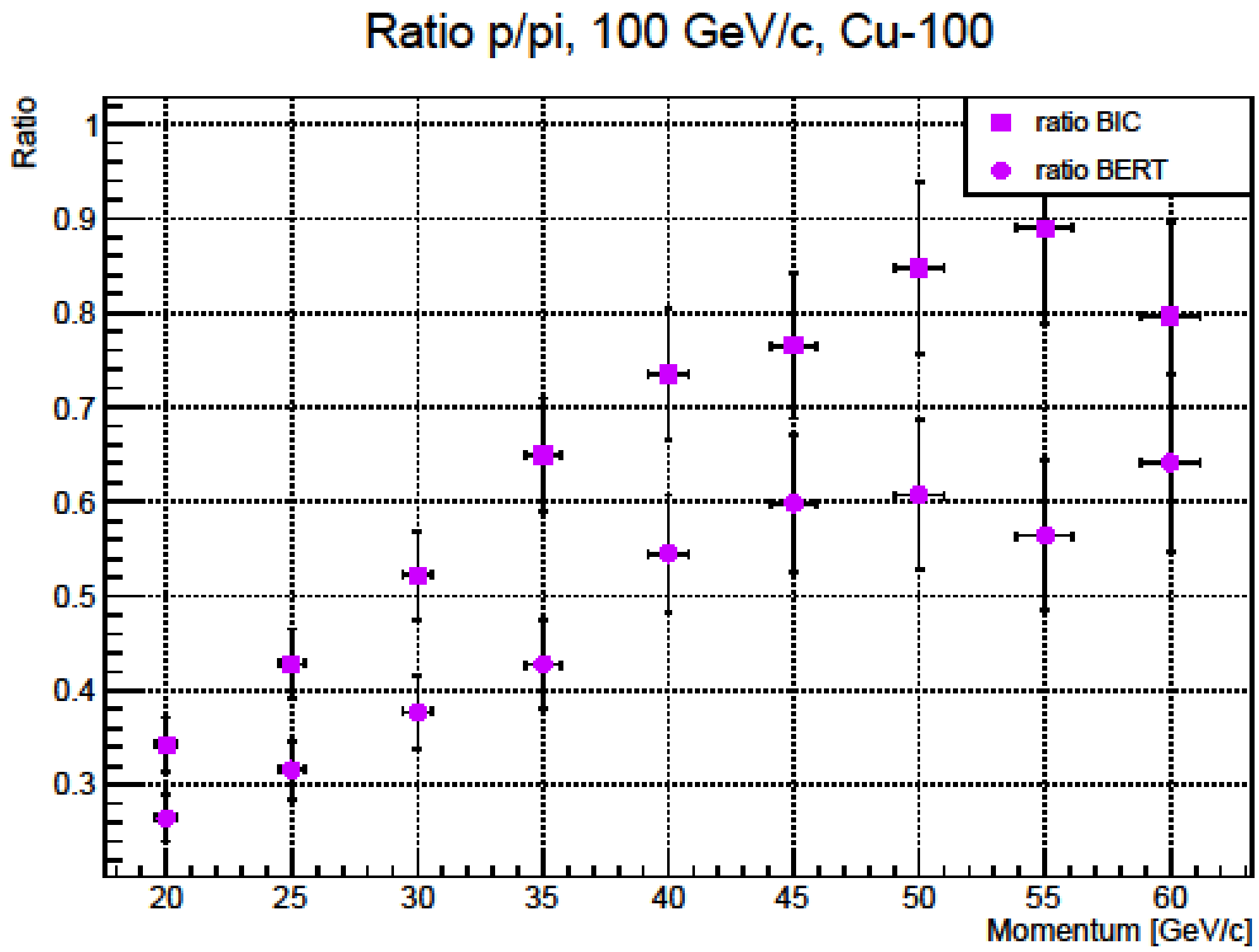
Ratio p/pi, 100 GeV/c, Cu-100
Ratio
Momentum [GeV/c]
ratio BIC
ratio BERT
1
0.9
0.8
0.7
0.6
0.5
0.4
0.3
20
25
30
35
40
45
50
55
60

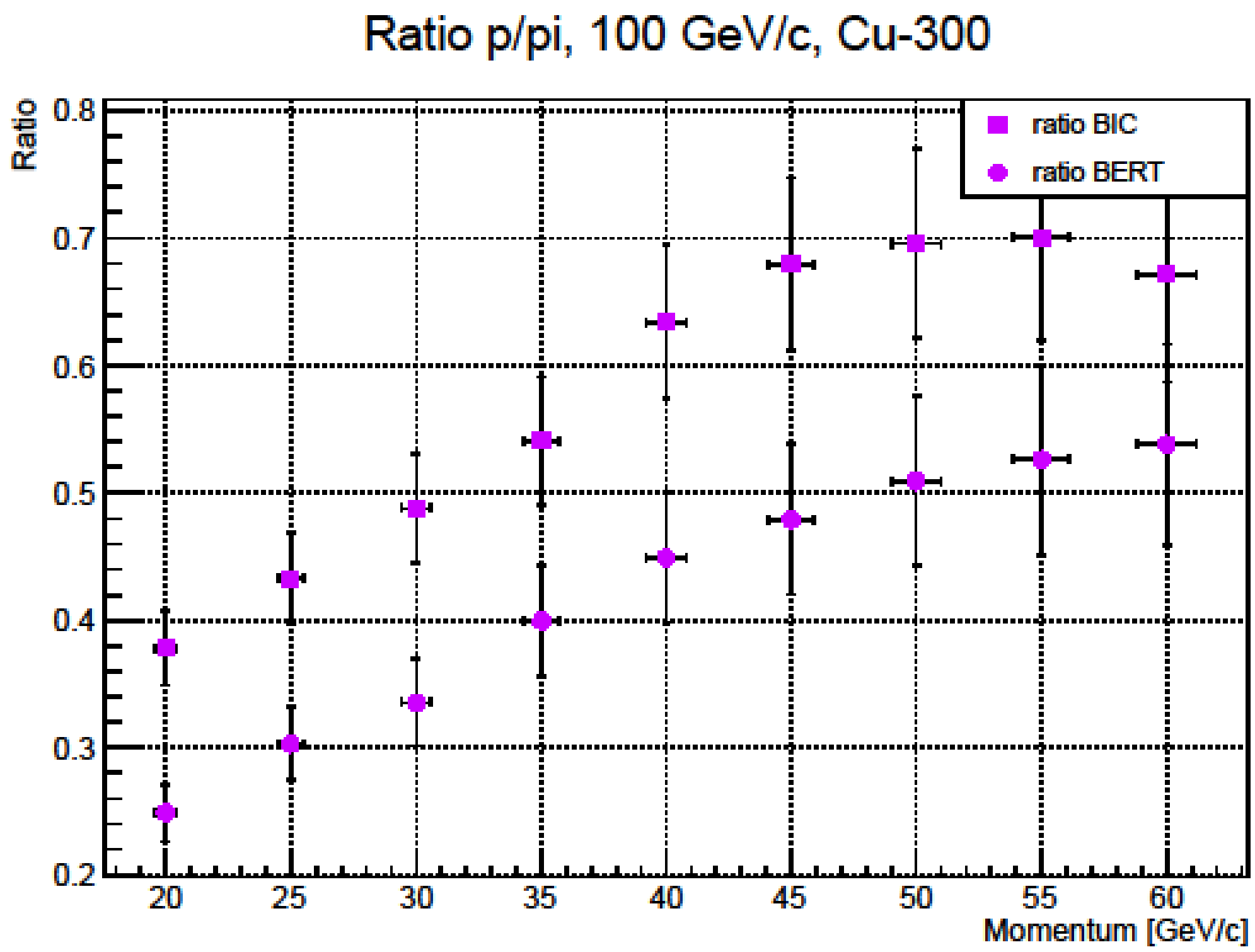
Ratio p/pi, 100 GeV/c, Cu-300
Ratio
Momentum [GeV/c]
ratio BIC
ratio BERT
0.8
0.7
0.6
0.5
0.4
0.3
0.2
20
25
30
35
40
45
50
55
60

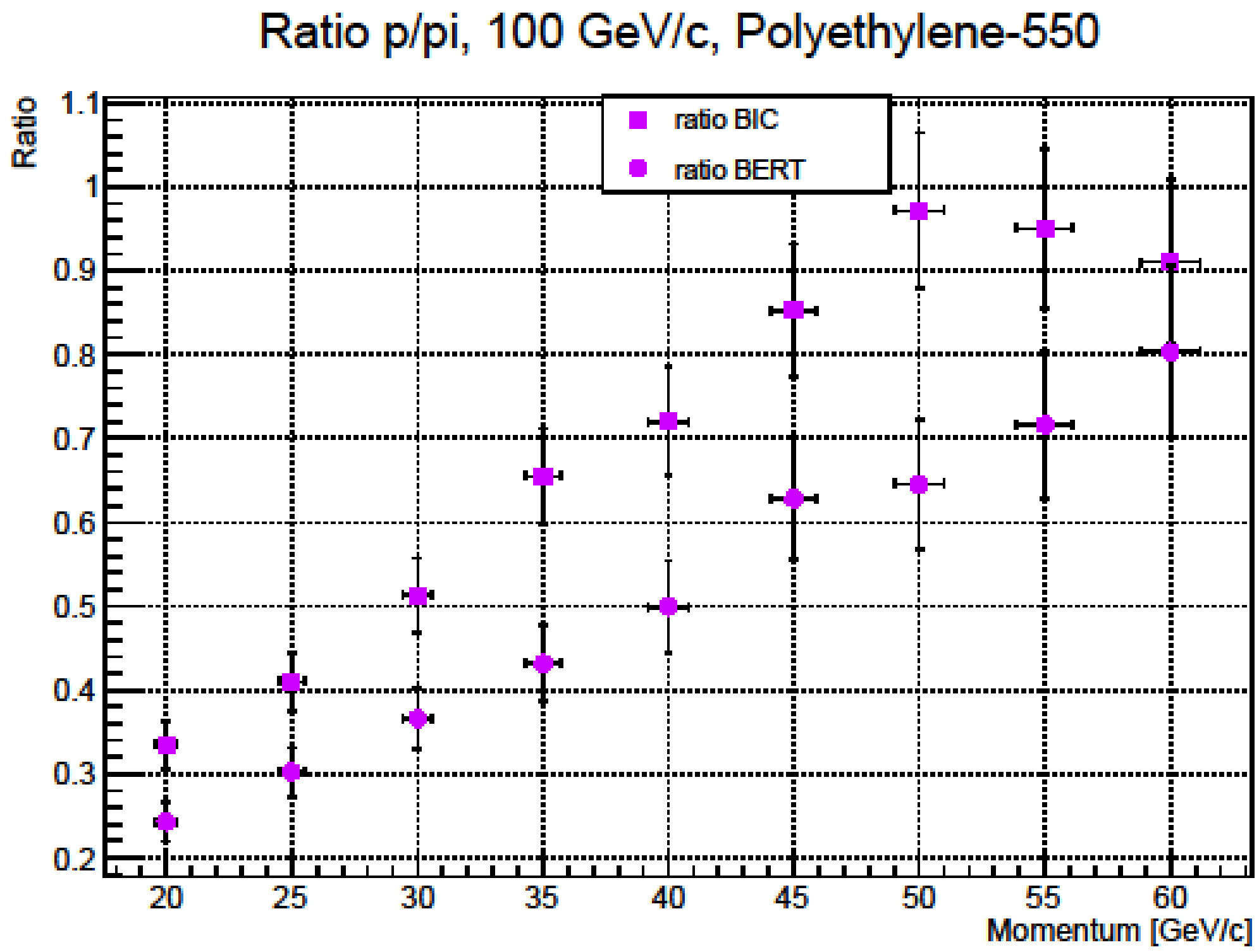
Ratio p/pi, 100 GeV/c, Polyethylene-550
Ratio
ratio BIC
ratio BERT
1.1
1
0.9
0.8
0.7
0.6
0.5
0.4
0.3
0.2
20
25
30
35
40
45
50
55
60
Momentum [GeV/c]

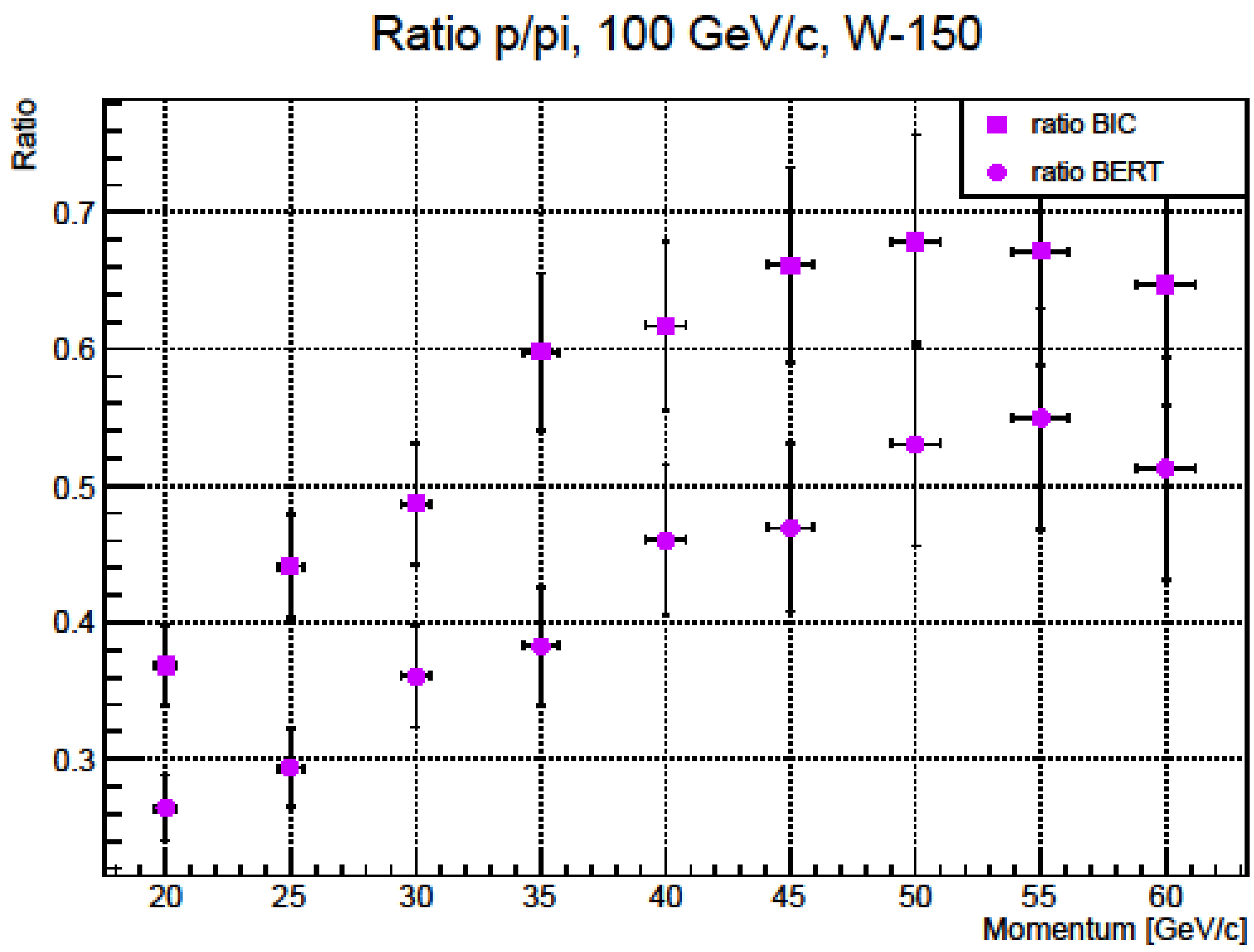
Ratio p/pi, 100 GeV/c, W-150
Ratio
ratio BIC
ratio BERT
0.7
0.6
0.5
0.4
0.3
20
25
30
35
40
45
50
55
60
Momentum [GeV/c]

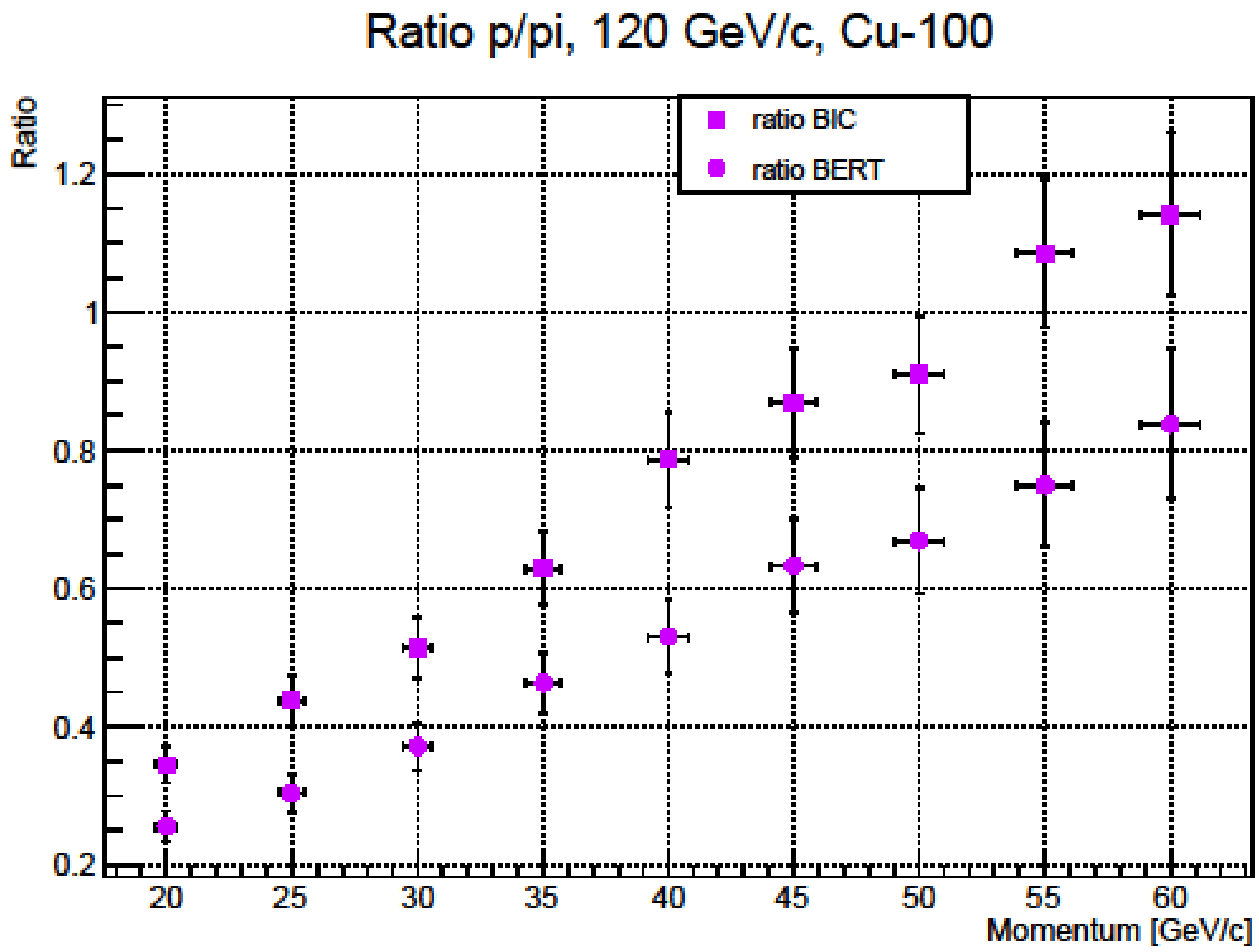
Ratio p/pi, 120 GeV/c, Cu-100
Ratio
ratio BIC
ratio BERT
1.2
1
0.8
0.6
0.4
0.2
20
25
30
35
40
45
50
55
60
Momentum [GeV/c]

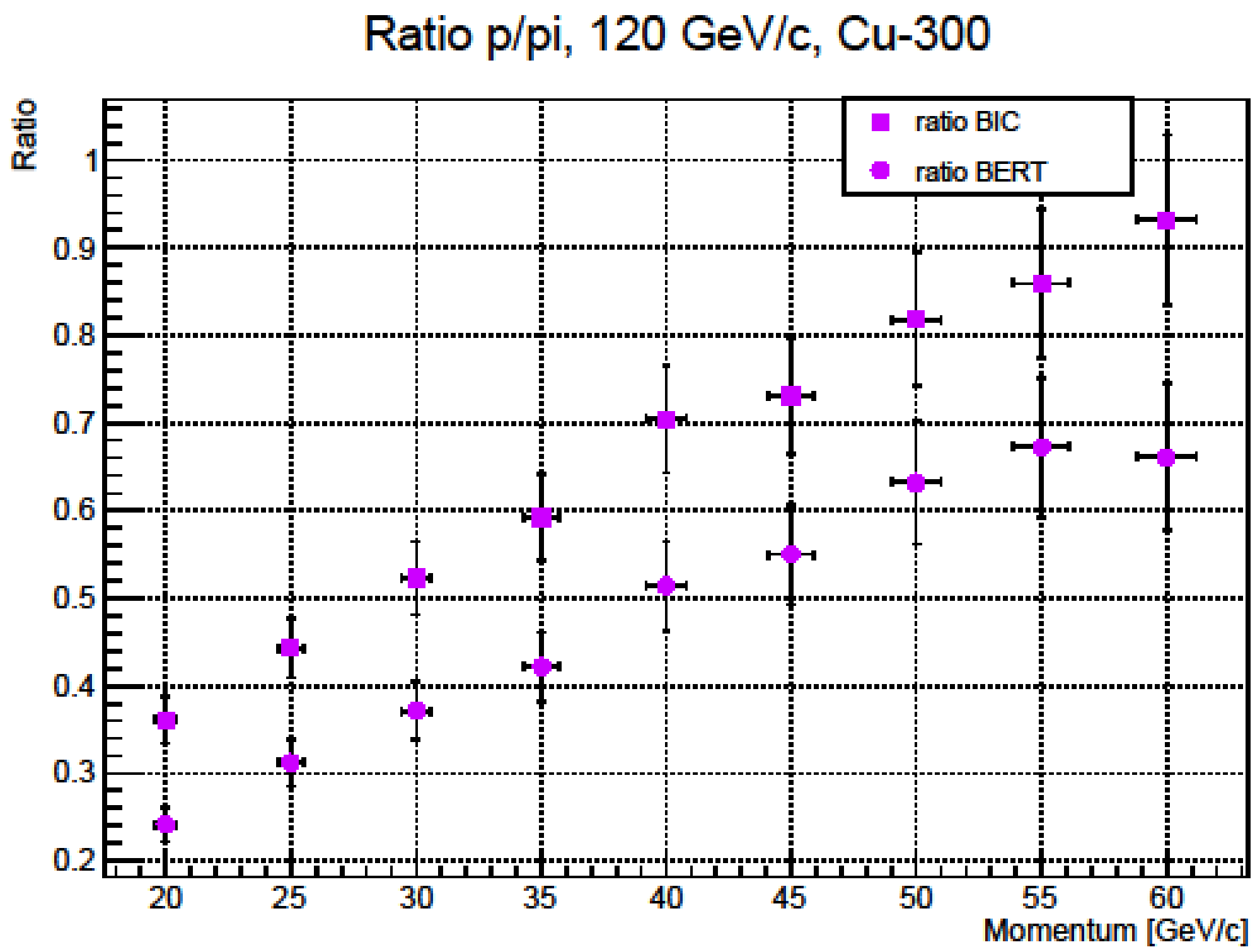
Ratio p/pi, 120 GeV/c, Cu-300
Ratio
ratio BIC
ratio BERT
1
0.9
0.8
0.7
0.6
0.5
0.4
0.3
0.2
20
25
30
35
40
45
50
55
60
Momentum [GeV/c]

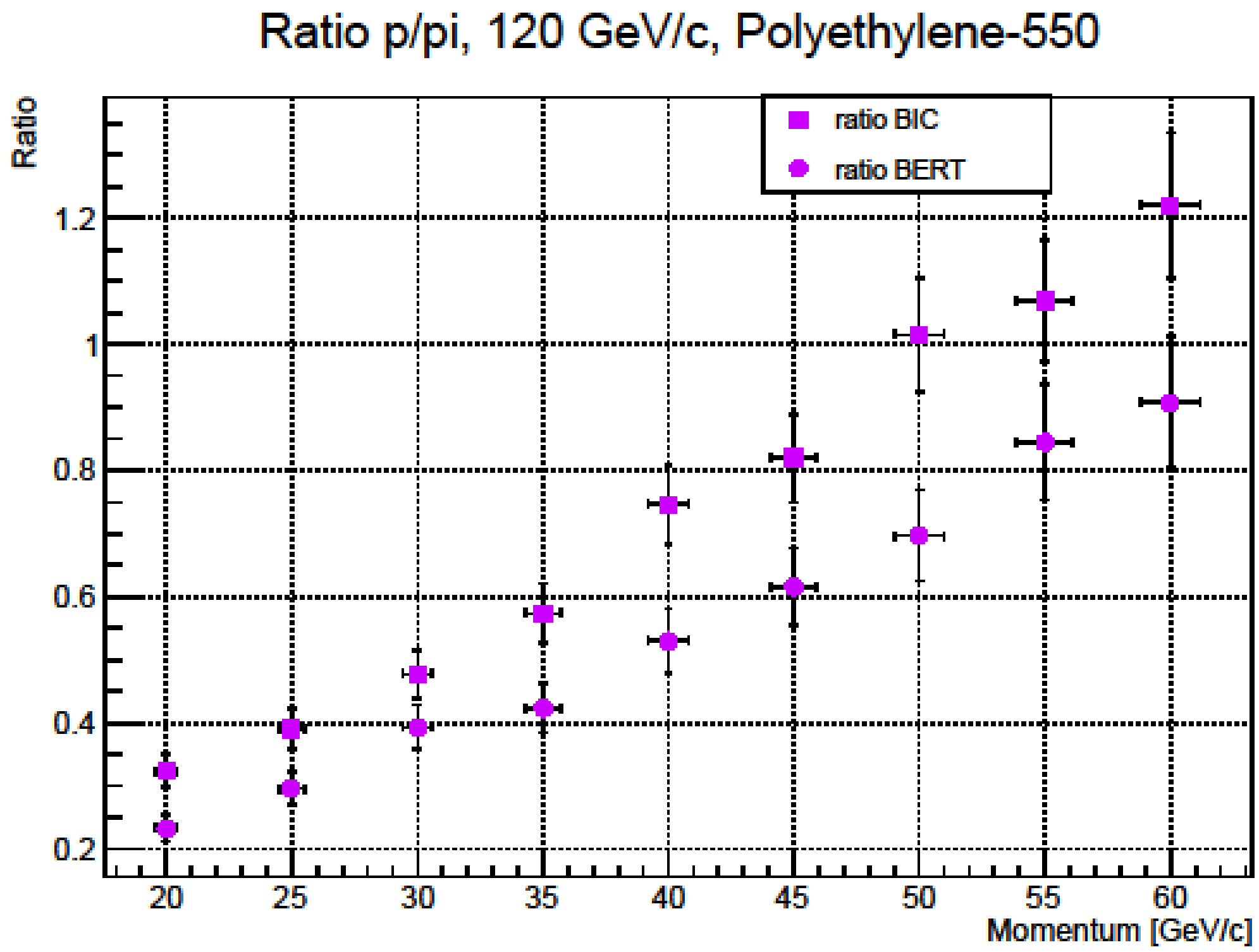
Ratio p/pi, 120 GeV/c, Polyethylene-550
ratio BIC
ratio BERT
Ratio
1.2
1
0.8
0.6
0.4
0.2
20
25
30
35
40
45
50
55
60
Momentum [GeV/c]

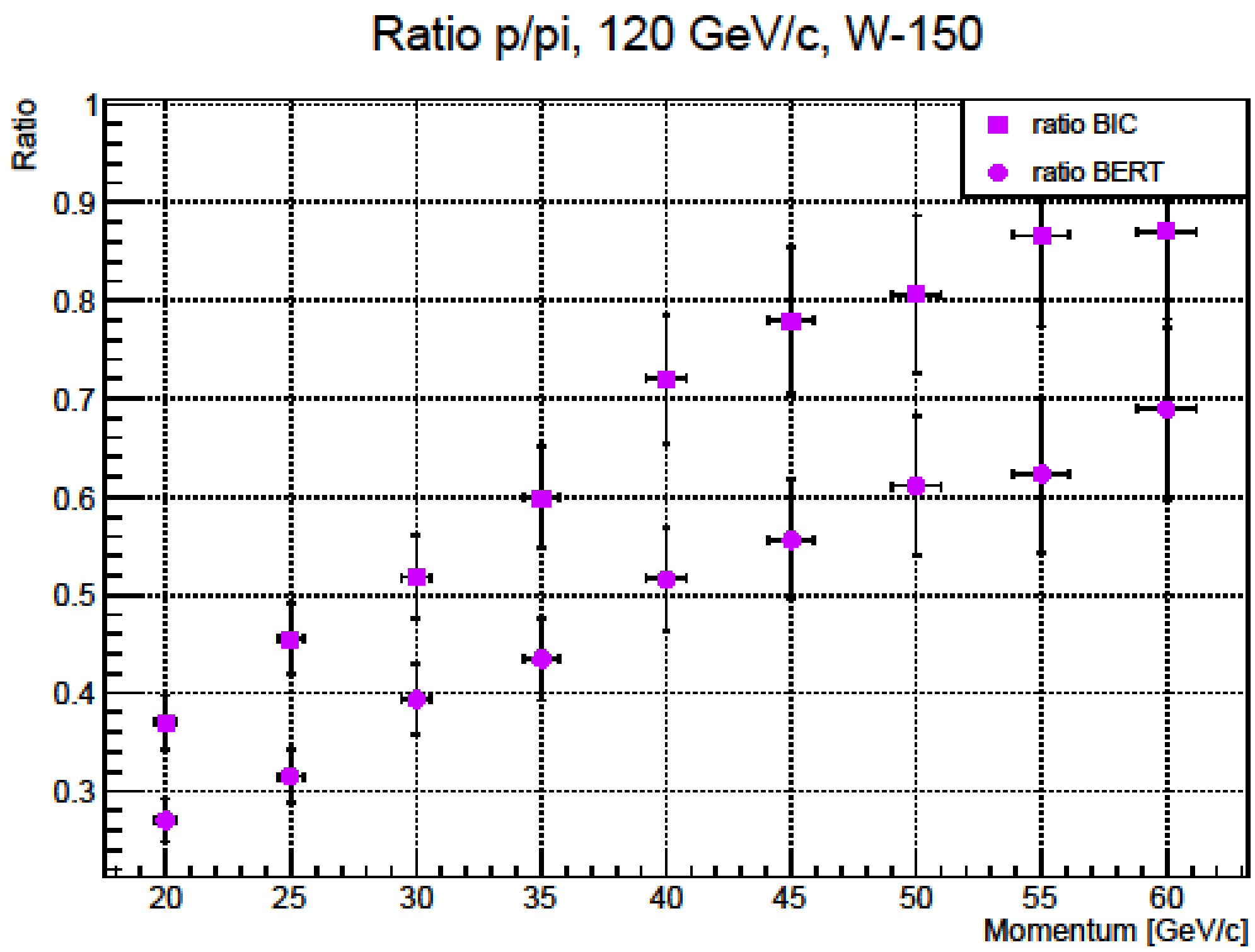
Ratio p/pi, 120 GeV/c, W-150
ratio BIC
ratio BERT
Ratio
1
0.9
0.8
0.7
0.6
0.5
0.4
0.3
20
25
30
35
40
45
50
55
60
Momentum [GeV/c]

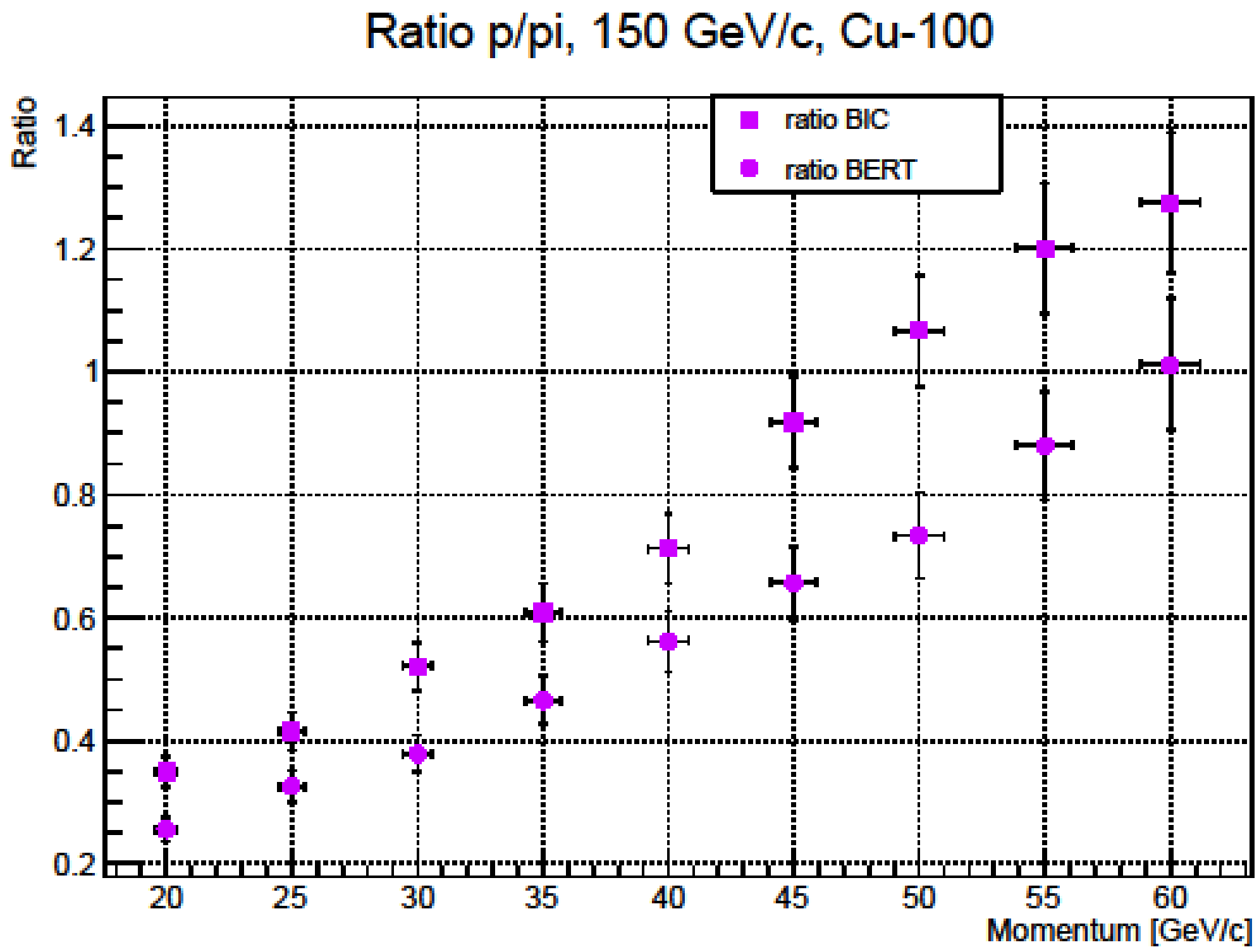
Ratio p/pi, 150 GeV/c, Cu-100
Ratio
ratio BIC
ratio BERT
1.4
1.2
1
0.8
0.6
0.4
0.2
20
25
30
35
40
45
50
55
60
Momentum [GeV/c]

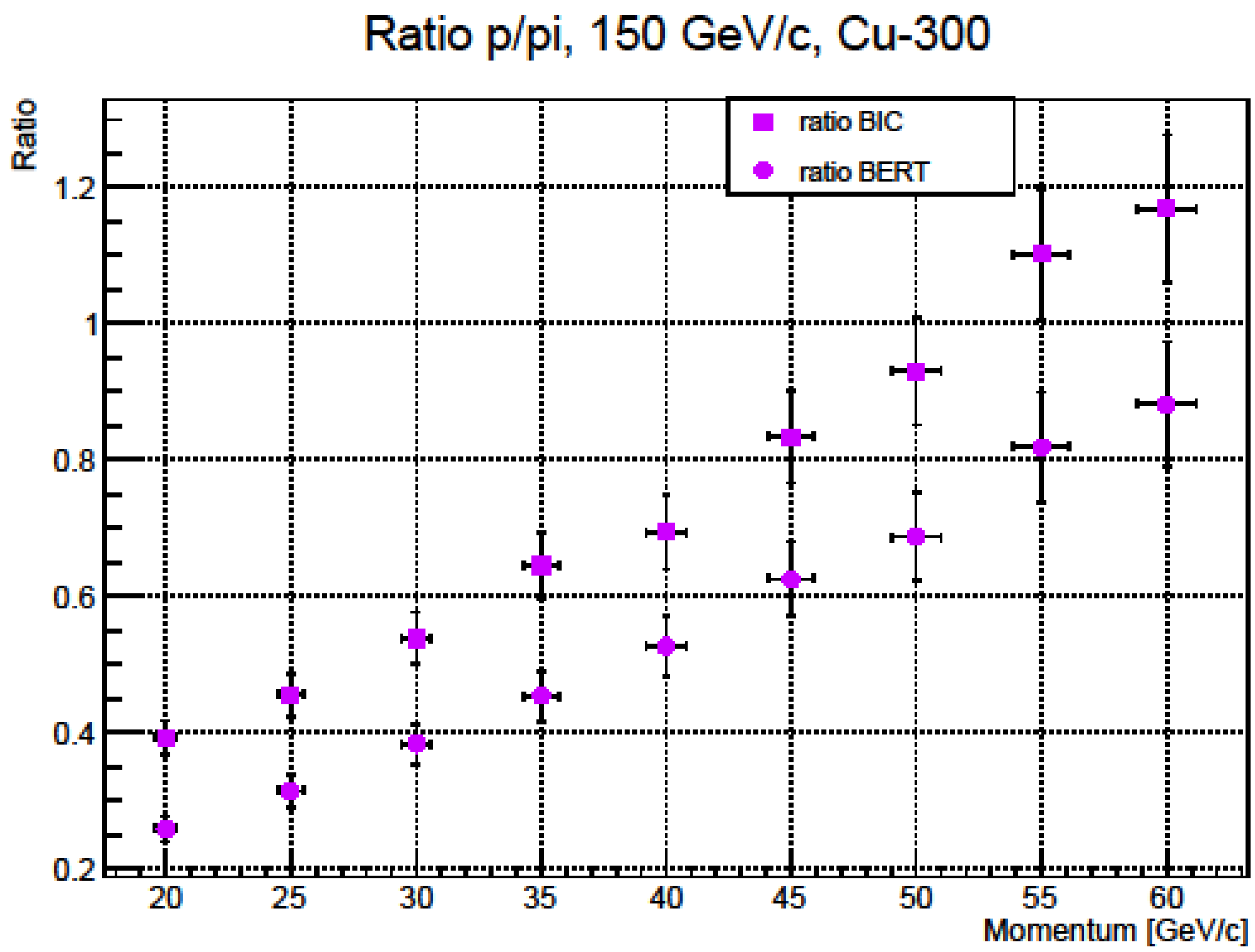
Ratio p/pi, 150 GeV/c, Cu-300
Ratio
ratio BIC
ratio BERT
1.2
1
0.8
0.6
0.4
0.2
20
25
30
35
40
45
50
55
60
Momentum [GeV/c]

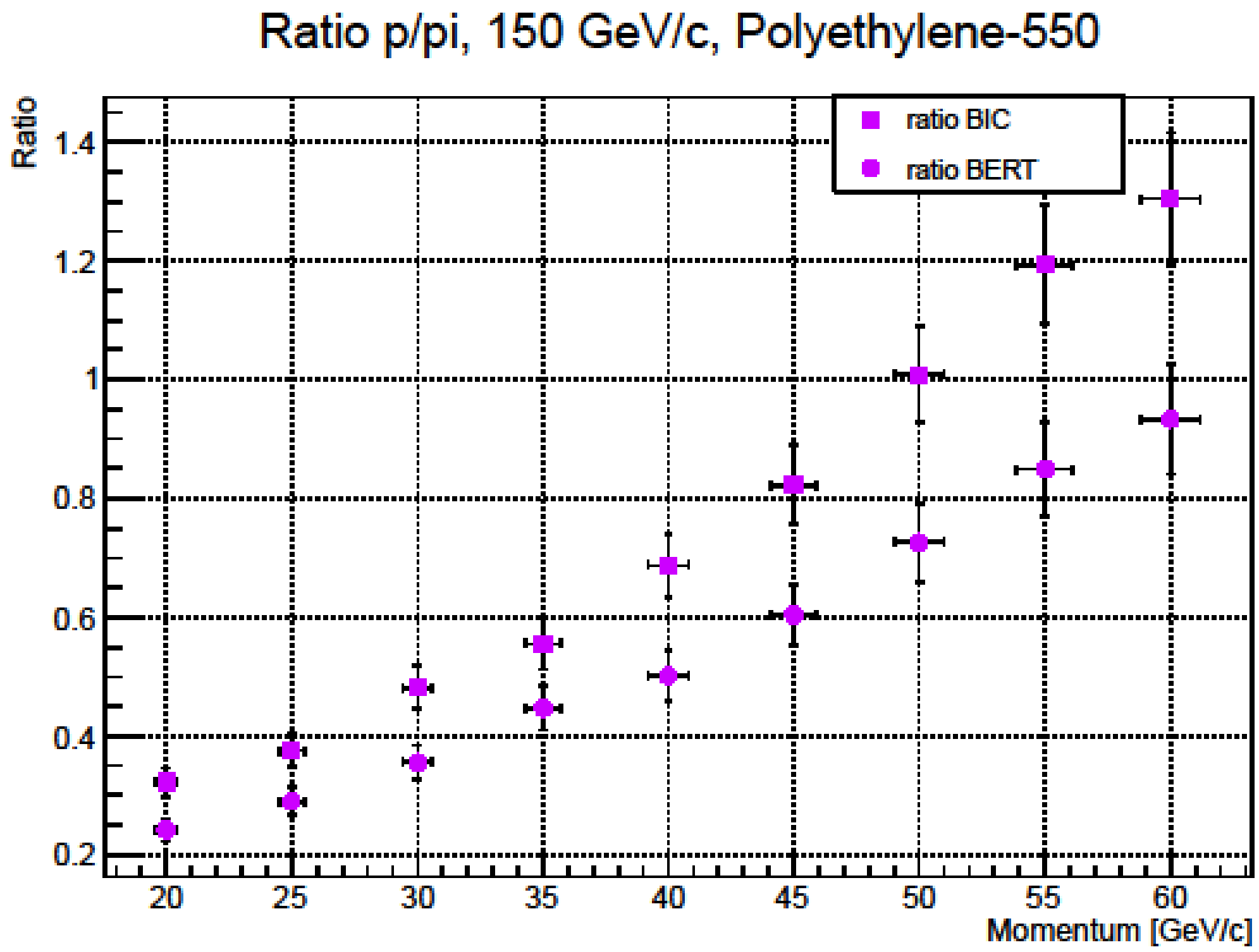

Ratio p/pi, 150 GeV/c, Polyethylene-550
Ratio
ratio BIC
ratio BERT
Momentum [GeV/c]

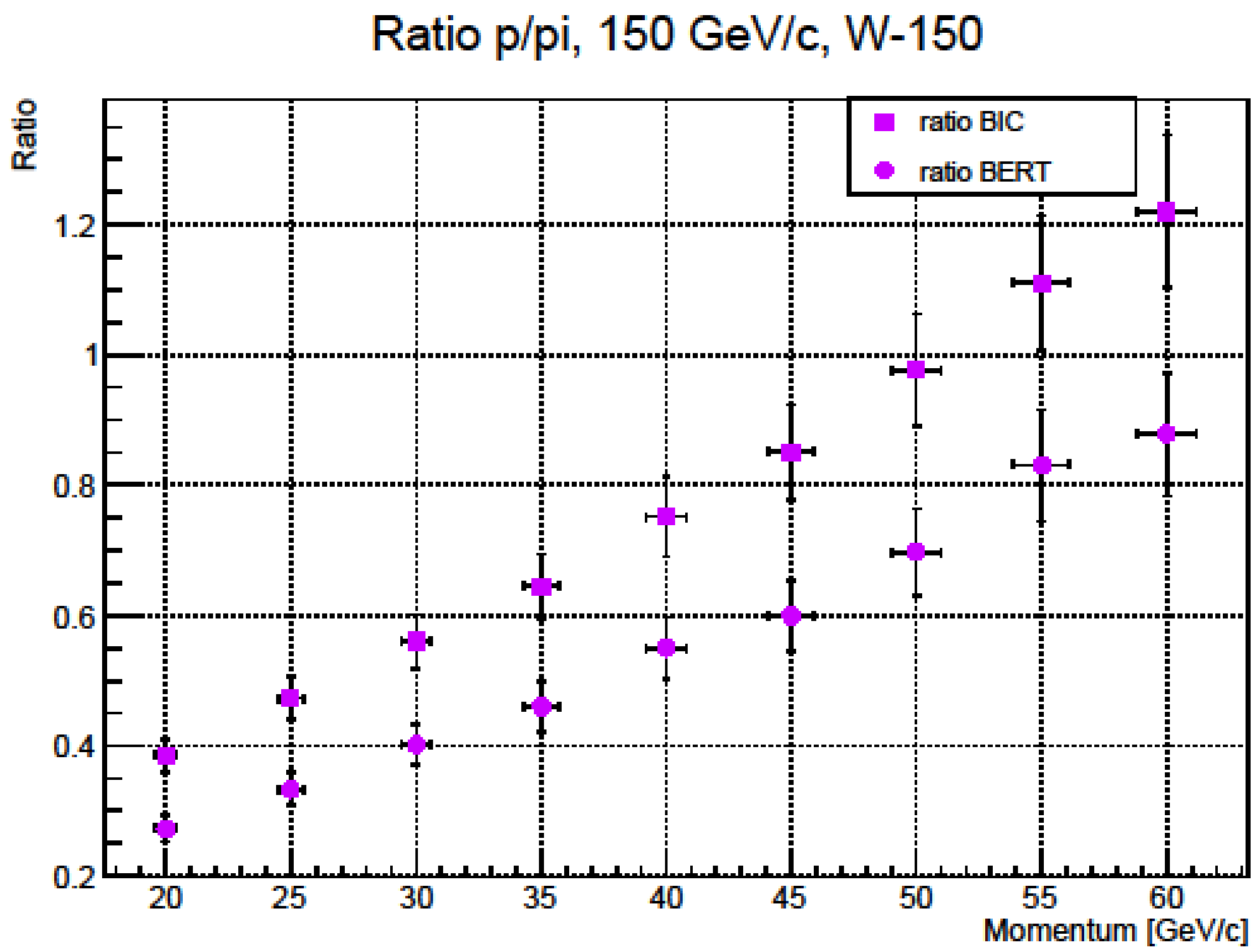

Ratio p/pi, 150 GeV/c, W-150
Ratio
ratio BIC
ratio BERT
Momentum [GeV/c]

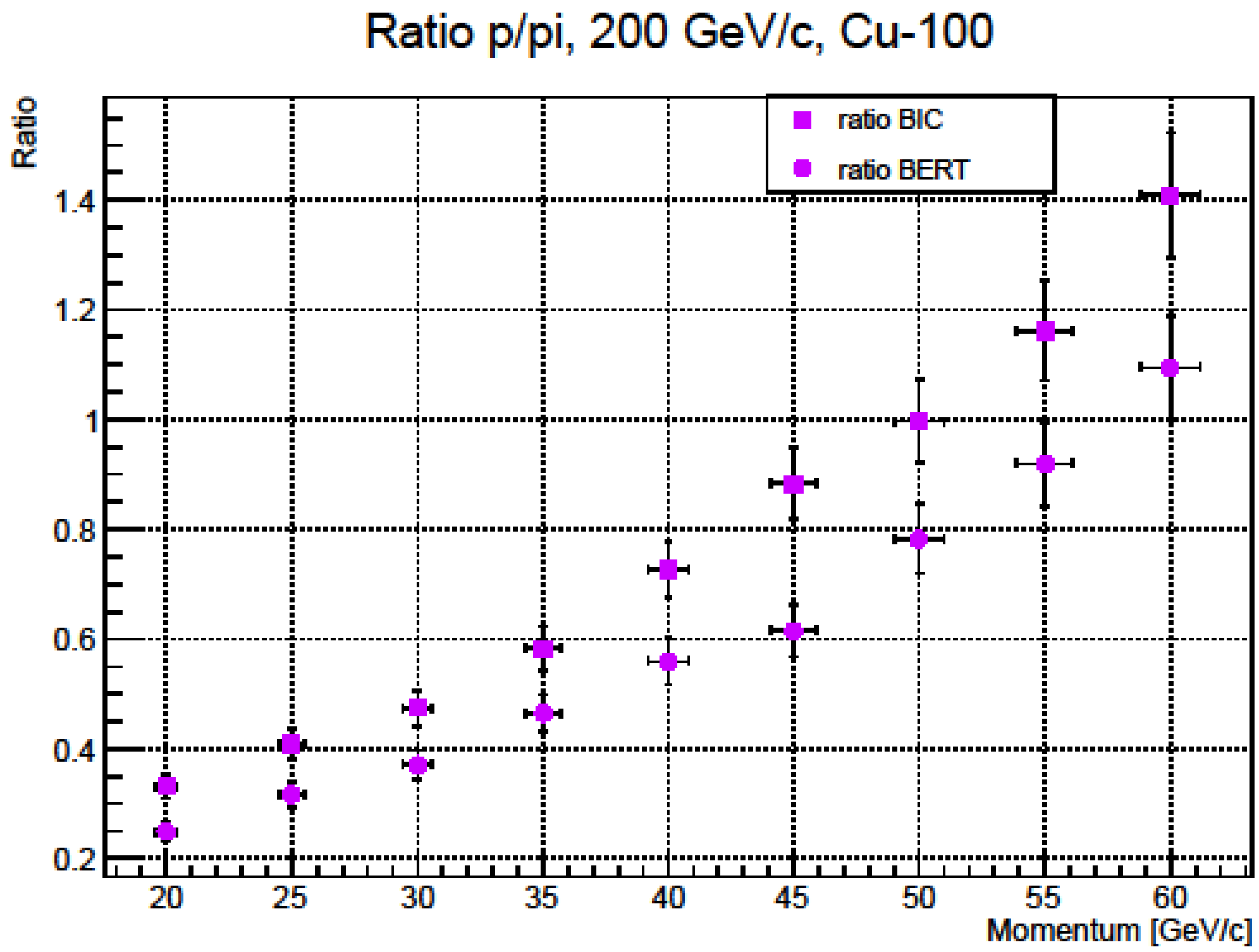

Ratio p/pi, 200 GeV/c, Cu-100
Ratio
Momentum [GeV/c]
ratio BIC
ratio BERT
1.4
1.2
1
0.8
0.6
0.4
0.2
20
25
30
35
40
45
50
55
60

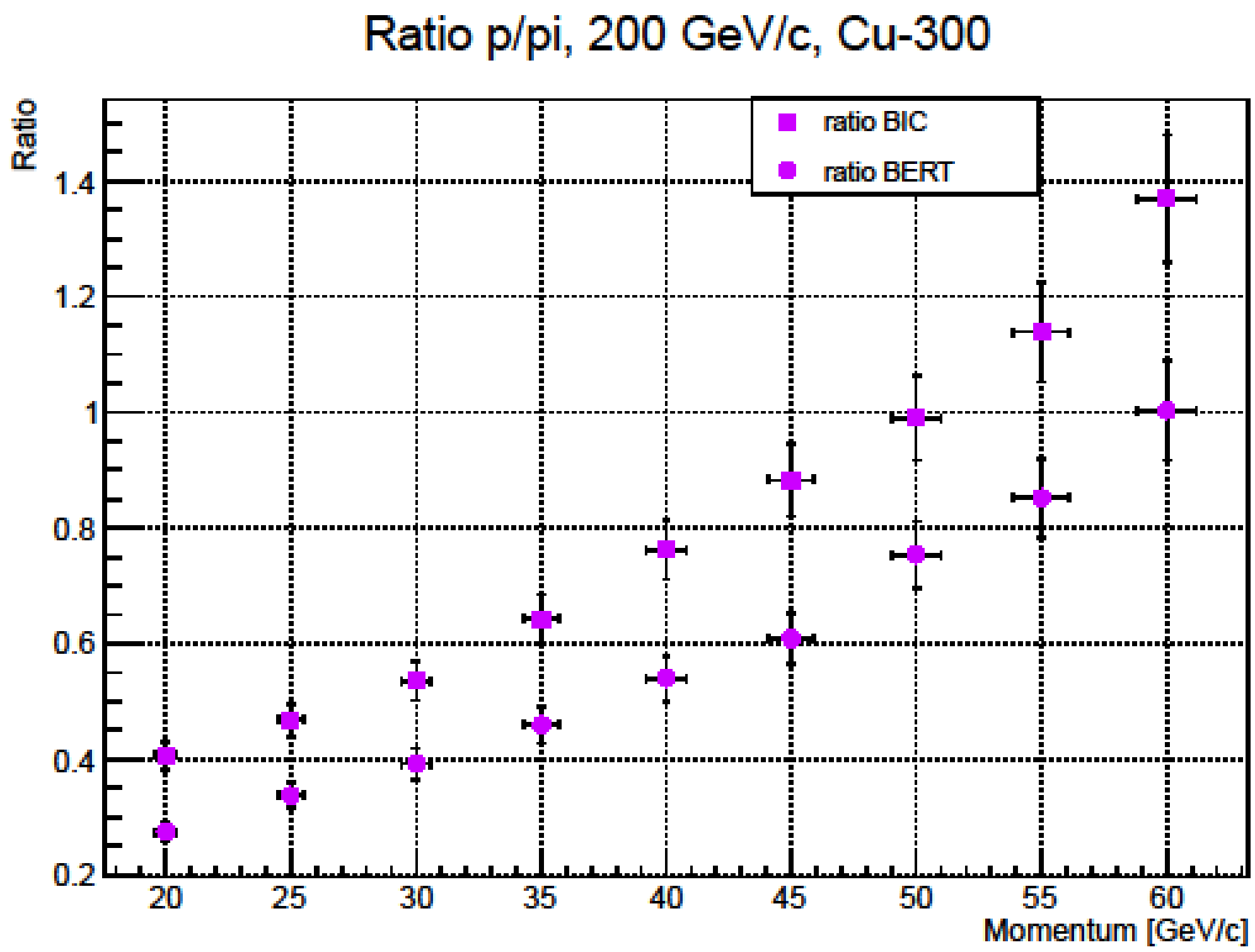

Ratio p/pi, 200 GeV/c, Cu-300
Ratio
Momentum [GeV/c]
ratio BIC
ratio BERT
1.4
1.2
1
0.8
0.6
0.4
0.2
20
25
30
35
40
45
50
55
60

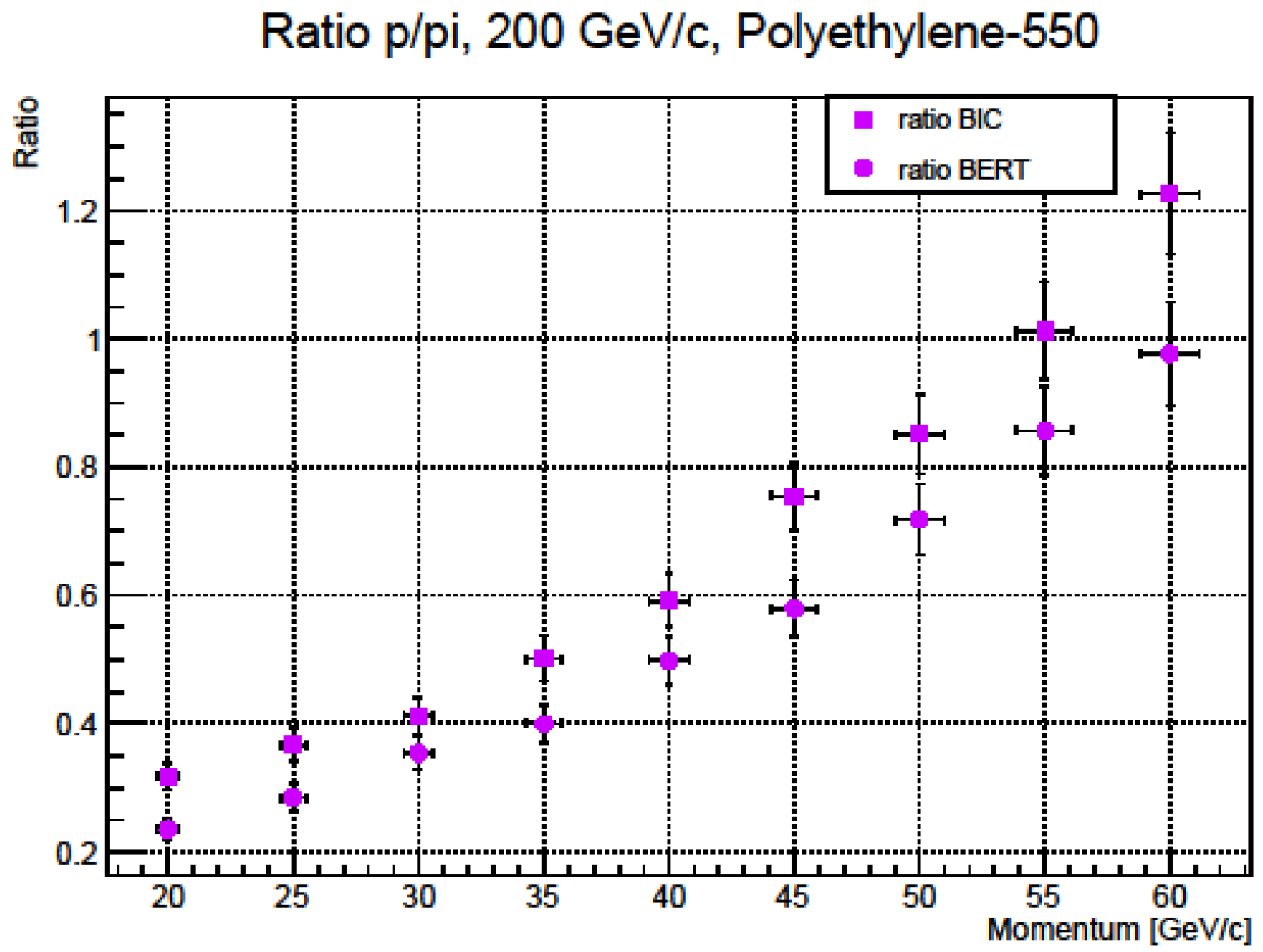

## 2.5 Ratio p over sum (p+pi) with FTFP_BERT

The following six plots summarize all the results for hadron production, i.e. section 2.1, 2.2 and 2.4 with FTFP_BERT. We show the ratio of protons over protons+pions. Each figure is for one specific target and we compare the results with two measurements of secondary particle production from beryllium (without the use of a filter). In the following figures, 'Sec. Atherton' points represent the data points and a fitted line from [1] and 'Sec. SPY' shows the same from [2]. Note that the proton content of the filtered tertiary beam has significant higher proton content. The highest proton content is obtained at 60 GeV/c with 100 mm Cu and a 200 GeV/c secondary beam.

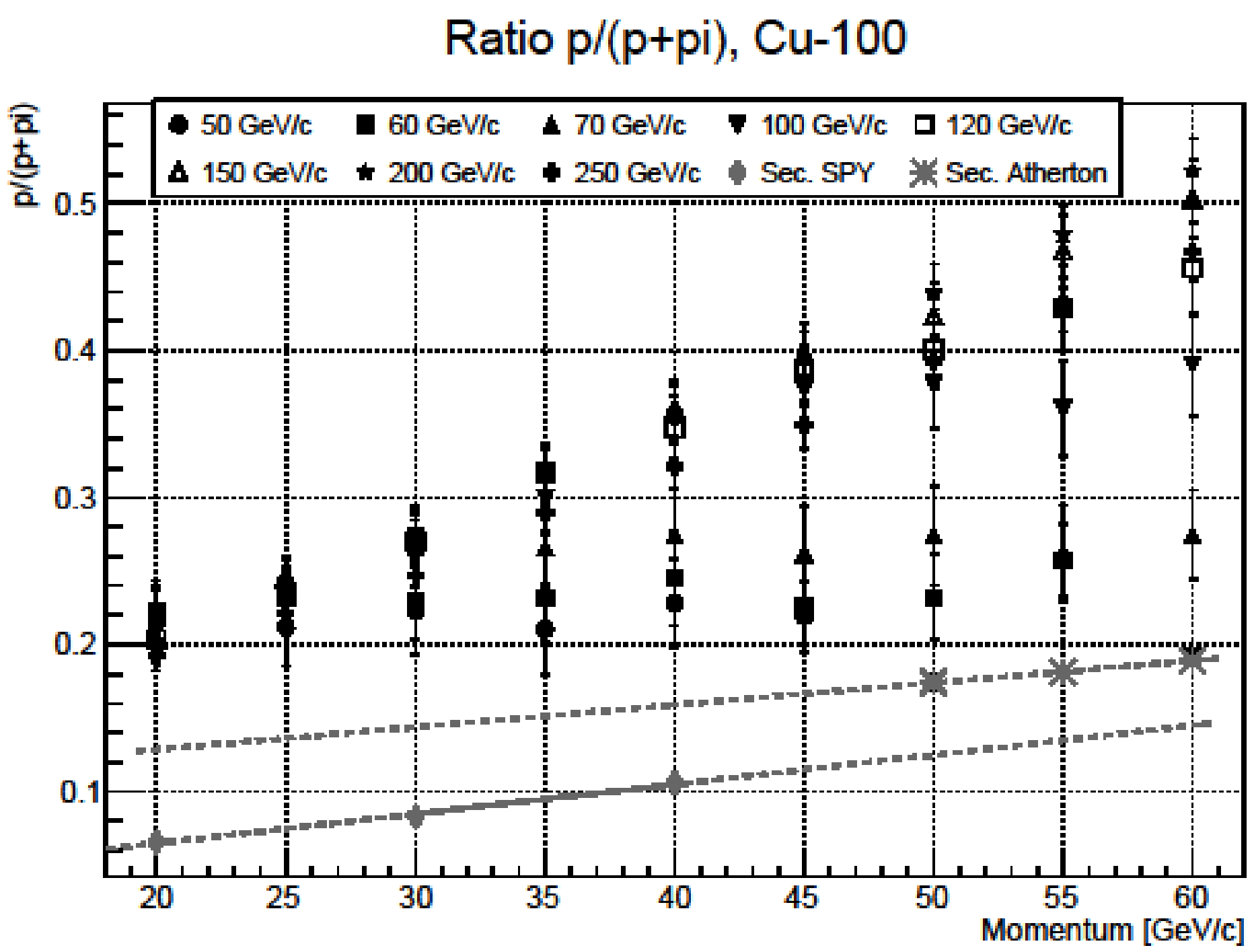

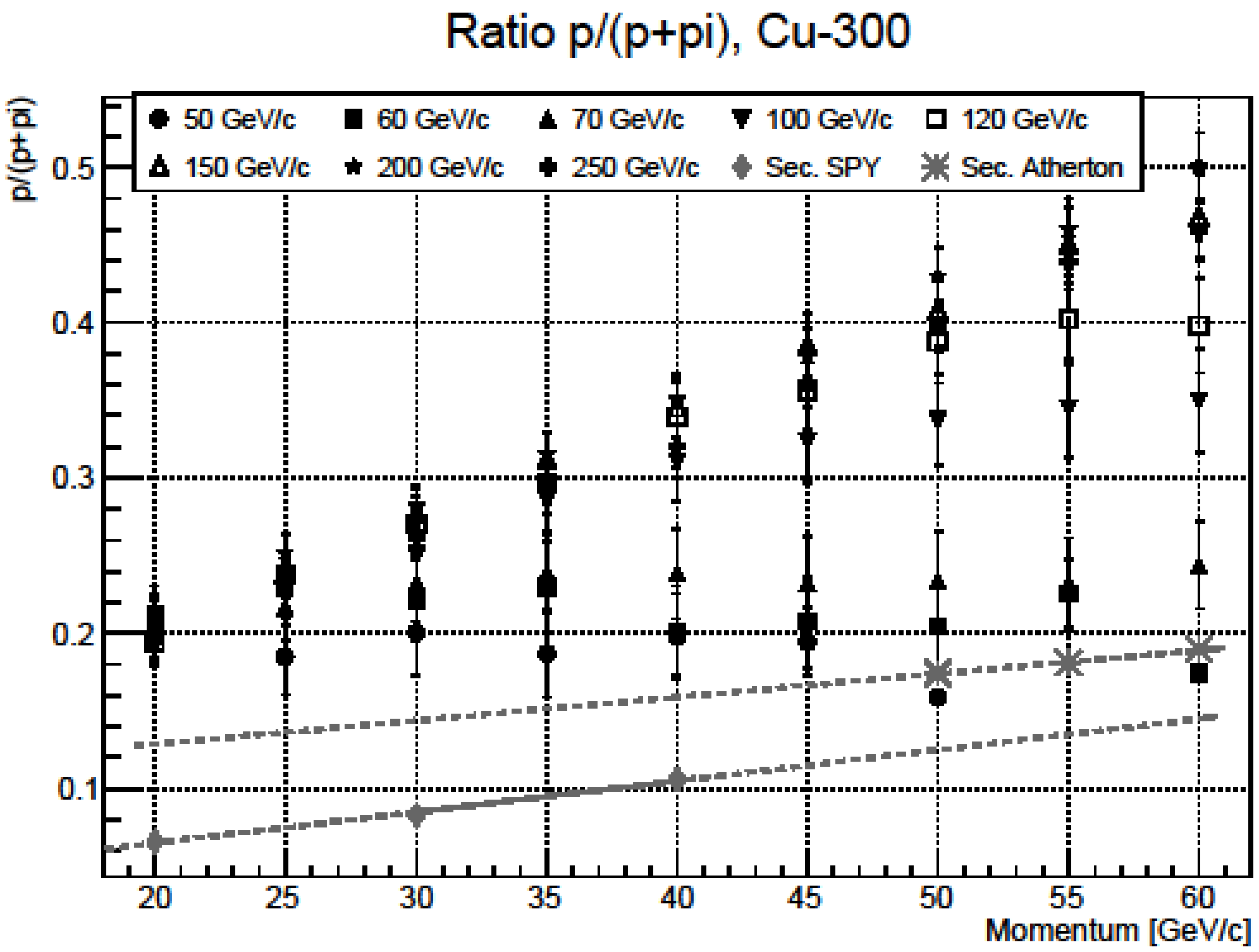
Ratio p/(p+pi), Cu-300
p/(p+pi)
50 GeV/c
60 GeV/c
70 GeV/c
100 GeV/c
120 GeV/c
150 GeV/c
200 GeV/c
250 GeV/c
Sec. SPY
Sec. Atherton
0.1
0.2
0.3
0.4
0.5
20
25
30
35
40
45
50
55
60
Momentum [GeV/c]

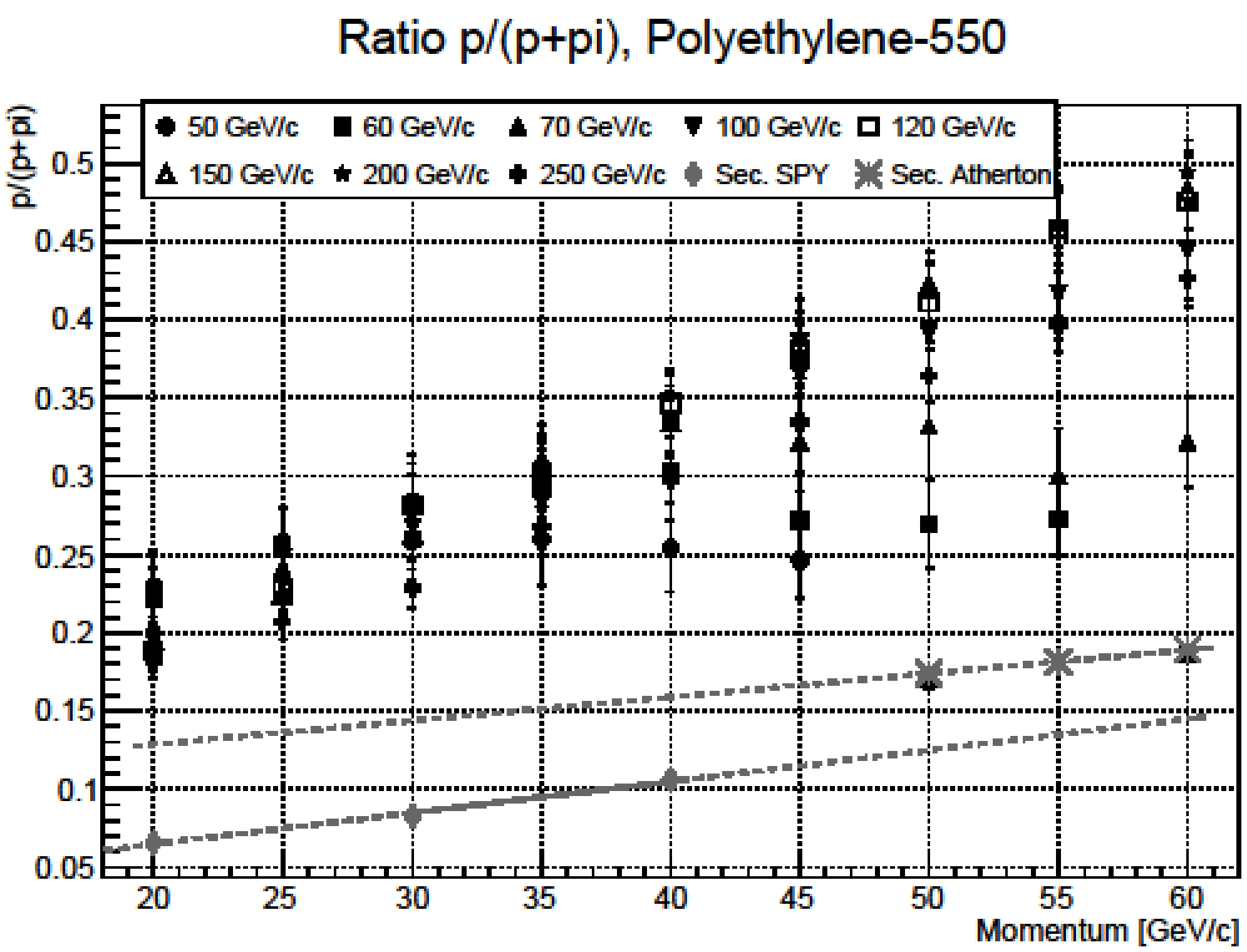
Ratio p/(p+pi), Polyethylene-550
p/(p+pi)
50 GeV/c
60 GeV/c
70 GeV/c
100 GeV/c
120 GeV/c
150 GeV/c
200 GeV/c
250 GeV/c
Sec. SPY
Sec. Atherton
0.05
0.1
0.15
0.2
0.25
0.3
0.35
0.4
0.45
0.5
20
25
30
35
40
45
50
55
60
Momentum [GeV/c]

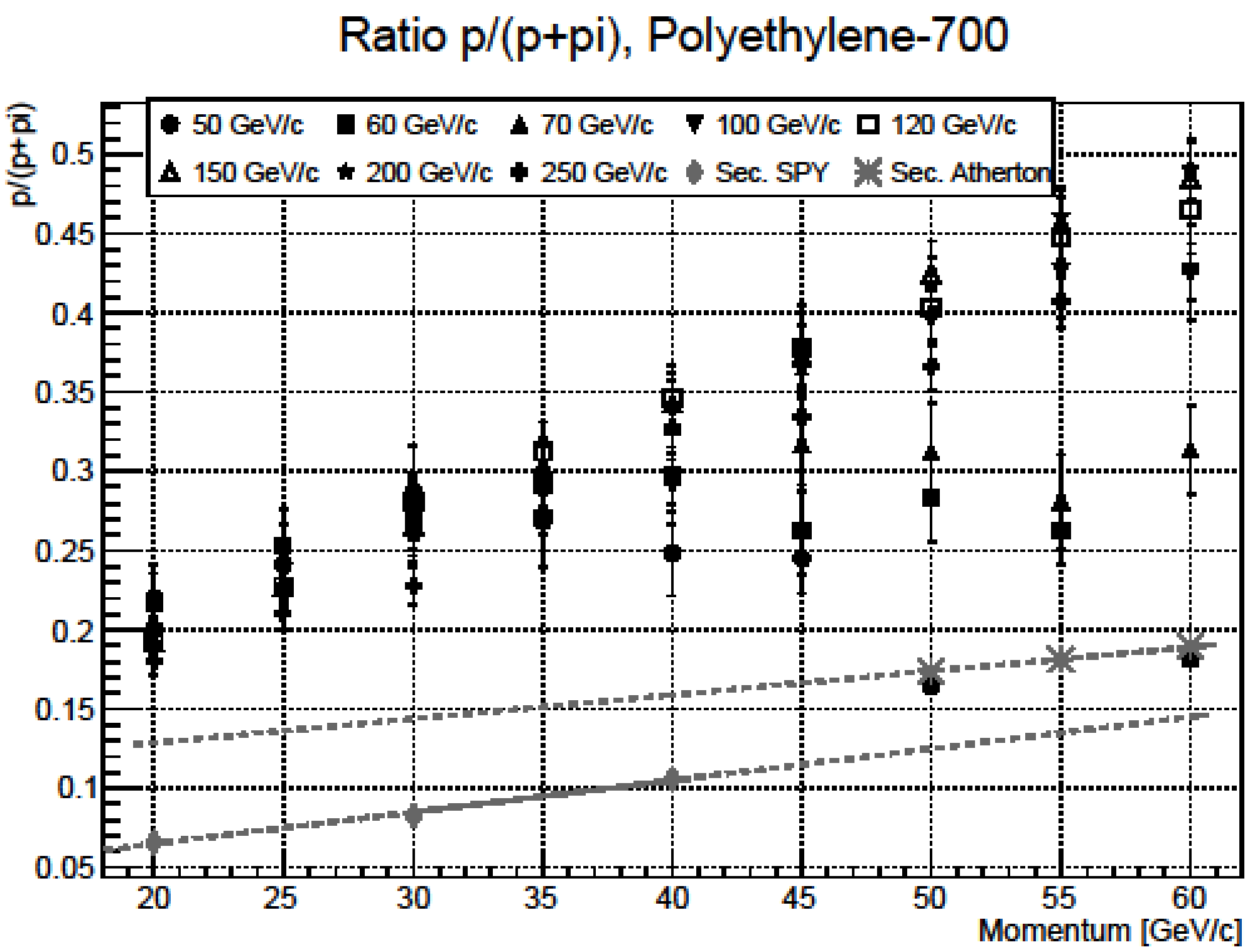
Ratio p/(p+pi), Polyethylene-700
p/(p+pi)
50 GeV/c
60 GeV/c
70 GeV/c
100 GeV/c
120 GeV/c
150 GeV/c
200 GeV/c
250 GeV/c
Sec. SPY
Sec. Atherton
0.5
0.45
0.4
0.35
0.3
0.25
0.2
0.15
0.1
0.05
20
25
30
35
40
45
50
55
60
Momentum [GeV/c]

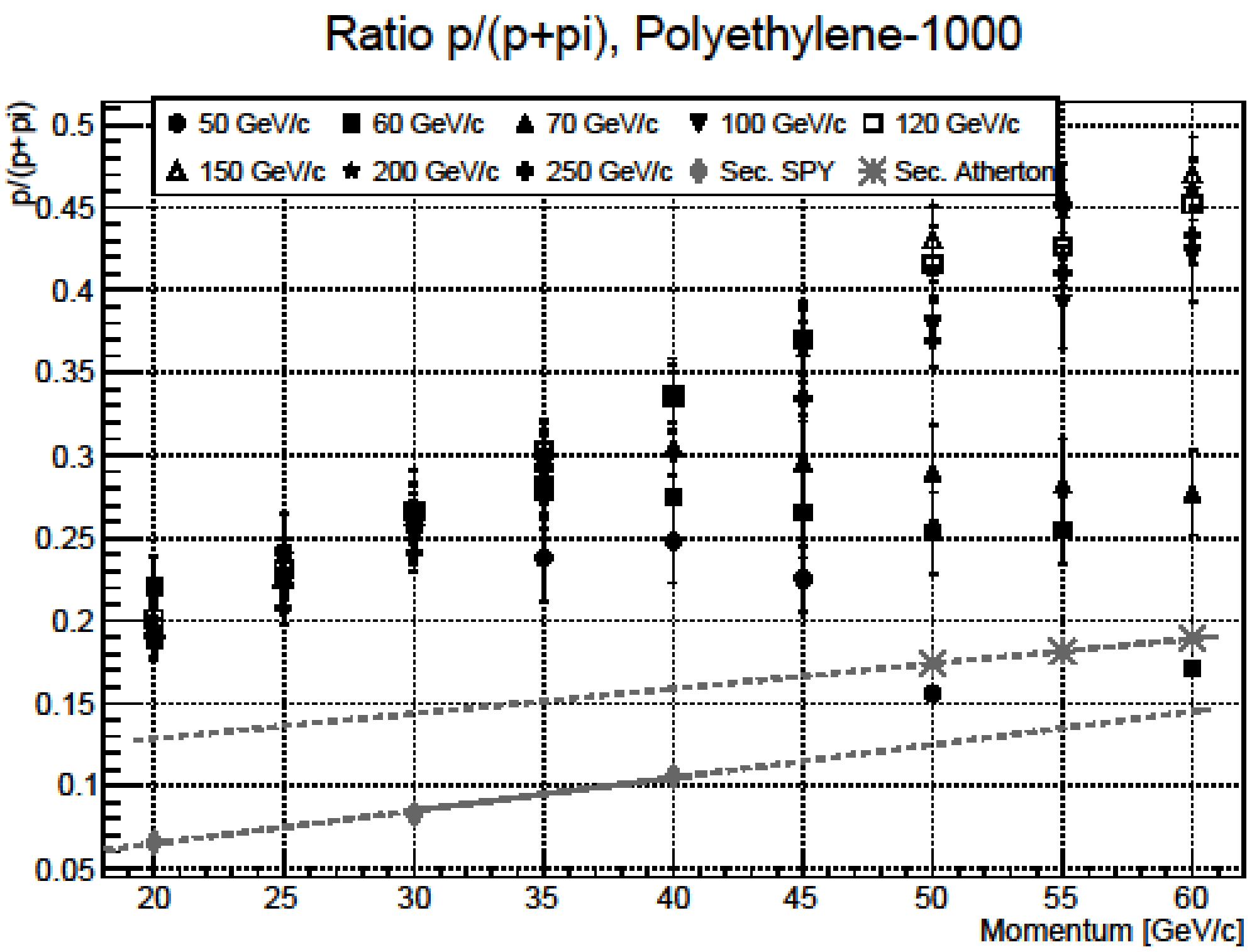
Ratio p/(p+pi), Polyethylene-1000
p/(p+pi)
50 GeV/c
60 GeV/c
70 GeV/c
100 GeV/c
120 GeV/c
150 GeV/c
200 GeV/c
250 GeV/c
Sec. SPY
Sec. Atherton
0.5
0.45
0.4
0.35
0.3
0.25
0.2
0.15
0.1
0.05
20
25
30
35
40
45
50
55
60
Momentum [GeV/c]

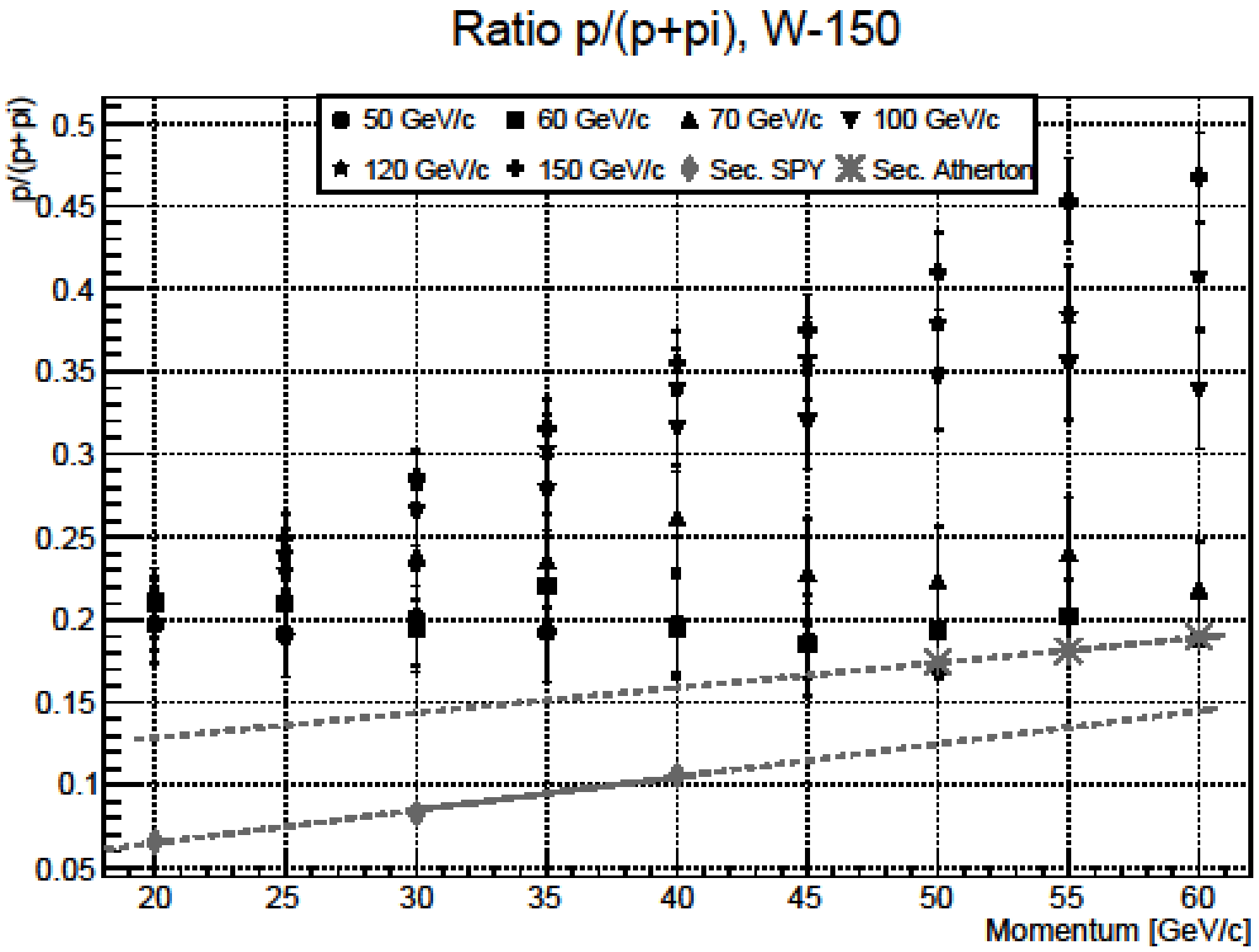

## 3. Acknowledgements

The authors would like to thank *A. Fabich* for useful discussions on the subject, and *A.E Rahmoun* for providing the data of the target's availabilities and geometries.